\documentclass[11pt]{article}
\usepackage{jheppub}

\usepackage{amssymb}
\usepackage{amsmath}
\usepackage{amsfonts}

\usepackage{graphicx, subfigure, placeins, float}

\usepackage{array} 
\usepackage{longtable} 
\usepackage{colortab} 
\usepackage{colortbl}
\usepackage{arydshln}

\usepackage{dsfont,bbm}
\usepackage{mathrsfs}  
\usepackage{xfrac}
\usepackage{slashed}
\usepackage{mathabx}
\usepackage{enumitem}
\usepackage{shuffle}
\usepackage{stmaryrd}

\makeatletter
\newcommand\xleftrightarrow[2][]{%
  \ext@arrow 9999{\longleftrightarrowfill@}{#1}{#2}}
\newcommand\longleftrightarrowfill@{%
  \arrowfill@\leftarrow\relbar\rightarrow}
\makeatother

\newcommand{\Rome}[1]{\uppercase\expandafter{\romannumeral#1}}

\newcommand{\itbf}[1]{\textbf{\textit{#1}}}

\newcolumntype{C}[1]{>{\centering\arraybackslash}m{#1}}

\makeatletter
\newcommand{\mathleft}{\@fleqntrue\@mathmargin0pt}
\newcommand{\mathcenter}{\@fleqnfalse}
\makeatother

\newcommand{\hlp}[1]{\hbox{\hspace{-#1pt}+}}
\newcommand{\hlm}[1]{\hbox{\hspace{-#1pt}$-$}}
\newcommand{\hlmath}[2]{\hbox{\hspace{-#1pt}$ \displaystyle #2 $}}

\newcommand{\pln}{\scriptscriptstyle{\text{PL}}}

\usepackage{extarrows}

\newcommand{\offW}{\mathbb{W}} 
\newcommand{\onW}{W} 
\newcommand{\onWbos}{{\cal W}} 
\newcommand{\PLff}{{\cal F}} 
\newcommand{\NPff}{\itbf{F}} 
\newcommand{\PLdensity}{{\cal I}} 
\newcommand{\NPdensity}{\itbf{I}} 

\title{Non-planar form factors of generic local operators via on-shell unitarity and color-kinematics duality}

\author[a]{Guanda Lin}
\emailAdd{linguandak@pku.edu.cn}
\author[b,c,d,e]{and Gang Yang}
\emailAdd{yangg@itp.ac.cn}
\affiliation[a]{School of Physics, Peking University, Beijing 100871, China}
\affiliation[b]{CAS Key Laboratory of Theoretical Physics, Institute of Theoretical Physics, \\Chinese Academy of Sciences, Beijing 100190, China}
\affiliation[c]{School of Fundamental Physics and Mathematical Sciences, Hangzhou Institute for Advanced Study, UCAS, Hangzhou 310024, China}
\affiliation[d]{International Centre for Theoretical Physics Asia-Pacific, Beijing/Hangzhou, China}
\affiliation[e]{School of Physical Sciences, University of Chinese Academy of Sciences, Beijing 100049, China}

\abstract{
Form factors, as quantities involving both local operators and asymptotic particle states, contain information of both the spectrum of operators and the on-shell amplitudes. So far the studies of form factors have been mostly focused on the large $N_{ c}$ planar limit, with a few exceptions of Sudakov form factors. In this paper, we discuss the systematical construction of full color dependent form factors with generic local operators. We study the color decomposition for form factors and discuss the general strategy of using on-shell unitarity cut method. As concrete applications, we compute the full two-loop non-planar minimal form factors for both half-BPS operators and non-BPS operators in SU(2) sector  in ${\cal N}=4$ SYM. Another important aspect is to investigate the color-kinematics (CK) duality for form factors of high-length operators. Explicit CK dual representation is found for the two-loop half-BPS minimal form factors with arbitrary number of external legs. The full-color two-loop form factor result provides an independent check of the infrared dipole formula for two-loop $n$-point amplitudes. By extracting the UV divergence, we also reproduce the known non-planar SU(2) dilatation operator at two loops. As for the finite remainder function, interestingly, the non-planar part is found to contain a new maximally transcendental part beyond the known planar result.
}

\begin{document}

\maketitle

\setcounter{footnote}{0}


\section{Introduction}

The large $N_{ c}$ limit (\emph{i.e.}~the planar limit) has been a useful approximation in the study of gauge theories \cite{tHooft:1973jz}. 
In this limit, the color degrees of freedom are decoupled and irrelevant. For example, for scattering amplitudes, it is enough to consider color-ordered planar amplitudes; for anomalous dimensions or correlation functions, only single-trace operators are involved in the computation. 
The planar limit also induces certain symmetries beyond the full theory. 
An example is the 
planar ${\cal N}=4$ super-Yang-Mills theory (${\cal N}=4$ SYM), in which, an infinite number of hidden symmetries (integrability) have been found which render the theory exactly solvable in the planar limit, see \emph{e.g.}~\cite{Beisert:2010jr} for a review.

On the other hand, it is important to go beyond the large $N_{ c}$ limit and consider the full-color contributions. 
In a full gauge theory, the mixing of different color structures as well as the correlation between color and spacetime degrees of freedom are expected to have significant physical effects. 
In the realistic QCD with $N_{ c}=3$, the color-subleading parts should be indispensable for understanding the mechanism of confinement. 
Besides, in the  phenomenological studies, non-planar contributions of amplitudes and their cross sections have also important effects.

As a useful testing ground for studying non-planar structures, we consider form factors in ${\cal N}=4$ SYM in this paper. 
They are defined as matrix elements between a local gauge invariant operator and $n$ on-shell outgoing particle states \cite{vanNeerven:1985ja, Alday:2007he, Maldacena:2010kp, Brandhuber:2010ad, Bork:2010wf, Brandhuber:2011tv, Bork:2011cj}:
\begin{equation}
\label{eq:def-FF}
{\cal F}_{{\cal O},n} = \int d^D x \, e^{-i q \cdot x} \langle \onW(p_1)\cdots \onW(p_n)|{\cal O}(x) | 0\rangle = (2\pi)^D\delta^{(D)}\Big(q-\sum_{i=1}^n p_i\Big)\langle 1\cdots n|{\cal O}(0) | 0\rangle ,
\end{equation}
where $\onW(p_i)$ represent asymptotic on-shell states, and ${\cal O}(x)$ is a local operator carrying an off-shell momentum $q = \sum_i p_i$ 
(via the Fourier transformation).
Since form factors involve both local operators and asymptotic on-shell states, 
they contain important information of both on-shell amplitudes and anomalous dimension of local operators. 
This has inspired many studies that generalize amplitudes techniques to form factors  \cite{Henn:2011by, Brandhuber:2012vm, Gao:2013dza, Penante:2014sza, Brandhuber:2014ica, Frassek:2015rka, Bork:2016hst, Bork:2016xfn, Bork:2017qyh, Bork:2014eqa, Koster:2016ebi, Koster:2016loo, Chicherin:2016qsf, Koster:2016fna, He:2016dol, Brandhuber:2016xue, He:2016jdg, Bolshov:2018eos, Bianchi:2018peu, Bianchi:2018rrj, Bork:2012tt, Johansson:2012zv, Huang:2016bmv,Huber:2019fea} and also use form factors to study the spectrum problem of generic operators \cite{Wilhelm:2014qua, Nandan:2014oga, Loebbert:2015ova, Brandhuber:2016fni, Loebbert:2016xkw, Ahmed:2016vgl, Brandhuber:2017bkg, Brandhuber:2018xzk, Caron-Huot:2016cwu,Ahmed:2019upm,Ahmed:2019yjt}.
A state-of-the-art pedagogical review for form factors in ${\cal N}=4$ SYM can be found in \cite{Yang:2019vag}. 

In the case of Sudakov form factors of the stress tensor supermultiplet in ${\cal N}=4$ SYM, the full non-planar corrections has been considered at four and five loops \cite{Boels:2012ew, Yang:2016ear} with the help of color-kinematics duality \cite{Bern:2008qj, Bern:2010ue}. On the other hand, for form factors of generic local operators of high dimensions, the studies have so far mostly focused on the planar limit. 
A concrete goal of this paper is to study the \emph{non-planar} form factors of generic high-length local operators in ${\cal N}=4$ SYM. 
This generalization is important in the sense that it will allow us to probe the non-planar spectrum of general local operators, as well as the non-planar IR structure of amplitudes with generic number of external legs. 

As mentioned in the beginning, the main new feature of non-planar form factors is that the color degrees of freedom must be taken into account. This generates several complications, comparing to the planar case. First, operators with multi-trace structures have to be considered. Second, the correlation between color and kinematic parts needs to be taken into account. Moreover, the kinematic part itself becomes more intricate: for example, new non-planar types of Feynman integrals will contribute.
To solve these problems, we will apply the color decomposition for form factors to first disentangle the color and kinematic parts. 
Similar strategy was also considered in loop amplitudes computation, see \emph{e.g.}~\cite{Edison:2011ta, Edison:2012fn, Feng:2011fja, Ochirov:2016ewn, Abreu:2018aqd, Badger:2019djh, Dalgleish:2020mof}.
For form factors of high dimension operators, an important difference comparing to amplitudes is that one can focus on a small set of density loop correction functions. 
Then for the non-planar kinematic parts, one can apply the on-shell unitarity method \cite{Bern:1994zx, Bern:1994cg, Britto:2004nc}. 
A generic computational strategy based on these will be described, and as concrete applications, we obtain two-loop non-planar form factors in both half-BPS and non-BPS SU(2) sectors.
The advantage of the above unitarity-based method is that it does not rely on any specific symmetry of a theory (such as supersymmetry or planar integrability), thus can be used in quite general context (either in non-planar sector, or in generic theories like QCD).

Another important tool for non-planar construction is the color-kinematics duality \cite{Bern:2008qj, Bern:2010ue} (see \cite{Bern:2019prr} for an extensive review of the developments). As mentioned above, this duality has been very useful for computing the Sudakov form factors in ${\cal N}=4$ SYM. It would be of great interest to see if this can be applied to form factors of high-length operators which involve multiple external legs.
In this paper, we make a concrete step and find a two-loop integral representation that manifests the color-kinematics duality for minimal form factors of half-BPS operators ${\rm tr}(\phi^L)$ for arbitrary $L$ (which equals to the number of external on-shell legs). It turns out that in order to satisfy the duality, certain scaleless integrals (which are zero after integration) are necessarily included.

Having the full two-loop form factor results, we are able to study several non-planar properties. First, we compute the two-loop full-color IR divergence, which provides an independent check for the dipole formula of amplitudes with arbitrary number of external legs \cite{Becher:2009cu, Gardi:2009qi}. 
From  the UV renormalization of non-BPS SU(2) form factors, we also obtain the two-loop non-planar SU(2) dilatation operator previously obtained in \cite{Beisert:2003jj}.
After subtracting the divergences, one can get the finite remainder function. Interestingly, we find out that, beyond the known planar results \cite{Brandhuber:2014ica, Loebbert:2015ova, Loebbert:2016xkw}, the leading transcendentality part receives a new non-planar correction. 

This paper is organized as follows. 
In Section~\ref{sec:review}, we provide a review of local gauge invariant operators and planar form factors. 
Section~\ref{sec:color} concerns the generic color structures of operators and form factors. Both the trace basis and trivalent basis are studied for decomposing the color structure of form factors. A brief review for color-kinematics duality and its generalization to generic form factors are also given.
In Section~\ref{sec:unitarity}, we explain the general strategy of computing full-color form factors with unitarity cut method, together with concrete examples. 
Section~\ref{sec:2l} provides the full construction and results for both two-loop half-BPS and SU(2) form factors. 
In Section~\ref{sec:CKBPS}, the representation satisfying color-kinematics duality is obtained for the two-loop half-BPS form factors. 
In Section~\ref{sec:iruv}, we discuss the non-planar properties of the two-loop IR structure, the finite remainders, and the SU(2) dilation operator.
Finally, Section~\ref{sec:disc} provides a summary and outlook.
Appendix~\ref{ap:ddm}-\ref{app:IRUVcancel} contain some further discussions and several technical details.
%

\section{Review: operators and form factors in the planar limit}
\label{sec:review}

In this section, we briefly review the gauge invariant operators and planar amplitudes and form factors. This section is mainly intended to set up the notation and provide some basic building block results that will be used in later sections. For simplicity, the discussion of this section will focus on the planar case and ignore the color degrees of freedom. The full-color structure will be discussed in detail in the next section. 

\subsection{Local gauge invariant operators}
\label{ssec:FandO}

A gauge invariant local operator ${\cal O}(x)$ is composed of trace of gauge covariant fields at same spacetime point $x$. In $\mathcal{N}=4$ SYM, the elementary fields are: gauge boson $A^{\mu}$, eight Weyl spinor $\Psi_{\alpha A},\bar{\Psi}_{\dot{\alpha}}^{A}$ with $\alpha,\dot{\alpha}=1,2$ and $A=1,2,3,4$, and six scalars  $\Phi_{I}$ with $I=1,\ldots,6$.  All the fields are in the adjoint representation of gauge group. In local operators, the gauge field $A^{\mu}$ shows up in the covariant derivative $D_{\mu} =\partial_{\mu} -i g_{\mathrm{YM}}\left[A_{\mu}, \cdot \right]$ and the field strength tensor $F^{\mu\nu}=(-{i}g)^{-1}[D^{\mu},D^{\nu}]$.  All the covariant fields $\offW_{i}$ appearing in local operators can be listed as
\begin{equation}
\label{offf}
\offW \in\left\{\Phi^{A B}, F^{\alpha \beta}, \bar{F}^{\dot{\alpha} \dot\beta}, \bar{\Psi}^{\dot{\alpha} A}, \Psi^{\alpha,ABC}\right\} ,
\end{equation} 
and each field can be dressed with arbitrary number of covariant derivatives $D^{\alpha\dot{\alpha}}$. 
In the planar limit, one only needs to consider single trace operators which is
\begin{equation}
\label{eq:SToperator}
{\cal O}_L = {\rm tr}(\offW_1^{(n_1)} \offW_2^{(n_2)} \ldots \offW_L^{(n_L)} ) \,, 
\end{equation}
where $\offW_i^{(n)} = D^n \offW_i$, and we also call $L$ the \emph{length} of the operator.

An important problem regarding the local operators is to compute their \emph{dimensions}. In general the canonical dimension of an operator receives quantum corrections which is called the anomalous dimensions. 
The anomalous dimension is captured by the dilatation operator, see \emph{e.g.}~\cite{Beisert:2003jj}.
Quantum correction of dilatation operator may be obtained from the computation of two-point correlation functions. In the planar limit, the eigenvalue of the dilation operators or the anomalous dimension can be computed non-perturbatively, via integrability techniques \cite{Beisert:2010jr}. 
Form factors provide an alternative way to compute the dilatation operator, and as we will discuss in this paper, the on-shell method also makes it straightforward to go beyond the planar limit.

\subsection{Tree-level amplitudes and form factors}
\label{subsec:treeAmpFF}

In contrast to correlation function of local operators, amplitudes and form factors in ${\cal N}=4$ SYM contain on-shell asymptotic states which can be conveniently represented by the on-shell superfield \cite{nair1988current}:
\begin{equation}
    \onW(p, \eta)=g_{+}(p)+\eta^{A} \bar{\psi}_{A}(p)+\frac{\eta^{A} \eta^{B}}{2 !} \phi_{A B}(p)+\frac{\epsilon_{A B C D} \eta^{A} \eta^{B} \eta^{C}}{3 !} \psi^{D}(p)+\eta^{1} \eta^{2} \eta^{3} \eta^{4} g_{-}(p) \,.
\end{equation}
To clarify our notation, we use $\offW_i$ to represent off-shell fields in operators, and $W_i(p_i,\eta_i)$ to denote on-shell super fields. We also denote $\onWbos_i(p_i)$ for certain component of $\onW_i(p_i,\eta_i)$, \emph{i.e.} $\onWbos_{i} \in \{\phi_i, \psi_i, \bar{\psi}_i, g_{i,\pm}\}$.

\subsubsection*{Tree-level amplitudes}

Amplitudes are gauge invariant quantities only depending on on-shell states and is given as $\langle \onW(p_1)\cdots \onW(p_n)| | 0\rangle$.
Here we briefly review some results of planar color-ordered tree amplitudes. See \emph{e.g.}~\cite{Henn:2014yza, Elvang:2015rqa} for more detailed introduction.

Firstly, the $n$-point tree level super MHV amplitudes take the well-known Parke-Taylor form \cite{Parke:1986gb}:  
\begin{equation}\label{eq:sua4}
    \hat{\mathcal{A}}_{n}^{(0),\text{MHV}}\big(1,\ldots,n\big)=\frac{\delta^{(4)}(p) \delta^{(8)}(q)}{\langle 12\rangle\langle 23\rangle \cdots \langle n  1\rangle} \,,
\end{equation}
where $p^{\alpha \dot{\alpha}}=\sum_{i=1}^{n}\lambda^{\alpha}\tilde{\lambda}^{\dot{\alpha}}$ and $q^{\alpha,A }=\sum_{i=1}^{n} \lambda_{i}^{\alpha} \eta_{i}^{A}$, and $\overline{\rm MHV}$ amplitudes can be derived by conjugate. Here we denote super amplitudes $\hat{\mathcal{A}}$ with a ``hat'' to stress their dependence on Grassmann variables, and amplitudes ${\mathcal{A}}$ without hat are bosonic components with only kinematic dependence. 
The four-point pure scalar amplitudes $\mathcal{A}^{(0)}_{4}(p_1^{X}, p_2^{Y}, p_3^{\bar{X}}, p_4^{\bar{Y}})$ can be obtained by extracting the $\eta_{1}^{1}\eta_{1}^{2}\eta_{2}^{1}\eta_{2}^{3}\eta_{3}^{3}\eta_{3}^{4}\eta_{4}^{2}\eta_{4}^{4}$ component from \eqref{eq:sua4}:
\begin{equation}
\label{eq:a4su2}
\begin{aligned}
   \mathcal{A}^{(0)}_{4}(p_1^{X}, p_2^{Y}, p_3^{\bar{X}}, p_4^{\bar{Y}})&=1 \,,
\end{aligned}
\end{equation}
where $X=\phi_{12},Y=\phi_{13}$ and $\bar{X}=\phi_{34},\bar{Y}=\phi_{42}$. And similarly,
\begin{equation}
\label{eq:a4su2b}
\begin{aligned}
   \mathcal{A}^{(0)}_{4}(p_1^{X}, p_2^{Y}, p_3^{\bar{Y}}, p_4^{\bar{X}})&=-\frac{\langle 13 \rangle \langle 24\rangle }{\langle 14 \rangle \langle 23\rangle }=-\left(1+\frac{s_{12}}{s_{14}}\right) \,.
\end{aligned}
\end{equation}

Furthermore, for higher point amplitudes, we also have N$^{k}$MHV amplitudes. They can be computed via BCFW on-shell recursion \cite{Britto:2005fq} or CSW vertex expansion \cite{Cachazo:2004kj} methods. In our two-loop computation, we encounter 6-point NMHV amplitudes of scalars. 
All of them can be conveniently expressed in a form with Mandelstam variables, for example,
\begin{equation}\label{6pexp}
    \mathcal{A}^{(0)}_{6}(p_1^{X},p_2^X,p_3^Y,p_4^{\bar{X}},p_5^{\bar{X}},p_6^{\bar{Y}})=-\frac{1}{s_{234}} \,.
\end{equation}
Other six-scalar amplitudes that are needed in the two-loop computation are listed in Appendix~\ref{ap:amp}.

\subsubsection*{Tree-level form factors}

Next we consider tree-level form factors. 
The simplest form factors are the minimal form factors, $\PLff_{{\cal O}_L,n}$ with $n=L$, in which the external legs are in one-to-one correspondence with the covariant fields in the operators, which are related via Wick contraction and LSZ reduction. The correspondence, or say a dictionary translating operators to minimal form factors in spinor helicity formalism,  can be summarized in Table \ref{tab:mtff} \cite{Beisert:2010jq, Zwiebel:2011bx,Wilhelm:2014qua}. 

\begin{table}[tbp]
    \centering
    \caption{Correspondence between local operators and minimal form factors.}
    \label{tab:mtff}
    \vskip .3 cm
    \begin{tabular}{|c|c|c|}
    \hline
     off-shell fields $\offW$ & on-shell states $\mathcal{W}_{o}$  & helicity spinor variables \\\hline
     $F^{\dot{\alpha} \dot{\beta}}$ & $\xrightarrow{\quad g_{+}\quad}$ & $\tilde{\lambda}^{\dot{\alpha}} \tilde{\lambda}^{\dot{\beta}}$\\
     $\bar{\Psi}^{\dot{\alpha} A}$ & $\xrightarrow{\quad\bar{\psi}_{ A}\quad}$ & $\tilde{\lambda}^{\dot{\alpha}} \eta^{A}$\\
     $\Phi^{A B}$& $ \xrightarrow{\quad \phi _{A B} \quad }$ & $\eta^{A} \eta^{B}$ \\
     $\Psi^{\alpha A B C}$ & $\xrightarrow{\quad \psi_{A B C} \quad }$ & $\lambda^{\alpha} \eta^{A} \eta^{B} \eta^{C}$ \\
     $F^{\alpha \beta}$ & $\xrightarrow{\quad g_{-} \quad }$ & $\lambda^{\alpha} \lambda^{\beta} \eta^{1} \eta^{2} \eta^{3} \eta^{4}$ \\
     $D^{\alpha \dot{\alpha}}$& $\xrightarrow{\quad p_{\alpha \dot{\alpha}} \quad }$ & $\lambda^{\alpha} \tilde{\lambda}^{\dot{\alpha}}$ \\\hline
\end{tabular}
\end{table}

To obtain a minimal form factor from a given operator, for example, for $\mathcal{O}=\operatorname{tr}(F^{3})=\operatorname{tr}\left(F_{\dot{\beta}}^{~\dot{\alpha}} F_{\dot{\gamma}}^{~\dot{\beta}} F_{\dot{\alpha}}^{~\dot{\gamma}}\right)$, the tree minimal form factor $\PLff^{(0)}_{\mathcal{O},3}$ reads
\begin{equation}
     \PLff^{(0)}_{\mathcal{O},3}\left(1^{g_+},2^{g_+},3^{g_+}\right)= \left(\tilde{\lambda}_{1}^{\dot{\alpha}} \tilde{\lambda}_{1 \dot{\beta}}\right)\left( \tilde{\lambda}_{2}^{\dot{\beta}} \tilde{\lambda}_{2 \dot{\gamma}}\right)\left( \tilde{\lambda}_{3}^{\dot{\gamma}} \tilde{\lambda}_{3 \dot{\alpha}}\right)+\text { cyc.perm.}(1,2,3) =3[12][23][31] \,.
\end{equation}

For more general non-minimal form factors, we introduce next$^k$-to-minimal form factor  $\PLff^{(0)}_{\mathcal{O}_L,n}$, which contains a length-$L$ operator and $n=L+k$ on-shell legs.
Non-minimal form factors also have simplicities such as the Parke-Taylor like structure and can be computed using amplitudes techniques \cite{Brandhuber:2010ad, Brandhuber:2011tv} (see also methods using twistor formalism \cite{Koster:2016ebi, Koster:2016loo, Chicherin:2016qsf, Koster:2016fna}). 
To construct two-loop form factors, one will need next-to-minimal form factors. In particular, next-to-minimal form factors of operators composed of pure scalar fields  is listed in  Table \ref{tab:nmff}.

\begin{table}[tbp]
     \centering
     \caption{Next-to-minimal tree level form factors as building blocks in unitarity \cite{Loebbert:2015ova}.}
     \label{tab:nmff}
    \vskip .3 cm     
     \begin{tabular}{|c|c|}
     \hline
        external states  & next to minimal form factors $\displaystyle\PLff_{L+1}^{(0)}$  \\\hline
        $\displaystyle\cdots \phi_{i,AB} g^{+}_{i+1} \phi_{i+2,CD}\cdots$  & $\displaystyle\displaystyle\cdots \eta_{i}^{A} \eta_{i}^{B} \frac{\langle i i+2\rangle}{\langle i i+1\rangle\langle i+1 i+2\rangle} \eta_{i+2}^{C} \eta_{i+2}^{D} \cdots$\\
        $\displaystyle\cdots \phi_{i,AB} g^{-}_{i+1,1234} \phi_{i+2,CD} \cdots$ & $\displaystyle\cdots \eta_{i}^{A} \eta_{i}^{B} \frac{[i+2]}{[i i+1][i+1 i+2]} \eta_{i+1}^{1} \eta_{i+1}^{2} \eta_{i+1}^{3} \eta_{i+1}^{4} \eta_{i+2}^{C} \eta_{i+2}^{D} \cdots$\\
        $\displaystyle\cdots \bar{\psi}_{i,A}\bar{\psi}_{i+1,B}\phi_{i+1,CD}\cdots$ & $\displaystyle\cdots\frac{1}{\langle i i+1\rangle}\left(\eta_{i}^{A} \eta_{i+1}^{B}-\eta_{i}^{B} \eta_{i+1}^{A}\right) \eta_{i+2}^{C} \eta_{i+2}^{D}\cdots$\\
         $\displaystyle\cdots \psi_{i}^{A}\psi_{i+1}^{B}\phi_{i+1,CD}\cdots$ & $\displaystyle\cdots\frac{-1}{[i i+1]}\left(\eta_{i,A^{\prime}} \eta_{i+1,B^{\prime}}-\eta_{i,B^{\prime}} \eta_{i+1, A^{\prime}}\right) \eta_{i+2}^{C} \eta_{i+2}^{D}\cdots$ 
         \\\hline
     \end{tabular}
 \end{table}


\subsection{Loop correction function and renormalization}\label{subsec:planarloopdensity}

In this subsection, we consider form factors at loop level. We introduce loop correction function and discuss its connection to the renormalization. 
The loop expansion for planar minimal form factors is defined as
\begin{equation}
\PLff = \sum_{l=0}^\infty g^{2 \ell} \PLff^{(\ell)} \,, \qquad \textrm{with} \quad g^{2}=\frac{g_{\mathrm{YM}}^{2} N_{\mathrm{c}}}{(4 \pi)^{2}}\left(4 \pi e^{-\gamma_{\mathrm{E}}}\right)^{\epsilon} \,.
\label{eq:loopexpPLff}
\end{equation}

\subsubsection*{Loop correction density function}

An $\ell$-loop form factor of generic operators may be expressed in terms of 
a loop correction function ${\mathcal{I}}^{(\ell)}$ acting on tree level form factors
\begin{equation}
\PLff_{\cal O}^{(\ell)}={\mathcal{I}}^{(\ell)}\cdot \PLff_{\cal O}^{(0)} \,.
\end{equation}

\begin{figure}
    \centering
    \subfigure[One-loop density.]{
        \includegraphics[width=0.34\linewidth]{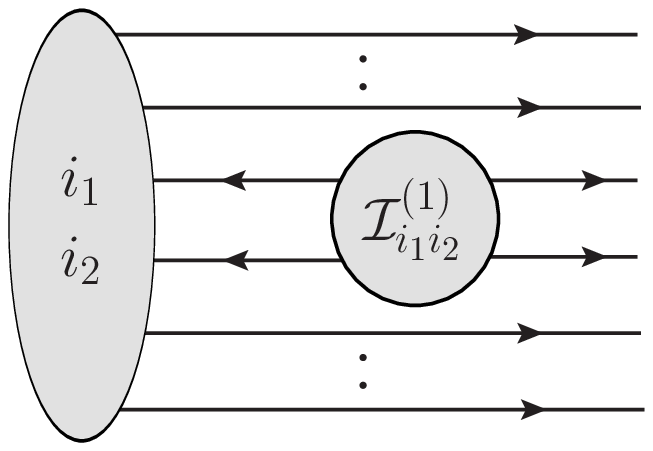}
         \label{fig:loopcorrectiondensity1}
        }
     \centering
   \subfigure[Higher-loop density functions.]{
        \includegraphics[width=0.34\linewidth]{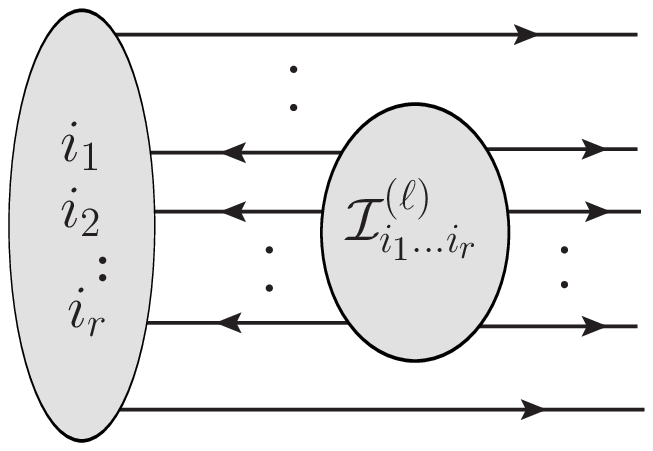}
        \label{fig:loopcorrectiondensity2}
        }
    \caption{Diagrammatic representation for loop correction density functions.}
    \label{fig:loopcorrectiondensity}
\end{figure}

The full loop correction function ${\mathcal{I}}^{(\ell)}$ can be decomposed in terms of \emph{loop density} functions, 
$\PLdensity^{(\ell)}_{i_1i_2\ldots i_r}$,
as represented by Figure~\ref{fig:loopcorrectiondensity}. 
The subindices ``$i_1 i_2 \ldots i_r$" denotes the sites of the operators that the density function acting on.
The number $r$ is named the \emph{interaction range}. At one loop, it is obvious that the interaction range is two. At $\ell$-loop, the maximal (connected) interaction range is $\ell+1$.
In the planar limit, $i_1\ldots i_r$ should be adjacent: $i,i+1,\ldots,i+(r-1)$.
The full form factor is given by summing over density functions of all external momenta plus summing over their insertion on all possible sites in the operators. For example, the one-loop form factor can be written as 
\begin{equation}
\PLff_{\cal O}^{(1)} = \sum_i \PLdensity^{(1)}_{i (i+1)} \cdot \PLff_{\cal O}^{(0)} \,.
\end{equation}

Since the operator may contain different type of fields, it should be clear that in general the loop correction functions are matrices acting in the field space. 
For example, the above one-loop density should be understood as in 
 \begin{align}
 \PLff^{(1)}_{{\cal O}_{\onWbos_i \onWbos_{i+1}}}(p_k^{\widetilde{\onWbos}_k}, p_{k+1}^{\widetilde{\onWbos}_{k+1}}) & = \Big( \PLdensity^{(1)}  \Big)_{\onWbos_i \onWbos_{i+1}}^{{\widetilde{\onWbos}_k} {\widetilde{\onWbos}_{k+1}}}(p_k, p_{k+1}) 
 \Big( {\widetilde{\onWbos}_k} {\delta \over \delta\onWbos_i}  {\widetilde{\onWbos}_{k+1}} {\delta \over \delta\onWbos_{i+1}} \Big) 
 \PLff^{(0)}_{{\cal O}_{\onWbos_i \onWbos_{i+1}}} \nonumber\\
 & =  \big( \PLdensity^{(1)} (p_k, p_{k+1}) \big)_{\onWbos_i \onWbos_{i+1}}^{{\widetilde{\onWbos}_k} {\widetilde{\onWbos}_{k+1}}} \  \PLff^{(0)}_{{\cal O}_{\widetilde{\onWbos}_k \widetilde{\onWbos}_{k+1}}}(p_k, p_{k+1}) \,,
 \label{eq:planarloopcorrection}
 \end{align}
where ${{\cal O}_{\onWbos_i \onWbos_{i+1}}}$ is to stress that the $i$-th and $(i+1)$-th field in operator $\mathcal{O}$ are $\mathcal{W}_i$ and $\mathcal{W}_{i+1}$,\footnote{Here we do not distinguish on-shell and off-shell fields, since they are related by the dictionary given in Table~\ref{tab:mtff}.}
and ``${\widetilde{\onWbos}_k, \widetilde{\onWbos}_{k+1}}$" label the fields of external on-shell legs.
A more precise definition for the full non-planar form factor will be given in the next section, see \eqref{eq:nplanarloopcorrection}.

Having this structure in mind, one can focus on the density functions in the computation of loop form factors, and it is enough to calculate all matrix elements such as $\big( \PLdensity^{(1)} \big)_{\onWbos_i \onWbos_j}^{\widetilde{\onWbos}_k \widetilde{\onWbos}_l}$ in the one-loop case.
For full-color form factors, one needs to dress loop corrections with color indices as well, which will be discussed in next section.
As we will discuss shortly, the matrix form is related to the fact that different operators are mixed with each other via quantum corrections. 
Through renormalization, one can extract the operator mixing information using form factors.

\subsubsection*{Renormalization, operator mixing and dilatation operator}

Loop correction functions of form factors of general operators suffer from both ultraviolet (UV) and infrared (IR) divergences. The IR structure is well understood and can be subtract systematically, while the UV divergences require renormalization and result in operator mixing. 

The renormalization of operators can be expressed as 
\begin{equation}
\mathcal{O}_{I}^{\mathrm{ren}}=\sum_{J} \mathcal{Z}_{I}^{~J} \mathcal{O}_{J}^{\mathrm{bare}} \,,
\end{equation}
where $\mathcal{Z}$ is the renormalization constant matrix. 
Renormalization of form factors can be expressed likewise
\begin{equation}
\PLff_{\mathcal{O}_{I}}^{\text {ren }}=\sum_{J} \mathcal{Z}_{I}^{~J} \PLff_{\mathcal{O}_{J}}^{\text {bare }} \,.
\end{equation}
More explicitly, we have
\begin{equation}\label{prenorm}
    \PLff_{\mathcal{O}_{I}}^{\text {ren }}=\sum_{J} \mathcal{Z}_{I}^{~J} \, \PLff_{\mathcal{O}_{J}}^{\text {bare }}
    =\sum_{J,K} \mathcal{Z}_{I}^{~J} \, (\PLdensity^{\text{bare}})_{J}^{~K} \, \PLff^{(0)}_{\mathcal{O}_{K}}
    \equiv \sum_{J} (\PLdensity^{\text{ren}})_{I}^{~J} \, \PLff_{\mathcal{O}_{J}} \,.
\end{equation}
The UV divergence of  $\mathcal{Z}_{I}^{~J} \, \mathcal{I}^{\text{bare}}_{J}$ should cancel order by order. Up to two-loop order, we have 
\begin{equation}
\begin{aligned}
   \mathcal{I}^{(1), \text {ren }}&=\mathcal{I}^{(1), \text {bare }}+\mathcal{Z}^{(1)} \,, \\
\mathcal{I}^{(2), \text {ren }}&=\mathcal{I}^{(2), \text {bare }}+\mathcal{Z}^{(2)}+\mathcal{Z}^{(1)}\mathcal{I}^{(1), \text {bare }} \,.
\end{aligned}
\end{equation}
We stress that the product of loop corrections and renormalization constants should be taken as matrix products. For example, the one-loop square should be understood as
\begin{equation}\label{prodop}
    \big(\PLdensity^{(1)}\big)^{2}\PLff^{(0)}_{\mathcal{O}_I}=\sum_{J,K}\big(\PLdensity^{(1)}\big)_{I}^{~J} \big(\PLdensity^{(1)}\big)_{J}^{~K}\PLff^{(0)}_{\mathcal{O}_{K}} \,.
\end{equation}

The renormalization can be expressed in the density form as well. Take the one-loop case as an example:
\begin{equation}
    \mathcal{Z}^{(1)}=\sum_{i}\mathcal{Z}_{i(i+1)}^{(1)}\,.
\end{equation}
By imposing UV cancellation at density level
\begin{equation}
    \mathcal{I}^{(1), \text {ren }}_{i(i+1)}=\mathcal{I}^{(1), \text {bare }}_{i(i+1)}+\mathcal{Z}^{(1)}_{i(i+1)}\,,
\end{equation}
one can extract the renormalization constant density $\big({\cal Z}^{(1)}\big)_{\onWbos_i \onWbos_j}^{\onWbos_k \onWbos_l}$.
Examples of the SO(6) and SL(2) sector can be found in \cite{Yang:2019vag}.

Given the renormalization constant $\mathcal{Z}$, it is straightforward to calculate the \emph{anomalous dilatation operator} $\delta \mathds{D}\equiv \mathds{D}-\mathds{D}_0$, where $\mathds{D}_0$ corresponds to classical dimensions, as\footnote{For special subsectors, such as the SU(2) sector, we only need to consider even order in the expansion. The odd power of $g$ expansion corresponds to the ``length changing effect", see \emph{e.g.}~\cite{Beisert:2003ys}. We will not consider such case in this paper.}
\begin{equation}
   \delta \mathds{D}=2\epsilon g^2 \frac{\partial}{\partial g^2} \mathcal{Z}= \sum_{\ell=1}^\infty g^{2\ell} \mathds{D}^{(\ell)}\,.
\end{equation}
Up to two-loops these $\mathds{D}^{(\ell)}$ can be derived from renormalization constant as 
\begin{equation}\label{eq:ZandD}
    \mathds{D}^{(1)}=2 \epsilon \mathcal{Z}^{(1)}, \quad \mathds{D}^{(2)}=4 \epsilon \Big[\mathcal{Z}^{(2)}-\frac{1}{2} \big(\mathcal{Z}^{(1)}\big)^{2}\Big] .
\end{equation}
Finally, by diagonalizing the dilatation operator, one can obtain the eigenstates and their eigenvalues are the anomalous dimensions.

\section{Full-color structure}\label{sec:color}

In this section, we consider the full-color structure of local operators and form factors. 
At non-planar level, the spacetime and color degrees of freedom are expected to be entangled.
To simplify the study,  we discuss color basis decomposition for form factors, such that the color and kinematic parts are separated in a well controlled way.
Besides, we also discuss and generalize the color-kinematics duality to form factors of high-length operators, which correlates the color and kinematics factors in a very nice way. 

To be concrete, we will mostly consider SU($N_{\mathrm{c}}$) gauge group for simplicity.
In ${\cal N}=4$ SYM, all fields are in the adjoint representation: $\offW = \offW^a T^a$,
where $T^{a}$,  $a=1,\dots,N_{\text{c}}^2-1$, are the generators of SU$(N_{ c})$ and satisfy commutation relations
\begin{equation}
[T^{a}, T^{b}]=i \sqrt{2} f^{a b c} T^{c}=\tilde{f}^{a b c} T^{c}, \qquad \tilde{f}^{a b c}=\operatorname{tr} (T^{a} T^{b} T^{c})-\operatorname{tr}(T^{a} T^{c} T^{b}) \,.
\end{equation}
When contracting color indices, we need the completeness relations: 
\begin{equation}{\label{contr}}
    \sum_{a=1}^{N_{\mathrm{c}}^2-1}\left(T^{a}\right)_{\text{x}_{1}}^{\text{y}_{1}}\left(T^{a}\right)_{\text{x}_{2}}^{\text{y}_{2}}=\delta_{\text{x}_{1}}^{\text{y}_{2}} \delta_{\text{x}_{2}}^{\text{y}_{1}}-\frac{1}{N_{ c}} \delta_{\text{x}_{1}}^{\text{y}_{1}} \delta_{\text{x}_{2}}^{\text{y}_{2}} \,,
\end{equation}
where $\text{x}_i, \text{y}_i = 1, \ldots N_c$ are (anti)fundamental indices.
We choose normalization: $\operatorname{tr}(T^{a}T^{b})=\delta^{ab}$.

For generic loop expansion at non-planar level, we use the color-independent coupling $\tilde{g} = g/\sqrt{N_{ c}}$. The loop expansion of full-color form factors is defined as
\begin{equation}
\NPff = \sum_{l=0}^\infty \tilde{g}^{2 \ell} \NPff^{(\ell)} \,, \qquad \textrm{with} \quad \tilde{g}^{2}=\frac{g_{\mathrm{YM}}^{2} }{(4 \pi)^{2}}\left(4 \pi e^{-\gamma_{\mathrm{E}}}\right)^{\epsilon} \,.
\label{eq:loopexpPLffnp}
\end{equation}

\subsection{Trace color basis}\label{ssec:cftr}

We first consider the color structure in terms of  trace basis. 

\subsubsection*{Color structure of operators}

For local operators, the color structures are naturally defined in terms of color traces. 
In the planar limit, one can confine oneself in single-trace operators as given in \eqref{eq:SToperator}, as interactions between fields of different traces are suppressed by $1/N_{\mathrm{c}}$. 
For full non-planar considerations,  it is necessary to consider multi-trace operators. We can define a generic operator as 
\begin{equation}
{\cal O}_L = C_{\cal O}(a_1, \ldots a_L) \offW_1^{a_1, (n_1)} \offW_2^{a_2, (n_2)} \ldots \offW_L^{a_L, (n_L)} \,, 
\end{equation}
where there color factor $C_{\cal O}$ in general contains multiple traces and is given as a product of the following terms:
\begin{equation}
\label{eq:tracepoly}
    \operatorname{tr}(\mathds{1}) = N_c,\ \operatorname{tr}(T^{a_1}T^{a_2}),\ \ldots,\  \operatorname{tr}(T^{a_1}T^{a_2} \ldots T^{a_m}), \ \ldots \,.
\end{equation}
One can characterize the color structure of an operator by the number of traces and the length of each trace.

\begin{figure}[t]
    \centering
    \includegraphics[width=0.45\linewidth]{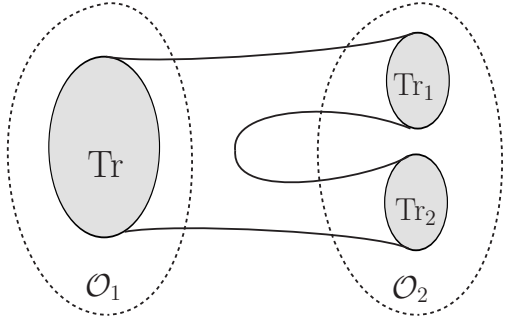}
    \caption{At non-planar level, operators of different number of traces can mix with each other. This diagram shows the interaction between a single trance operator and a double trace operator.}
    \label{fig:operatormixing}
\end{figure}

Loop corrections can merge or split traces, as illustrated in Figure~\ref{fig:operatormixing}, which lead to the mixing between operators with different color structure.
Only after considering operators of all color structures can one determines the anomalous dimensions and eigenstates of the mixing matrix. 
When increasing the length (\emph{i.e.}~the number of fields) of the operators, the number of inequivalent trace factors can grow very fast and the non-planar sectors may be expected to eventually dominate the contribution.

\subsubsection*{Trace basis for form factors }

The trace basis can be also used for form factors, and  the corresponding color factors can be given as products of terms in \eqref{eq:tracepoly} as well.
It should be clear that different products of  \eqref{eq:tracepoly} are linear independent.
Here it is important to note the difference between the trace structure of local operators and form factors:
the operators are always color singlet, 
while the color factor of a form factor is a SU$(N_{\mathrm{c}})$ tensor with free color indices carried by the external on-shell states. 

We denote the trace basis by $\mathcal{T}_k$, and one can make a color decomposition of form factors as:
\begin{equation}\label{ffcdec}
    \NPff^{(\ell)}_{\mathcal{O},n}=\sum_k \mathcal{T}_k \left[ \mathcal{F}^{(\ell)}_{\mathcal{O},n} \right]_{\mathcal{T}_k} \,,
\end{equation}
where $\mathcal{T}_{k}$ are product of traces with total length $n$,
and the color-stripped form factors are denoted as $\big[ \mathcal{F}^{(\ell)}_{\mathcal{O},n} \big]_{\mathcal{T}_k}$.
We use bold italic character $\NPff (\itbf{A})$ to denote the full-color form factors (amplitudes).
The linear independence of $\mathcal{T}_k$ also implies an important fact: each $\big[ \mathcal{F}^{(\ell)}_{\mathcal{O},n}\big]_{\mathcal{T}_k}$ should be a gauge invariant quantity and  have appropriate physical factorization and proper IR/UV structure. 

As an example for the trace color factors, for the minimal form factor of a length-4 operator, and the trace color basis $\{\mathcal{T}_k\}$ can be chosen as
\begin{equation}
    N_{\mathrm{c}}^{m}\operatorname{tr}(T^{a_1}T^{a_{\sigma(2)}}T^{a_{\sigma(3)}}T^{a_{\sigma(4)}}), \quad N_{\mathrm{c}}^{n}\operatorname{tr}(T^{a_1}T^{a_{\tau(2)}})\operatorname{tr}(T^{a_{\tau(2)}}T^{a_{\tau(2)}}),
    \quad \text{ with } \sigma\in S_3,  \tau\in Z_3 \,, \nonumber
\end{equation}
where powers of $N_c$ depend on the number of loops in general.

\subsubsection*{Definition of full-color loop correction}

Following Section~\ref{subsec:planarloopdensity}, we discuss color-dressed loop correction functions. Here we use trace basis to illustrate the definition and concepts, but the discussion will also apply to other color bases such as DDM basis given in next subsection.

As in the planar case, we can write the full-color loop form factor as 
\begin{equation}
\NPff^{(\ell)}={\NPdensity}^{(\ell)}\cdot \NPff^{(0)} \,.
\end{equation}
Furthermore, $\NPdensity^{(\ell)}$ can be decomposed in terms density functions. We label $\ell$-loop density for color-dressed loop correction as $\NPdensity^{(\ell)}_{i_1i_2\cdots i_r}$, where the subscript $i_i i_2 \cdots i_r$ denote the \emph{sites} of the operator on which the density acting on, as in Figure~\ref{fig:loopcorrectiondensity}.
Importantly, the sites are not required to be adjacent with each other in the full-color result, unlike the planar case.

Now let us consider a one-loop example in detail. The full one-loop form factor can be expressed in terms of density function as:
\begin{align}
\NPff^{(1)}_{{\cal O}({\onWbos_1 .. \onWbos_L})}\Big(p_1^{\widetilde{\onWbos}_1}, .., p_L^{\widetilde{\onWbos}_L}\Big) & =
\sum_{k\neq l, i \neq j}
\big( \NPdensity^{(1)} \big)_{\onWbos_i \onWbos_j}^{\widetilde{\onWbos}_k \widetilde{\onWbos}_l}(p_k, p_l)
\Big( {\widetilde{\onWbos}_k} {\delta \over \delta\onWbos_i}  {\widetilde{\onWbos}_l} {\delta \over \delta\onWbos_j} \Big) 
\Big( {p_k} {\delta \over \delta q_i}  {p_l} {\delta \over \delta q_j} \Big) 
 \label{eq:nplanarloopcorrection}
\\ 
& \quad {1\over S_{(ij)}^{(kl)}} \sum_{\{{m,n\}}} \prod_{m\neq k,l;  n\neq i,j} 
\Big( {p_m} {\delta \over \delta q_n} \delta^{\widetilde{\onWbos}_m}_{\onWbos_n}   \Big) 
\cdot \NPff^{(0)}_{{\cal O}({\onWbos_1 \ldots \onWbos_L})}\Big(q_1^{\onWbos_1}, .., q_L^{\onWbos_L}\Big) \,.
\nonumber
\end{align}
The first line comes from the contribution from intrinsic one-loop corrections (which is a range-2 interaction). And the second line is the trivial tree factors that involve remaining $(L-2)$ legs, where the $(L-2)$ states ($\{{\widetilde{\onWbos}_m}\}$) should be in one-to-one correspondence with the $(L-2)$ fields ($\{{{\onWbos}_n}\}$) in the operator, and $S_{(ij)}^{(kl)}$ accounts for the possible symmetry factors in the case when there are identical fields. Obviously, to compute one-loop form factor, one only needs to focus on the density function $( \NPdensity^{(1)} )_{\onWbos_i \onWbos_j}^{\widetilde{\onWbos}_k \widetilde{\onWbos}_l}$. This discussion should be straightforward to be generalized to higher loop form factors.

\subsubsection*{Color structure of loop density functions}
Similar to \eqref{ffcdec}, one can also perform a color decomposition for the density function and expand $\NPdensity^{(\ell)}$ as 
\begin{equation}
\label{eq:colordecomforI}
    \NPdensity^{(\ell)}_{i_1i_2\cdots i_r}(p_{j_1},\ldots,p_{j_r})=\sum_{\alpha}  \check{\mathcal{C}}_{\alpha} \left[\left( \mathcal{I}^{(\ell)}\right)^{\widetilde{\mathcal{W}}_{j_1}\cdots \widetilde{\mathcal{W}}_{j_r} }_{\mathcal{W}_{i_1}\cdots \mathcal{W}_{i_r}}(p_{j_1},\ldots,p_{j_r})\right]_{\check{\mathcal{C}}_{\alpha}} \,,
\end{equation}
where $\check{\mathcal{C}}_{\alpha}$ is the color factor which should be understood as an "operator" acting on color space of sites $i_1 \cdots i_r$. A one-loop example of color decomposition of $\itbf{I}_{ij}^{(1)}(p_k,p_l)$ is 
\begin{equation}
	\itbf{I}^{(1)}_{ij}(p_k,p_l)=\sum_{\alpha}\check{\mathcal{C}}_{\alpha} \left[\left( \mathcal{I}^{(1)}\right)^{\widetilde{\mathcal{W}}_{k}\widetilde{\mathcal{W}}_{l}}_{\mathcal{W}_i\mathcal{W}_j}(p_k,p_l)\right]_{\check{\mathcal{C}}_{\alpha}}\,.
\end{equation}
And $\check{\mathcal{C}}_{\alpha}$ should be understood as
\begin{equation}\label{eq:ccheck1l}
\check{\mathcal{C}}_{\alpha} = \check{\mathcal{C}}_{\alpha} \Big(T^{a_k}, T^{a_l}, {\delta \over \delta T^{a_{i}}}, {\delta \over \delta T^{a_{j}}} \Big) \,,
\end{equation}
where it plays a role of annihilating two color charges $a_i, a_j$ in the operator (or say tree level form factor) and regenerate them for the external legs. 
As an analogy, in the field space the operator form is given in 
\eqref{eq:nplanarloopcorrection}:  
$\Big( {\widetilde{\onWbos}_k} {\delta \over \delta\onWbos_i} \Big) \Big( {\widetilde{\onWbos}_l} {\delta \over \delta\onWbos_j} \Big) \cdot \NPff^{(0)}_{{\cal O}_{\onWbos_i \onWbos_j}}$. 
The action in color space should be understood in a similar way as $\check{\mathcal{C}}_{k} \cdot \NPff^{(0)}_{{\cal O}_{\onWbos_i \onWbos_j}}$.

Let us explain these notations with concrete examples in trace basis. 
First, we define variation in color space operator, or $\emph{color variation}$,  as $\displaystyle \Check{T}^{a}\equiv \frac{\delta}{\delta T^{a}}$, and its action on color structure reads\footnote{Similar definition has been used in \emph{e.g.}~\cite{Beisert:2003jj}. Here we do not consider normal order because we forbid $\check{a}$ act on $T^{a}$ with in the same trace.}
\begin{equation}
    (\Check{T}^{a})_{\text{z}}^{\text{w}}\operatorname{tr}(T^{a_1}T^{a_2}\cdots T^{a_L})=\sum_{i=1}^{L}\delta^{a a_i}(T^{a_{i+1}})_{\text{z}}^{\text{x}_{i+2}}(T^{a_{i+2}})_{\text{x}_{i+2}}^{\text{x}_{i+3}}\cdots (T^{a_{i-1}})_{\text{x}_{i-2}}^{\text{w}} \,,
\end{equation}
where  repeated upper and lower indices means summation form 1 to $N_{\mathrm{c}}$. 
We can construct color trace by inserting variation operators, for example, $\operatorname{tr}(\check{T}^{a}T^{b}T^{c})\equiv(T^{b})_{\text{y}_{2}}^{\text{y}_{3}}(T^{c})_{\text{y}_{3}}^{\text{y}_{1}}(\check{T}^{a})_{\text{y}_1}^{\text{y}_2}$. 
As another example of the product of two traces
\begin{equation}
\begin{aligned}
    \operatorname{tr}(\check{T}^{a}T^{b}T^{c})\operatorname{tr}(T^{a_1}T^{a_2}\cdots T^{a_L})&=(T^{b})_{\text{y}_{2}}^{\text{y}_{3}}(T^{c})_{\text{y}_{3}}^{\text{y}_{1}}\sum_{i=1}^{L}\delta^{a a_i}(T^{a_{i+1}})_{\text{y}_1}^{\text{x}_{i+2}}\cdots (T^{a_{i-1}})_{\text{x}_{i-2}}^{\text{y}_2}\\
    &=\sum_{i=1}^{L}\delta^{a a_i} \operatorname{tr}(T^{a_1}\cdots T^{a_{i-1}}T^{b}T^{c}T^{a_{i+1}}\cdots T^{a_L}) \,.
\end{aligned}
\end{equation}
For simplicity of notation, we abbreviate $\operatorname{tr}(T^{a_1}\cdots T^{a_{L}})$ as $\operatorname{tr}(a_1\cdots a_{L})$, while $\operatorname{tr}(\check{a}bc)$ is short for $\operatorname{tr}(\check{T}^{a}T^{b}T^{c})$.

After clarifying the notations, we see that there are six color factors in the trace basis and the density function can be expanded as:
\begin{equation}
\label{eq:1loopNPdensityinTracebasis}
\NPdensity^{(1)}_{i(i+1)}(p_1,p_2)=
\sum_{k=1}^6 \check{\mathcal{T}}_{k}\left[\left( \mathcal{I}^{(1)}\right)^{\widetilde{\mathcal{W}}_1\widetilde{\mathcal{W}}_2}_{\mathcal{W}_{i}\mathcal{W}_{i+1}}(p_1,p_2)\right]_{\check{\mathcal{T}}_k} \,,
\end{equation}
where  the color basis $\{\check{\mathcal{T}}_{k}\}$ can be chosen as 
\begin{equation}
    \sigma \cdot \operatorname{tr}({a_1}{a_{2}}{\check{a}_{l_2}}{\check{a}_{l_1}})\equiv  \operatorname{tr}({a_1}\sigma(a_{2})\sigma(\check{a}_{l_2})\sigma(\check{a}_{l_1})), 
    \quad \text{ with } \sigma\in S_3 \,.
\end{equation}
with $\check{a}_{l_1},\check{a}_{l_2}$ acting on color indices in the $i$-th and $i+1$-th sites, for example, 
\begin{equation}
\label{eq:1loopcolorexample}
\check{\mathcal{T}}_{1} \cdot {\cal C}_{\cal O} = \operatorname{tr}(a_1 a_2 \check{a}_{l_2} \check{a}_{l_1}) {\rm tr}(\cdots \underbrace{a_{l_1}a_{l_2}}_{\mathrm{sites} \ i,i+1} \cdots) = N_c \, {\rm tr}(\cdots a_1 a_2 \cdots) + \cdots \,,
\end{equation}
which contains a planar color factor $N_c \, {\rm tr}(\cdots a_1 a_2 \cdots)$. We stress that
Note that to clearly distinguish internal and external lines, we will re-name the internal lines as $l_1,\ldots,l_r$, so that the $\delta/\delta T^{a_i}$ in \eqref{eq:ccheck1l} should be replaced by $\delta/\delta T^{a_{l_1}}$. 
In later discussion, for the convenience of notation, we usually omit the label of fields for $\mathcal{I}$ for simplicity and make an abbreviation as: 
\begin{equation}
     \mathcal{I}^{(1)}(p_1,p_2)\equiv \left( \mathcal{I}^{(1)}\right)^{\widetilde{\mathcal{W}}_1\widetilde{\mathcal{W}}_2}_{\mathcal{W}_{i}\mathcal{W}_{i+1}}(p_1,p_2) \,,
\end{equation}
and similar abbreviation for general case.

Before discussing other color basis, let us make two important comments:
\begin{enumerate}
\item 
We have introduced two kinds of \emph{pictures} to deal with loop level color structure for form factors. In the first picture, one can decompose the full loop form factor to a set of color basis as \eqref{ffcdec}, which is similar to the usual treatment for amplitudes. In the second picture, we introduce  the loop density functions and consider the color decomposition at density level as \eqref{eq:colordecomforI}.
One can always translate one picture to another, but they are suitable in different cases. 
For short length operators, such as $L=2,3$, the simplicity of color structure implies that the first picture is better. 
When considering form factors of generic high-length operators, the second picture is certainly much more convenient. 

\item
As previously mentioned, in the non-planar discussion, one has to includes operators with different color structures, such as multiple-trace operators. 
Using the picture of loop density functions, one makes a ``decoupling" between loop corrections and the detail structure of operators. In the computation of density functions, the structure of operator is not important (and therefore one can use single trace operator for simplicity). 
It is only when computing the complete form factor, one needs to act the density functions on certain particular operator (which is a combinatoric computation);
and comparing to single trace operators, the extra complexity of multiple-trace operators is only at the combinatoric level (but not in a quantum loop computation).

\end{enumerate}

\subsection{Trivalent (DDM) color basis}\label{ssec:cfddm}

The trace basis we discussed in the last subsection has the important advantage that the corresponding kinematic coefficients are form factors with particular ordering of on-shell particles. Consequently, they have much simpler kinematic properties and are easier to construct via unitarity cuts.

On the other hand, it is also clear that the trace basis for form factors (and also for amplitudes) are usually \emph{overcomplete}, namely, the physical form factors or amplitudes live in a smaller subspace of the space spanned by trace basis. 
This is obvious from the Feynman diagram origin of the amplitudes or form factors, where the color factors are given by product of trivalent structure constants ${f}^{abc}$.
By solving linear Jacobi relations for all trivalent graphs, a minimal set of color basis in terms of $f$'s can be obtained \cite{DelDuca:1999rs}.  We will call this alternative basis as trivalent color basis, or DDM basis.

Since the trace basis is larger than the DDM basis, this implies that the kinematic factors in the trace basis are not all independent but satisfy certain linear relations. At tree level, this precisely explains the origin of tree-level U(1) decoupling and KK relations \cite{Kleiss:1988ne}. 
The advantage of the trivalent color basis is that it can reduce the number of independent kinematic parts significantly, thus making the reconstruction of full-color quantities simpler.

To distinguish different bases, we use following different notations: 
\begin{equation}
\textrm{Trace basis}:  \ \  \NPff = \sum_{i=1}^{d(\mathcal{T})} \mathcal{T}_i \times [ \mathcal{F} ]_{\mathcal{T}_i} \,, \qquad
\textrm{DDM basis}:  \ \  \NPff = \sum_{i=1}^{d(\mathcal{D})} \mathcal{D}_i \times [ \mathcal{F} ]_{\mathcal{D}_i} \,, 
\end{equation}
where $[ \mathcal{F} ]_{\mathcal{C}_i}$ is the kinematic coefficient of color factor ${\mathcal{C}_i}$, and $d(\mathcal{C})$ is the dimension of the color basis.

Below we first review the amplitude case and then generalize to form factors.

\subsubsection*{A review for amplitudes}

Consider four-point tree amplitude $\itbf{A}_{4}^{(0)}$ as an example. In the trace basis, there are six elements:
\begin{equation}\label{tree4}
    \big\{\mathcal{T}_{\sigma}\big\}=\big\{\operatorname{tr}(T^{a_1}T^{a_{\sigma(2)}}T^{a_{\sigma(3)}}T^{a_{\sigma(4)}})\big\}_{\sigma\in S_3} .
\end{equation}
On the other hand, the trivalent DDM basis contains only two elements, and 
\begin{equation}
    \itbf{A}_{4}^{(0)}=\mathcal{D}_{1}\left[\mathcal{A}_{4}^{(0)}\right]_{\mathcal{D}_1}+\mathcal{D}_{2}\left[\mathcal{A}_{4}^{(0)}\right]_{\mathcal{D}_{2}} , 
\end{equation}
where
\begin{equation}
\big\{\mathcal{D}_{1},\mathcal{D}_{2}\big\}=\big\{\tilde{f}^{a_1 a_2 \text{x}}\tilde{f}^{\text{x} a_3 a_4},\tilde{f}^{a_1 a_3 \text{x}}\tilde{f}^{\text{x} a_2 a_4}\big\}. 
\end{equation}
Expressing the DDM basis in trace basis one obtains
\begin{equation}
\begin{aligned}\label{a4ddf}
    \itbf{A}_{4}^{(0)}&=\left[\mathcal{A}_{4}^{(0)}\right]_{\mathcal{D}_{1}}\big(\operatorname{tr}\left(a_{1}a_{2}a_{3}a_{4}\right)+\text{rev.}\big)+\left[\mathcal{A}_{4}^{(0)}\right]_{\mathcal{D}_{2}}\big(\operatorname{tr}\left(a_{1}a_{3}a_{2}a_{4}\right)+\text{rev.}\big)\\
    &+\left(-\left[\mathcal{A}_{4}^{(0)}\right]_{\mathcal{D}_{1}}-\left[\mathcal{A}_{4}^{(0)}\right]_{\mathcal{D}_{2}}\right)\big(\operatorname{tr}\left(a_{1}a_{2}a_{4}a_{3}\right)+\text{rev.}\big) ,
\end{aligned}   
\end{equation}
where `rev.' means reverse of trace ordering. 
From \eqref{a4ddf}, it is obvious that the six color-ordered amplitudes has only two independent components, and one can easily read out the reflection symmetry and U(1) decoupling relation (which is equivalent to the KK-relation at four points):
\begin{align}
	&\mathcal{A}^{(0)}_{n}\left(1,\cdots, n\right)=(-1)^{n}\mathcal{A}^{(0)}_{n}\left(n,\cdots ,1\right) \,, \quad \textrm{for} \ n=4\,, \nonumber\\
	&\mathcal{A}_{4}^{(0)}\left(i,j,k,l\right)+\mathcal{A}_{4}^{(0)}\left(i,k , l,j\right)+\mathcal{A}_{4}^{(0)}\left(i,k,j , l\right)=0 \,,
	\label{eq:A4KK}
\end{align} 
in which $\mathcal{A}_{4}^{(0)}(i,j,k,l)$ is a color-ordered amplitude with color factor ${\rm tr}(a_i a_j a_k a_l)$.

More generally, for $n$-point tree amplitudes the DDM basis can be given as  \cite{DelDuca:1999rs}
\begin{equation}\label{DDMb}
   \big\{\mathcal{D}_{\scriptscriptstyle\text{DDM}}(\sigma) \big\}_{\sigma\in S_{n-2}}= \big\{\sigma \cdot \tilde{f}^{a_{1} a_{{2}} \text{x}_{1}} \tilde{f}^{\text{x}_{1} a_{{3}} \text{x}_{2}} ...  \tilde{f}^{\text{x}_{n-3} a_{{n-1}}a_{n}} \big\} \equiv \big\{ \tilde{f}^{a_{1} a_{\sigma(2)} \text{x}_{1}} \tilde{f}^{\text{x}_{1} a_{\sigma(3)} \text{x}_{2}} ... \tilde{f}^{\text{x}_{n-3} a_{\sigma(n-1)}a_{n}} \big\}.
\end{equation}
As a result, there are only $(n-2)!$ linear independent partial amplitudes, which leads to linear relation between partial amplitudes in trace basis. 
By rewriting the DDM basis into trace basis, one can reproduce the KK relations for color-ordered amplitudes  \cite{Kleiss:1988ne}:
\begin{equation}\label{kk}
    \mathcal{A}_{n}^{(0)}(1,\{\alpha\}, n,\{\beta\})=(-1)^{n_{\beta}} \sum_{\sigma \in \{\alpha\}\shuffle\{\beta\}^{T}} \mathcal{A}_{n}^{(0)}\left(1, \sigma, n\right) \,,
\end{equation}
where $\{\beta\}^{T}$ is $\{\beta\}$ with the ordering reversed, and $\{\alpha\}\shuffle\{\beta\}^{T}$ is the shuffle product of $\{\alpha\}$ and $\{\beta\}^{T}$, \emph{i.e.} the ordered product that preserves the order of $\{\alpha\}$ and $\{\beta\}^{T}$ within the merged list. 

In summary, the idea of DDM basis is to express color factors using (linear) independent chain of structure constant. If we consider the constraints from group theory only, then DDM basis is the minimal basis for color factors in general gauge theory with fields in adjoint representation. This basis reduces the number of matrix elements that one needs to calculate, and can be used to derive KK-like relations for partial amplitudes in the trace basis. 
Similar relations have been considered at higher loop level and applied for computing amplitudes, see \emph{e.g.}~\cite{Edison:2011ta, Edison:2012fn, Feng:2011fja, Ochirov:2016ewn, Abreu:2018aqd, Badger:2019djh, Dalgleish:2020mof}.

\subsubsection*{Trivalent basis and KK-relations for form factors}

We now consider the trivalent basis for form factors. Comparing to amplitudes, the main new feature of form factors is the appearance of a local operator.  

We consider first the tree level case. The minimal tree form factors have the same trace structure as the operator and the color decomposition is trivial. Thus, only for non-minimal tree form factors one needs to consider trivalent basis.  
The construction of tree-level trivalent basis is straightforward by sewing trivalent basis for tree-level amplitudes on tree-level form factors. 

Basing on trivalent basis, one can derive KK-like relation for general color-ordered form factors. The essential difference between tree-level KK relations for general form factors and for amplitudes is that the order of fields is fixed for a given operator, and partial form factors appearing in KK relations must obey this order. For example, for $\mathcal{O}=\operatorname{tr}\big(\mathbf{W}_1\ldots \mathbf{W}_{L} \big)$ and external states $\left\{\mathcal{W}_{1}(p_1),\ldots,\mathcal{W}_{L}(p_L),g_{+}(p_{L+1}),g_{+}(p_{L+2})\right\}$, the order of $\mathcal{W}_{i}$ must be $(1,2,\ldots,L)$ in partial form factors (in trace basis). Corresponding KK relations for the color-ordered form factors then read
\begin{align}
        \sum_{\sigma \in \{p_3,\ldots,p_{L} \}\shuffle\{p_{L+2},p_{L+1}\}}\mathcal{F}_{L+2}(p_2,\sigma,p_1)=&\mathcal{F}_{L+2}\left(p_2,\{p_3,\ldots,p_{L}\},p_1,\{p_{L+1},p_{L+2}\}\right) \,,\\
        \sum_{\sigma\in \{p_{L+2}\}\shuffle \{p_1,\ldots,p_{i},p_{L+1},p_{i+1},\ldots,p_{L}\}}\mathcal{F}_{L+2}(\sigma)=&0 \,.
\end{align}
%

\begin{figure}[t]
    \centering
    \subfigure[Loop correction density.]{
        \includegraphics[width=0.35\linewidth]{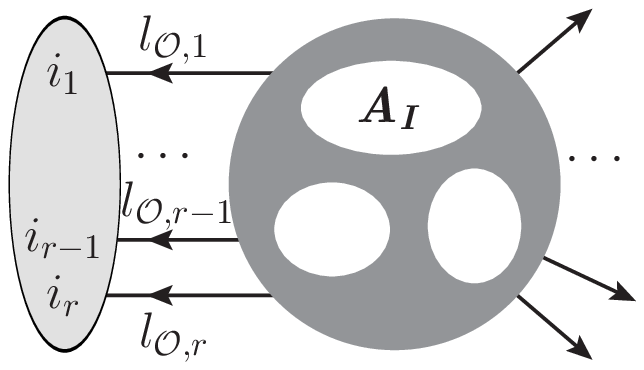}
         \label{fig:ckden1}
        }
     \centering
   \subfigure[Amplitudes from setting $l_{\mathcal{O},i}$ on-shell.]{
        \includegraphics[width=0.38\linewidth]{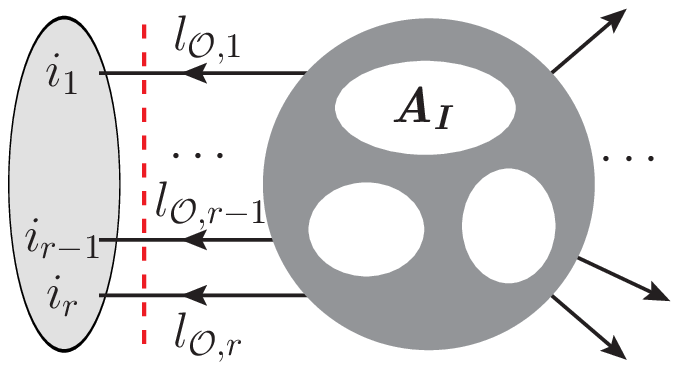}
        \label{fig:ckden2}
        }
    
    \caption{Diagrammatic representation for loop correction density and corresponding amplitudes.}
    \label{fig:ckden}
\end{figure}

When it comes to loop-level form factors, the trivalent part of the color factor is provided by an amplitude part that is sewed to the operator, such as shown in Figure~\ref{fig:ckden}.
Therefore, it is natural to consider the density picture.\footnote{In the context of loop-level form factors, we only focus on those "connected" densities, \emph{i.e.} those can not be trivially separated as  two disconnected lower-loop densities. }
At $\ell$ loops, the possible interaction range $r$ is $2 \leq r \leq(\ell+1)$. 
Consider an $\ell$-loop minimal form factor,
the range-$r$ density is given by sewing a $(\ell+1-r)$-loop $(2r)$-point amplitude to the minimal tree form factor. 
To construct trivalent basis for the form factor density, the problem is reduced to select proper linear independent color factors for this cut amplitude. 

Consider for example the one-loop density. The interaction range is $r=2$. For a minimal form factor, the cut amplitude is a four-point tree amplitude. 
Clearly, one can choose two trivalent basis for the one-loop density function:
\begin{equation}\label{eq:1lbasis}
    \Big\{\check{\mathcal{D}}_{1}=\tilde{f}^{a_{1}\check{a}_{{l}_{1}}\text{x}}\tilde{f}^{\text{x}a_{2}\check{a}_{{l}_{2}}}, \ \check{\mathcal{D}}_{2}= \tilde{f}^{a_{2}\check{a}_{{l}_{1}}\text{x}}\tilde{f}^{\text{x} a_{1}\check{a}_{{l}_{2}}} \Big\} \,.
\end{equation}
so that the minimal one-loop correction density has the form
\begin{equation}\label{1lddm}
    \NPdensity_{i_1i_2}^{(1)}(p_1,p_2)=\sum_{i=1}^2\check{\mathcal{D}}_{i}\left[\mathcal{I}^{(1)}(p_1,p_2)\right]_{\check{\mathcal{D}}_{i}}\,.
\end{equation}

It is natural to map between this trivalent basis and the trace basis in the previous section. For example, $\check{\mathcal{D}}_1= \left(\operatorname{tr}\left(a_{1} a_{2} \check{a}_{l_2} \check{a}_{l_1}\right)+\operatorname{tr}\left(a_{1} \check{a}_{l_1} \check{a}_{l_2} a_{2}\right)-\operatorname{tr}\left(a_{1} \check{a}_{l_2} a_{2} \check{a}_{l_1}\right)-\operatorname{tr}\left(a_{1} \check{a}_{l_2} a_{2} \check{a}_{l_1}\right)\right)$ and one can easily find out relations between kinematic parts, such as
\begin{equation}
    \left[\mathcal{I}^{(1)}\right]_{\check{\mathcal{D}}_{1}}=\left[\mathcal{I}^{(1)}\right]_{\check{\mathcal{T}}_{1}}\,.
\end{equation}

The above discussion can be generalized to higher loops in a similar way.
We will discuss two-loop minimal form factors in detail in Section~\ref{sec:2l}, see also  Appendix~\ref{ap:ddm}.

\subsection{Color-Kinematics duality for form factors}\label{ssec:ck}

A remarkable observation made by Bern, Carrasco and Johansson \cite{Bern:2008qj, Bern:2010ue} reveals that the kinematic numerators of perturbative amplitudes may satisfy similar Jacobi relations of color factors. 
An extensive review of  the developments can be found in \cite{Bern:2019prr}. 
Here we first briefly review this duality between color and kinematics in the context of amplitudes, then we generalize to the loop density functions of form factors. 

The color-kinematics duality (CK duality) states that it is possible to rearrange the perturbative amplitudes, such that the Jacobi relations satisfied by color factors of cubic graphs also apply for the kinematic numerators of same diagrams.
More concretely, an $\ell$-loop amplitude with full-color dependence can be given in the form
\begin{equation}\label{ckorigin}
    \itbf{A}_{\mathcal{O},n}^{(\ell)}=\sum_{\sigma_n}\sum_{\Gamma_{i}}\int \prod_{\alpha=1}^{\ell} \mathrm{d}^{D} l_{\alpha} \frac{1}{S_{i}} \frac{\mathcal{C}_{i} N_{i}}{\prod_{a} d_{i, a}} \,,
\end{equation}
where for a given trivalent graph denoted by $\Gamma_i$, $S_{i}$ is the symmetry factor of the graph, $\mathcal{C}_i$ is the color factor, and $d_{i,a}$ are the propagators. 
The kinematic numerators, $N_i$, are the only objects that are not determined by $\Gamma_i$ and contain the intrinsic physical information of the amplitudes. The CK duality requires the numerators $N_i$ to take a special form so that they satisfy the same Jacobi relations for the color factors:
\begin{equation}
\mathcal{C}_i=\mathcal{C}_j+\mathcal{C}_k \quad \rightarrow \quad N_i=N_j+N_k \,.
\end{equation}
We will refer the kinematic Jacobi relation as dual Jacobi relations.
  
Color-kinematics duality is very powerful in computing full-color amplitudes at high loops, see \emph{e.g.}~\cite{Bern:2010ue, Carrasco:2011mn, Bern:2012uf, Bern:2013qca, Bern:2014sna, Bern:2017yxu, Johansson:2017bfl, Bern:2017ucb, Kalin:2018thp, Boels:2013bi, Bern:2013yya, Mogull:2015adi}. 
Dual Jacobi relations, together with other constraints, result in significant simplification: if the CK duality form exists, only a small subset of numerators, called \emph{master numerators}, are necessary to consider. One can start with making an ansatz for the master numerators which are enough to determine the full integrand, and then employ various physical constrains such as unitarity check to determine the ansatz. 

To apply the similar procedure to calculate full-color form factors, one should take into account the effect of local operators.
For the case of Sudakov form factors of short length operators, CK duality has been known up to five loops \cite{Boels:2012ew, Yang:2016ear}. On the other hand, whether CK duality is true for form factors of high-length operators is still unknown. 

Analogue to the two pictures mentioned in the end of Section~\ref{ssec:cftr}, one can either directly deal with color factors of full form factors, or, consider color factors of loop correction densities. 
Since we want to study operators with general length, it is convenient to consider the second picture.
In this case, color factors for form factors are determined in terms of two aspects: the trivalent topology of the sub-amplitude, and the local operator  where the trivalent topology ``springs" from. 
More concretely, following Figure~\ref{fig:ckden}, for an operator $\mathcal{O}=\operatorname{tr}(\mathbf{W}_{1}\ldots\mathbf{W}_{L})$,\footnote{We use a single trace operator to clarify our notation, and the principle applies for general operators.} a range-$r$ interaction includes trivalent topology connected to the $i_{1},i_{2},\ldots,i_{r}$-th fields, with $i_{\alpha}\in 1,\ldots,L$,  
so the color factor from a given loop correction density $\NPdensity^{(\ell)}_{i_1i_2\cdots i_r} = \NPdensity^{(\ell)}_{\{i_{\alpha}\}}$ is given as 
\begin{equation}
    \mathcal{C}(\{i_{\alpha}\},\Gamma^{\itbf{A}})= {\check{\mathcal{C}}^{\{i_{\alpha}\}}_{\Gamma^{\itbf{A}}}} \cdot \mathcal{C}_{\mathcal{O}}
    = \Big( \prod \tilde{f} \prod_{\alpha} \tilde{f}^{\text{x}_{\alpha}\text{y}_{\alpha}\check{a}_{i_\alpha}} \Big) \cdot \operatorname{tr}(a_{1}\ldots a_{i_{\alpha}}\ldots a_{i_{L}}) \,,
\end{equation}
where $\Gamma^{\itbf{A}}$ are the trivalent topologies of sub-amplitudes.
It is worthwhile stressing that for general operators, color factors $\mathcal{C}(\{i_{\alpha}\},\Gamma^{\itbf{A}})$ with different $\{i_{\alpha}\}$ are linearly independent with each other. 
Therefore, we do not expect there are Jacobi relations between color factors of different $\{i_{\alpha}\}$. As a corollary, interaction with different ranges should be viewed as decoupled in the context of dual Jacobi relations.

\begin{figure}
    \centering
    \subfigure[$s$-channel]{
        \includegraphics[width=0.31\linewidth]{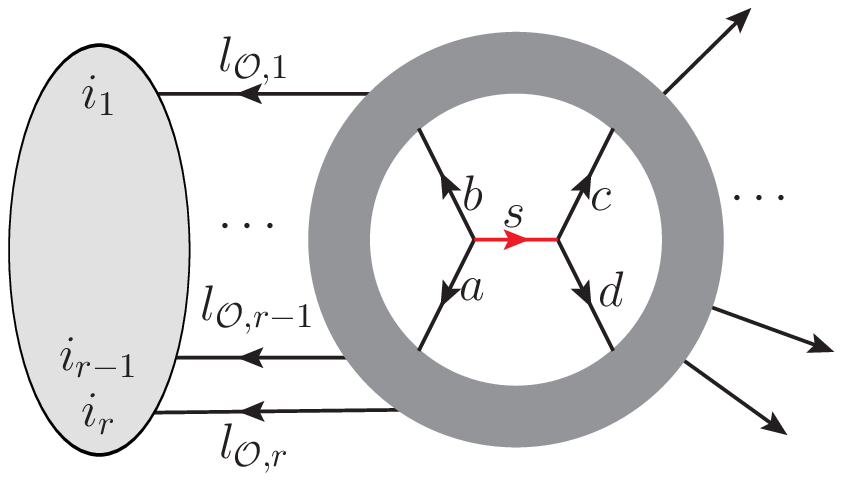}
         \label{fig:cks}
        }
    \centering
    \subfigure[$t$-channel]{
        \includegraphics[width=0.31\linewidth]{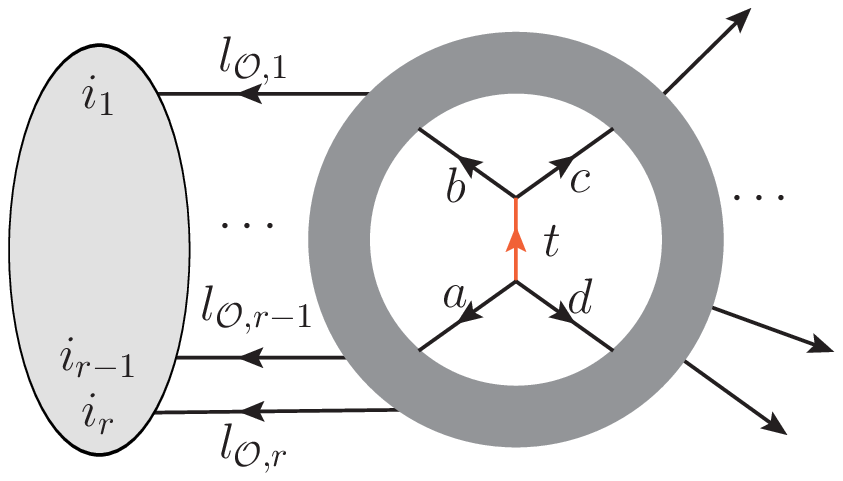}
        \label{fig:ckt}
        }
    \centering
    \subfigure[$u$-channel]{
        \includegraphics[width=0.31\linewidth]{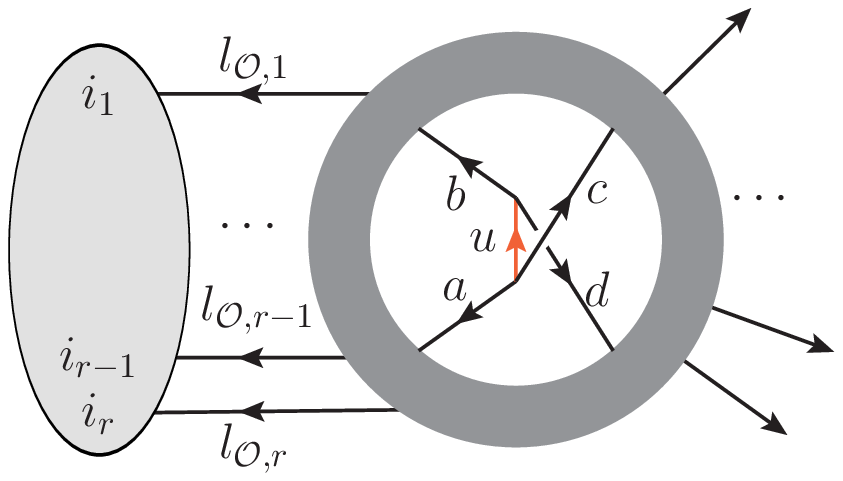}
        \label{fig:cku}
        }
    \caption{Three topologies related by Jacobi identities.}
    \label{fig:ckgra}
\end{figure}

The above discussion shows that for form factor densities, one can first classify the loop corrections according the interaction ranges, and then for each range $r$, consider Jacobi identities between color factors with the same range interaction $\{i_{\alpha}\}$. Similar to \eqref{ckorigin}, one can write the loop density as
\begin{equation}\label{eq:densityCK}
    \NPdensity^{(\ell)}_{\{i_{\alpha}\}}=\sum_{\sigma}\sum_{\Gamma^{\itbf{A}}_{j}}\int \prod_{\alpha=1}^{\ell} \mathrm{d}^{D} l_{\alpha} \frac{1}{S^{\{i_{\alpha}\}}_{j}} \frac{\check{\mathcal{C}}^{\{i_{\alpha}\}}_{j} N^{\{i_{\alpha}\}}_{j}}{\prod_{a} d_{j, a}} \,,
\end{equation}
where ${\check{\mathcal{C}}^{\{i_{\alpha}\}}_{j}}$ is the color factor from the sub-amplitudes topology ${\Gamma^{\itbf{A}}_{j}}$.

For any propagators not directly connected to the operators, labeled by red line in Figure~\ref{fig:cks}, one can  rotate it into the other two channels as in Figure~\ref{fig:ckt} and \ref{fig:cku}, which we call $s,t$ and $u$ channel respectively. The color factors of these three diagrams are 
\begin{equation}
    \mathcal{C}^{\{i_{\alpha}\}}_{s}=\tilde{f}^{a b s} \tilde{f}^{c d s} \prod \tilde{f} \cdot \mathcal{C}_{\mathcal{O}}, \quad 
    \mathcal{C}^{\{i_{\alpha}\}}_{t}=\tilde{f}^{b c t} \tilde{f}^{d a t} \prod \tilde{f}\cdot \mathcal{C}_{\mathcal{O}}, \quad 
    \mathcal{C}^{\{i_{\alpha}\}}_{u}=\tilde{f}^{a c u} \tilde{f}^{b d u} \prod \tilde{f}\cdot \mathcal{C}_{\mathcal{O}} \,,
\end{equation}
where $\prod\tilde{f}$ comes from other vertices which is identical in these three diagrams,  $\mathcal{C}_{\mathcal{O}}$ is color factor of operator $\mathcal{O}$. The color factors of densities ${\check{\mathcal{C}}^{\{i_{\alpha}\}}_{j}}$, defined in $\mathcal{C}^{\{i_{\alpha}\}}_{s}= {\check{\mathcal{C}}^{\{i_{\alpha}\}}_{j}} \cdot \mathcal{C}_{\mathcal{O}}$, satisfy the corresponding Jacobi relation 
\begin{equation}\label{cjacob}
    \check{\mathcal{C}}^{\{i_{\alpha}\}}_{s}= \check{\mathcal{C}}^{\{i_{\alpha}\}}_{t}+ \check{\mathcal{C}}^{\{i_{\alpha}\}}_{u} \,.
\end{equation}
Thus, the kinematic numerators are expected to satisfy the dual Jacobi relation:
\begin{equation}\label{djacob}
    N^{\{i_{\alpha}\}}_s=N^{\{i_{\alpha}\}}_t+N^{\{i_{\alpha}\}}_u \,,
\end{equation}
resulting in a large set of linear relations between kinematic numerators.

For supersymmetric theories, such as $\mathcal{N}=4$ SYM, it is possible to impose further constraints on numerators. For quantities with good UV behavior, such as scattering amplitudes or BPS form factors, supersymmetric cancellation may lead to following simplifications:
(1) one may eliminate topology with sub-bubble/triangle, except when those sub-bubble/triangle are directly connected to the operators, 
and (2) one may impose constraint on the power of loop momenta which is useful to reduce the size of ansatz.
Though, one should be cautious that these conditions may not always be satisfied and may be relaxed according to specific circumstances, see  Section~\ref{sec:CKBPS} for more discussions.

\section{General discussion of non-planar unitarity}\label{sec:unitarity}

In this section, we describe the general strategy of computing full-color form factors with (generalized) unitarity cut method \cite{Bern:1994zx, Bern:1994cg, Britto:2004nc}.  
Unitarity cut serves to construct perturbative amplitudes or form factors from on-shell ``building blocks" which have lower loops or points. 
Here \emph{cut} refers to setting certain (off-shell) internal lines to be on-shell
\begin{equation}
    \frac{i}{l^{2}} \stackrel{\text { cut }}{\longrightarrow} 2 \pi \delta_{+}\left(l^{2}\right) ,
\end{equation}
such that the loop quantities are factorized as product of simpler on-shell building blocks. 
This cut process captures physical singularities of kinematic functions, and one can reconstruct the analytic function via the information given by its singularities. 

Below we first outline the general strategy of performing full-color non-planar unitarity in Section~\ref{ssec:NPcutstrategy}, then we provide concrete examples in Section~\ref{ssec:unitarityeg} to illustrate the strategy. The results of examples will also be used in later calculation in Section~\ref{sec:2l}. Some further details and features about (non-planar) unitarity are also provided in Appendix~\ref{ap:npuni}.

\subsection{General strategy}
\label{ssec:NPcutstrategy}

From the discussion in the previous section, the full-color form factor takes the color decomposition form as
\begin{equation}
\label{eq:colordecomp-general}
\NPff^{(\ell)} = \sum_i {\cal C}_i  \, \left[ \PLff^{(\ell)} \right] _{{\cal C}_i} \,,
\end{equation}
where ${\cal C}_i$ are the color factors, and the color-stripped kinematic coefficients are
\begin{equation}
\left[ {\cal F}^{(\ell)} \right] _{{\cal C}_i} = \sum \text{\big(momentum integrals\big)} \,,
\end{equation}
which are what we want to compute using unitarity cut method.

It should be helpful to first review the planar unitarity. The planar form factors, denoted as $\left[\mathcal{F}^{(\ell)}\right]_{\scriptscriptstyle \rm PL}$, satisfies the following planar cut:\footnote{We consider only tree building blocks here, but sub-loop building blocks are also possible.}
\begin{equation}\label{eq:pl-unitarity}
    \left[\mathcal{F}^{(\ell)}\right]_{\scriptscriptstyle \rm PL}\Big|_{\rm cut}= \int  {\rm dPS}_{\{l_a^2\}} \text{ \big(color-ordered tree products\big)}\,, 
\end{equation}
where the phase space measure is understood as ${\rm dPS}_{\{l_a^2\}} = \prod_a d^D l_a \delta_+(l_a^2)$. 
The planar unitarity is simple in two aspects: (1) the tree blocks are simple color-ordered quantities, and (2) the ordering within each blocks and the connection between blocks are completely fixed by planar topology. 

Planar form factors can be regarded as special (planar) components of full-color form factors, or in other words, they are the kinematic components associated to planar color factors 
$\mathcal{T}_{\scriptscriptstyle \rm PL}$ that are leading in the large $N_c$ limit.
The choosing of the tree blocks is determined by the fact that the product of their trace color factors ${\cal T}_{\textrm{tree-block}}$ contribute the planar color factor:
\begin{equation}
\prod \mathcal{T}_{\textrm{tree-block}} = \mathcal{T}_{\scriptscriptstyle \rm PL} + \big(\textrm{subleading terms in $N_c$}\big) \,.
\end{equation}

This understanding allows us to generalize unitarity cut to kinematic component $\left[ {\cal F}^{(\ell)} \right] _{{\cal C}_i}$ with a general color factor ${{\cal C}_i}$. 
Let us first consider ${{\cal C}_i} = \mathcal{T}_i$ as trace basis, and we will comment on the cases with other basis shortly.
For a generic color factor $\mathcal{T}_i$ (which can be multi-trace for example), one can apply unitarity in a similar way as 
\begin{equation}
\label{eq:unitaryastreeproduct}
\left[\mathcal{F}^{(\ell)}\right]_{\mathcal{T}_i} \Big|_{\rm cut}
= 
c_i \int  {\rm dPS}_{\{l_a^2\}} \text{\big(color-ordered tree products\big)}  \,,
\end{equation}
where the tree products are chosen such that 
\begin{equation}
\label{eq:Ttreeproduct}
\prod \mathcal{T}_{\textrm{tree-block}} = c_i \mathcal{T}_i +  \big(\textrm{non-$\mathcal{T}_i $ factors}\big) ,
\end{equation}
and $c_i$ is a possible numerical factor. 
Here it is important that tree building blocks are still color-ordered, thus taking advantage of the simplicity of color-ordered quantities. The RHS of  \eqref{eq:unitaryastreeproduct} are specified by the ordering of each tree blocks. 
We define each such configuration as a \emph{cut channel}. 

A single cut channel typically only probes partial physical singularities of the integrand, therefore,
to determine the full kinematic integrand of $\left[\mathcal{F}^{(\ell)}\right]_{\mathcal{T}_i}$, one needs to consider a spanning set of cut channels having color factor ${\mathcal{T}_i}$.
For planar form factors, usually only a small number of cut channels are needed which can be easily determined by planarity. 
For the full-color case, a more careful analysis is needed based on \eqref{eq:unitaryastreeproduct}.
Different ways of sewing tree blocks as well as different planar ordering of each tree block should be taken into account.
This will be explained with explicit examples in Section~\ref{ssec:unitarityeg}.  
In Appendix~\ref{ap:npuni} we also describe some systematic way of classifying cut channels and also discuss some further features of nonplanar cuts.

Finally, to obtain the full form factor, one needs to consider all color factors in the basis. 
Let us  now comment on the choice of other color basis.
If the color basis ${\cal C}_j$ of full-color form factor $\itbf{F}^{(\ell)}$ is not trace basis, for example is trivalent basis (DDM basis or CK ansatz form), such that ${\cal C}_j = \prod \tilde{f}$, one can expand the $\mathcal{C}_j$ in terms of trace basis and perform a basis change. For example, for ${\cal C}_j =  b_{j}^{(i)} {\mathcal{T}_i} + \big(\textrm{non-$\mathcal{T}_i $ factors})$, the kinematic parts from different basis are related as
\begin{equation}
    \left[\mathcal{F}^{(\ell)}\right]_{\mathcal{T}_i}=\sum_{j}\left[\mathcal{F}^{(\ell)}\right]_{\mathcal{C}_{j}} b_{j}^{(i)} \,. 
    \end{equation}
Thus knowing $\left[\mathcal{F}^{(\ell)}\right]_{\mathcal{T}_i}$ will be able determine $\left[\mathcal{F}^{(\ell)}\right]_{\mathcal{C}_{j}}$.
Practically it is often convenient to start with trivalent DDM basis, 
which has relatively small number of color basis and can also manifest certain symmetry structure of the results. 

The above discussion also applies at the level of density functions $\itbf{I}^{(\ell)}$ defined in Section~\ref{sec:color}. In our computation of two-loop form factors, thanks to universal structure of density functions, one can select a small subset of $\mathcal{T}_i$, such that their kinematic parts are enough to fix all kinematic functions in $\itbf{I}$. A full application to two-loop form factors will be given in Section~\ref{sec:2l}.
Similar strategy also applies to the CK dual ansatz which will be discussed in Section~\ref{sec:CKBPS}.

The above strategy may be summarize as following graph:
%
%
$$
\begin{tabular}{| c |}
\hline  color-ordered \\ tree blocks  \\ \hline
\end{tabular}
\xlongleftrightarrow[\mbox{all channels}]{\mbox{unitarity-cut}}
\begin{tabular}{| c |}
\hline  Kinematic \\ part $[\mathcal{F}^{(\ell)}]_{\mathcal{T}_i} $  \\ \hline
\end{tabular}
\xlongrightarrow[\mbox{all ${\mathcal{T}_i}$}]{\mbox{collect}}
\begin{tabular}{| c |}
\hline  $\NPff^{(\ell)}$ in \\ trace basis  \\ \hline
\end{tabular}
\leftrightarrow
\begin{tabular}{|c|}
\hline  $\NPff^{(\ell)}$ in DDM basis \\
or CK dual form  \\ \hline
\end{tabular}
$$

\subsection{Examples for unitarity construction}
\label{ssec:unitarityeg}

We now apply the above strategy to construct some two-loop form factor examples.
For simplicity, we choose operator as half-BPS operators ${\rm tr}(X^L)$.
These examples will be useful in later full-color form factor construction in Section~\ref{sec:2l}.

\subsubsection*{Example 1: the planar case}

\begin{figure}
    \centering
    \subfigure[]{\includegraphics[width=0.33\linewidth]{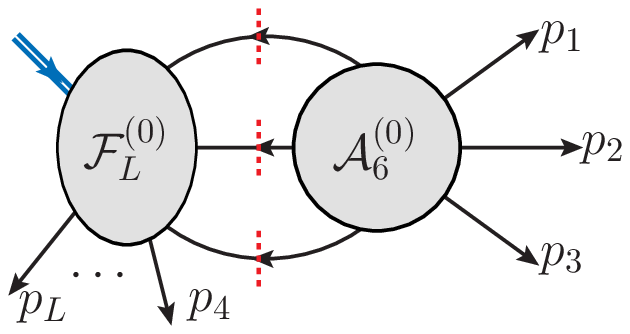}\label{fig:pa6cut}}
    \subfigure[]{\includegraphics[width=0.3\linewidth]{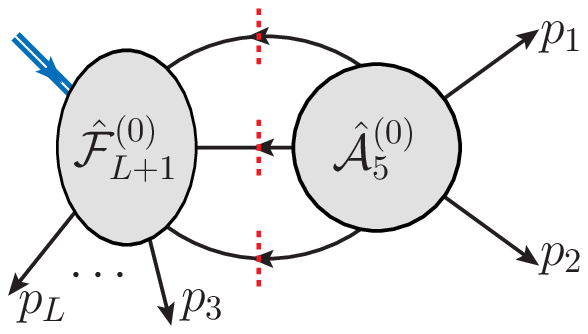}\label{fig:pa5cut}}
    \subfigure[]{\includegraphics[width=0.3\linewidth]{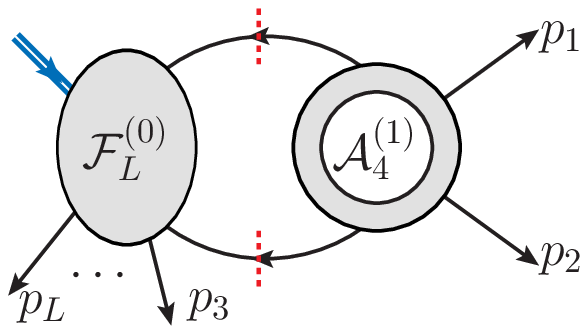}\label{fig:pa4cut}}
    \caption{Planar cut channels at two loops. Without loss of generality, we choose the external lines as $p_1,p_2,p_3$. }
    \label{fig:pcut}
\end{figure}

We briefly review the the planar limit case. 
In this case, our targeting color factor is 
\begin{equation}
\mathcal{T}_{\scriptscriptstyle \rm PL}=N_{ c}^2\operatorname{tr}(a_1\cdots a_{L}) \,.
\end{equation}
To determine the integrand, it is enough to consider three types of cut channels as shown in Figure~\ref{fig:pcut}. 

We start from $s_{123}$ triple-cut (TC) with planar ordering in Figure~\ref{fig:pa6cut}.
The product of two trace factors from tree building blocks gives 
\begin{equation}
\operatorname{tr}_{\mathcal{A}}(\check{a}_{l_3}\check{a}_{l_2}\check{a}_{l_1}a_1a_2 a_3)\times \operatorname{tr}_{{\mathcal{F}}}(a_{l_1} a_{l_2} a_{l_3} a_4 \cdots a_L) =1 \times \mathcal{T}_{\scriptscriptstyle \rm PL} + \big(\textrm{subleading terms in $N_c$}\big) \,,
\end{equation}
which indeed contains the planar color factor.
The kinematic part under this cut is\footnote{Terms proportional to $l_i^2$ are possible contributions which can not be fixed by this cut. Other cuts can detect them so they are contained in ``terms fixed by other channels" in \eqref{eq:bpspl3}. }
\begin{align}\label{tcfull1}
\nonumber 
\left.\mathcal{F}_{\scriptscriptstyle \rm PL}^{(2)}(p_1,p_2,p_3,\ldots)\right|_{s_{123}\textrm{-TC}} & = \int \mathrm{\mathrm{dPS}}_{3,l} \mathcal{F}^{(0)}_{L}\left(-l_{1},-l_{2},-l_{3},p_4,\ldots,p_L\right) \mathcal{A}^{(0)}_{6}(l_{3},l_{2},l_{1},p_{1},p_{2},p_{3}) \\
    &=\begin{aligned}
        \includegraphics[width=0.15\linewidth]{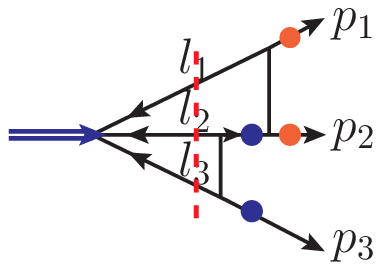}
         \end{aligned} \hlp{5} 
         \begin{aligned}
        \includegraphics[width=0.15\linewidth]{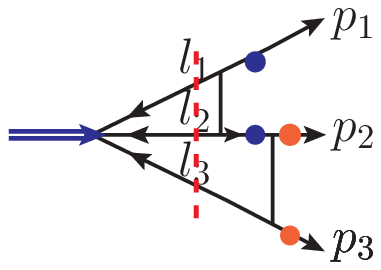}
        \end{aligned}
        \hlm{5}
        \begin{aligned}
        \includegraphics[width=0.15\linewidth]{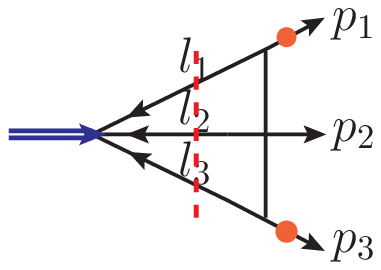}
    \end{aligned} ,
\end{align}
where the six-scalar tree amplitude expression (given in Appendix~\ref{ap:amp}) and expression for the tree form factor (here $\mathcal{F}_{L}^{(0)}=1$)  are used.
In the second line of \eqref{tcfull1}, the tree product is reorganized as three cut integrals (see \cite{Brandhuber:2014ica} for more details).
Here for convenience, we introduce in \eqref{eq:bpspl3} a graphic notation to use colored dots to represent numerators: two dots with the same color on lines with momenta $q_{i}$ and $q_{j}$ means $2(q_i\cdot q_j)$. 
For example, the numerators for three integrals  in \eqref{eq:bpspl3} are respectively:
\begin{equation}
4 (p_1 \cdot p_2) [p_3 \cdot (l_1 + p_1 + p_2)] \,, \quad 
4 (p_2 \cdot p_3) [p_1 \cdot (l_3 + p_2 + p_3)] \,, \quad 
2 (p_1 \cdot p_3) \,. 
\end{equation}
It should be obvious that $s_{i(i+1)(i+2)}$ triple cuts contribute similar terms as \eqref{eq:bpspl3} by changing $\{p_1,p_2,p_3\}$ to $\{p_i,p_{i+1},p_{i+2}\}$. 
Removing the cuts, the form factor can be expressed as
\begin{equation}\label{eq:bpspl3}
\begin{aligned}
    \mathcal{F}_{\scriptscriptstyle \rm PL}^{(2)}(p_1,p_2,p_3,\ldots)
    =&\begin{aligned}
        \includegraphics[width=0.15\linewidth]{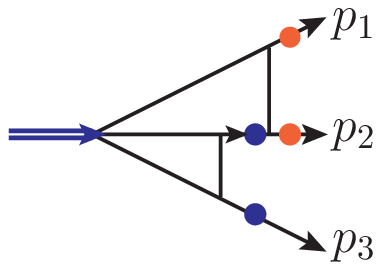}
         \end{aligned} \hlp{5} 
         \begin{aligned}
        \includegraphics[width=0.15\linewidth]{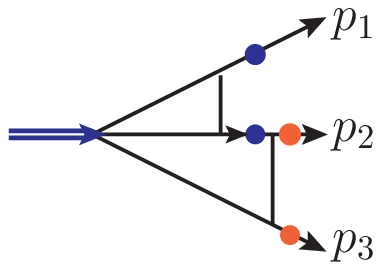}
        \end{aligned}
        \hlm{5}
        \begin{aligned}
        \includegraphics[width=0.15\linewidth]{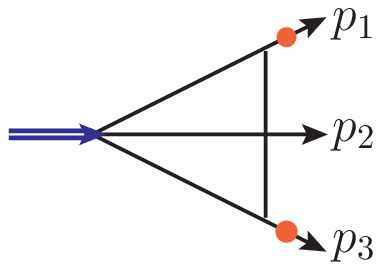}
    \end{aligned}  \\
            &+ \big(\text{terms fixed by other channels}\big) \,.
\end{aligned}
\end{equation}


Next we consider the $s_{12}$ triple-cut (TC) channels in Figure~\ref{fig:pa5cut}. As a check, the product of tree color factors gives
\begin{equation}
    \operatorname{tr}_{\mathcal{A}}(\check{a}_{l_3}\check{a}_{l_2}\check{a}_{l_1}a_1a_2)\times \operatorname{tr}_{{\mathcal{F}}}(a_{l_1} a_{l_2} a_{l_3}  a_3 \cdots a_L) =1 \times \mathcal{T}_{\scriptscriptstyle \rm PL} + \big(\textrm{subleading terms in $N_c$}\big) \,.
\end{equation}
The cut of the kinematic part is:
\begin{equation}
\mathcal{F}_{\scriptscriptstyle \rm PL}^{(2)}(p_1,p_2,p_3,\ldots,p_L) \big|_{s_{12}\textrm{-TC}
}=\int \mathrm{\mathrm{d\hat{PS}}}_{3,l} \hat{\mathcal{F}}^{(0)}_{L+1}\left(-l_{1},-l_{2},-l_{3}, p_3, \ldots,p_L\right) \hat{\mathcal{A}}^{(0)}_{5}(l_{3},l_{2},l_{1},p_{1},p_{2}) \,.
\label{eq:cut2TC}
\end{equation}
The expressions of five-point tree amplitudes and next-to-minimal form factors are reviewed in Section~\ref{subsec:treeAmpFF}, and ``hats" in \eqref{eq:cut2TC} remind that one should take super form factor/amplitudes or equivalently, summing over all possible internal state configurations.
Without going into details, we point out that in this channel only integrals with external momenta $\{p_L,p_1,p_2,p_3\}$ are possibly probed, because the tree blocks only contain these momenta. Given the knowledge from previous $s_{123}$-TC (and similar $s_{L12}$-TC), the new contribution from this cut reads:
\begin{align}
\mathcal{F}_{\scriptscriptstyle \rm PL}^{(2)}(p_1,p_2,p_3,\ldots,p_L) = &
    \begin{aligned}
         \includegraphics[width=0.15\linewidth]{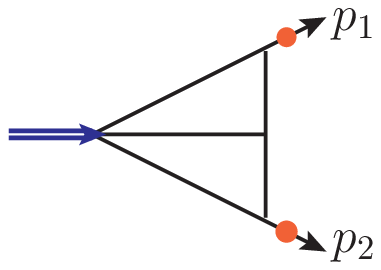}
        \end{aligned}
        \hlp{5}
        \begin{aligned}
           \includegraphics[width=0.17\linewidth]{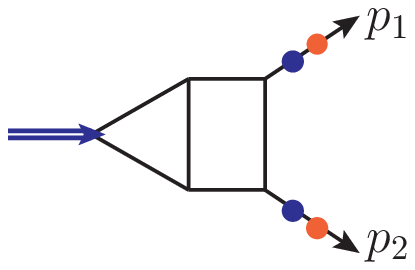}
        \end{aligned}
        \hskip -10pt + \big(\text{terms from $s_{123}$- and $s_{L12}$-TC}\big) \nonumber \\
        &+ \big(\text{terms fixed by other channels}\big) \,. \label{eq:bpspl2}
\end{align}

Finally, one can check that the $s_{12}$ double-cut in Figure~\ref{fig:pa4cut} does not give any new contribution. Thus the kinematic function in \eqref{eq:bpspl2} (plus their permutations) satisfies all required cut channels and determines the  complete planar form factor results.

The full planar form factor $\mathcal{F}_{\scriptscriptstyle \rm PL}^{(2)}(p_1,p_2,p_3,\ldots,p_L)$ can be written in a compact form as
\begin{equation}
    \mathcal{F}_{\scriptscriptstyle \rm PL}^{(2)}(p_1,p_2,p_3,\ldots,p_L)=\sum_{i=1}^{L}\widetilde{\mathcal{F}}_{\scriptscriptstyle \text{ PL}}^{(2)}(p_i, p_{i+1}, p_{i+2}) \,,
\end{equation}
where we define a density function (specified by a tilde):
\begin{align}\label{eq:planarfinaldensity}
&\widetilde{\mathcal{F}}_{\scriptscriptstyle \rm PL}^{(2)}(p_1, p_2, p_3)=\\
\nonumber 
&   \begin{aligned}
        \includegraphics[width=0.15\linewidth]{fig/tribox2top.eps}
         \end{aligned} \hlp{5} 
         \begin{aligned}
        \includegraphics[width=0.15\linewidth]{fig/boxtri2top.eps}
        \end{aligned}
        \hlm{5}
        \begin{aligned}
        \includegraphics[width=0.15\linewidth]{fig/tritri2top.eps}
    \end{aligned} \hskip -5pt +{1\over2}\Big( \hskip -3pt  \begin{aligned}
           \includegraphics[width=0.15\linewidth]{fig/tritri2pt1top.eps}
        \end{aligned}
        \hlp{5}
        \begin{aligned}
           \includegraphics[width=0.17\linewidth]{fig/pladderm1top.eps}
        \end{aligned}
        \hlp{5} (s_{12}\shortrightarrow s_{23})\Big) \,.
\end{align}
Other density functions are related to it as:
\begin{equation}\label{eq:pl-dens-cyc}
    \widetilde{\mathcal{F}}_{\scriptscriptstyle \text{ PL}}^{(2)}(p_i, p_{i+1}, p_{i+2})=
    \widetilde{\mathcal{F}}_{\scriptscriptstyle \text{ PL}}^{(2)}(p_1, p_2, p_3)\big|_{\{ 1\shortrightarrow i, 2\shortrightarrow (i+1), 3\shortrightarrow (i+2) \}} \,. 
\end{equation}
Note that the 1/2 factor in \eqref{eq:planarfinaldensity} shows that we have evenly distributed the range-2 integrals to two density functions.

\subsubsection*{Example 2: a non-planar case}

Next we consider a non-trivial non-planar example with a triple trace color factor as
\begin{equation}\label{eq:cft2}
\mathcal{T}_{\scriptscriptstyle \rm NP}=\operatorname{tr}(B a_1)\operatorname{tr}(C a_3a_2)\operatorname{tr}(A) \,,
\end{equation} 
where we assume $A,B,C$ are all non-empty and assign $A=\{a_4,\ldots,a_8\},B=\{a_7,\ldots,a_6\},C=\{a_5,\ldots\}$, representing other sets of color indices in the traces.
As in the previous planar case, we will mainly focus on the cut channels that involve  external momenta $\{p_1,p_2,p_3\}$.

\begin{figure}\label{fig:npcut}
    \centering
    \subfigure[$s_{123}$ triple-cut.]{
        \includegraphics[width=0.31\linewidth]{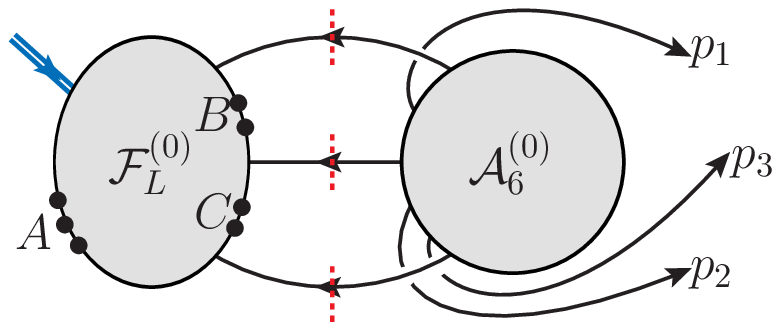} 
        \label{fig:a6}
        }
    \centering
    \subfigure[$s_{124}$ triple-cut.]{
        \includegraphics[width=0.31\linewidth]{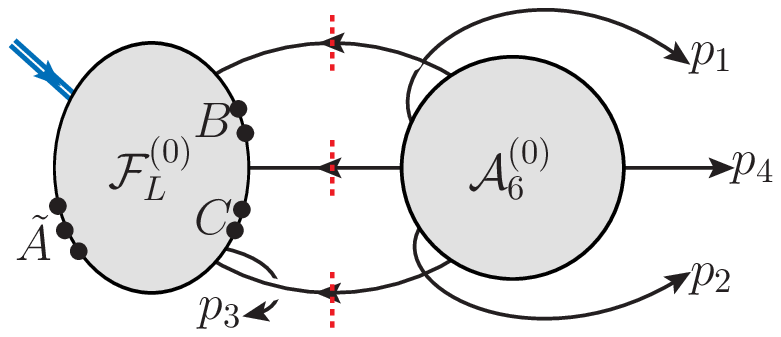} 
        \label{fig:a6another}
        }
    \centering
    \subfigure[$s_{12}$ triple-cut.]{
        \includegraphics[width=0.28\linewidth]{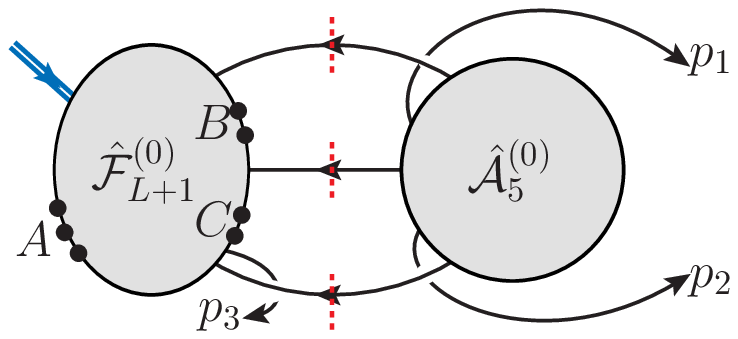}
        \label{fig:a5}
        }
	 \caption{Selected non-planar cut channels at two loops for $\mathcal{F}_{\scriptscriptstyle \rm NP}^{(2)}$. These cuts serves to calculate kinematic dependence of $p_1$ to $p_4$ for  $\mathcal{F}^{(2)}_{\scriptscriptstyle \rm NP}$. } 
\end{figure}

Following the strategy in Section~\ref{ssec:NPcutstrategy}, we need to identify the cut channels that contribute to the color factor $\mathcal{T}_{\scriptscriptstyle \rm NP}$. 
Let us first consider the $s_{123}$ triple-cut. The allowed configuration is shown in Figure~\ref{fig:a6}. The product of tree tree factors gives the required color factor:
\begin{align}
 \operatorname{tr}_{\mathcal{A}}(\check{a}_{l_3}a_{3} a_2\check{a}_{l_2} a_{1}\check{a}_{l_1})\times\operatorname{tr}_{\mathcal{F}}(A a_{l_{1}} B a_{l_{2}}C a_{l_{3}}) = 1\times \mathcal{T}_{\scriptscriptstyle \rm NP}+\cdots   \,,
\end{align}
where `$\cdots$' represents other factors.
The kinematic part under this cut is: 
\begin{equation}
\mathcal{F}_{\scriptscriptstyle \rm NP}^{(2)}(p_1,p_2,p_3,\ldots) \Big|_{s_{123}\textrm{-TC}} = \int \mathrm{\mathrm{dPS}}_{3,l} \mathcal{F}^{(0)}_{L}\left(A,-l_{1},B,-l_{2},C,-l_{3}\right) \mathcal{A}^{(0)}_{6}(l_{3},p_{3},p_{2},l_{2},p_{1},l_{1}) .
\end{equation}

Plugging in the expressions for the tree form factor (here $\mathcal{F}_{L}^{(0)}=1$) and the six-scalar tree amplitude ${\cal A}^{(0)}(l^{\bar X}_{3},p^X_{3},p^X_{2},l^{\bar X}_{2},p^X_{1},l^{\bar X}_{1})$ (see Appendix~\ref{ap:amp}), and after removing the cut, one has 
\begin{equation}
\label{tc1s0}
\mathcal{F}_{\scriptscriptstyle \rm NP}^{(2)}(p_1,p_2,p_3,\ldots)= -\mathcal{K}_{\scriptscriptstyle \rm NP}(123) + \big(\text{terms determined by other channels}\big) \,,
\end{equation}
where
\begin{align}
    \mathcal{K}_{\scriptscriptstyle \rm NP}(123)&=\Big(
    \hskip -3pt
    \begin{aligned}
       \includegraphics[height=0.10\linewidth]{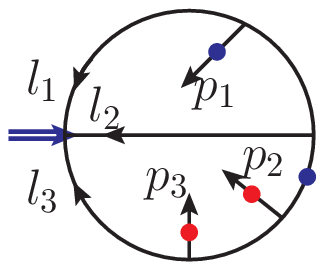}
    \end{aligned} 
   \hskip -3pt - \hskip -5pt 
    \begin{aligned}
       \includegraphics[height=0.10\linewidth]{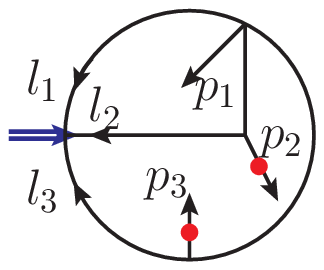}
    \end{aligned}
   \hskip -3pt - \hskip -5pt 
    \begin{aligned}
       \includegraphics[height=0.10\linewidth]{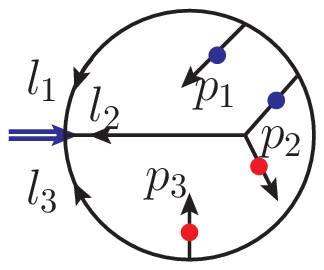}
    \end{aligned}
    \hskip -5pt \Big)
    +\Big(\hskip -3pt
    \begin{aligned}
       \includegraphics[height=0.10\linewidth]{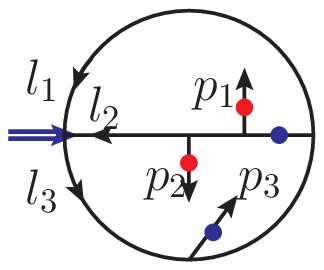}
    \end{aligned} 
    \hskip -3pt - \hskip -5pt 
    \begin{aligned}
       \includegraphics[height=0.10\linewidth]{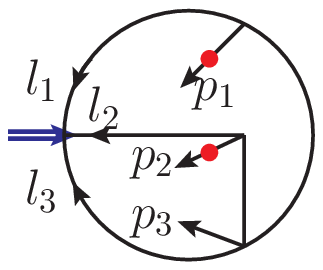}
    \end{aligned}
     \hskip -3pt - \hskip -5pt 
    \begin{aligned}
       \includegraphics[height=0.10\linewidth]{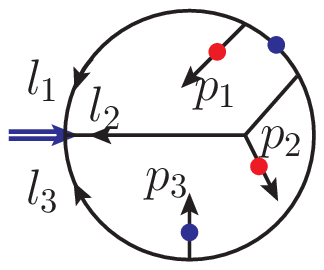}
    \end{aligned}
    \hskip -6pt \Big) \nonumber\\
    &+\Big(
    \hskip -1pt 
    \underline{\begin{aligned}
       \includegraphics[height=0.10\linewidth]{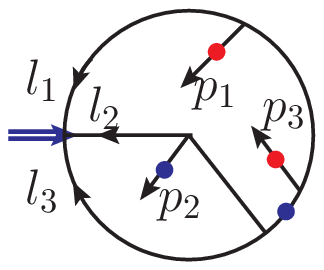}
    \end{aligned}}
    \hskip -3pt - \hskip -5pt 
    \begin{aligned}
       \includegraphics[height=0.10\linewidth]{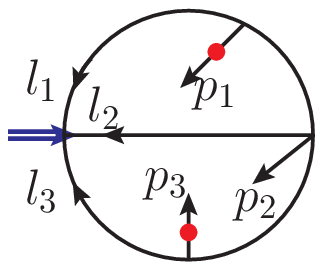}
    \end{aligned}
    \hskip -3pt - \hskip -5pt 
    \begin{aligned}
       \includegraphics[height=0.10\linewidth]{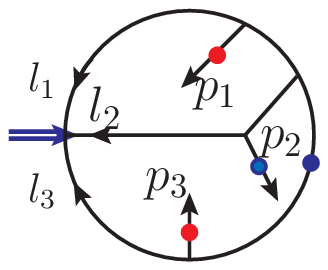}
    \end{aligned}
    \hskip -5pt \Big)
    +\Big(
    \hskip -3pt 
    \begin{aligned}
       \includegraphics[height=0.10\linewidth]{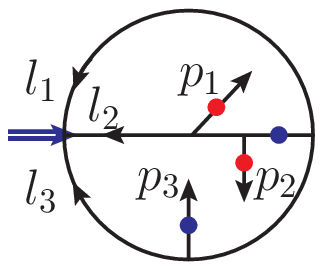}
    \end{aligned}
     \hskip -3pt + \hskip -5pt 
    \begin{aligned}
       \includegraphics[height=0.10\linewidth]{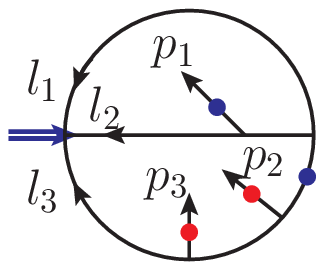}
    \end{aligned}
    \hskip -3pt - \hskip -5pt 
    \begin{aligned}
       \includegraphics[height=0.10\linewidth]{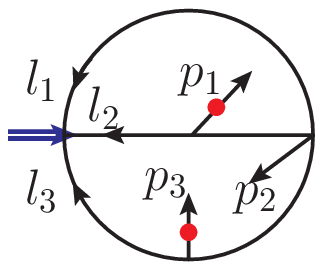}
    \end{aligned}\hskip -2pt \Big) \nonumber\\
    &-\Big(
    \hskip -1pt 
    \underline{\begin{aligned}
       \includegraphics[height=0.10\linewidth]{fig/triboxn5.eps}
    \end{aligned}}
     \hskip -3pt + \hskip -5pt 
    \begin{aligned}
       \includegraphics[height=0.10\linewidth]{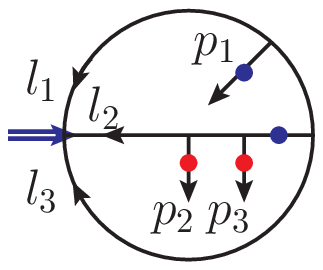}
    \end{aligned}
     \hskip -3pt - \hskip -5pt 
    \begin{aligned}
       \includegraphics[height=0.10\linewidth]{fig/tritrin1.eps}
    \end{aligned}\hskip -1pt \Big) .
    \label{tc1s}
\end{align}
The above expression is straightforward to obtain since the RHS expression is in one-to-one correspondence with the six-scalar amplitude expression when cutting $l_i$ legs.
Note that the two graphs underlined cancel between each other, which are topologically actually forbidden regarding the ordering from  $ \mathcal{A}^{(0)}_{6}(l_{3},p_{3},p_{2},l_{2},p_{1},l_{1})$. 
They are purposely inserted and we arrange $\mathcal{F}_{\scriptscriptstyle \rm NP}^{(2)}(p_1,p_2,p_3,\ldots)$ into this form because it will be helpful for discussion in Section~\ref{ssec:2lbps}.

Next we consider a $s_{12}$ triple-cut  (TC) channel, 
and our choice of ordering for blocks in this channel  as shown in Figure~\ref{fig:a5}: the $p_3$ leg in Figure~\ref{fig:a6} moving to the lower side of  $C$. One can check that color factor is correct:
\begin{align}
\operatorname{tr}_{\mathcal{A}}(\check{a}_{l_3}\check{a}_{l_2}a_{2}a_{1}\check{a}_{l_1})\times\operatorname{tr}_{\mathcal{F}}(A a_{l_{1}} B a_{l_{2}} {C}  a_{3} a_{l_{3}}) =1\times \mathcal{T}_{\scriptscriptstyle \rm NP} + \ldots  \,.
\end{align}
The kinematic parts under cut $s_{12}$-TC satisfy:
\begin{align} 
\label{tc3p0-0}
\mathcal{F}_{\scriptscriptstyle \rm NP}^{(2)}(p_1,p_2,p_3,p_4,...) \Big|_{s_{12}\textrm{-TC}} =\int \mathrm{\mathrm{d\hat{PS}}}_{3,l}\hat{\mathcal{F}}_{L+1}^{(0)}(-l_1,B,-l_2,C,p_3,-l_3,p_4,\tilde{A})  \hat{\mathcal{A}}_{5}^{(0)}\left(l_1,l_3,l_2,p_2,p_1\right),
\end{align}
in which the tree five-point amplitude and next-to-minimal form factors are reviewed in Section~\ref{subsec:treeAmpFF}.
And we define that $\{p_4,\tilde{A}\}=A$. 

To simplify the discussion, we note that the next-to-minimal form factor $\hat{\mathcal{F}}_{L+1}^{(0)}$ in \ref{tc3p0-0} can be splitted into three (gauge invariant) parts according to kinematic dependence\footnote{
Note that no fermions can be exchanged between internal lines because of non-adjacency of $l_i$ in $\hat{\mathcal{F}}_{L+1}^{(0)}(-l_1,B,-l_2,C,p_3,-l_3,A)$. }
\begin{equation}\label{eq:tcfuldens}
\begin{aligned}
    \hat{\mathcal{F}}_{L+1}^{(0)}&(-l_1,B,-l_2,C,p_3,-l_3,p_4,\tilde{A})\\
    &=\hat{\mathcal{F}}_{L+1,\text{dens}}^{(0)}(p_3,-l_3,p_4)+\hat{\mathcal{F}}_{L+1,\text{dens}}^{(0)}(p_5,-l_2,p_6) +\hat{\mathcal{F}}_{L+1,\text{dens}}^{(0)}(p_7,-l_1,p_8)\,.
\end{aligned}
\end{equation}
Since the three terms within  \eqref{eq:tcfuldens} are gauge invariant respectively and depend on different external momenta, one can calculate them separately. 
We thus decompose \eqref{tc3p0-0} correspondingly as
\begin{align} \label{tc3p0}
& \mathcal{F}_{\scriptscriptstyle \rm NP}^{(2)}(p_1,p_2,p_3,p_4,\ldots) \Big|_{s_{12}\textrm{-TC}}\\
& =\int \hat{\mathrm{dPS}}_{3,l}\hat{\mathcal{F}}_{L+1,\text{dens}}^{(0)}(p_3,-l_3,p_4)  \hat{\mathcal{A}}_{5}^{(0)}\left(l_1,l_3,l_2,p_2,p_1\right)+
\big(\text{terms dependent on $p_5$ to $p_8$}\big) . \nonumber
\end{align}
Let us focus on the first term on the RHS of \eqref{tc3p0}. Similar to the planar example, this part also probes terms from $s_{123}$ and $s_{124}$ triple-cut.
For example, the cut contribution from the integrand $-\mathcal{K}_{\scriptscriptstyle \rm NP}(123)$ in \eqref{tc1s}  is 
\begin{equation}
\label{eq:tc1cut}
\begin{aligned}
    -&\Big(
    \hskip -2pt 
    \begin{aligned}
       \includegraphics[height=0.105\linewidth]{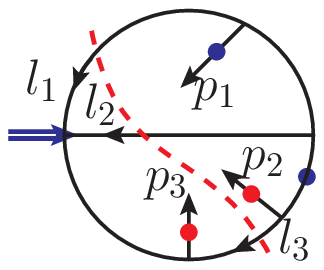}
    \end{aligned} 
    \hlm{3} 
    \begin{aligned}
       \includegraphics[height=0.105\linewidth]{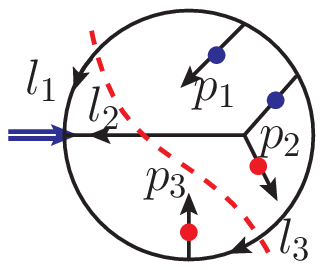}
    \end{aligned}
    \hlm{3} 
    \begin{aligned}
       \includegraphics[height=0.105\linewidth]{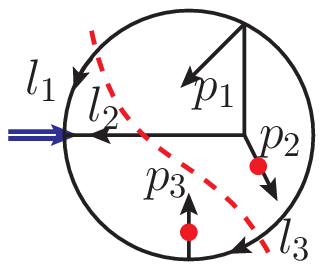}
    \end{aligned}
    \hskip -3pt \Big)
    -\Big(
    \hskip -2pt \begin{aligned}
       \includegraphics[height=0.105\linewidth]{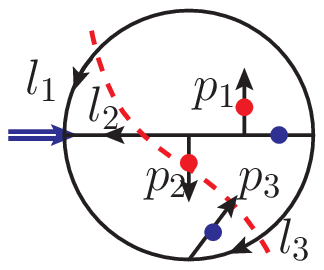}
    \end{aligned}
     \hlm{3} 
    \begin{aligned}
       \includegraphics[height=0.105\linewidth]{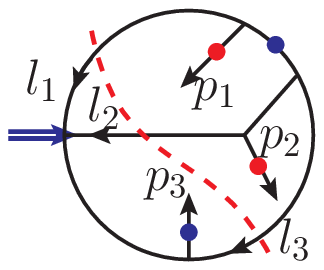}
    \end{aligned}
    \hskip -3pt \Big)\\
    -&\Big(
    \hskip -2pt \begin{aligned}
       \includegraphics[height=0.105\linewidth]{fig/triboxn3Cut.eps}
    \end{aligned}
    \hlm{3} 
    \begin{aligned}
       \includegraphics[height=0.105\linewidth]{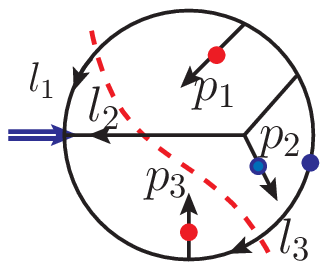}
    \end{aligned}
    \hskip -3pt \Big)
    -\Big(
    \hskip -2pt 
    \begin{aligned}
       \includegraphics[height=0.105\linewidth]{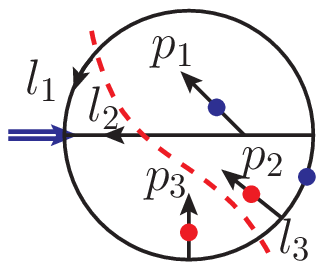}
    \end{aligned}
    \hlp{3}
    \begin{aligned}
       \includegraphics[height=0.105\linewidth]{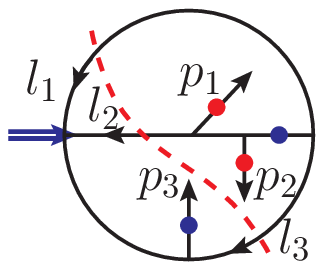}
    \end{aligned}
     \hlm{3}
    \begin{aligned}
       \includegraphics[height=0.105\linewidth]{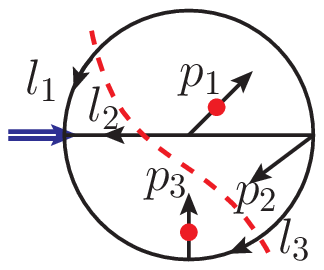}
    \end{aligned}
    \hskip -3pt \Big).
\end{aligned}
\end{equation}
The contribution from $s_{124}$-cut-detectable terms is similar 
and can be obtained from $s_{123}$-cut result \eqref{eq:tc1cut} using the following relation:
\begin{equation}
    \text{$s_{124}$-cut-detectable terms}=\text{$s_{123}$-cut-detectable terms}\big|_{p_3\shortrightarrow p_4, l_1\leftrightarrow l_2,p_1\leftrightarrow p_2}\,.
\end{equation}
Besides these cut contributions, there appears a new contribution detected by the $s_{12}$-triple cut:
\begin{equation}\label{eq:tc1cor}
   -\frac{s_{12}}{s_{1l_1}s_{2 l_2}}=-\frac{1}{2}
   \hskip -3pt
   \begin{aligned}
    \includegraphics[height=0.105\linewidth]{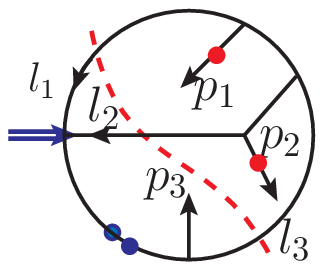}
   \end{aligned}\hskip -7pt =  -\hskip -2pt \begin{aligned}
    \includegraphics[height=0.105\linewidth]{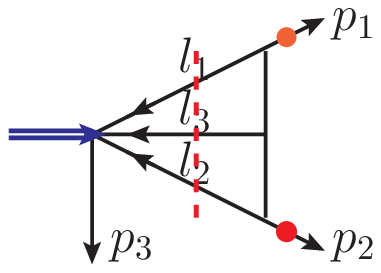}
   \end{aligned} \,.
\end{equation}
Note that this is a range-2 integral and is obviously vanishing under the $s_{123}$-triple cut.
Similar to the calculation of the first term of \eqref{tc3p0}, the kinematic dependence on $p_5$ to $p_8$ can be fixed in a similar way and we will not reproduce the details here.

So far we have considered $s_{12}$ triple-cut channel. To determine the integrand that involves $\{p_1,p_2,p_3\}$,  the remaining cuts to consider are $s_{23}$ triple-cut and $s_{13}$ triple-cut. They can be considered in the same way as for the $s_{12}$ triple-cut, and one finds two similar range-2 functions as \eqref{eq:tc1cor}, with kinematic dependence on $s_{13}$ and $s_{23}$ respectively. 

Including these range-2 contribution to \eqref{tc1s},  one can introduce a density function $\widetilde{\mathcal{F}}_{\scriptscriptstyle \rm NP}^{(2)}(p_1, p_2, p_3)$, similar to planar density in \eqref{eq:planarfinaldensity}, as 
\begin{equation}\label{tcfull2}
     \widetilde{\mathcal{F}}_{\scriptscriptstyle \rm NP}^{(2)}(p_1, p_2, p_3)=-
  \mathcal{K}_{\scriptscriptstyle \rm NP}(123)
     -\frac{1}{2}\bigg(
     \hskip -2pt
     \begin{aligned}
      \includegraphics[height=0.10\linewidth]{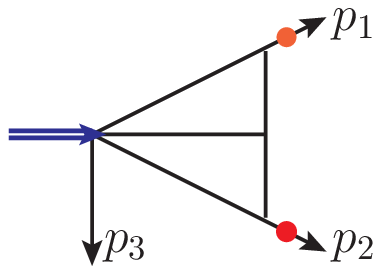}
  \end{aligned}\hlp{5}
  \begin{aligned}
      \includegraphics[height=0.10\linewidth]{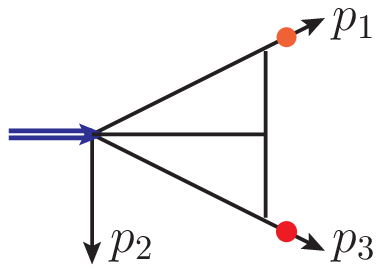}
  \end{aligned}\hlp{5}
  \begin{aligned}
      \includegraphics[height=0.10\linewidth]{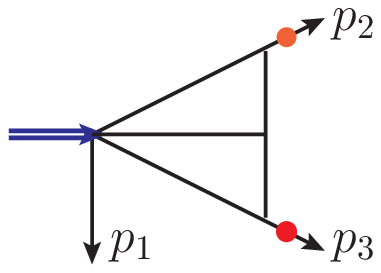}
  \end{aligned}\hskip -6pt \bigg)\,.
\end{equation}
so that the full form factor can be given by summing all density functions as
\begin{equation}
    \mathcal{F}^{(2)}_{\scriptscriptstyle \rm NP}=\sum_{i<j<k} \widetilde{\mathcal{F}}^{(2)}_{\scriptscriptstyle \mathrm{NP}}(p_i, p_j, p_k)\,.
\end{equation} 
The difference, comparing to the planar case, is that $\widetilde{\mathcal{F}}^{(2)}_{\scriptscriptstyle \mathrm{NP}}(p_i, p_j, p_k)$ is no longer a universal function as \eqref{eq:pl-dens-cyc}: 
\begin{equation}
    \widetilde{\mathcal{F}}^{(2)}_{\scriptscriptstyle \mathrm{NP}}(p_i, p_j, p_k) \neq \widetilde{\mathcal{F}}^{(2)}_{\scriptscriptstyle \mathrm{NP}}(p_1, p_2, p_3)\big|_{\{1\shortrightarrow i, 2\shortrightarrow j, 3\shortrightarrow k \} }\,,
\end{equation}
so in principle, we need to calculate every $(i,j,k)$ separately, following the same procedure from \eqref{eq:cft2} to \eqref{tcfull2}. However, as will be shown in next section, with a proper analysis of color basis, knowing $\widetilde{\mathcal{F}}^{(2)}_{\scriptscriptstyle \mathrm{NP}}(p_1, p_2, p_3)$ is actually enough. 
Finally, we flatten integrals in a more convenient form:
\begin{align}
 \widetilde{\mathcal{F}}_{\scriptscriptstyle \rm NP}^{(2)}(p_1, p_2, p_3)
  =-&\Big(
    \begin{aligned}
      \includegraphics[height=0.10\linewidth]{fig/tribox2top.eps}
    \end{aligned} 
    \hlm{5}
    \begin{aligned}
      \includegraphics[height=0.10\linewidth]{fig/tritri2top.eps}
    \end{aligned}
    \hlm{5}
    \begin{aligned}
      \includegraphics[height=0.10\linewidth]{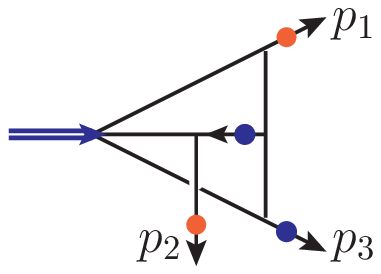}
    \end{aligned}
    \hlmath{5}{+\frac{1}{2}}
    \begin{aligned}
     \includegraphics[height=0.10\linewidth]{fig/pirbcomtop.eps}
  \end{aligned}
  \hskip -8pt\Big)+\text{cyc}(p_1,p_2,p_3) \nonumber\\
  -&\Big(\begin{aligned}
    \includegraphics[height=0.10\linewidth]{fig/boxtri2top.eps}
  \end{aligned}
  \hlp{5}
  \begin{aligned}
    \includegraphics[height=0.10\linewidth]{fig/tribox2top.eps}
  \end{aligned}
  \hlm{5}
  \begin{aligned}
    \includegraphics[height=0.10\linewidth]{fig/tritri2top.eps}
  \end{aligned}
  \Big) \nonumber\\
  +&\Big(\begin{aligned}
    \includegraphics[height=0.10\linewidth]{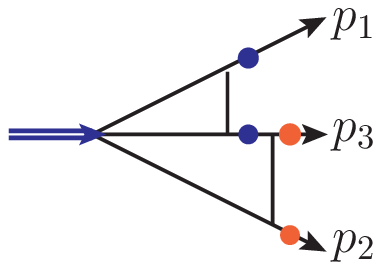}
  \end{aligned}
  \hlp{5}
  \begin{aligned}
    \includegraphics[height=0.10\linewidth]{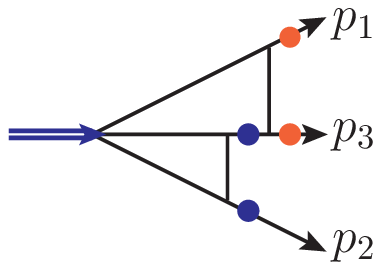}
  \end{aligned}
  \hlm{5}
  \begin{aligned}
    \includegraphics[height=0.10\linewidth]{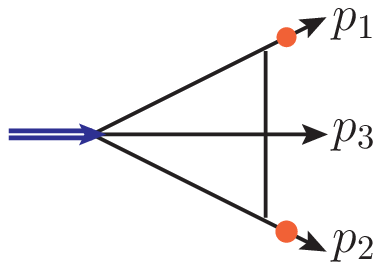}
  \end{aligned}
  \Big) \,.
  \label{tcfull3}
\end{align}
%

\section{Full-color two-loop form factors}\label{sec:2l}

In this section, we apply the strategy described in the previous section to construct the full-color form factor $\itbf{F}^{(2)}$, or more precisely, $\itbf{F}^{(2)}$ expressed with loop correction $\itbf{F}^{(2)}=\itbf{I}^{(2)}\itbf{F}^{(0)}$. 
Before going into the details, let us first clarify the general structure of two-loop form factors. 
The two-loop form factors have the following schematic structure (a more precise form can be obtained following \eqref{eq:nplanarloopcorrection} which is not important for later discussions):
\begin{equation}\label{eq:fc-2l-loop}
\begin{aligned}
    \itbf{F}^{(2)}&=\itbf{I}^{(2)}\itbf{F}^{(0)} =\Big(\sum_{i<j<k}\itbf{I}^{(2)}_{ijk}+\sum_{i<j}\itbf{I}^{(2)}_{ij}+ \sum_{ij \neq kl}  \itbf{I}^{(1)}_{ij} \itbf{I}^{(1)}_{kl}   \Big)\itbf{F}^{(0)}\,.
\end{aligned}
\end{equation}
The last part $\sum_{ij \neq kl}  \itbf{I}^{(1)}_{ij} \itbf{I}^{(1)}_{kl}$ correspond to the product of two decoupled one-loop corrections, which is trivial and we ignore it in the following discussion. 
The first two parts are intrinsic two-loop corrections, and we separate them as range-3 and range-2 densities respectively. 
To calculate full form factors, one only needs to compute the following density matrix elements: $\big( \NPdensity^{(2)}_{ijk}  \big)_{\onWbos_i \onWbos_j \onWbos_k}^{\widetilde{\onWbos}_1 \widetilde{\onWbos}_2 \widetilde{\onWbos}_3}$ and   $\big( \NPdensity^{(2)}_{ij}  \big)_{\onWbos_i \onWbos_j }^{\widetilde{\onWbos}_1 \widetilde{\onWbos}_2 }$.
It should be clear that these quantities, as function of momenta, depend also on the transition of states (\emph{i.e.}~fields). 
As will be shown later, by means of properly selecting $\mathcal{T}_i$ and $\big[\mathcal{F}^{(2)}\big]_{\mathcal{T}_i}$, 
these functions can be determined by a small set of equations.

Below we will first consider the color decomposition for $\itbf{I}^{(2)}$ which will help to identify a set of kinematic functions that are enough to determine $\itbf{I}^{(2)}$. Then, we will compute two classes of form factors: (1) the half-BPS form factors with ${\rm tr}(X^L)$; (2) the form factor with non-BPS operators in the SU(2) sector.  
The similar construction should apply to more general form factors.

\subsection{Color decomposition of two-loop form factors}
\label{ssec:nbasis2l}

In this subsection we consider the color decomposition for the density function as in \eqref{eq:colordecomforI}:
\begin{align}
\big( \itbf{I}^{(2)} \big)_{\onWbos_i \onWbos_j \onWbos_k}^{\widetilde{\onWbos}_1 \widetilde{\onWbos}_2 \widetilde{\onWbos}_3} (123)
& =\sum_{\alpha} \check{\mathcal{C}}_\alpha \left[ \big( \mathcal{I}^{(2)} \big)_{\onWbos_i \onWbos_j \onWbos_k}^{\widetilde{\onWbos}_1 \widetilde{\onWbos}_2 \widetilde{\onWbos}_3}(123)\right]_{\check{\mathcal{C}}_\alpha}, 
\\
\big( \itbf{I}^{(2)} \big)_{\onWbos_i \onWbos_j }^{\widetilde{\onWbos}_1 \widetilde{\onWbos}_2} (12)
& =\sum_{\beta} \check{\mathcal{C}}_\beta \left[ \big( \mathcal{I}^{(2)} \big)_{\onWbos_i \onWbos_j}^{\widetilde{\onWbos}_1 \widetilde{\onWbos}_2}(12)\right]_{\check{\mathcal{C}}_\beta}.
\end{align}
Without loss of generality, we choose the external momenta to be $\{p_1,p_2,p_3\}$ and set labeling of $\itbf{I}^{(2)}_{ijk}$ as  $\itbf{I}^{(2)}_{ijk}(123)$ and $\itbf{I}^{(2)}_{ij}$ as  $\itbf{I}^{(2)}_{ij}(12)$. Meanwhile, the color indices of external states are $a_1,a_2,a_3$.  When there is no ambiguity, the 
momenta label (123) and (12) can be omitted for simplicity.

\subsubsection*{DDM basis}

We start with choosing a set of trivalent DDM basis, where
${\check{\mathcal{C}}_{\alpha}} = {\check{\mathcal{D}}_{\alpha}}$ are trivalent color factors in terms of $\tilde{f}^{abc}$. 

\begin{figure}
    \centering
    \subfigure[Cubic Graph: $\check{\mathcal{D}}_{\Delta_1}$]{
    \includegraphics[width=0.225\linewidth]{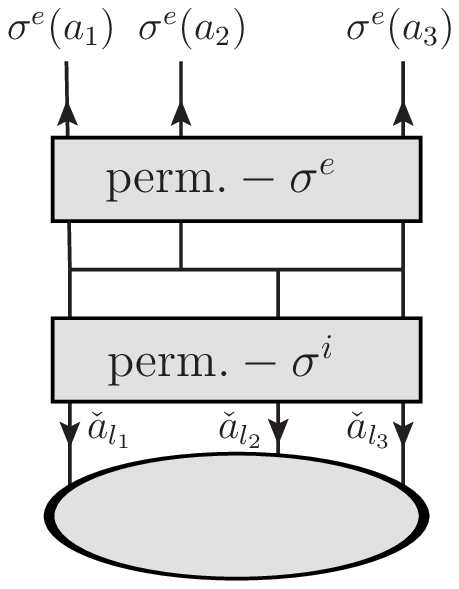}
    \label{fig:ddm2lm}
    }
    \subfigure[Cubic Graph: $\check{\mathcal{D}}_{\Delta_2}$]{
    \includegraphics[width=0.225\linewidth]{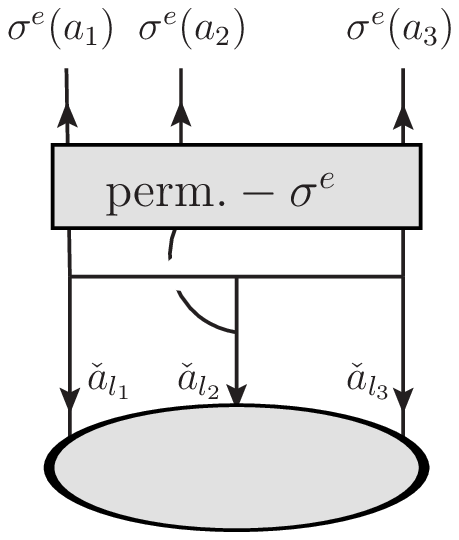}
    \label{fig:ddm2lm1}
    }
    \subfigure[Cubic Graph: $\check{\mathcal{D}}_{\Delta_3}$]{
        \includegraphics[width=0.225\linewidth]{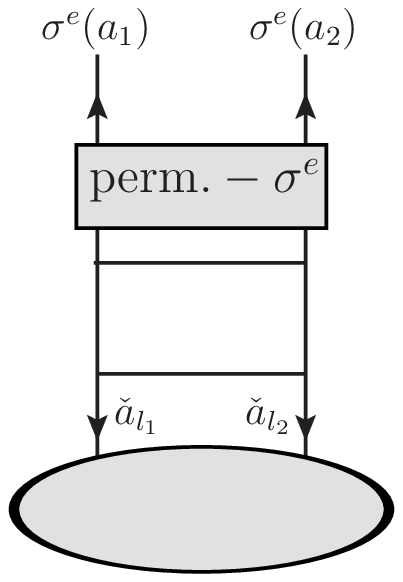}
         \label{fig:ddm2lm2}
        }
   \subfigure[Cubic Graph: $\check{\mathcal{D}}_{\Delta_4}$]{
        \includegraphics[width=0.225\linewidth]{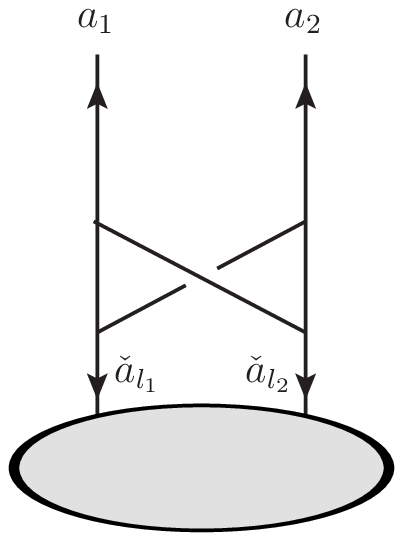}
        \label{fig:ddm2lm3}
        }
    \caption{Cubic Graphs for minimal two-loop correction density.}
    \label{fig:tribas2l}
\end{figure}

Consider first the range-3 color basis. The trivalent topologies can be obtained by sewing six-point tree-level amplitudes to minimal form factors. One can choose two independent configurations whose diagrammatically representation are given in Figure~\ref{fig:ddm2lm} and \ref{fig:ddm2lm1}, respectively. They correspond to the following two sets of color structures:
\begin{align}
    &\text{Set}_{1}: \qquad \left\{\check{\mathcal{D}}_{\Delta_1}(\sigma^{e},\sigma^{i})\right\}=\left\{\sigma^{i}\sigma^{e} \cdot f^{\check{a}_{l_1}a_{1}\text{x}_{1}}f^{\text{x}_{1}a_{2}\text{x}_{2}}f^{\text{x}_{2}\text{x}_{3}\check{a}_{l_2}}f^{\text{x}_{3}a_{3}\check{a}_{l_3}}\right\}, 
    \label{eq:set1} \\
    &\text{Set}_{2}: \qquad \left\{\check{\mathcal{D}}_{\Delta_2}(\sigma^{e})\right\}=\left\{\sigma^{e} \cdot f^{\check{a}_{l_1}a_{1}\text{x}_{1}}f^{\check{a}_{l_2}a_{2}\text{x}_{2}}f^{\check{a}_{l_3}a_{3}\text{x}_{3}}f^{\text{x}_{1}\text{x}_{2}\text{x}_{3}}\right\} \,,
\label{eq:set2}
\end{align}
where the permutations $\sigma_3^{e}\in S_{3}$ and they act on color indices of external states $\{a_1, a_2, a_3\}$, while $\sigma_3^{i}\in \mathbb{Z}_{3}$ (cyclic permutation) act on internal color variation $\{\check{a}_{l_1}, \check{a}_{l_2}, \check{a}_{l_3}\}$ (which is to act on three sites $\{i,j,k\}$ in the operator).
As a check, the number of elements of the above range-3 basis is $3!\times 3 + 3! = 24$,  which is the same as the dimension of DDM basis for six-point amplitude $\itbf{A}_{6}^{(0)}$.

The range-2 cubic graphs can be understood as sewing a one-loop four-point cubic graph on minimal form factors. There are also two independent configurations as shown in Figure~\ref{fig:ddm2lm2} and \ref{fig:ddm2lm3}. Thus, range-2  color basis for minimal two-loop correction density can be given as 
\begin{align}
    \text{Set}_{3}&: \qquad  \left\{\check{\mathcal{D}}_{\Delta_3}(\sigma^{e})\right\}=
    \left\{\sigma^{e} \cdot f^{\text{x}_{1}\check{a}_{l_2}\text{x}_{2}}f^{\text{x}_{2}\check{a}_{l_1}\text{x}_{3}}f^{\text{x}_{3} a_{1}\text{x}_{4}}f^{\text{x}_{4} a_{2} \text{x}_{1}}\right\} \,, 
    \label{eq:set3}\\
    \text{Set}_{4}&: \qquad  \left\{\check{\mathcal{D}}_{\Delta_4}(\mathbf{1})\right\}=\left\{ f^{\text{x}_{1}\check{a}_{l_2}\text{x}_{2}}f^{\text{x}_{2}a_{2}\text{x}_{3}}f^{\text{x}_{3} a_{1}\text{x}_{4}}f^{\text{x}_{4} \check{a}_{l_1} \text{x}_{1}}\right\} \,,
\label{eq:set4}
\end{align}
where the permutations $\sigma_2^{e}\in S_{2}$ and they act on color indices of external states $\{a_1, a_2\}$.

For the convenience of notation, we will use an explicit set of color factors $\check{\mathcal{D}}_{i}$ to represent the elements in $\check{\mathcal{D}}_{\Delta_a}$: 
(1) $\check{\mathcal{D}}_{i}$, $i=1,..,18$,  for 18 elements $\check{\mathcal{D}}_{\Delta_{1}}$ in  \eqref{eq:set1}, (2) $\check{\mathcal{D}}_{j}$,  $j=19,\ldots,24$,  for 6 elements $\check{\mathcal{D}}_{\Delta_{2}}$ in  \eqref{eq:set2},   
(3) $\check{\mathcal{D}}_{25}, \check{\mathcal{D}}_{26}$ for 2 elements $\check{\mathcal{D}}_{\Delta_{3}}$ in  \eqref{eq:set3}, 
and (4) $\check{\mathcal{D}}_{27}$ for the single element $\check{\mathcal{D}}_{\Delta_{4}}$ in  \eqref{eq:set4}. 
Explicit form of $\check{\mathcal{D} }_{i}$ are listed in Appendix~\ref{ap:ddm}.
Four of the them will appear frequently in the following discussion, and we give them explicitly here:
\begin{equation}
\check{\mathcal{D}}_{1} = \check{\mathcal{D}}_{\Delta_1}(\mathbf{1},\mathbf{1})\,, \quad
\check{\mathcal{D}}_{19} = \check{\mathcal{D}}_{\Delta_2}(\mathbf{1})\,, \quad
\check{\mathcal{D}}_{25} = \check{\mathcal{D}}_{\Delta_3}(\mathbf{1})\,, \quad
\check{\mathcal{D}}_{27} = \check{\mathcal{D}}_{\Delta_4}(\mathbf{1}) \,.
\end{equation}

The permutational relations for the color factors imply that the kinematic parts also satisfy certain permutational relations:
\begin{align}
\label{eq:densitypermrels1}
\left[ \big( \mathcal{I}^{(2)} \big)_{\onWbos_i \onWbos_j \onWbos_k}^{\widetilde{\onWbos}_1 \widetilde{\onWbos}_2 \widetilde{\onWbos}_3}(123)\right]_{\Delta_{i}(\sigma^{e},\sigma^{i})} & =
\left[ \big( \mathcal{I}^{(2)}\big)_{ \sigma^{i}\scriptscriptstyle({\onWbos_i \onWbos_j \onWbos_k})}^{ \sigma^{e} \scriptscriptstyle ({\widetilde{\onWbos}_1 \widetilde{\onWbos}_2 \widetilde{\onWbos}_3})}(\sigma^{e}(123)) \right]_{\Delta_i(\mathbf{1},\mathbf{1})}\,,\\
\label{eq:densitypermrels2}
\left[ \big( \mathcal{I}^{(2)} \big)_{\onWbos_i \onWbos_j }^{\widetilde{\onWbos}_1 \widetilde{\onWbos}_2 }(12)\right]_{\Delta_{i}(\sigma^{e})} & =
\left[ \big( \mathcal{I}^{(2)}\big)_{ \scriptscriptstyle({\onWbos_i \onWbos_j })}^{ \sigma^{e} \scriptscriptstyle ({\widetilde{\onWbos}_1 \widetilde{\onWbos}_2 })}(\sigma^{e}(12)) \right]_{\Delta_i(\mathbf{1})}\,.
\end{align}
This observation implies that one can focus on the following correction functions: 
\begin{equation}
\label{eq:masterkine}
\left[ \big( \mathcal{I}^{(2)} \big)_{\onWbos_i \onWbos_j \onWbos_k}^{\widetilde{\onWbos}_1 \widetilde{\onWbos}_2 \widetilde{\onWbos}_3}\right]_{\check{\mathcal{D}}_{1}} \,,\ 
\left[ \big( \mathcal{I}^{(2)} \big)_{\onWbos_i \onWbos_j \onWbos_k}^{\widetilde{\onWbos}_1 \widetilde{\onWbos}_2 \widetilde{\onWbos}_3}\right]_{\check{\mathcal{D}}_{19}} \,,\ 
\left[ \big( \mathcal{I}^{(2)} \big)_{\onWbos_i \onWbos_j }^{\widetilde{\onWbos}_1 \widetilde{\onWbos}_2 }\right]_{\check{\mathcal{D}}_{25}} \,,\ 
\left[ \big( \mathcal{I}^{(2)} \big)_{\onWbos_i \onWbos_j}^{\widetilde{\onWbos}_1 \widetilde{\onWbos}_2 }\right]_{\check{\mathcal{D}}_{27}} \,,\ 
\end{equation}
 taking into account all possible field configurations of $\onWbos$ and $\widetilde{\onWbos}$.

\subsubsection*{Equations for solving correction functions}

As discussed in Section~\ref{ssec:NPcutstrategy}, to apply unitarity, it is convenient to relate the above trivalent basis to trace basis.
In the following, we discuss their relations and select proper $\mathcal{T}_i$ and corresponding $\big[\mathcal{F}^{(2)}\big]_{\mathcal{T}_i}$ that are enough to determine the four types of correction functions defined in \eqref{eq:masterkine}. 

The first relation comes from planar form factor with color factor as 
\begin{equation}
\mathcal{T}_{t_1}\equiv \mathcal{T}_{\scriptscriptstyle \rm PL} =N_c^2 \operatorname{tr}(a_1\cdots a_L) \,.
\end{equation} 
From the trivalent topologies in Table~\ref{fig:tribas2l},  it is not hard to see that only $\check{\mathcal{D}}_{1}$ and $\check{\mathcal{D}}_{25}$ contribute to the planar topology. 
Indeed, one can obtain
\begin{align}\label{eq:ieqns1}
	    \left[\mathcal{F}^{(2)}\right]_{\mathcal{T}_{t_1}}&=\Big(\sum_{i<j<k}\itbf{I}^{(2)}_{ijk}+\sum_{i<j}\itbf{I}^{(2)}_{ij}\Big)\itbf{F}^{(0)}\Big|_{\mathcal{T}_{t_1}}\\
    \nonumber
    &=
    \Big(\big[  \mathcal{I}^{(2)} (123)\big]_{\check{\mathcal{D}}_{1}}  + \big[ \mathcal{I}^{(2)}(12) \big]_{\check{\mathcal{D}}_{25}}  +\cdots \Big)  \mathcal{F}^{(0)}(p_1,\cdots,p_L) \,,
\end{align}
which involves only the first and third type of correction functions in \eqref{eq:masterkine}. 
Let us explain \eqref{eq:ieqns1} (and similar equations below) in some more detail:
\begin{itemize}
	\item For simplicity, we temporarily hide the field index ${\onWbos_i }$ and ${\widetilde{\onWbos}_i }$ and 
	the `$\ldots$' in the bracket represents similar density functions depending on other momenta.
	\item  To reach the expansion of \eqref{eq:ieqns1}, a careful analysis of color structure is required. Concretely, only $\check{\mathcal{D}}_{1}$ and $\check{\mathcal{D}}_{25}$ acting on adjacent sites can give non-vanishing contribution to $\mathcal{T}_{t_1}$:
	\begin{align}
	\check{\mathcal{D}}_{1} \cdot \operatorname{tr}(l_1 l_2 l_3 a_4 \cdots a_L)=1\times \mathcal{T}_{t_1}+\cdots \,, \\
	\check{\mathcal{D}}_{25} \cdot \operatorname{tr}(l_1 l_2 a_3 a_4 \cdots a_L)=1\times \mathcal{T}_{t_1}+\cdots \,.
	\end{align} 
As a result, 
	\begin{equation}
	\begin{aligned}
		& \itbf{I}_{123}^{(2)}(123)\cdot \Big(\operatorname{tr}(a_{l_1}a_{l_2}a_{l_3}\cdots a_L)\mathcal{F}^{(0)}(p_1,p_2,p_3,\ldots,p_L)\Big) \Big|_{\mathcal{T}_{t_1}} \\
		= & \check{\mathcal{D}}_{1} \cdot \operatorname{tr}(a_{l_1}a_{l_2}a_{l_3}\cdots a_L) \Big|_{\mathcal{T}_{t_1}}\times \left[\mathcal{I}^{(2)}\right]_{\check{\mathcal{D}}_1} \cdot \mathcal{F}^{(0)}(p_1,p_2,p_3,\ldots,p_L)\\
		= &  \left[\mathcal{I}^{(2)}(123)\right]_{\check{\mathcal{D}}_1} \times \mathcal{F}^{(0)}(p_1,p_2,p_3,\ldots,p_L) \,.
	\end{aligned}
	\end{equation}
	\item The reason we only explicitly show the three terms in \eqref{eq:ieqns1} is mainly because they are relevant to cuts to consider below.
	Basing on the unitarity discussion in last section, our practical use of equations like \eqref{eq:ieqns1} is to put them under cuts. Hence, consider only those $\mathcal{I}$ functions involved in certain unitarity channel is sufficient. 
\end{itemize}

The second equation comes from a triple-trace color factor\footnote{Note that we focus on single trace operator for simplicity.} 
\begin{equation}
\mathcal{T}_{t_2}
\equiv \mathcal{T}_{\scriptscriptstyle \rm NP}
=\operatorname{tr}(B a_1)\operatorname{tr}(C a_3 a_2) \operatorname{tr}(A)
\end{equation} 
discussed in the second example of Section~\ref{ssec:unitarityeg}. This triple-trace color is intentionally selected because (1) range-2 color factors $\check{\mathcal{D}}_{\Delta_3}$ and $\check{\mathcal{D}}_{\Delta_4}$ do not contribute, one can just focus on range-3 densities consequently; (2) to split the trace of a tree-level form factor (single-trace) into a triple-trace color factor, the sites on which the loop correction density acts must be non-adjacent, separated by $A,B,C$ in $\mathcal{T}_{t_2}$. 
More detailed examination shows that only $\check{\mathcal{D}}_{i}$ with $i=11,18,19$ contributes:
\begin{equation*}
	\check{\mathcal{D}}_{11}\cdot \operatorname{tr}(A l_1 B l_2 C l_3)\Big|_{\mathcal{T}_{t_2}}=-1, \ \check{\mathcal{D}}_{18}\cdot \operatorname{tr}(A l_1 B l_2 C l_3)\Big|_{\mathcal{T}_{t_2}}=1,\ \check{\mathcal{D}}_{19}\cdot \operatorname{tr}(A l_1 B l_2 C l_3)\Big|_{\mathcal{T}_{t_2}}=1\,,
\end{equation*}
where explicit expressions of $\check{\mathcal{D}}_{11, 18}$ can be found in Appendix~\ref{ap:ddm}.
As a result, the kinematic part satisfies the following equation
\begin{equation}\label{eq:ieqns2}
    \left[\mathcal{F}^{(2)}\right]_{\mathcal{T}_{t_2}}=\Big(\left[  \mathcal{I}^{(2)}(123)\right]_{\check{\mathcal{D}}_{11}}-\left[  \mathcal{I}^{(2)}(123)\right]_{\check{\mathcal{D}}_{18}}+\left[  \mathcal{I}^{(2)}(123)\right]_{\check{\mathcal{D}}_{19}} +\cdots \Big) \mathcal{F}^{(0)}(A,p_1,B,p_2,C,p_3) ,
\end{equation}
where `$\cdots$' in brackets represents functions depending on other momenta. Since $\left[\mathcal{I}^{(2)}\right]_{\check{\mathcal{D}}_{i}}$ with $i=1,\ldots,18$ can be regarded as known from previous relation \eqref{eq:ieqns1}, the new relation \eqref{eq:ieqns2} can be used to determine $\left[\mathcal{I}^{(2)}\right]_{\check{\mathcal{D}}_{19}}$. 

Finally, we need an equation to determin the last term in \eqref{eq:masterkine}. We consider the color factor: 
\begin{equation}
\mathcal{T}_{t_3}=N_{\rm c}\operatorname{tr}(a_1 B)\operatorname{tr}(a_2 A) \,,
\end{equation} 
with $A=\{a_4,\ldots,a_5\}\equiv \{a_4,\tilde{A}\}$ and $B=\{a_6,\ldots,a_3\}\equiv \{\tilde{B},a_3\}$. A similar analysis can be performed basing on the feature of targeting color factor, and we observe that any range-3 density contributing to such color factor must act on sites with two of them adjacent. 
One find the equation for the kinematic factors as
\begin{align}\label{eq:ieqns3}
    \left[\mathcal{F}^{(2)}\right]_{\mathcal{T}_{t_3}}=&\Big(\left[\mathcal{I}^{(2)}(12)\right]_{\check{\mathcal{D}}_{27}}+\left[\mathcal{I}^{(2)}(123)\right]_{\check{\mathcal{D}}_{19}}+\left[\mathcal{I}^{(2)}(123)\right]_{\check{\mathcal{D}}_{11}}-\left[\mathcal{I}^{(2)}(123)\right]_{\check{\mathcal{D}}_{18}}\\
    \nonumber
    &+\left[\mathcal{I}^{(2)}(124)\right]_{\check{\mathcal{D}}_{21}}+\left[\mathcal{I}^{(2)}(124)\right]_{\check{\mathcal{D}}_{5}}-\left[\mathcal{I}^{(2)}(124)\right]_{\check{\mathcal{D}}_{15}}+\cdots \Big)\cdot \mathcal{F}^{(0)}(p_1,A,p_2,B) \,,
\end{align}
which involves the wanted function $\big[\mathcal{I}^{(2)}\big]_{\check{\mathcal{D}}_{27}}$. One can check the color factors are consistent, for example:
\begin{equation*}
\check{\mathcal{D}}_{27} \cdot \operatorname{tr}(l_1A l_2 B)\big|_{\mathcal{T}_{t_3}}=1 \,.
\end{equation*}

The three equations \eqref{eq:ieqns1}, \eqref{eq:ieqns2}, \eqref{eq:ieqns3} provide the minimal set of equations for determining density $\mathcal{I}$ functions, which means that after solving these three equations, we can find a solution satisfying all unitarity cuts for $\itbf{F}$ (expressed via $\itbf{I}$). One might notice that there are only three equations while we have four functions to solve. 
This is due to that there is a degree of freedom in defining the density functions, which is expected as a redistribution property for form factors.
The full form factor is certainly independent of such definition. We will also see in the end of Section~\ref{sec:CKBPS} an explicit formula to explain how the redistribution works.

\subsection{Two-loop BPS loop correction}\label{ssec:2lbps}

In this subsection, we apply the above strategy to compute the half-BPS form factors of operators ${\rm tr}(X^L)$. Since all the fields are identical, we omit the labeling of fields $\mathcal{W}_i$.
Discussion here provides details for \eqref{eq:ieqns1}-\eqref{eq:ieqns3}. Besides the computation for $\mathcal{F}_{\mathcal{T}_{t_{1,2,3}}}$ mentioned in previous subsection, we will also perform another non-trivial check for  $\mathcal{F}_{\mathcal{T}_{t_{4}}}$ with a different color factor.

\subsubsection*{Kinematic parts of BPS loop correction in trivalent basis}

To begin with, we consider \eqref{eq:ieqns1}. This equation involves $\mathcal{I}_{\check{\mathcal{D}}_{1}}$ and $\mathcal{I}_{\check{\mathcal{D}}_{25}}$. Comparing \eqref{eq:bpspl2} with \eqref{eq:ieqns1}, a natural choice is
\begin{equation}\label{eq:ID25}
    \left[\mathcal{I}^{(2)}(12)\right]_{\check{\mathcal{D}}_{25}}=\begin{aligned}
         \includegraphics[height=0.10\linewidth]{fig/tritri2pt1top.eps}
        \end{aligned}
        \hlp{5}
        \begin{aligned}
         \includegraphics[height=0.10\linewidth]{fig/pladderm1top.eps}
        \end{aligned}\,,
\end{equation}
and 
\begin{equation}\label{eq:ID1}
    \Big[\mathcal{I}^{(2)}(123)\Big]_{\check{\mathcal{D}}_{1}}=\begin{aligned}
        \includegraphics[height=0.10\linewidth]{fig/tribox2top.eps}
         \end{aligned} \hlp{5} \begin{aligned}
        \includegraphics[height=0.10\linewidth]{fig/boxtri2top.eps}
        \end{aligned}
        \hlm{5}
        \begin{aligned}
        \includegraphics[height=0.10\linewidth]{fig/tritri2top.eps}
    \end{aligned}\,.
\end{equation}
Note that there is some freedom of distributing the range-2 integrals to the range-3 result, so the above choice is not unique. 
As we will discuss  in Section~\ref{sec:iruv}, \eqref{eq:ID25}-\eqref{eq:ID1} and other results derived later also provide a convenient form for the IR subtraction.

As for the second relation \eqref{eq:ieqns2}, one can directly apply \eqref{tcfull3} and \eqref{eq:ieqns2}. 
The form of \eqref{tcfull3} is intentionally arranged such that every row maps to one of the terms in \eqref{eq:ieqns2}. 
Terms with color factor $\check{\mathcal{D}}_{11}$ and $\check{\mathcal{D}}_{18}$ are related to $\check{\mathcal{D}}_{1}$ (via permutation) and correspond to the last two rows in \eqref{tcfull3}. For example, $\check{\mathcal{D}}_{11}=\check{\mathcal{D}}_{\Delta_1}((p_1,p_3),(l_3,l_1,l_2))$ and
\begin{equation}
    \left[\mathcal{I}^{(2)}\right]_{\check{\mathcal{D}}_{11}}=\left[\mathcal{I}^{(2)}\right]_{\check{\mathcal{D}}_{1}}\Big|_{p_1\shortrightarrow p_2\shortrightarrow p_3}=\Big(\begin{aligned}
    \includegraphics[height=0.10\linewidth]{fig/boxtri5.eps}
  \end{aligned}
  \hlp{5}
  \begin{aligned}
    \includegraphics[height=0.10\linewidth]{fig/tribox5.eps}
  \end{aligned}
  \hlm{5}
  \begin{aligned}
    \includegraphics[height=0.10\linewidth]{fig/tritri5.eps}
  \end{aligned}
  \Big).
\end{equation}
Consequently, 
\begin{equation}
\label{eq:ID29}
     \left[\mathcal{I}^{(2)}(123)\right]_{\check{\mathcal{D}}_{19}}=-\Big(
    \begin{aligned}
      \includegraphics[height=0.10\linewidth]{fig/tribox2top.eps}
    \end{aligned} 
    \hlm{5}
    \begin{aligned}
      \includegraphics[height=0.10\linewidth]{fig/tritri2top.eps}
    \end{aligned}
    \hlm{5}
    \begin{aligned}
      \includegraphics[height=0.10\linewidth]{fig/pirbtop.eps}
    \end{aligned}
    \hlmath{5}{+\ \frac{1}{2}}
    \begin{aligned}
     \includegraphics[height=0.10\linewidth]{fig/pirbcomtop.eps}
  \end{aligned}\hskip -5pt \Big)+\text{cyc}(p_1,p_2,p_3).
\end{equation}
This form apparently satisfy the symmetries as expected. 

\begin{figure}
    \centering
    \subfigure[$s_{12}$ double-cut.]{
        \includegraphics[width=0.31\linewidth]{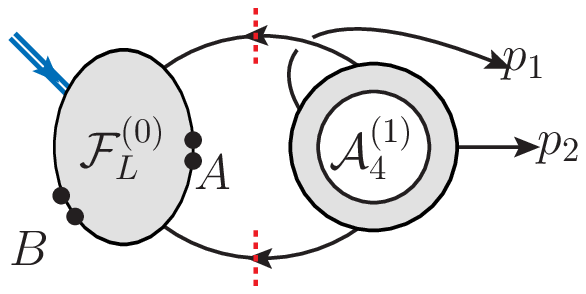} 
        \label{fig:d27a4}
        }
    \centering
    \subfigure[a $s_{12}$ triple-cut.]{
        \includegraphics[width=0.31\linewidth]{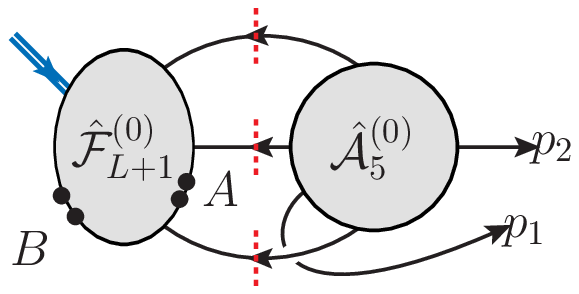} 
        \label{fig:d27a5}
        }
    \centering
    \subfigure[another $s_{12}$ triple-cut.]{
        \includegraphics[width=0.31\linewidth]{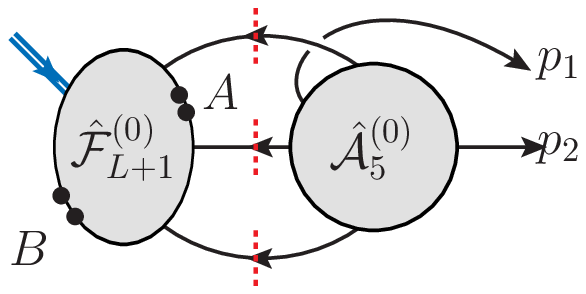}
        \label{fig:d27a52}
        }
	 \caption{Selected non-planar cut channels at two loops for  for $\mathcal{F}_{\mathcal{T}_{t_3}}^{(2)}$. These cuts are sufficient to determine $\big[\mathcal{I}\big]^{(2)}_{\check{\mathcal{D}}_{27}}$} 
	 \label{fig:anothernpcut}
\end{figure}

Finally, we have the third relation \eqref{eq:ieqns3}. The cuts we will use to determine part of $\mathcal{F}$ are presented in Figure~\ref{fig:anothernpcut}. 
Since our purpose is to fix the range-2 density function, a simple starting point is the two-particle double-cut (DC), as shown in Figure~\ref{fig:d27a4}. One can first check the color factor:
\begin{equation}
    N_c \operatorname{tr}_{\mathcal{A}}(\check{a}_{l_2}a_1\check{a}_{l_1}a_{2})\times\operatorname{tr}_{\mathcal{F}}(a_{l_{1}} A  a_{l_{2}} B ) =1\times \mathcal{T}_{t_3} + \cdots  \,. \nonumber
\end{equation}
The kinematic part under cut is:
\begin{equation}
    \left[\mathcal{F}^{(2)}\right]_{s_{12}\textrm{-DC}}=\int \mathrm{dPS}_{2,l}\mathcal{F}^{(0)}_{L}(-l_1,A,-l_2,B)\mathcal{A}_{4}^{(1)}(l_2,p_1,l_1,p_2)\,,
\end{equation}
where it is easy to derived the RHS as 
$\begin{aligned}
    \includegraphics[height=0.10\linewidth]{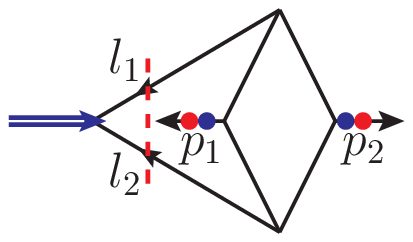}
\end{aligned}$. 
Since no other terms in \eqref{eq:ieqns3} possibly contribute to this cross-ladder integral, it has to come from $\big[\mathcal{I}^{(2)}(12)\big]_{\check{\mathcal{D}}_{27}}$.

Furthermore, we also consider a $s_{12}$ triple-cut (TC) as shown in Figure~\ref{fig:d27a5}. The color factor is:
\begin{equation}
    \operatorname{tr}_{\mathcal{A}}(\check{a}_{l_2}\check{a}_{l_1}a_{2}\check{a}_{l_3}a_{1})\times\operatorname{tr}_{\mathcal{F}}(a_{l_{1}} a_{l_2} A  a_{l_{3}} B ) =1\times \mathcal{T}_{t_3} + \cdots  \,,  \nonumber
\end{equation}
and the kinematic part under the cut is:
\begin{equation}
    \left[\mathcal{F}^{(2)}\right]_{\mathcal{T}_{t_3}}\Big|_{s_{12}\textrm{-TC}}=\int \mathrm{dPS}_{2,l}\hat{\mathcal{F}}^{(0)}_{L+1}(-l_1,-l_2,A,-l_3,B)\hat{\mathcal{A}}_{5}^{(0)}(l_2,l_1,p_2,l_3,p_1)\,.
\end{equation}
Similar to the discussion in \eqref{eq:tcfuldens}, here we can also split the next-to-minimal form factor $\hat{\mathcal{F}}_{L+1}^{(0)}$ according to their kinematic dependence as 
\begin{equation}
    \hat{\mathcal{F}}^{(0)}_{L+1}(-l_1,-l_2,A,-l_3,B)=\hat{\mathcal{F}}^{(0)}_{L+1,\scriptscriptstyle \rm dens}(p_3,-l_1,-l_2,p_4)+\hat{\mathcal{F}}^{(0)}_{L+1,\scriptscriptstyle \rm dens}(p_5,-l_3,p_6)\,,
\end{equation}
and we focus on the first part:\footnote{ Within this part, the super-density $\hat{\mathcal{F}}^{(0)}_{L+1,\scriptscriptstyle \rm dens}(p_3,-l_1,-l_2,p_4)$ allows exchanging gluons in $l_1$  or $l_2$ line, or  a pair of fermions in both lines.}
\begin{equation}
\begin{aligned}
	\left[\mathcal{F}^{(2)}\right]_{\mathcal{T}_{t_3}}\Big|_{s_{12}\textrm{-TC}}&=\int \mathrm{dPS}_{3,l}\hat{\mathcal{F}}^{(0)}_{L+1,\scriptscriptstyle \rm dens}(p_3,-l_1,-l_2,p_4)\hat{\mathcal{A}}_{5}^{(0)}(l_2,l_1,p_2,l_3,p_1)\\
	& \quad \ + \big(\text{kinematic dependence on  $p_5$ and $p_6$}\big) \,. 
\end{aligned}
\end{equation}
The RHS of \eqref{eq:ieqns3} contains kinematic dependence on $\{p_1,p_2,p_3,p_4\}$, and except $\big[\mathcal{I}^{(2)}(12)\big]_{\check{\mathcal{D}}_{27}}$, all of them have been derived from previous discussion. After subtracting them, we have
\begin{equation}
\label{eq:ID27}
\begin{aligned}
    \left[\mathcal{I}^{(2)}(12)\right]_{\check{\mathcal{D}}_{27}}=& \begin{aligned}
        \includegraphics[width=0.17 \linewidth]{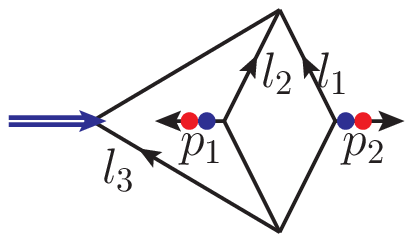}
    \end{aligned}
    \hlm{5}\Big(
   \hskip +2pt \frac{1}{2}\begin{aligned}
       \includegraphics[height=0.10\linewidth]{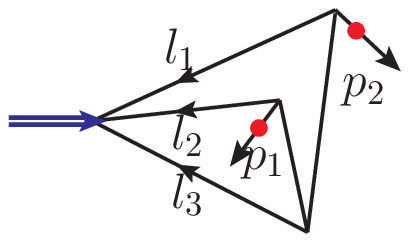}
    \end{aligned} \hlmath{5}{+} \begin{aligned}
       \includegraphics[height=0.10\linewidth]{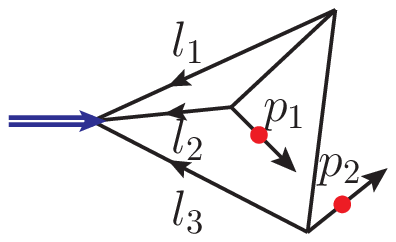}
    \end{aligned}\hlmath{5}{+\frac{1}{2}} \begin{aligned}
       \includegraphics[height=0.10\linewidth]{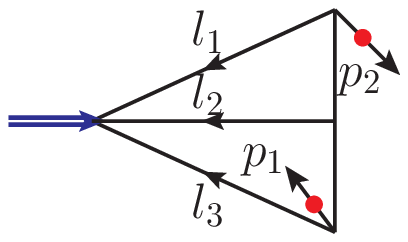}
    \end{aligned}\hskip -5pt \Big) \\
    &+ \big(\text{terms detected in other channels}\big).
\end{aligned}
\end{equation}
Integrals in \eqref{eq:ID27} are put in the form such that the cut structure (Figure~\ref{fig:d27a5}) is manifest. 
There is another $s_{12}$ triple-cut (Figure~\ref{fig:d27a52}) which can be considered similarly. Details of the calculation can be found in Appendix~\ref{ap:npuni}, where some further features for non-planar unitarity are discussed.
After considering this channel,  $\left[\mathcal{I}^{(2)}(12)\right]_{\check{\mathcal{D}}_{27}}$ is now complete.  The final integral form of $\left[\mathcal{I}^{(2)}(12)\right]_{\check{\mathcal{D}}_{27}}$ is 
\begin{equation}\label{eq:sid27}
	\left[\mathcal{I}^{(2)}(12)\right]_{\check{\mathcal{D}}_{27}}= -4 \begin{aligned}
          \includegraphics[height=0.10\linewidth]{fig/tritri2pt1top.eps}
        \end{aligned}\hlmath{5}
        + \begin{aligned}
           \includegraphics[height=0.10\linewidth]{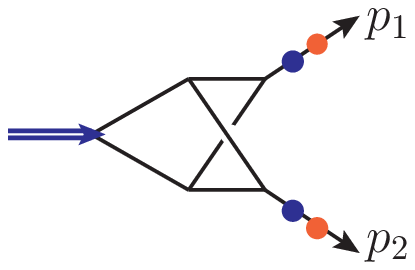}
        \end{aligned} .
\end{equation}

\subsubsection*{Check of other channels}
As mentioned above, there are other unitarity channels that we have not used and they can also be checked. Here we present an example. The targeting color factor is  
\begin{equation}
\mathcal{T}_{t_4}=N_{ c} \operatorname{tr}(a_{1}a_{2})\operatorname{tr}(a_{3}\cdots a_{L}) \,.
\end{equation} 
By extracting $\mathcal{T}_{t_4}$ component in \eqref{eq:fc-2l-loop}, one finds:
\begin{equation}
    \left(\itbf{I}^{(2)}_{12}+\cdots \right)\itbf{F}^{(0)}\Big|_{\mathcal{T}_{t_4}}= \Big( 3\left[\mathcal{I}^{(2)}(12)\right]_{\check{\mathcal{D}}_{25}}+3\left[\mathcal{I}^{(2)}(12)\right]_{\check{\mathcal{D}}_{26}}+2 \left[\mathcal{I}^{(2)}(12)\right]_{\check{\mathcal{D}}_{27}}+\cdots \Big) \mathcal{F}^{(0)} \,,
    \label{eq:t4cutexample}
\end{equation}
where the factor 3 and 2 comes from color structure, 
\begin{equation*}
\check{\mathcal{D}}_{25}\operatorname{tr}(l_{1}l_{2}a_3\cdots a_{L})\big|_{\mathcal{T}_{t_4}}=3,\quad 
\check{\mathcal{D}}_{26}\operatorname{tr}(l_{1}l_{2}a_3\cdots a_{L})\big|_{\mathcal{T}_{t_4}}=3, \quad
\check{\mathcal{D}}_{27}\operatorname{tr}(l_{1}l_{2}a_3\cdots a_{L})\big|_{\mathcal{T}_{t_4}}=2\,.
\end{equation*}

\begin{figure}
    \centering
    \subfigure[a $s_{12}$ triple-cut.]{
        \includegraphics[width=0.31\linewidth]{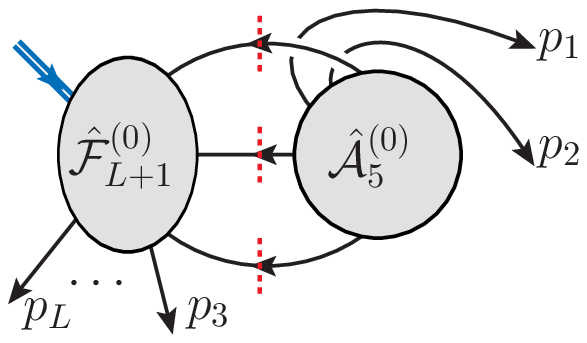} 
        \label{fig:check1}
        }
    \centering
    \subfigure[another $s_{12}$ triple-cut.]{
        \includegraphics[width=0.31\linewidth]{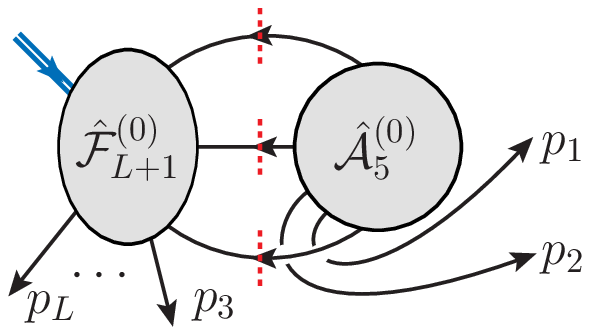}
        \label{fig:check2}
        }
    \caption{Selected cuts for $\big[\mathcal{F}^{(2)}\big]_{\mathcal{T}_{t_4}}$ for check. }
    \label{fig:check}
\end{figure}
We focus on the most constraining cut channels: the $s_{12}$ triple-cut as shown in Figure~\ref{fig:check}:\footnote{There are two more cut channels not shown in Figure~\ref{fig:check} which can be obtained by exchanging $p_1$ and $p_2$ in channel-(a) and (b). There is no need to compute them because both color factor $\mathcal{T}_{t_4}$ and kinematic coefficients, the range-2 integrals, are symmetric between $p_1$ and $p_2$ so that we just count in an extra 2 factor.\label{fn:factor2} }
\begin{equation}
\begin{aligned}
 &\text{Cut in Figure~\ref{fig:check1}}: \quad \mathcal{T}_{c}=\operatorname{tr}_{\mathcal{A}}(\check{a}_{l_1}\check{a}_{l_3}\check{a}_{l_2}a_1 a_2 )\times \operatorname{tr}(a_{l_1}a_{l_2}a_{l_3}a_{3}\cdots a_{L}) \,, \\
   & \left[\mathcal{F}_{L,12}^{(2)}\right]_{\mathcal{T}_c}\bigg|_{s_{12}\textrm{-TC(a)}}=\int \mathrm{dPS}_{3,l}\hat{\mathcal{F}}^{(0)}_{L+1}(-l_1,-l_2,-l_3,p_3,\ldots,p_{L}) \hat{\mathcal{A}}_{5}^{(0)}(l_1,l_3,l_2,p_1,p_2) \,, \\
   &\text{Cut in Figure~\ref{fig:check2}}: \quad \mathcal{T}_{c}=\operatorname{tr}_{\mathcal{A}}(\check{a}_{l_1}\check{a}_{l_3}a_1 a_2\check{a}_{l_2} )\times \operatorname{tr}_{\mathcal{F}}(a_{l_1}a_{l_2}a_{l_3}a_{3}\cdots a_{L}) \,, \\
   & \left[\mathcal{F}_{L,12}^{(2)}\right]_{\mathcal{T}_c}\bigg|_{s_{12}\textrm{-TC(b)}}=\int \mathrm{dPS}_{3,l}\hat{\mathcal{F}}^{(0)}_{L+1}(-l_1,-l_2,-l_3,p_3,\ldots,p_{L}) \hat{\mathcal{A}}_{5}^{(0)}(l_1,l_3,p_1,p_2,l_2) \,.
\end{aligned}
\end{equation}

The complete expression is a little lengthy and we only mention that after subtracting cuts from range-3 contributions, one obtains range-2 integrals as:
\begin{equation}\label{eq:finalcheck}
  3 \begin{aligned}
        \includegraphics[height=0.10\linewidth]{fig/pladderm1top.eps} 
    \end{aligned}\hskip -5pt + \begin{aligned}
        \includegraphics[height=0.10\linewidth]{fig/npladdertop.eps} 
    \end{aligned} \hskip -5pt - \begin{aligned}
        \includegraphics[height=0.10\linewidth]{fig/tritri2pt1top.eps} 
    \end{aligned}.
\end{equation}
The factor 3 in the first term can be understood from cuts: triple-cut-(a) contains three type of twists of the ``planar-ladder" integral 
\begin{equation}
    \begin{aligned}
        \includegraphics[height=0.085\linewidth]{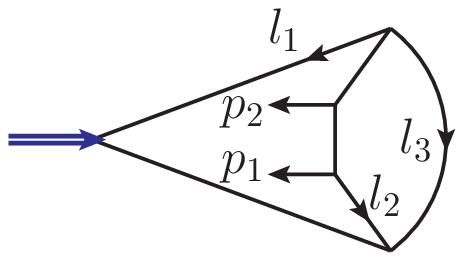} 
    \end{aligned}\hskip -5pt + \begin{aligned}
        \includegraphics[height=0.085\linewidth]{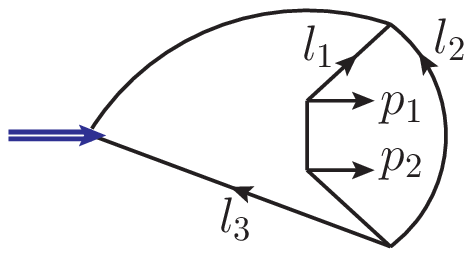} 
    \end{aligned} \hskip -5pt +\begin{aligned}
        \includegraphics[height=0.085\linewidth]{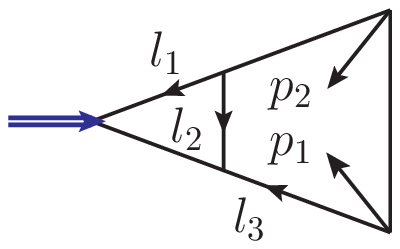} 
    \end{aligned}\,,
\end{equation}
which explains the factor 3 in \eqref{eq:finalcheck}. 
See also the discussion of a similar example in Appendix~\ref{ap:npuni}.

Together with the factors 2 in LHS is explained in Footnote\,\ref{fn:factor2}, one has
\begin{equation}
    2 \times  \eqref{eq:finalcheck}=3\left[\mathcal{I}^{(2)}(12)\right]_{\check{\mathcal{D}}_{25}}+3\left[\mathcal{I}^{(2)}(12)\right]_{\check{\mathcal{D}}_{26}}+2 \left[\mathcal{I}^{(2)}(12)\right]_{\check{\mathcal{D}}_{27}}\,,
\end{equation}
which \eqref{eq:ID25} and \eqref{eq:sid27} are used. This is indeed consistent with the wanted result in \eqref{eq:t4cutexample}. 

\subsubsection*{Summarizing the BPS result}

Results for complete correction density functions are summarized in the following form:
\begin{itemize}[leftmargin=*]
\item[$\square$] \textbf{Range-2}
    \begin{subequations}
    \begin{align}
        &\left[\mathcal{I}^{(2)}(12)\right]_{\mathcal{D}_{25}}=\begin{aligned}
           \includegraphics[height=0.10\linewidth]{fig/tritri2pt1top.eps}
        \end{aligned}\hlp{5}\begin{aligned}
           \includegraphics[height=0.10\linewidth]{fig/pladderm1top.eps}
        \end{aligned} 
        \label{bpsr21} , \\
        &\left[\mathcal{I}^{(2)}(12)\right]_{\mathcal{D}_{27}}= \begin{aligned}
           \includegraphics[height=0.10\linewidth]{fig/npladdertop.eps}
        \end{aligned}\hlmath{5}{-4} \begin{aligned}
          \includegraphics[height=0.10\linewidth]{fig/tritri2pt1top.eps}
        \end{aligned}
        \label{bpsr22} , \\
        & \itbf{I}^{(2)}=\sum_{\sigma \in S_{2}}\left( \check{\mathcal{D}}_{\Delta_{3}(\sigma)} \left[\mathcal{I}^{(2)}(\sigma(p_1,p_2))\right]_{\check{D}_{25}} \right) + \check{\mathcal{D}}_{\Delta_{4}(\mathbf{1})} \left[\mathcal{I}^{(2)}(p_1,p_2)\right]_{\check{D}_{27}} \,. 
    \end{align}
    \end{subequations}

\item[$\square$]\textbf{Range-3}
    \begin{subequations}
    \begin{align}
    &\left[\mathcal{I}^{(2)}(123)\right]_{\mathcal{D}_{1}}=\Big(\begin{aligned}
        \includegraphics[height=0.10\linewidth]{fig/tribox2top.eps}
        \end{aligned}\hlp{5}\begin{aligned}
        \includegraphics[height=0.10\linewidth]{fig/boxtri2top.eps}
        \end{aligned}\hlm{5}\begin{aligned}
        \includegraphics[height=0.10\linewidth]{fig/tritri2top.eps}
    \end{aligned}\Big) 
    \label{bpsr31} , \\
    \nonumber
    &\left[\mathcal{I}^{(2)}(123)\right]_{\mathcal{D}_{19}}=-\Big(
    \begin{aligned}
      \includegraphics[height=0.10\linewidth]{fig/tribox2top.eps}
    \end{aligned} 
    \hlm{5}
    \begin{aligned}
      \includegraphics[height=0.10\linewidth]{fig/tritri2top.eps}
    \end{aligned}
    \hlm{5}
    \begin{aligned}
      \includegraphics[height=0.10\linewidth]{fig/pirbtop.eps}
    \end{aligned}
    \hlmath{5}{+\frac{1}{2}}
    \begin{aligned}
     \includegraphics[height=0.10\linewidth]{fig/pirbcomtop.eps}
  \end{aligned}\hskip -6pt \Big) \hskip -2pt + \text{cyc}(p_1,p_2,p_3)
  \label{bpsr32}\,,\\
  & \itbf{I}_{123}^{(2)}=\sum_{
  \sigma^{e} \in S_{3}} \bigg(\sum_{\sigma^{i} \in \mathbb{Z}_{3}} \check{\mathcal{D}}_{\Delta_{1}(\sigma^e,\sigma^i)}\left[\mathcal{I}^{(2)}(\sigma^e(p_1,p_2,p_3))\right]_{\check{D}_{1}} + \check{\mathcal{D}}_{\Delta_{2}(\sigma^e)} \left[\mathcal{I}^{(2)}(\sigma^{e}(p_1,p_2,p_3))\right]_{\check{D}_{19}}\bigg) \,. 
\end{align}
\end{subequations}
\end{itemize}
Below we show that the above result can be given in a much simpler form. In particular, the range-3 non-planar integrals can all be reduced to planar ones.

\subsubsection*{Cancellation of range-3 non-planar integrals}

Results in \eqref{bpsr31}-\eqref{bpsr32} can actually be simplified further with an integral relation. This is  inspired by  \cite{Aybat:2006mz} where it was found that all non-planar range-3 Wilson line integrals do not exist in the final result. Although the objects considered there are Wilson lines, one may expect the similar pattern to appear also in form factors. Indeed, we find that the range-3 non-planar integrals, although having non-trivial integrand, can be reduced to a linear combination of planar integrals. 

The three non-planar integrals in \eqref{bpsr32} can be written together as 
\begin{equation}
     \mathcal{I}_{\rm 3, np}=  \left(s_{13} ~ 2k_1\cdot p_2 + s_{12} ~ 2k_2\cdot p_3 + s_{23} ~ 2k_3\cdot p_1\right)\hskip -5pt
    \begin{aligned}
        \includegraphics[height=0.10\linewidth]{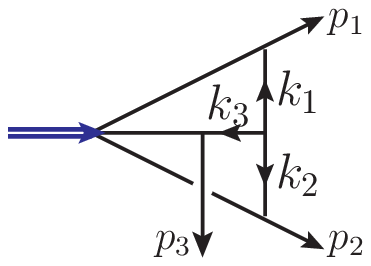}
    \end{aligned} .
\end{equation}
As proved in  Appendix~\ref{ap:bps}, $\mathcal{I}_{\rm 3,np}$ has an implicit parity symmetry, and it is then easy to show that
\begin{equation}
\label{eq:npr3detop}
\begin{aligned}
2\mathcal{I}_{\rm 3, np}   
    &=N_{3,\rm np}\hskip -2pt \begin{aligned}
        \includegraphics[height=0.10\linewidth]{fig/pirade.eps}
    \end{aligned}\hskip -8pt =\begin{aligned}
        \includegraphics[height=0.10\linewidth]{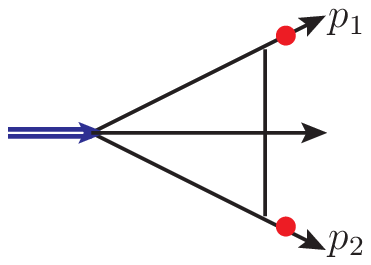}
    \end{aligned}\hlmath{5}{-}\begin{aligned}
        \includegraphics[height=0.10\linewidth]{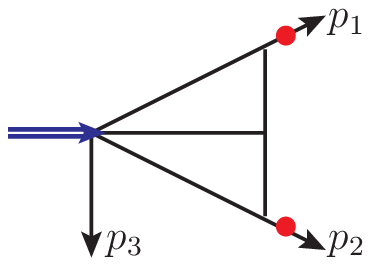}
    \end{aligned}\hlp{5}\text{cyc}(p_1,p_2,p_3)\,,
\end{aligned}
\end{equation}
in which only planar integrals are left. And 
\begin{equation}
    N_{3,\rm np}=\left(s_{12} (k_3+p_3)^2- s_{12} k_3^2 + s_{13} (k_2+p_2)^2- s_{13} k_2^2 + s_{23} (k_1+p_1)^2- s_{23} k_1^2 \right)\,.
\end{equation}

Using \eqref{eq:npr3detop}, one can rewrite \eqref{bpsr32} in a much simpler form as
\begin{equation}\label{eq:bpsr32new}
    \left[\mathcal{I}^{(2)}\right]_{\check{\mathcal{D}}_{19}}=-\Big(\begin{aligned}
       \includegraphics[height=0.10\linewidth]{fig/tribox2top.eps}
    \end{aligned}
    \hlmath{5}{-\frac{1}{2}}
    \begin{aligned}
       \includegraphics[height=0.10\linewidth]{fig/tritri2top.eps}
    \end{aligned}
    \Big)+ \text{cyc}(p_1,p_2,p_3) \,.
\end{equation}
As a comparison, let us recall \eqref{bpsr31}
\begin{equation}\label{eq:bpsr31new}
    \left[\mathcal{I}^{(2)}\right]_{\check{\mathcal{D}}_{1}}=\Big(\begin{aligned}
        \includegraphics[height=0.10\linewidth]{fig/tribox2top.eps}
        \end{aligned}\hlmath{5}{-\frac{1}{2}}\begin{aligned}
        \includegraphics[height=0.10\linewidth]{fig/tritri2top.eps}
    \end{aligned}\Big) +(p_1\leftrightarrow p_3) \,.
\end{equation}
We can see that range-3 BPS densities can be expressed via only one building block after using integration relations aforementioned.%
\footnote{The similarity of the above two forms, together with color identity given in \eqref{eq:d1md1a},  allow us to express the full-color range-3 density as
\begin{equation*}
    \itbf{I}_{123}^{(2)}=\sum_{\sigma^e, \sigma^i\in S_3}\check{\mathcal{D}}_{\Delta_1}(\sigma^e,\sigma^i) \times \mathcal{I}_{\text{BPS,a}}^{(2)}(\sigma^{e}(p_1,p_2,p_3))\,,
\end{equation*}
where $\mathcal{I}_{\text{BPS,a}}^{(2)}$ are the terms in parenthesis (see \eqref{eq:IBPSa}).
}

\subsection{Two-loop SU(2) loop correction}
\label{ssec:2lsu2}

In this subsection, we employ the method aforementioned to non-BPS form factors with operators in the SU(2) sector. 
The color decomposition in Section~\ref{ssec:nbasis2l} applies here, and one can still focus on the $\{ \big[\mathcal{I}_{123}^{(2)}\big]_{\check{\mathcal{D}}_1}, \ \big[\mathcal{I}_{123}^{(2)}\big]_{\check{\mathcal{D}}_{19}} , \ \big[\mathcal{I}_{123}^{(2)}\big]_{\check{\mathcal{D}}_{25}}, \ \big[\mathcal{I}_{123}^{(2)}\big]_{\check{\mathcal{D}}_{27}} \}$. 

Unlike the BPS case, in SU(2) sector there are two types of fields $X$ and $Y$,  so the corresponding density functions have more than one field configurations. One should therefore take into account all inequivalent configurations. 
In the concrete calculation, the Super-Ward-Identity with respect to SU(2) R-symmetry can provide a useful simplification, which have been observed in \cite{Loebbert:2016xkw} in the planar case and will be generalized to full-color case in our calculation. Using these relations, we can identify a small set of kinematic functions with specific field configurations that one needs to compute.

\subsubsection*{SU(2) supersymmetric relations}
We first review the planar case. 
All possible field configurations can be given as follows:\footnote{The density involving two $Y$ fields can be simply obtained by exchanging $X$ and $Y$ fields, and we will not consider them.}
\begin{itemize}
	\item range-2: $YX \rightarrow YX$, $YX \rightarrow XY$ and reverting the order of the fields. The total number of configurations is 4.
	\item  range-3: $YXX \rightarrow YXX$, $XYX \rightarrow XYX$, $YXX \rightarrow XYX$,  $XYX \rightarrow YXX$ and $YXX \rightarrow XXY$ and reverting the order of the fields.  The total number is 9.
\end{itemize}

Supersymmetry provides further symmetric constraint on form factors via Super-Ward-Identity:
\begin{equation}
0=\langle\Phi(1) \cdots \Phi(n)[\mathfrak{S}, \mathcal{O}]\rangle+\sum_{i=1}^{n}\langle\Phi(1) \cdots[\mathfrak{S}, \Phi(i)] \cdots \Phi(n) \mathcal{O}\rangle\,.
\end{equation}
where $\Phi$ 
can be general on-shell superfield or component fields and  $\mathfrak{S}$  is a supersymmety generator, which can be taken as R-charge generators $R^{A}_{B}$ in our particular discussion on SU(2) form factors. 
As a result, planar SU(2) relations have the following form in general
\begin{equation}\label{eq:fc-su2-st}
\begin{aligned}
    \sum_{\tau \in \mathbb{Z}_2} \left[\left(\mathcal{I}^{(2)}\right)^{ \tau(\scriptscriptstyle{YX})}_{\scriptscriptstyle{YX}}\right]_{\scriptscriptstyle \rm PL }
    &=\left[\left(\mathcal{I}^{(2)}\right)^{ \scriptscriptstyle{XX}}_{\scriptscriptstyle{XX}}\right]_{\scriptscriptstyle \rm PL }
    \,,\\
    \sum_{\tau_a \in \mathbb{Z}_3} \left[\left(\mathcal{I}^{(2)}\right)^{ \tau_{b}(\scriptscriptstyle{YXX})}_{\tau_{a}(\scriptscriptstyle{YXX})}\right]_{\scriptscriptstyle \rm PL }
    &= \left[\left(\mathcal{I}^{(2)}\right)^{ \scriptscriptstyle{XXX}}_{\scriptscriptstyle{XXX}}\right]_{\scriptscriptstyle \rm PL }
    , \quad \forall  \tau_{b}\in \mathbb{Z}_3\,,\\
     \sum_{\tau_b \in \mathbb{Z}_3} \left[\left(\mathcal{I}^{(2)}\right)^{ \tau_{b}(\scriptscriptstyle{YXX})}_{\tau_{a}(\scriptscriptstyle{YXX})}\right]_{\scriptscriptstyle \rm PL }
     &= \left[\left(\mathcal{I}^{(2)}\right)^{ \scriptscriptstyle{XXX}}_{\scriptscriptstyle{XXX}}\right]_{\scriptscriptstyle \rm PL }
     , \quad \forall  \tau_{a}\in \mathbb{Z}_3\,.
\end{aligned}
\end{equation}
Note that unlike in \eqref{eq:densitypermrels1}-\eqref{eq:densitypermrels2}, here the permutation $\tau$ acts \emph{only} on fields (but not momenta). 
Also, planar densities satisfy the reflection property:
\begin{equation}
    \left[\left(\mathcal{I}^{(2)}\right)^{\scriptscriptstyle ABC}_{\scriptscriptstyle A^{\prime} B^{\prime} C^{\prime}}\right]_{\scriptscriptstyle \rm PL}(p_1,p_2,p_3)=\left[\left(\mathcal{I}^{(2)}\right)^{\scriptscriptstyle CBA}_{\scriptscriptstyle C^{\prime} B^{\prime} A^{\prime}}\right]_{\scriptscriptstyle \rm PL}(p_3,p_2,p_1)\,.
\end{equation}
In the end, it turns out that one only needs to compute
\begin{itemize}
	\item range-2: $YX \rightarrow YX$,
	\item  range-3: $YXX \rightarrow YXX$, $YXX \rightarrow XXY$, 
\end{itemize}
to get all planar results, see also \cite{Loebbert:2016xkw} for more details.

Next we consider to the full-color case. It is straightforward to generalize \eqref{eq:fc-su2-st} as
\begin{equation}\label{eq:fc-su2-all}
\begin{aligned}
    \sum_{\tau \in \mathbb{Z}_2} \left(\itbf{I}^{(2)}_{12}\right)^{ \tau(\scriptscriptstyle{YX})}_{\scriptscriptstyle{YX}}&=\left(\itbf{I}^{(2)}_{12}\right)^{ \scriptscriptstyle{XX}}_{\scriptscriptstyle{XX}}\,,\\
    \sum_{\tau_a \in \mathbb{Z}_3} \left(\itbf{I}^{(2)}_{123}\right)^{ \tau_{b}(\scriptscriptstyle{YXX})}_{\tau_{a}(\scriptscriptstyle{YXX})}&=\left(\itbf{I}^{(2)}_{123}\right)^{ \scriptscriptstyle{XXX}}_{\scriptscriptstyle{XXX}}, \quad \forall  \tau_{b}\in \mathbb{Z}_3\,,\\
     \sum_{\tau_b \in \mathbb{Z}_3} \left(\itbf{I}^{(2)}_{123}\right)^{ \tau_{b}(\scriptscriptstyle{YXX})}_{\tau_{a}(\scriptscriptstyle{YXX})}&=\left(\itbf{I}^{(2)}_{123}\right)^{ \scriptscriptstyle{XXX}}_{\scriptscriptstyle{XXX}}, \quad \forall  \tau_{a}\in \mathbb{Z}_3\,.
\end{aligned}
\end{equation}
By expanding \eqref{eq:fc-su2-all} to a set of color basis, for example our trivalent basis, one also has the SU(2) relations for color-stripped densities|just replace $\itbf{I}$ with $\big[\mathcal{I}\big]_{\check{\mathcal{D}}_i}$ for some $\check{\mathcal{D}}_i$.

As mentioned, we only need to focus on the densities with ${\check{\mathcal{D}}_i}, i=1,19,25,27,$ which we describe respectively below:
\begin{enumerate}
    \item For $\big[\mathcal{I}^{(2)}\big]_{\check{\mathcal{D}}_1}$, our experience in BPS computation shows that they can be mapped to the planar case, and the independent matrix elements can be also chosen as the planar case. 
    \item For $\big[\mathcal{I}^{(2)}\big]_{\check{\mathcal{D}}_{19}}$, because of the cyclic symmetry of color factor $\check{D}_{19}$, we have
    \begin{equation}\label{eq:su2id19cyc}
        \left[\left(\mathcal{I}^{(2)}\right)^{\sigma \tau(\scriptscriptstyle{YXX})}_{\sigma(\scriptscriptstyle{YXX})} \big(\sigma(p_1,p_2,p_3)\big) \right]_{\check{\mathcal{D}}_{19}}=\left[\left(\mathcal{I}^{(2)}\right)^{ \tau(\scriptscriptstyle{YXX})}_{\scriptscriptstyle{YXX}} (p_1,p_2,p_3) \right]_{\check{\mathcal{D}}_{19}}, \quad \forall\tau,\sigma\in \mathbb{Z}_3\,.
    \end{equation}
    Thus, the nine density functions for $\big[\mathcal{I}^{(2)}\big]_{\check{\mathcal{D}}_{19}}$ are reduced to only three ones in RHS of \eqref{eq:su2id19cyc}. And the SU(2) relation from \eqref{eq:fc-su2-all} involving only these three functions is
    \begin{equation}
        \sum_{\tau\in \mathbb{Z}_3}\left[\left(\mathcal{I}^{(2)}\right)^{ \tau(\scriptscriptstyle{YXX})}_{\scriptscriptstyle{YXX}}\right]_{\check{\mathcal{D}}_{i}}=\left[\left(\mathcal{I}^{(2)}\right)^{ \scriptscriptstyle{XXX}}_{\scriptscriptstyle{XXX}}\right]_{\check{\mathcal{D}}_{i}}\,. 
    \end{equation}
    So we can further select two of these three functions as independent basis. 
    \item For$\big[\mathcal{I}^{(2)}\big]_{\check{\mathcal{D}}_{25}}$ or $\big[\mathcal{I}^{(2)}\big]_{\check{\mathcal{D}}_{27}}$, invariance of density function under flipping $X$ and $Y$ fields reduces four configurations to two. And the following SU(2) equations 
    \begin{equation}
        \left[\left(\mathcal{I}^{(2)}\right)^{ \scriptscriptstyle{YX}}_{\scriptscriptstyle{YX}}\right]_{\check{\mathcal{D}}_{j}}+\left[\left(\mathcal{I}^{(2)}\right)^{ \scriptscriptstyle{YX}}_{\scriptscriptstyle{YX}}\right]_{\check{\mathcal{D}}_{j}}=\left[\left(\mathcal{I}^{(2)}\right)^{ \scriptscriptstyle{XX}}_{\scriptscriptstyle{XX}}\right]_{\check{\mathcal{D}}_{j}}\,, \quad j=25,27,
    \end{equation}
     lead to the conclusion that $YX\rightarrow YX$ is sufficient. 
\end{enumerate}
To summarize, the independent matrix elements for SU(2) form factors can be chosen as 
\begin{equation}
\begin{aligned}
    \Big[{\mathcal{I}}^{(2)}\Big]_{\check{\mathcal{D}}_{1}}&: \quad  YXX \rightarrow YXX, \ YXX \rightarrow XXY \,, \\
\Big[{\mathcal{I}}^{(2)}\Big]_{\check{\mathcal{D}}_{19}}&: \quad  YXX \rightarrow YXX, \ YXX \rightarrow XYX \,, \\
    \Big[{\mathcal{I}}^{(2)}\Big]_{\check{\mathcal{D}}_{25}}& \text{ and }  \Big[{\mathcal{I}}^{(2)}\Big]_{\check{\mathcal{D}}_{27}}: \quad  YX \rightarrow YX \,.
\end{aligned}
\end{equation}

\subsubsection*{Unitarity construction}

Having identified the target density functions, we can use unitarity method to construct them. The unitarity computation for these building blocks is totally parallel to the BPS case. The only change is that the tree building blocks contain both $X, Y$ as external states. The six-point NMHV scalar amplitudes with $X,Y,\bar{X}\text{ and }\bar{Y}$ as external states are major building blocks, which are explicitly given in Appendix~\ref{ap:amp}.  We will not reproduce the details of the calculation and just summarize the final results below:
\begin{itemize}[leftmargin=*]
\item[$\square$]\textbf{Range-2}
    \begin{subequations}\label{su2r2}
    \begin{align}
    &\left[\left({\mathcal{I}}^{(2)}\right)^{\scriptscriptstyle YX}_{\scriptscriptstyle YX}\right]_{\mathcal{D}_{25}}=\Big(\hskip -2pt \begin{aligned}
           \includegraphics[height=0.10\linewidth]{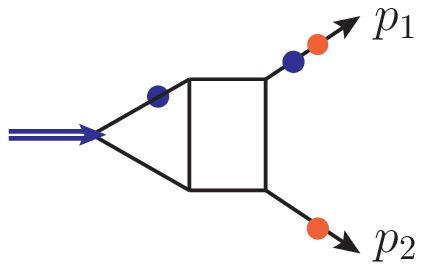}
        \end{aligned}
        \hlmath{8}{+}
        \begin{aligned}
           \includegraphics[height=0.10\linewidth]{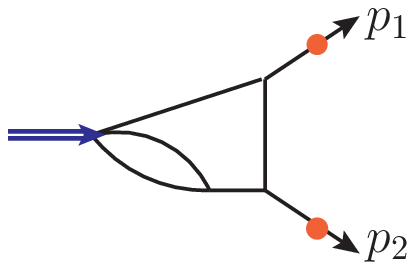}
        \end{aligned}\hskip -5pt \Big)
        +2 
        \begin{aligned}
           \includegraphics[height=0.10\linewidth]{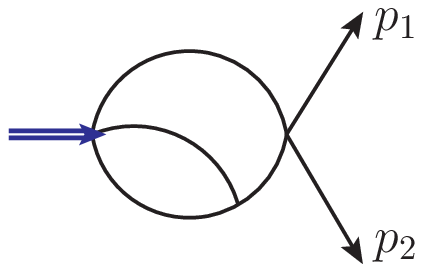}
        \end{aligned} \hskip -10pt + \eqref{bpsr21}
        , \label{su2r21} \\
        &\left[\left({\mathcal{I}}^{(2)}\right)^{\scriptscriptstyle YX}_{\scriptscriptstyle YX}\right]_{\mathcal{D}_{27}}={-\frac{1}{2}}\begin{aligned}
           \includegraphics[height=0.10\linewidth]{fig/npladdertop.eps}
        \end{aligned}\hlmath{10}{-}\begin{aligned}
           \includegraphics[height=0.10\linewidth]{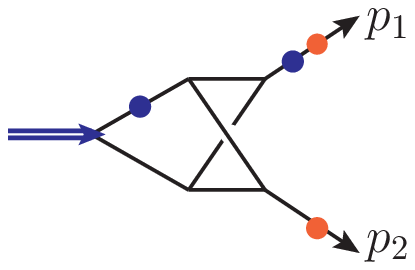}
        \end{aligned} \hskip -10pt +  \eqref{bpsr22}=  \eqref{bpsr22},
         \label{su2r22}
    \end{align}
    \end{subequations}
  where the second equation in \eqref{su2r22} results from 
    \begin{equation}
        \begin{aligned}
           \includegraphics[height=0.10\linewidth]{fig/npladdersu2.eps}
        \end{aligned}\hlmath{5}{=}\begin{aligned}
           \includegraphics[height=0.10\linewidth]{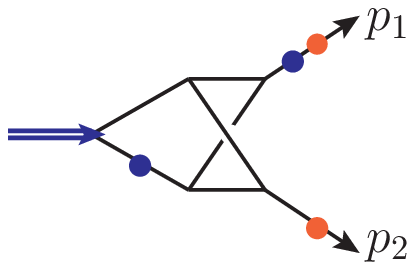}
        \end{aligned}\hlmath{5}{=}-\frac{1}{2}\begin{aligned}
           \includegraphics[height=0.10\linewidth]{fig/npladdertop.eps}
        \end{aligned}\,.
    \end{equation}
%

\item[$\square$]\textbf{Range-3}
    \begin{subequations}
    \label{su2r3}
    \begin{align}
        \left[\left({\mathcal{I}}^{(2)}\right)^{\scriptscriptstyle YXX}_{\scriptscriptstyle YXX}\right]_{\mathcal{D}_1}=&\Big(\begin{aligned}
           \includegraphics[height=0.10\linewidth]{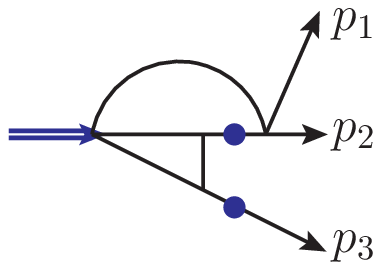}
        \end{aligned}\hlp{5}\begin{aligned}
           \includegraphics[height=0.10\linewidth]{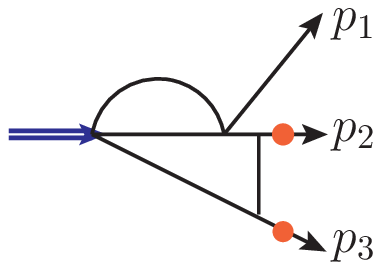}
        \end{aligned}\Big)+\eqref{bpsr31}
        \label{su2r31} \,,\\
        \left[\left({\mathcal{I}}^{(2)}\right)^{\scriptscriptstyle XXY}_{\scriptscriptstyle YXX}\right]_{\mathcal{D}_1}=&\Big(\begin{aligned}
           \includegraphics[height=0.10\linewidth]{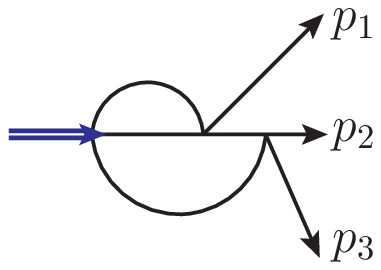}
        \end{aligned}\Big) \label{su2r32} \,,\\
         \left[\left({\mathcal{I}}^{(2)}\right)^{\scriptscriptstyle XYX}_{\scriptscriptstyle YXX}\right]_{\mathcal{D}_{19}}=&-\Big(\begin{aligned}
           \includegraphics[height=0.10\linewidth]{fig/tritrimasstop.eps}
        \end{aligned}\hlp{5}\begin{aligned}
           \includegraphics[height=0.10\linewidth]{fig/bubboxtop.eps}
        \end{aligned}\Big)-\Big(\begin{aligned}
          \includegraphics[height=0.10\linewidth]{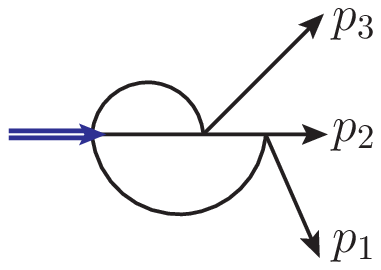}
        \end{aligned}\hlp{5}\begin{aligned}
          \includegraphics[height=0.10\linewidth]{fig/bubtrimass23.eps}
        \end{aligned}\Big)
        \label{su2r34} \,, \\
        \notag
        \left[\left({\mathcal{I}}^{(2)}\right)^{\scriptscriptstyle YXX}_{\scriptscriptstyle YXX}\right]_{\mathcal{D}_{19}}=&\Big(\begin{aligned}
           \includegraphics[height=0.10\linewidth]{fig/tritrimasstop.eps}
        \end{aligned}\hlp{5}\begin{aligned}
           \includegraphics[height=0.10\linewidth]{fig/bubboxtop.eps}
        \end{aligned}\hlp{5}\begin{aligned}
           \includegraphics[height=0.10\linewidth]{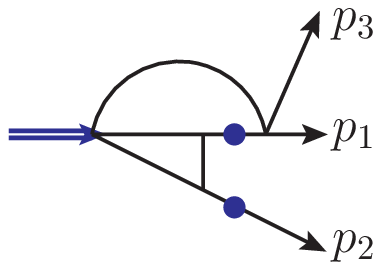}
        \end{aligned}\hlp{5}\begin{aligned}
           \includegraphics[height=0.10\linewidth]{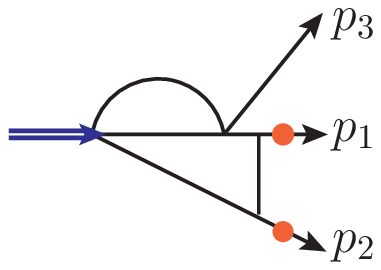}
        \end{aligned}\Big)\\
        &+\Big(\begin{aligned}
           \includegraphics[height=0.10\linewidth]{fig/bubtrimasstop.eps}
        \end{aligned}\Big)+\eqref{bpsr32} 
        \label{su2r35} \,.
    \end{align}
\end{subequations}
\end{itemize}
From \eqref{su2r21}-\eqref{su2r35}, one can see that it is possible to separate the BPS part as 
    \begin{equation}\label{eq:isu2tilde}
        \left[\left(\tilde{\mathcal{I}}^{(2)}\right)^{\scriptscriptstyle \mathbf{W}_i^{\prime}}_{\scriptscriptstyle \mathbf{W}_i}\right]_{\check{\mathcal{D}}_{i}}=
        \left[\left({\mathcal{I}}^{(2)}\right)^{\scriptscriptstyle \mathbf{W}_i^{\prime}}_{\scriptscriptstyle \mathbf{W}_i}\right]_{\check{\mathcal{D}}_{i}}
        - \delta_{\scriptscriptstyle \mathbf{W}_i}^{\scriptscriptstyle \mathbf{W}_{i}^{\prime}} \bigg[\left({\mathcal{I}}^{(2)}_{\rm BPS}\right)^{\scriptscriptstyle X\cdots X}_{\scriptscriptstyle X \cdots X}\bigg]_{\check{\mathcal{D}}_{i}}\,,
    \end{equation}
with $i=1,19,25,27$ as our focused elements.
As will be discussed in Section~\ref{ssec:uv}, it is possible to further separate the IR and UV divergences in a manifest way.

\section{Color-kinematics duality for two-loop BPS form factors}
\label{sec:CKBPS}
In this section, we study non-planar form factors based on the color-kinematics duality. As a concrete result, we will reconstruct the two-loop BPS form factor and obtain a representation satisfying CK duality at the loop density level. 

\begin{figure}
    \centering
    \subfigure[]{\includegraphics[width=0.18\linewidth]{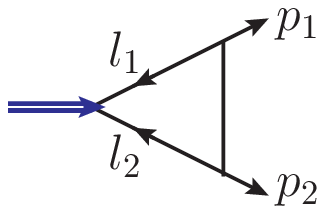}}
    \subfigure[]{\includegraphics[width=0.18\linewidth]{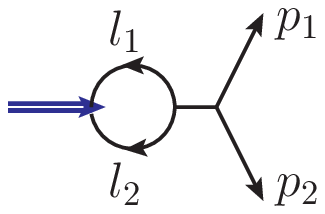}}
    \caption{Trivalent topology for one-loop minimal BPS form factors. 
    }
    \label{fig:ck1l}
\end{figure}

Before discussing two loops, it should be helpful to have a simple check at one-loop level. There are two types of trivalent topologies at one-loop level as shown in Figure~\ref{fig:ck1l}.
For BPS case, the existence of one-loop CK-dual representation is easy to see,
where the planar kinematic part is (see Appendix~\ref{ap:1l} for discussions and notations)
\begin{equation}
    \mathcal{I}_{\rm BPS}^{(1)}(12)=-\mathrm{I}_{\rm tri}[s_{12}]\,,
\end{equation}
and the full-color one-loop correction is
\begin{equation}
    \itbf{I}^{(1)}_{12,\rm BPS}=-\tilde{f}^{\check{a}_{l_1}a_{1}\text{x}}\tilde{f}^{\text{x}a_{2}\check{a}_{l_2}}s_{12}\begin{aligned}
    \includegraphics[width=0.14\linewidth]{fig/trinew.eps}
    \end{aligned}-\tilde{f}^{\check{a}_{l_1}a_{2}\text{x}}\tilde{f}^{\text{x} a_{1}\check{a}_{l_2}}s_{12}\begin{aligned}
    \includegraphics[width=0.14\linewidth]{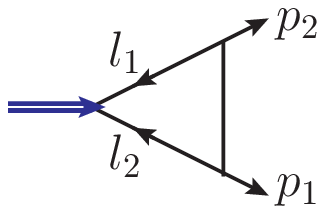}
    \end{aligned}\,.
\end{equation}
Obviously, such quantity satisfies CK duality with $N_{\rm tri}=s_{12}$ and $N_{\rm bub}=0$.
Similar construction can be also generalized to non-BPS operators, and the explicit CK-dual form for one-loop SU(2) form factors is provided in Appendix~\ref{ap:1l}. We leave more general discussions to another paper.

As we will show in this section,  two-loop generalization of CK duality is also possible. Following Section~\ref{ssec:ck}, below we first use a six-point tree amplitude as an example to illustrate the construction of CK duality. Then we go on to construct the two-loop BPS form factors. 

\subsection{Six-scalar tree amplitude}

As a warm up, we describe the calculation of CK-dual representation for the full-color six-point amplitude $\itbf{A}^{(0)}_{6}(p_1^X, p_2^X, p_3^X, q_1^{\bar X}, q_2^{\bar X}, q_3^{\bar X})$. Via unitarity cut, this six-point amplitude is also a building block  in the two-loop BPS form factor, which will facilitate the construction for the two-loop form factor.
Following \eqref{ckorigin}, we expect the six scalar amplitude to be given in the following form:
\begin{equation}
        \itbf{A}^{(0)}_{6}(p_1^X, p_2^X, p_3^X, q_1^{\bar X}, q_2^{\bar X}, q_3^{\bar X})
        = \sum_{\sigma} \sum_{\Gamma_{i}}\frac{1}{S_i}\frac{\mathcal{C}_{i}N_{i}}{\prod_{a}d_{i,a}} \,.
\end{equation}
%

\begin{figure}
        \centering
        \includegraphics[width=0.9\linewidth]{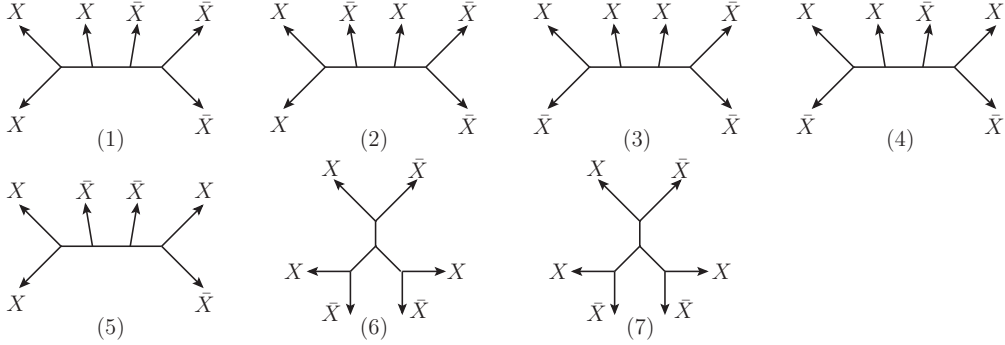}
        \caption{Topology with scalar amplitudes with $X$ and $\bar{X}$ external states, labeled as  $\Gamma_{i}$. Only $\Gamma_4$ and $\Gamma_6$ give non-vanishing numerators. }
        \label{fig:6pttree}
\end{figure}

We can construct this form in the following steps:
\begin{enumerate}[leftmargin=*]
    \item \emph{Generating trivalent diagrams $\Gamma_i$.} 
    
    The trivalent topologies of inequivalent external scalar configurations are given in Figure~\ref{fig:6pttree}. 
    Note that we do not need to specify the momenta in this step; hence permutations of $\{p_1,p_2,p_3\}$ will not lead to different trivalent diagrams $\Gamma_i$.

    \item \emph{Dual Jacobi relation and master topology. }
    
    Dual Jacobi relations like \eqref{djacob} provide connections between numerators, for example: 
    \begin{align}
        N\Big(\begin{aligned}
           \includegraphics[width=0.2\linewidth]{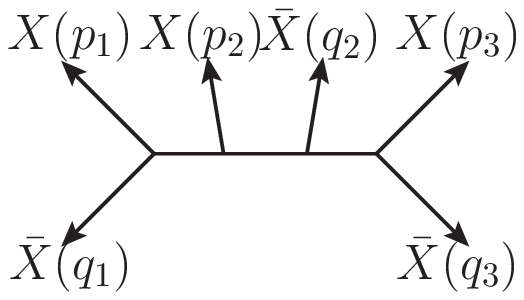}
        \end{aligned}\Big)-N\Big(\begin{aligned}
           \includegraphics[width=0.2\linewidth]{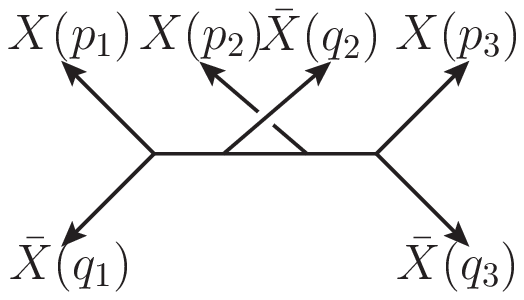}
        \end{aligned}\Big)&=N\Big(\begin{aligned}
           \includegraphics[width=0.2\linewidth]{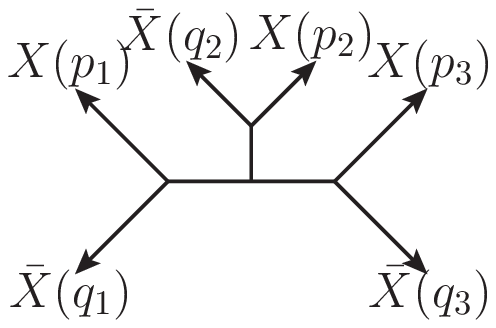}
        \end{aligned}\Big) \nonumber \\
        \quad N_{4}(p_1,p_2,p_3,q_1,q_2,q_3)-N_{4}(p_3,p_2,p_1,q_3,q_2,q_1)&=N_{6}(p_1,p_2,p_3,q_1,q_2,q_3),
    \end{align}
    \begin{align}
        N\Big(\begin{aligned}
           \includegraphics[width=0.2\linewidth]{fig/6pttree1.eps}
        \end{aligned}\Big)-N\Big(\begin{aligned}
           \includegraphics[width=0.2\linewidth]{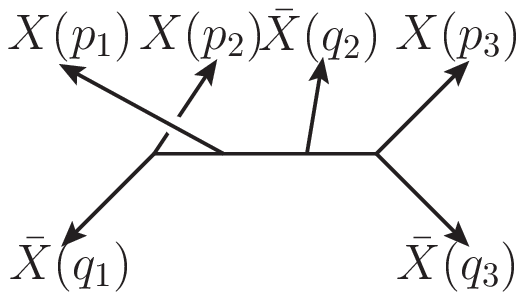}
        \end{aligned}\Big)&=-N\Big(\begin{aligned}
           \includegraphics[width=0.2\linewidth]{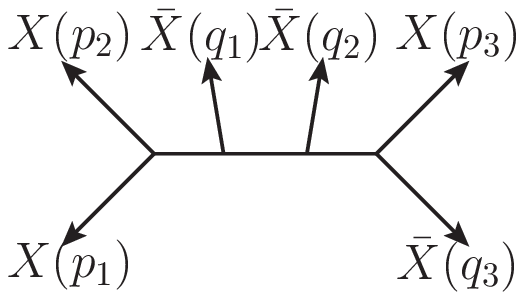}
        \end{aligned}\Big) \nonumber \\
        \quad N_{4}(p_1,p_2,p_3,q_1,q_2,q_3)-N_{4}(p_2,p_1,p_3,q_1,q_2,q_3)&=-N_{5}(p_1,p_2,p_3,q_1,q_2,q_3).
    \end{align}
    Using these identities, one finds out that Figure~\ref{fig:6pttree}(4) can be chosen as the master topology and other numerators can be expressed by linear combination of $N_{4}$.

    \item \emph{Ansatz and constraint for master numerator} 
    
     Before writing an ansatz for the master numerator $N_4$, let us consider the following two constrains:
     
        (a) We assume that one can set all numerators of diagrams with two external $X$ or $\bar{X}$ connected to the same vertex zero. 
        This directly means that except the numerators of (4) and (6) in Figure~\ref{fig:6pttree}, all other numerators are zero.
        Under this assumption, $N_4$ should be symmetric by shifting the first two or the last two arguments, such as  $N_4(p_1,p_2,p_3,q_1,q_2,q_3)=N_4(p_1,p_2,p_3,q_1,q_3,q_2)$, due to the fact that $N_3=0$. 
        
        (b) Then we use a simple cut condition, by setting the center propagator $(q_1+p_1+p_2)^2 \rightarrow 0$. This cut channel can actually detects topology (2)-(5), but as we mentioned that only (4) is non-vanishing. Under this cut, six point amplitudes factorize into product of two four point amplitudes, with $P\equiv p_1+p_2+q_1$
        \begin{equation}
           \left(P^{2}\itbf{A}^{(0)}_{6}(p_1^X, p_2^X, p_3^X, q_1^{\bar X}, q_2^{\bar X}, q_3^{\bar X})\right)\Big|_{P^{2}\shortrightarrow 0}=\itbf{A}^{(0)}_{4}(q_1^{\bar{X}},p_1^{X},p_2^{X},P^{\bar{X}})\itbf{A}^{(0)}_{4}(q_2^{\bar{X}},q_3^{\bar{X}},p_3^{X},-P^{{X}})\,,
        \end{equation}
        with this condition, we see that
        \begin{equation}
            N_{4}(p_1,p_2,p_3,q_1,q_2,q_3)\Big|_{P^2 \shortrightarrow 0} = -s_{p_1 p_2}s_{q_2 q_3}\,,
        \end{equation}
        where the minus sign on RHS comes from the color factor.

        Given the above two constraints, an ansatz for $N_4$ can be given as\footnote{In this ansatz we also make the local numerator assumption, \emph{i.e.}~the numerators are only polynomials of Mandelstam variables.}
        \begin{equation}
        \begin{aligned}
        \label{eq:ansatzN4tree}
            N_{4}(p_1,p_2,p_3,q_1,q_2,q_3)=&-s_{p_1 p_2}s_{q_2 q_3}+s_{q_1p_1p_2}\sum_{i=1}^{4}\beta_i u_i\,,
        \end{aligned}
        \end{equation}
        where $\beta_i$ are numerical parameters to be further determined and
        $$u_i \in \{s_{p_1p_2}+s_{q_2q_3},s_{p_3 q_1},s_{p_1 p_2 p_3},s_{p_1 p_2 q_1}\}$$
        is a set of independent lorentz product which is symmetric under $(p_1\leftrightarrow p_2)$ and $(q_2\leftrightarrow q_3)$.

    \item \emph{Symmetry of numerators.} 
    
    From the master ansatz \eqref{eq:ansatzN4tree}, all other numerators can be obtained through dual Jacobi relations.  We then impose the condition that the numerators should have symmetries inherited from symmetries of corresponding topology.
    
    For the two non-vanishing diagram, one can see that diagram Figure~\ref{fig:6pttree}(4) has no intrinsic symmetry, while  diagram Figure~\ref{fig:6pttree}(6) has cyclic and parity symmetry and its symmetric factor is $6$.
As a result, one has 
\begin{equation}
N_{6}(p_1,p_2,p_3,q_1,q_2,q_3) = \pm N_{6}(\sigma\{p_1,p_2,p_3\},\sigma\{q_1,q_2,q_3\}) \,,  \quad {\rm with} \ \sigma\in S_{3} \,.
\end{equation} 
To determine the sign, we use the following color convention:
    \begin{equation}
        \mathcal{C}_6(p_1,p_2,p_3,q_1,q_2,q_3) =(-1)^{\sigma} \mathcal{C}_6(\sigma\{p_1,p_2,p_3\},\sigma\{q_1,q_2,q_3\}), \quad \sigma\in S_{3}\,. 
    \end{equation}
    Thus, we have the following relation for $N_6$
    \begin{equation}
         N_{6}(p_1,p_2,p_3,q_1,q_2,q_3) =(-1)^{\sigma} N_{6}(\sigma\{p_1,p_2,p_3\},\sigma\{q_1,q_2,q_3\}), \quad \sigma\in S_{3}\,.
    \end{equation}
    
    This symmetry condition turns out to be enough to fix all the parameters $\beta_i$, and the only non-vanishing parameter is
    \begin{equation}
        \beta_3=\frac{1}{2}\,. 
    \end{equation}
    
    Thus, the master numerator \eqref{eq:ansatzN4tree} is
    \begin{equation}
        -s_{p_1 p_2}s_{q_2 q_3}+\frac{1}{2}s_{p_1 p_2 p_3}s_{p_1 p_2 q_1}\,.
    \end{equation}
%

\begin{table}[t!]
    \caption{Various coefficients for the CK-dual tree scalar amplitudes. The ``cyc."  in the second numerator represents cyclically permuting $\{p_1,p_2,p_3\}$ and $\{q_1,q_2,q_3\}$ simultaneously.  The color factors $\mathcal{C}_i$ can be directly read out from diagrams, given the rule that each vertex gives a $f^{abc}$ with legs $a,b,c$ arranged in clockwise order.}
    \label{tab:cktab-tree}
   \vskip .4 cm
    \centering
    \begin{tabular}{|C{3.5cm}|C{5cm} |C{2cm}|}
    \hline
    $\Gamma_{i}$ & $N_i$ & $S_i$ \\ \hline
        \includegraphics[width=0.8\linewidth]{fig/6pttree1.eps} & $-s_{p_1 p_2}s_{q_2 q_3}+\frac{1}{2}s_{p_1 p_2 p_3}s_{p_1 p_2 q_1}$ & 1 \\\hline
        \includegraphics[width=0.8\linewidth]{fig/6pttree3.eps} & 
        $-{1\over2} s_{p_1 p_2} (s_{p_2q_2}-s_{p_1q_1} + s_{p_3 q_2} - s_{p_3 q_1}) $ 
        + cyc. & 6
        \\\hline
    \end{tabular}
\end{table}

    \item \emph{Analytic check.} 
    
    Finally, we write down the full tree amplitude as 
    \begin{equation}
    \label{ckbpsres}
        \itbf{A}^{(0)}_{6}=\sum_{\sigma_1,  \sigma_2}\sum_{\Gamma_i}\frac{1}{S_i}\frac{\mathcal{C}_{i}(\sigma_{1},\sigma_{2})N_{i}(\sigma_{1},\sigma_{2})}{\prod_{a}d_{i,a}(\sigma_{1},\sigma_{2})} \,,
    \end{equation}
    where $\sigma_{1},\sigma_{2}$ are $S_3$ permutations of three $X$ and three $\bar{X}$ respectively. 
    The various factors are summarized in Table~\ref{tab:cktab-tree}. 
   
    We have checked that \eqref{ckbpsres} is indeed equivalent to the result from direct computations. For example, one can expand the full-color amplitudes to color-ordered amplitudes and compare with the expressions in Appendix~\ref{ap:amp}. 
 
\end{enumerate}

It is worthwhile to mention that under the conditions that we have imposed, the solution is unique (up to terms proportional to $p_i^2$ which are zero). 
Through unitarity cut, the six-point amplitude will be an important building block in the two-loop BPS form factor, thus \eqref{ckbpsres} will serve as a useful input when we consider range-3 density.

\subsection{Two-loop CK-dual form of BPS loop correction}
Now we construct the CK-dual representation for the two-loop half-BPS form factor of ${\rm tr}(X^L)$. As explained in Section~\ref{ssec:ck}, one only needs to focus on the loop density functions. The strategy of construction is similar to the aforementioned one, and we will sketch the steps and highlight the difference.

\begin{figure}[t]
    \centering
    \subfigure[]{\includegraphics[width=0.22\linewidth]{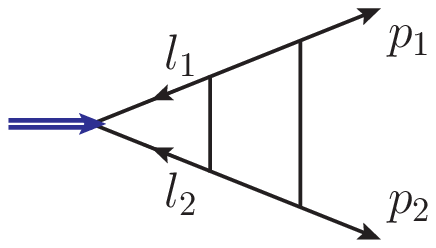}}
    \subfigure[]{\includegraphics[width=0.22\linewidth]{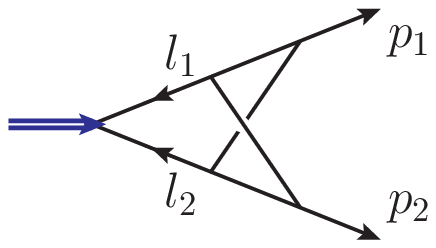}}
    \subfigure[]{\includegraphics[width=0.22\linewidth]{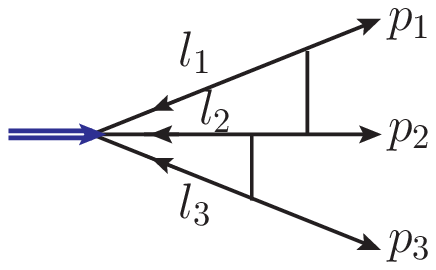}}
    \subfigure[]{\includegraphics[width=0.22\linewidth]{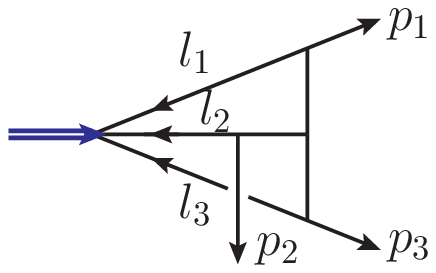}}
    \subfigure[]{\includegraphics[width=0.18\linewidth]{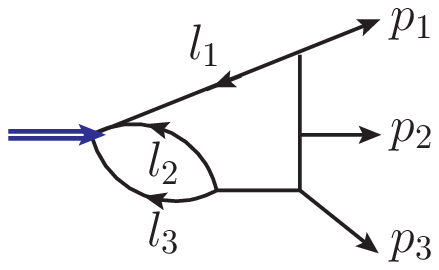}}
    \subfigure[]{\includegraphics[width=0.18\linewidth]{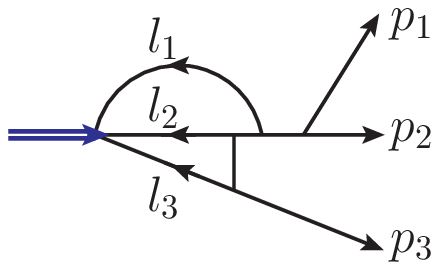}}
    \subfigure[]{\includegraphics[width=0.18\linewidth]{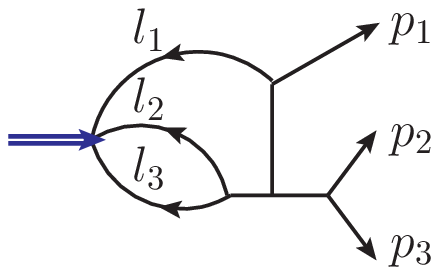}}
    \subfigure[]{\includegraphics[width=0.18\linewidth]{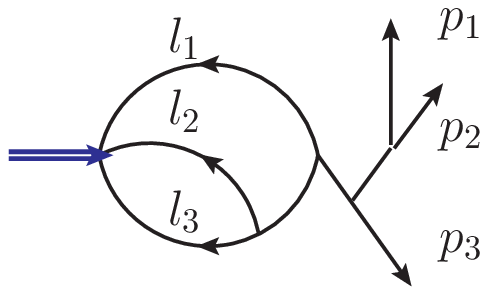}}
    \subfigure[]{\includegraphics[width=0.18\linewidth]{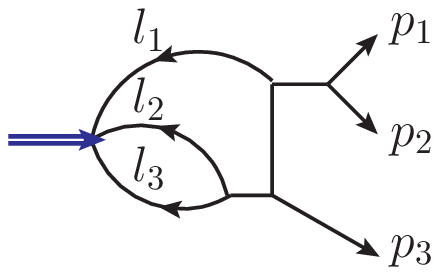}}

    \caption{Trivalent topologies that contribute to BPS form factors. The numerator functions are defined according the labeling of momenta as functions $N_i(p_1,p_2,l_1,l_2)$ for $i\in$\{a,b\}, and $N_i(p_1,p_2, p_3, l_1,l_2,l_3)$ for $i\in$\{c,d,e,f,g,h,i\}. 
    Color factors are defined with the rule that each vertex gives a $\tilde{f}^{abc}$ with legs $a,b,c$ arranged in clockwise order. The four topologies in the first row represent, essentially, the four types of non-vanishing contributions while the topologies in the second row will disappear after integration. }
    \label{fig:CKfinal}
\end{figure}

\begin{enumerate}[leftmargin=*]
    \item \emph{Generating trivalent topologies.} 
    It is convenient to separate the two-loop topologies into two classes: the range-2 integrals and range-3 integrals. Due the difference of color structures, these two classes are decoupled with each other under Jacobi relations, see the discussion in Section~\ref{ssec:ck}.
    Samples of trivalent topologies are given in  Figure~\ref{fig:CKfinal}. Within these figures, only graphs with finally non-vanishing numerators are listed. 
    \item \emph{Dual Jacobi relation and master topology.} 
    
After generating and applying the dual Jacobi relations, the master topologies can be chosen as Figure~\ref{fig:CKfinal}(a) and Figure~\ref{fig:CKfinal}(c). 
    
    A set of dual Jacobi relations are:
    \begin{equation}\label{extrajacob}
    \begin{aligned}
     (1): \quad &  N_{\rm  c}(p_1,p_2,p_3,l_1,l_2,l_3)-N_{\rm  c}(p_3,p_2,p_1,l_3,l_2,l_1)=N_{\rm d}(p_1,p_2,p_3,l_1,l_2,l_3)\\
     (2): \quad &  N_{\rm  c}(p_1,p_2,p_3,l_1,l_2,l_3)-N_{\rm  c}(p_1,p_2,p_3,l_1,l_3,l_2)=N_{\rm e}(p_1,p_2,p_3,l_1,l_2,l_3)\\
     (3): \quad &  N_{\rm  c}(p_1,p_2,p_3,l_1,l_2,l_3)-N_{\rm  c}(p_2,p_1,p_3,l_1,l_2,l_3)=N_{\rm f}(p_1,p_2,p_3,l_1,l_2,l_3)\\
     (4): \quad &  N_{\rm e}(p_1,p_2,p_3,l_1,l_2,l_3)-N_{\rm e}(p_1,p_3,p_2,l_1,l_3,l_2)=N_{\rm g}(p_1,p_2,p_3,l_1,l_2,l_3)\\
     (5): \quad &  N_{\rm e}(p_1,p_2,p_3,l_1,l_2,l_3)-N_{\rm e}(p_2,p_1,p_3,l_1,l_3,l_2)=N_{\rm i}(p_1,p_2,p_3,l_1,l_2,l_3)\\
     (6): \quad & N_{\rm g}(p_3,p_1,p_2,l_1,l_2,l_3)-N_{\rm i}(p_1,p_2,p_3,l_1,l_3,l_2)=N_{\rm h}(p_1,p_2,p_3,l_1,l_2,l_3).
    \end{aligned}
    \end{equation}
   Diagrammatic representation of these relations are given in Figure~\ref{fig:extrajacobi}.
    \begin{figure}
        \centering
        \includegraphics[width=1\linewidth]{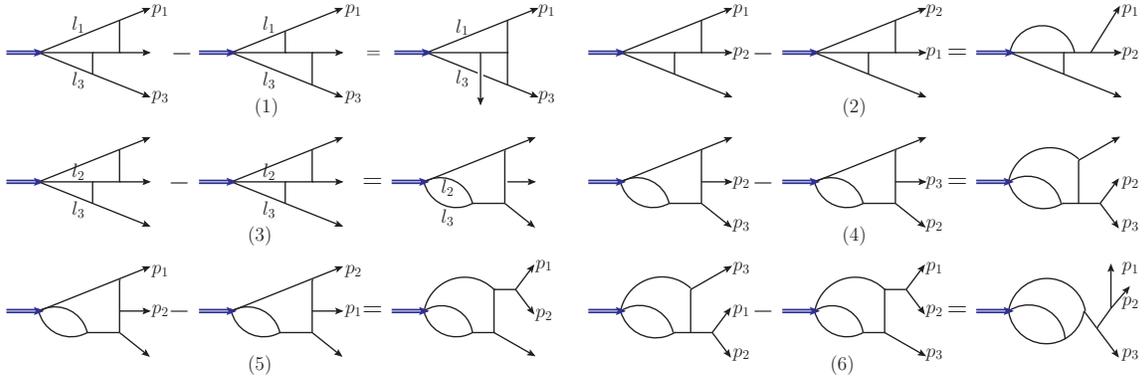}
        \caption{Diagrammatic representation of dual Jacobi relations in \eqref{extrajacob}. }
        \label{fig:extrajacobi}
    \end{figure}
    
    We would like to mention that, among these equations, the last five equations involve certain special integrals as will be discussed later: diagrams (e)-(i) have non-vanishing numerators but actually have zero integrals and hence do not contribute to the final form factor.

    \item \emph{Constraint and ansatz for master numerators.} 
    
    Since the half-BPS form factor considered by us has good UV behavior, one may expect the following power counting properties for the numerators:
    \begin{itemize}
     \item For any $n$-point one-loop sub-diagram, if this sub-diagram does not contains $q^2$ vertex, then the power of loop momenta should not exceed $n-4$.\footnote{Note that we regard $l_i^2$, together with $l_i\cdot p_j$, as power $1$ of $l_i$.}
     \item On the other hand, if this one-loop sub-diagram is a form factor containing the $q^2$ vertex, then the naive power counting shows that power of loop momenta should be less than or equal to $n-3$. 
 \end{itemize} 
    
    This argument can directly exclude sub-bubble or sub-triangle diagrams, except that     the triangle is connected to $q^2$ vertex.  
    Such a naive constraint will set numerators of (e)-(i) in Figure~\ref{fig:CKfinal} to be zero. However, as will be seen later, this constraint is too strong. 
To relax the constraint, we enlarge this ansatz to include terms proportional to $l_k^2$. 
This allows all of the diagrams explicitly listed in Figure~\ref{fig:CKfinal} to be non-vanishing.
    
With these considerations, we propose the following ansatz form for two master numerators: for range-2, the master numerator can be nothing but 
    \begin{equation}
        N_{\rm a}(p_1,p_2)=s_{12}^{2}\,,
    \end{equation}
     and for range-3, we write the numerator as 
    \begin{equation}
    \label{eq:ansatzRange3}
        N_{\rm  c}(p_1,p_2,p_3,l_1,l_2,l_3)=s_{p_1 p_2}s_{q_2 q_3}-\frac{1}{2}s_{p_1 p_2 p_3}s_{p_1 p_2 q_1}+\sum_{i=j}^{9}c_{j} v_{j}\,,
    \end{equation}
where $c_j$ are numerical parameters to be fixed and
\begin{equation}
v_i = \{s_{12} l_1^2 \,, \ s_{13} l_1^2 \,, \ s_{23} l_1^2 \,, \ s_{12} l_2^2 \,, \ s_{13} l_2^2 \,, \ s_{23} l_2^2 \,, \ s_{12} l_3^2 \,, \ s_{13} l_3^2 \,, s_{23} l_3^2  \} \,.
\end{equation}
Note that the ansatz form \eqref{eq:ansatzRange3} is constructed basing on CK-dual form of tree-level six point amplitudes in Table~\ref{tab:cktab-tree}, which is the building block of three-particle triple cut. In this triple-cut channel, only missing terms are terms proportional to $l_{k}^{2}$. 
 
 Through dual Jacobi relations, all other numerators can be obtained.

    \item \emph{Symmetry of numerators.}
        
    We impose the condition that the numerator should have the symmetry inherited from the corresponding topology.
    For range-2 diagrams, the numerators have no free parameter and they also automatically satisfy the symmetry constraint. 

    For the range-3 case, the symmetries of topologies Figure~\ref{fig:CKfinal}(d) and Figure~\ref{fig:CKfinal}(e) are similar to the symmetries of tree-level amplitudes.
    For topology  Figure~\ref{fig:CKfinal}(d) we impose the condition 
    \begin{equation}
    N_{\rm d}(p_1,p_2,p_3,l_1,l_2,l_3)=(-1)^{\sigma}N_{\rm d}(\sigma\{p_1,p_2,p_3\},\sigma\{l_1,l_2,l_3\}), \quad \textrm{for } \forall \sigma \in S_3 \,.
    \end{equation} 
    Symmetry equations for other numerators can also be written down, such as $N_{\rm e}(l_1,l_2,l_3,\ldots)=-N_{\rm e}(l_1,l_3,l_2,\ldots)$, with the minus sign from the minus sign of exchanging two legs connected to the same trivalent vertex.
    
    The above symmetry constraints give three equations as 
    \begin{equation}
    \label{eq:paraeq1}
        2 c_1 = 1 + 2 c_8, \quad c_2 + c_3 = c_7 + c_9, \quad c_1 + c_2 + c_5 = c_4 + c_7 + c_8 \,.
    \end{equation}

    \item \emph{Determining remaining parameters via unitarity.} 
    
    Using another unitarity channel|the planar two-particle triple-cut channel,  we can further  give more constraint on these parameters.
    
    \begin{table}[t]
        \caption{Non-vanishing diagrams $\Gamma$ and their numerators under planar cut to determine remaining parameters. Note that there are two cuts from the last ladder diagram. The sign from color factor have been absorbed into the numerators $N_{(i)}$.}
        \label{tab:ckcut}
   \vskip .4 cm
        \centering
        \begin{tabular}{|c|m{2.4cm}|m{2.4cm}|m{2.4cm}|m{2.4cm}|m{2.4cm}|m{2.4cm}|}
        \hline 
         $\Gamma_{i}$ & $\begin{aligned}
           \includegraphics[width=\linewidth]{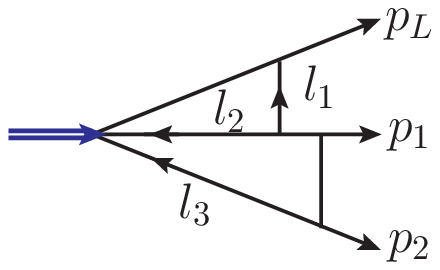}
           \end{aligned}$  & $\begin{aligned}\includegraphics[width=\linewidth]{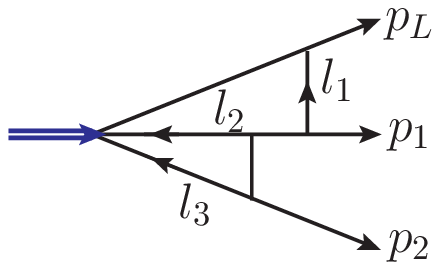}\end{aligned}$ & $\begin{aligned}
           \includegraphics[width=\linewidth]{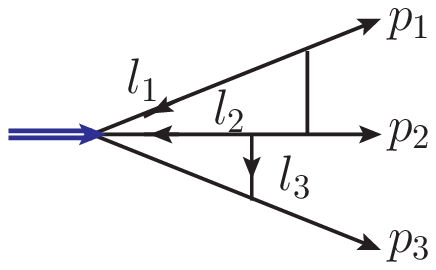}
           \end{aligned}$ & $\begin{aligned}
           \includegraphics[width=\linewidth]{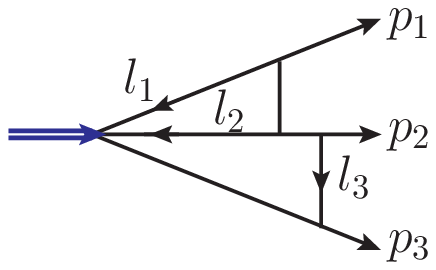}
           \end{aligned}$ & $\begin{aligned}
           \includegraphics[width=\linewidth]{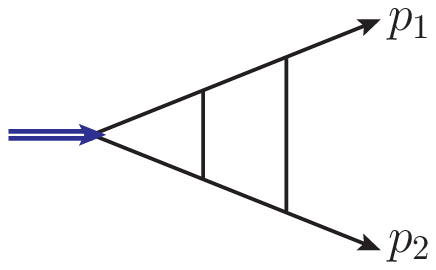}
           \end{aligned}$\\ 
           \hline
           $N_{(i)}$  & \small $+N_{\rm  c}(p_2,p_1$,$ p_L$,$l_3$, $l_2$,$l_1-p_L)$ & \small $ +N_{\rm  c}(p_L,p_1$,$p_2$, $l_1-p_L$, $l_2,l_3)$ & \small $+N_{\rm  c}(p_1,p_2$,$p_3$,$l_1$, $l_2$,$l_3-p_3)$& \small$+N_{\rm  c}(p_3,p_2$,$p_1$, $l_3-p_3$,$l_2$,$l_1)$&  \quad $N_{\rm a}=s_{12}^2$  \\\hline
        \end{tabular}
    \end{table}

    Concretely, we consider planar $s_{12}$ triple-cut (TC) channel as in  Figure~\ref{fig:pa5cut}. 
    Sample diagrams contribute to this cut channel are listed in Table~\ref{tab:ckcut}.
Basing on our discussion in Section~\ref{sec:unitarity}, we act the loop correction with trivalent color factor on a single trace operator and expand the color factor of $\itbf{F}^{(2)}=\itbf{I}^{(2)}\itbf{F}^{(0)}$ via trace basis. Our targeting color factor is $N_{ c}^{2}\operatorname{tr}(a_1 a_2 \cdots a_{L})$, and this cut channel can detect the following terms:
    \begin{equation}
    \begin{aligned}
        &\int \hat{\mathrm{dPS}_{3,l}}\hat{\mathcal{F}}^{(0)}_{L+1}(-l_1,-l_2,-l_3,p_3,\ldots,p_L) \hat{\mathcal{A}}^{(0)}_{5}(l_3,l_2,l_1,p_1,p_2)\\
        &=\left[\left(\itbf{I}_{12}^{(2)}+\itbf{I}_{L12}^{(2)}+\itbf{I}_{123}^{(2)}\right)\operatorname{tr}(a_1 a_2 \cdots a_{L})\right]_{N_{ c}^{2}\operatorname{tr}(a_1 a_2 \cdots)}\Big|_{s_{12}\textrm{-TC}} =\sum_{i=1}^{6} \frac{N_{(i)}}{D_i}\Big|_{s_{12}\textrm{-TC}}\,.
    \end{aligned}
    \end{equation}
     Using this cut constraint, one finds four more equations for these parameters:
     \begin{equation}
     \label{eq:paraeq2}
     c_2=0, \quad c_3=0, \quad 0= -\frac{1}{2} + c_1,\quad \frac{1}{2}  - c_4 + c_5 =0\,.
     \end{equation}
   
    After using \eqref{eq:paraeq1} and \eqref{eq:paraeq2}, there are still two parameters, chosen as $c_4,c_6$. It turns out that
    kinematic terms of these parameters do not change the final results of form factors. Below we will discuss the role of these parameters in more detail.
    
    \item \emph{Final results and discussion}
    
    The final results for the range-3 master numerator are
   \begin{equation}\label{bpsckr3}
       N_{\rm c}(p_1,p_2,p_3,l_1,l_2,l_3)=s_{l_2 l_3}s_{12}-\frac{1}{2}s_{l_1 12 }s_{123} -\frac{1}{2}l_2^2 s_{12}+\frac{1}{2}l_1^2 s_{23} + l_2^2(c_4 s_{13}+c_6 (s_{12}+  s_{23})) \,.
    \end{equation}

    An interesting fact is that numerators $N_j$ with $j \in$ $\{$e,f,g,h,i$\}$ are not equal to zero, as required by the CK duality, but all of the them have actually zero contribution to the form factor.     

     First,  any term proportional to $c_4$ or $c_6$ gives a scaleless integral  that are zero in dimensional regularization. For example, the $l_2^2$ terms for $N_{\rm c}$ in \eqref{bpsckr3} correspond to massless bubbles; and similarly for those terms proportional to $c_{4,6}$ in $N_j$ with $j \in$ $\{$e,f,g,h,i$\}$ ($c_{4,6}$ are cancelled in $N_{\rm b}$). If we set $c_4=c_6=0$, then $N_{\rm i}=0$ for the last diagram in Figure~\ref{fig:CKfinal}. 
    In the following discussion, for simplicity and without changing the form factor results, we present results with $c_{4,6}=0$. 

    Second, for the remaining integrals of topology $j \in$ $\{$e,f,g,h$\}$, they are also all equivalent to scaleless integrals. Alternatively, one may also argue the vanishing of $\{$e,g,h$\}$ by using integral symmetries. For example, for $N_{\rm e}$, the integral is proportional to:
    \begin{equation}\label{eq:bubbps}
        \propto \int {\mathrm{d}^{D}l_{2}}{\mathrm{d}^{D}l_{3} } \frac{(l_3^2-l_2^2)}{l_1^2 l_2^2 l_3^2 (l_2+l_3)^{2}(l_1+p_1)^{2}(l_1+p_1+p_2)^{2}}=0 \,,
    \end{equation}
    which is zero because the numerator $N_{\rm e}$ is anti-symmetric when exchanging $l_2$ and $l_3$ but the denominator and measure does not change under this permutation. The dropping out of topology $\Gamma_{\rm i}$, however, relies solely on zero scaleless integral. 
    
    \end{enumerate}
    
To summarize, the full-color two-loop correction, as given in \eqref{eq:densityCK}, reads 
\begin{equation}\label{eq:compck}
\begin{aligned}
        &\itbf{I}^{(2)}_{123} =\sum_{\sigma\in S_3\times S_3 }\sum_{i={\rm c}}^{\rm h}\int \prod_{j=1}^{2} {\mathrm{d}^{D}l_{i}\over i(\pi)^{D\over2}} \frac{1}{S_{i}^{123}} \sigma \cdot  \frac{\check{\mathcal{C}}^{123}_{i} N_{i}^{123}}{\prod_{a} d_{i, a}} \,,\\
        &\itbf{I}^{(2)}_{12} =\sum_{\sigma \in S_2\times S_2 }\sum_{i={\rm a}}^{\rm b}\int \prod_{j=1}^{2} {\mathrm{d}^{D}l_{i}\over i(\pi)^{D\over2}} \frac{1}{S_{i}^{12}} \sigma \cdot  \frac{\check{\mathcal{C}}^{12}_{i} N_{i}^{12}}{\prod_{a} d_{i, a}} \,,\\
\end{aligned}
\end{equation}
where we summarize the trivalent topology in Figure~\ref{fig:CKfinal} and the corresponding factors in Table~\ref{tab:cktab}. 
Note that the permutations here are fully permuting both internal and external fields, \emph{i.e.}~full permutations of $\{p_1,p_2,p_3\}$ and $\{l_1,l_2,l_3\}$,
considering both internal and external fields are identical.

\begin{table}[t!]
    \caption{Various coefficients for the CK-dual BPS form factor. Trivalent topologies $\Gamma_i$ are given in  Figure~\ref{fig:CKfinal}. Color factors $\mathcal{C}_i$ can be directly read out from diagrams, given the rule that each vertex gives a $f^{abc}$ with legs $a,b,c$ arranged in clockwise order. Note that the last topology (e) actually has zero contribution, we keep it here in order to manifest the CK duality of the numerators. 
    }
    \label{tab:cktab}
   \vskip .4 cm
    \centering
    \begin{tabular}{|l | c| c|}
    \hline
    $\Gamma_i$ &  $N_i$ (with $c_4=c_6=0$) & 
    $S_i$  \\\hline
    (a) \begin{tabular}{cc}   \\ ~  \end{tabular}    & $s_{12}^2$ 
    & 2  \\\hline
    (b) \begin{tabular}{cc}  ~ \\  ~ \end{tabular}    & $s_{12}^2$ 
    & 4 \\\hline
    (c) \begin{tabular}{cc}  ~ \\  ~ \end{tabular}    & $s_{12}s_{l_2 l_3}-\frac{1}{2}s_{123}s_{12 l_1}-\frac{1}{2}l_2^2 s_{12}+\frac{1}{2}l_1^2 s_{23} $ 
    & 1 \\\hline
    (d) \begin{tabular}{cc}  ~ \\  ~ \end{tabular}    & 
    ${1\over2}s_{12}[(s_{2l_2}-s_{1l_1})  + 2p_3\cdot(l_2-l_1)]$+cyc. 
    & 6 \\\hline
    (e) \begin{tabular}{cc}  ~ \\  ~ \end{tabular}    & $\frac{1}{2}s_{12}(l_2^2-l_3^2)$ 
    & 2 \\\hline
    (f) \begin{tabular}{cc}  ~ \\  ~ \end{tabular}    & $\frac{1}{2}(s_{13}-s_{12})l_1^{2}$ 
    & 2 \\\hline
    (g) \begin{tabular}{cc}  ~ \\  ~ \end{tabular}    & $\frac{1}{2}(s_{12}-s_{13})(l_2^{2}-l_3^{2})$ 
    & 2 \\\hline
    (h) \begin{tabular}{cc}  ~ \\  ~ \end{tabular}    & $N_{g}$
    & 4 \\\hline
    \end{tabular}
\end{table}

Last but not least, let us compare the above result obtained using CK duality with the unitarity result in Section~\ref{ssec:2lbps}. An obvious difference is that for range-2 densities, the CK result in \eqref{eq:compck} contains only ladder integrals while in \eqref{bpsr21} and \eqref{bpsr22} there are also another type of integral $\mathcal{I}^{\prime}(12) \equiv \hskip -3pt
\begin{aligned}
\includegraphics[height=0.08\linewidth]{fig/tritri2pt1top.eps}
\end{aligned}
$.  
Calculation of the difference between the CK  and the unitarity results shows that the difference $\Delta\itbf{I}\equiv \itbf{I}^{\rm CK}-\itbf{I}^{\rm unitarity}$ can be given as:\footnote{To get this result, one may need to expand CK results on the trivalent basis in Section~\ref{ssec:nbasis2l} using the equations like \eqref{eq:d1md1a} in Appendix~\ref{ap:ddm}. Note that $N_{\rm d}$ is actually equivalent to the one we obtained in Sect.~\ref{ssec:2lbps}.}
\begin{align}
    &\Delta\itbf{I}=\nonumber \\
    &\sum_{i<j<k}\bigg\{\sum_{\substack{\sigma^{i}\in \mathbb{Z}_3\\ \sigma^{e}\in S_3}} \check{\mathcal{D}}_{\Delta_{1}}(\sigma^{e},\sigma^{i})\ \sigma^{e}\cdot \left(\mathcal{I}^{\prime} ({ij})+\mathcal{I}^{\prime} ({jk})\right)- \sum_{ \sigma^{e}\in S_3} \check{\mathcal{D}}_{\Delta_{2}}(\sigma^{e}) \left(\mathcal{I}^{\prime}({ij})+\mathcal{I}^{\prime} ({jk})+\mathcal{I}^{\prime} ({ik})\right)\bigg\}\nonumber \\
    &+ 2 \sum_{i<j} \sum_{\sigma^{e}\in S_2}\Big(\check{\mathcal{D}}_{\Delta_{3}}(\sigma^{e})-2 \check{\mathcal{D}}_{\Delta_{4}}( \mathbf{1})\Big) \mathcal{I}^{\prime} ({ij})\,, \label{eq:redistribution}
\end{align}
where $\mathcal{I}^{\prime}({ij})$ is a range-2 function and origins from a redistribution between range-2 and range-3 densities by comparing $\itbf{I}^{\rm CK}$ and $\itbf{I}^{\rm unitarity}$:
\begin{equation}
    \int \frac{\mathrm{d}^{D}l_2}{i(\pi)^{D\over 2}}\frac{\mathrm{d}^{D}l_3}{i(\pi)^{D\over 2}}\frac{ s_{23}}{ l_1^2 l_2^2 l_3^2 (l_1+p_i)^2(l_3+p_j)^2}
   =\frac{1}{2}\begin{aligned}
       \includegraphics[width=0.15\linewidth]{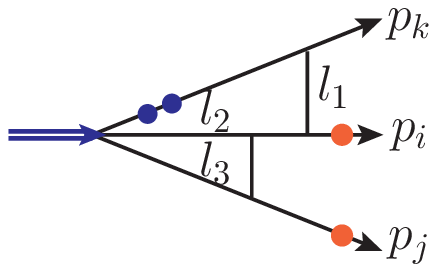}
    \end{aligned}=\begin{aligned}
       \includegraphics[width=0.15\linewidth]{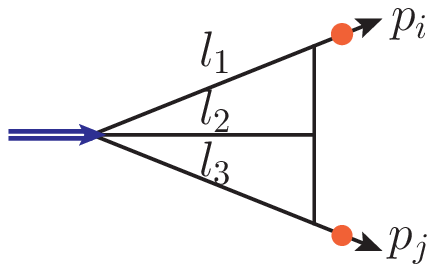}
    \end{aligned} \,.
\end{equation}
One can check that
\begin{equation}
   \Delta \itbf{I} \cdot \itbf{F}^{(0)}=0 \,,
\end{equation}
with arbitrary $\itbf{F}^{(0)}$, regardless of the length and structure of corresponding (BPS) operator. In conclusion, we see that the two forms obtained in this section and the last section are equivalent.
Similar identities will also be used in next section when we check IR divergence cancellation.

\section{IR, UV divergences and finite remainder}\label{sec:iruv}

Having obtained the full-color form factor integrands in the last two sections, one can study the properties of integrated results. 
The bare form factors contain infrared (IR) and ultraviolet (UV) divergences. Since we consider the full-color results, both IR and UV structures depend on the color structure. 
The similarity and difference between planar and non-planar results will be a main focus of our discussion in this section. 

As a brief outline, we will first review the planar IR structure and then generalize it to full-color case. 
Then we discuss the finite remainder functions of the BPS form factor, where we find a new non-planar contribution to the maximally transcendentality part. Third, we consider the SU(2) non-BPS form factor, with an emphasis on the new feature of UV divergences and the non-planar dilatation operator. In the end, we highlight those new non-planar features and discuss their origination as a summary of our non-planar discussion. 
Some technical details are given in Appendix~\ref{app:IRUVcancel}.

\subsection{Infrared structure with full-color}\label{ssec:ir}

Infrared divergences originate from integral configurations where loop momenta become soft or collinear. Such singularities are universal in the sense that they only depend on the external on-shell particles and are irrelevant to the type of operators in the form factors. Since the BPS form factor contains no UV divergences, it provides a clean object to understand the IR structure, which we focus on in this subsection.

In the planar limit, the color structure is trivial and IR divergences can be simply captured by the BDS ansatz part \cite{Bern:2005iz}, introduced originally for planar ${\cal N}=4$ amplitudes. On the other hand, for full-color form factors, the color mixing effect generates new interesting structure. In particular, there are new IR contributions at ${\cal O}(1/\epsilon)$ order which involve three external legs at the same time.
Such terms are the so-called $\itbf{H}^{(2)}$ functions first obtained in the four-point amplitudes \cite{Bern:2002tk,Bern:2003ck} 
and then derived for general $n$-point amplitudes in \cite{Bern:2004cz}. They satisfy the color-dipole formula which determines the complete IR singularities up to two loops \cite{Becher:2009cu, Gardi:2009qi}. Our computation of $n$-point form factors will provide an independent concrete check for the dipole formula of amplitudes and form factors that involve arbitrary number of external legs.

Below we first briefly review the planar IR structure and then consider the generalization to full-color dependence. 

\subsubsection*{Planar IR structure and BDS ansatz formula}

At planar level, IR structure has a very simple form because all the Feynman diagrams are planar so that all the internal lines are confined to the wedges formed between two adjacent hard lines. As a result, planar IR divergences can be simply captured by the two-point Sudakov form factor \cite{Mueller:1979ih,Collins:1980ih,Sen:1981sd}. 
Analysis via renormalization group shows that IR divergence can be given as a simple exponential form \cite{Magnea:1990zb, Bern:2005iz}:
\begin{equation}
\label{bds}
\log \underline{\mathcal{I}}=-\sum_{l=1}^{\infty} g^{2 l}\left[\frac{\gamma_{\mathrm{cusp}}^{(l)}}{(2 l \epsilon)^{2}}+\frac{\mathcal{G}_{\mathrm{coll}}^{(l)}}{2 l \epsilon}\right] \sum_{i=1}^{n}\left(-s_{i i+1}\right)^{-l \epsilon} \cdot \mathbf{1}+\mathcal{O}\left(\epsilon^{0}\right) ,
\end{equation}
where $\underline{\mathcal{I}}^{(l)}$ is $l$ loop renormalized planar loop correction (for BPS form factor, $\mathcal{I}=\underline{\mathcal{I}}$), $\gamma_{\rm cusp}$ is cusp anomalous dimension \cite{Korchemsky:1985xj, Korchemsky:1988si}
and $\mathcal{G}_{\rm coll}$ is collinear anomalous dimension \cite{Cachazo:2007ad, Dixon:2017nat}.\footnote{The non-planar corrections of cusp and collinear anomalous dimensions only start at fourth loop order. The analytic expression of the non-planar ${\cal N}=4$ cups anomalous dimension were obtained based on both Wilson loop \cite{Henn:2019swt} and form factor \cite{Huber:2019fxe, Boels:2017skl} computations. The ${\cal N}=4$ non-planar collinear anomalous dimension is so far only known numerically \cite{Boels:2017ftb}.}
Note that the trivial $N_{\mathrm{c}}$ dependence are included in $g^{2l}$; for the non-planar case, we will use $g_{\rm YM}$ rather than $g$ and explicitly write down the $N_{\mathrm{c}}$ dependence.  

It is convenient to rewrite \eqref{bds} as the BDS form  \cite{Bern:2005iz}, and at two-loop level:
\begin{equation}\label{bds2l}
\underline{\mathcal{I}}^{(2)} 
= \frac{1}{2}\left(\underline{\mathcal{I}}^{(1)}(\epsilon)\right)^{2} + f^{(2)}(\epsilon) \underline{\mathcal{I}}^{(1)}(2 \epsilon) + \mathcal{R}^{(2)}+\mathcal{O}(\epsilon) ,
\end{equation}
where the divergences are all captured in the first two terms and are determined by the one-loop correction $\mathcal{I}^{(1)}$, together with a universal kinematic independent quantity 
\begin{equation}
f^{(2)}(\epsilon)=-2 \zeta_{2}-2 \zeta_{3} \epsilon-2 \zeta_{4} \epsilon^{2} \,,
\end{equation}
which contains the information of two-loop cusp and collinear anomalous dimensions.
The remaining function $\mathcal{R}^{(2)}$ is two-loop finite remainder function.


As a pure side remark, one may compare the BDS subtraction with the Catani IR subtraction \cite{Catani:1998bh} (see also \cite{Sterman:2002qn}): 
\begin{align}
{\cal F}^{(1)} &= I^{(1)}(\epsilon) {\cal F}^{(0)} + {\cal F}^{(1),{\rm fin}} + {\cal O}(\epsilon) \,,  \\
{\cal F}^{(2)} &= I^{(2)}(\epsilon) {\cal F}^{(0)} +  I^{(1)}(\epsilon) {\cal F}^{(1)} + {\cal F}^{(2),{\rm fin}} + {\cal O}(\epsilon)   \,,
\end{align}
where
\begin{align}
I^{(1)}(\epsilon) = & - {e^{\gamma_E \epsilon} \over \Gamma(1-\epsilon)}  \frac{1}{\epsilon^2} \sum_{i=1}^n \Big(-{s_{i,i+1} \over \mu^2} \Big)^{-\epsilon} \,, \\
I^{(2)}(\epsilon) = & - {1\over2} \big[ I^{(1)}(\epsilon) \big]^2 - {e^{-\gamma_E \epsilon} \Gamma(1-2\epsilon) \over \Gamma(1-\epsilon)}  {\pi^2\over3} I^{(1)}(2\epsilon) + n {e^{\gamma_E \epsilon} \over \epsilon \Gamma(1-\epsilon)} \frac{\zeta_3}{2}  \,.
\end{align}
Although the original Catani subtraction is used in QCD, it can be applied here to ${\cal N}=4$ SYM theory by simply dropping lower transcendental pieces (including the beta function contribution).
Note that BDS and Catani subtractions are equivalent for the divergence parts, but they have a (scheme-changing) difference in the finite remainders.

\subsubsection*{Non-planar IR structure and dipole formula}

By dressing color factors, a generalization of \eqref{bds2l} is \cite{Bern:2002tk,Bern:2003ck, Bern:2004cz}
\begin{equation}\label{eq:bds2lnp}
\underline{\itbf{I}}^{(2)}
=\frac{1}{2}\Big(\underline{\itbf{I}}^{(1)}(\epsilon) \Big)^{2} + \tilde{f}^{(2)}(\epsilon) \underline{\itbf{I}}^{(1)}(2 \epsilon) + \itbf{H}^{(2)}(\epsilon) + \itbf{R}^{(2)}+\mathcal{O}(\epsilon) \,,
\end{equation}
which contains a new function  $\itbf{H}^{(2)}$ with only simple pole in $\epsilon$.
In the planar limit, \eqref{eq:bds2lnp} should reproduce the planar IR structure \eqref{bds}, therefore, the new extra contribution $\itbf{H}^{(2)}$ should be a pure non-planar effect.
This form can be understood based on the \emph{dipole-formula} (see \cite{Becher:2009cu, Gardi:2009qi} for further discussion): 
\begin{equation}\label{dip}
\Gamma_{n}^{\mathrm{dip.}}\left(\left\{p_{i}\right\}, \mu\right)=-\frac{1}{2} \gamma_{\text{cusp}}(g_{\rm YM};N_{\mathrm{c}})\sum_{i<j} \log \left(\frac{-s_{i j}}{\mu^{2}}\right) \hat{\mathbf{T}}_{i} \cdot \hat{\mathbf{T}}_{j}+\sum_{i=1}^{n} \mathcal{G}_{\text{coll},i}(g_{\rm YM};N_{\mathrm{c}})\,.
\end{equation}

Note that the coupling constant before $\itbf{I}^{(2)}$ should be $\tilde{g}^{4}$ rather than $g^4$ (see also \eqref{eq:loopexpPLffnp}):
\begin{equation}
\tilde{g}^{2}=\frac{g_{\mathrm{YM}}^{2} }{(4 \pi)^{2}}\left(4 \pi e^{-\gamma_{\mathrm{E}}}\right)^{\epsilon} \,, \qquad \tilde{g}^2 = g^2/ N_{ c} \,.
\end{equation} 
Correspondingly, $\tilde{f}^{(2)}= N_{\mathrm{c}} f^{(2)}$ where the $N_{\mathrm{c}}$ factor comes from $g^2/\tilde{g}^{2}$. 
Below we justify this formula and derive the explicit form of $\itbf{H}^{(2)}$  using the BPS form factor results.

\subsubsection*{One-loop square}

To perform IR subtraction as in \eqref{eq:bds2lnp}, one needs to compute the square of full-color one-loop correction
$\itbf{I}^{(1)} = \sum_{i_1<i_2}\itbf{I}^{(1)}_{i_1 i_2}$ which is reviewed in Appendix~\ref{ap:1l}.
The one-loop square has the following structure 
\begin{equation}\label{1lsqdef}
\begin{aligned}
    \left(\itbf{I}^{(1)}\right)^{2}=&\sum_{i_1<i_2,i_3<i_4}\itbf{I}^{(1)}_{i_1 i_2}\itbf{I}^{(1)}_{i_3 i_4}
    =    \left.\big(\itbf{I}^{(1)}\big)^{2}\right|_{\textrm{range-2}}+\left.\big(\itbf{I}^{(1)}\big)^{2}\right|_{\textrm{range-3}}+\left.\big(\itbf{I}^{(1)}\big)^{2}\right|_{\textrm{range-4}} \,,
\end{aligned}
\end{equation}
where the range is defined in terms of length of $\{i_1,i_2\}\bigcup\{i_3,i_4\}$. For example: 
\begin{equation}
\begin{aligned}
    \left.\big(\itbf{I}^{(1)}\big)^{2}\right|_{\textrm{range-2}}=\sum_{i_{1}<i_{2}}\big(\itbf{I}^{(1)}\itbf{I}^{(1)}\big)_{i_1 i_2} \,,\qquad 
    \left.\big(\itbf{I}^{(1)}\big)^{2}\right|_{\text{range-3}}=\sum_{i_{1}<i_{2}<i_{3}}\big(\itbf{I}^{(1)}\itbf{I}^{(1)}\big)_{i_1 i_2 i_3} \,.
\end{aligned}
\end{equation}
The color factor of range-2 and range-3 interactions here can also be expanded on the trivalent basis of loop correction densities, \emph{i.e.}~$\check{\mathcal{D}}_{j}$ defined in \eqref{eq:set1}-\eqref{eq:set4}. Although the detailed definition of one-loop square is a little bit subtle which will be clarified in Appendix~\ref{ap:bps}, the final result is clear and simple:
\begin{itemize}[leftmargin=*]
        \item[$\square$] \textbf{Range-2}\\
    \begin{equation}\label{1lsqr2}
    \begin{aligned}
        &\big(\itbf{I}^{(1)}\itbf{I}^{(1)}\big)_{12}=\sum_{j=25}^{27}\check{\mathcal{D}}_{j}\left[\big(\mathcal{I}^{(1)}\mathcal{I}^{(1)}\big)(p_1,p_2)\right]_{\check{\mathcal{D}}_{j}} ,\\
        &\left[\big(\mathcal{I}^{(1)}\mathcal{I}^{(1)}\big)(p_1,p_2)\right]_{\check{\mathcal{D}}_{25}}=\mathcal{I}^{(1)}(12)^2 \,, \qquad \left[\big(\mathcal{I}^{(1)}\mathcal{I}^{(1)}\big)(p_1,p_2)\right]_{\check{\mathcal{D}}_{27}}=0  \,.
    \end{aligned}
    \end{equation}
    \item[$\square$] \textbf{Range-3}\\
    \begin{align}\label{1lsqr3}
    \nonumber
        &\big(\itbf{I}^{(1)}\itbf{I}^{(1)}\big)_{123}=\sum_{j=1}^{24}\check{\mathcal{D}}_{j}\left[\big(\mathcal{I}^{(1)}\mathcal{I}^{(1)}\big)(p_1,p_2,p_3)\right]_{\check{\mathcal{D}}_{j}} , \\
        &\left[\big(\mathcal{I}^{(1)}\mathcal{I}^{(1)}\big)(p_1,p_2,p_3)\right]_{\check{\mathcal{D}}_{1}}=2\  \mathcal{I}^{(1)}(12)\mathcal{I}^{(1)}(23) \,, \\ 
        \nonumber
        &\left[\big(\mathcal{I}^{(1)}\mathcal{I}^{(1)}\big)(p_1,p_2,p_3)\right]_{\check{\mathcal{D}}_{19}}=-\mathcal{I}^{(1)}(12)\mathcal{I}^{(1)}(23) - \mathcal{I}^{(1)}(23)\mathcal{I}^{(1)}(13) - \mathcal{I}^{(1)}(13)\mathcal{I}^{(1)}(12) \,.
    \end{align}
\end{itemize}
Note that the range-4 part is straightforward to write down and we will not give them here. Also, in the context of BPS form factors $\mathcal{I}^{(1)}(ij)=\text{I}_{\rm tri}[s_{ij}]$ with $\text{I}_{\rm tri}$ in Appendix~\ref{ap:1l}.

\subsubsection*{Verification of IR structure}

Using two-loop form factor results \eqref{bpsr21}-\eqref{bpsr32}, and the expression for one-loop-square \eqref{1lsqdef},  
we can now apply \eqref{eq:bds2lnp} to check the cancellation of IR singularity.\footnote{The computation of the integrated result is standard: one can first perform IBP reduction for the integrand using public packages, (see \emph{e.g.}~\cite{Smirnov:2008iw, Lee:2013mka, Maierhoefer:2017hyi}), and the master integrals in our problem are all known in terms of 2d harmonic polylogarithms \cite{Gehrmann:2000zt, Gehrmann:2001jv}.}
The dipole formula implies that the divergent part of the following quantity 
\begin{equation}\label{eq:partsubtract}
    {\underline{\itbf{I}}}^{(2)}-\frac{1}{2}\left({\underline{\itbf{I}}}^{(1)}(\epsilon)\right)^{2}-\tilde{f}^{(2)}(\epsilon) {\underline{\itbf{I}}}^{(1)}(2 \epsilon)
\end{equation}
is expected to have $\epsilon$ simple pole only, which we will verify by direct calculation.

The remaining $1/\epsilon$ divergent terms are summarized in the function $\itbf{H}^{(2)}$, whose kinematic dependence is only polynomial of $\log s_{ij}$. More precisely, we find the following form:
\begin{equation}\label{nir}
\begin{aligned}
    \itbf{H}^{(2)}&=\sum_{i<j<k} \itbf{H}^{(2)}_{ijk} \,, \\
    \itbf{H}^{(2)}_{123}&=\bigg(\sum_{\sigma \in S_{3} }(-1)^{\sigma}\check{\mathcal{D}}_{\Delta_{2}}(\sigma)\bigg)\frac{1}{4\epsilon}\mathrm{log}\left(\frac{-s_{12}}{-s_{23}}\right)\mathrm{log}\left(\frac{-s_{23}}{-s_{13}}\right)\mathrm{log}\left(\frac{-s_{13}}{-s_{12}}\right)\,,
\end{aligned}
\end{equation}
where $\sigma$ represents permutation of $p_1,p_2,p_3$, and these permutations do not change the triple-log term up to a sign. Note that these terms have only components on the non-planar set of color factors $\check{\mathcal{D}}_{\Delta_2}(\sigma)$. This is equivalent to the known result, see \emph{e.g.}  \cite{Aybat:2006mz}.
More details of the cancellation and derivation are given in Appendix~\ref{ap:bps}, and we will try to understand the $\itbf{H}^{(2)}$ functions in Section~\ref{ssec:hterm}.

\subsection{BPS remainder}\label{ssec:2lrem}

After subtracting IR divergences, one obtains the finite remainder functions of full-color BPS  form factors. 
As we will see, the non-planar BPS remainder receives a  new contribution in maximal transcendentality part.

\subsubsection*{A brief review of planar remainder}

We first review the planar remainder, which is discussed in detail in \cite{Brandhuber:2014ica}. Planar remainder can be expressed as summing remainder density functions as
    \begin{equation}\label{eq:rembasis}
        \mathcal{R}^{\pln}=\sum_{i} \mathcal{R}^{\pln}_{i (i+1) (i+2)}\,,
    \end{equation}
and we recall the explicit form of remainder remainder here:
    \begin{equation}
        \mathcal{R}^{\pln}_{123}=\mathcal{R}_{\rm basis}(u,{v},{w})+(u\leftrightarrow {v})+\zeta_4\,,
    \end{equation}
where
\begin{equation}
    u=\frac{s_{12}}{s_{123}},\quad {v}=\frac{s_{23}}{s_{123}},\quad {w}=\frac{s_{13}}{s_{123}}\,,
\end{equation}
and $\mathcal{R}_{\rm basis}(u,{v},{w})$ is 
    \small
    \begin{equation}\label{eq:bpsrem}
    \begin{aligned}
    &G\left(\left\{1-u, 1-u, 1,0\right\}, {v}\right)-\mathrm{Li}_{4}\left(1-u\right)-\mathrm{Li}_{4}\left(u\right)+\mathrm{Li}_{4}\left(\frac{u-1}{u}\right)\\
    &-\log \left(\frac{1-u}{{w}}\right)\left[\operatorname{Li}_{3}\left(\frac{u-1}{u}\right)-\mathrm{Li}_{3}\left(1-u\right)\right]-\log \left(u\right)\left[\mathrm{Li}_{3}\left(\frac{{v}}{1-u}\right)+\mathrm{Li}_{3}\left(-\frac{{w}}{{v}}\right)+\mathrm{Li}_{3}\left(\frac{{v}-1}{{v}}\right)\right] \\
    &-\mathrm{Li}_{2}\left(\frac{u-1}{u}\right) \mathrm{Li}_{2}\left(\frac{{v}}{1-u}\right)+\mathrm{Li}_{2}\left(u\right)\left[\log \left(\frac{1-u}{{w}}\right) \log \left({v}\right)+\frac{1}{2} \log ^{2}\left(\frac{1-u}{{w}}\right)\right] \\
    &+\frac{1}{24} \log ^{4}\left(u\right)-\frac{1}{6} \log ^{3}\left(u\right) \log \left({w}\right) +\frac{1}{3} \log \left(u\right) \log ^{3}\left({v}\right)+\frac{1}{3} \log \left(u\right) \log ^{3}\left(1-u\right) \\
    &-\frac{1}{8} \log ^{2}\left(u\right) \log ^{2}\left({v}\right)-\frac{1}{2} \log ^{2}\left(1-u\right) \log \left(u\right) \log \left(\frac{{w}}{{v}}\right) -\frac{1}{2} \log \left(1-u\right) \log ^{2}\left(u\right) \log \left({v}\right)\\
    &-\zeta_{2}\left[\log \left(u\right) \log \left(\frac{1-{v}}{{v}}\right)+\frac{1}{2} \log ^{2}\left(\frac{1-u}{{w}}\right)-\frac{1}{2} \log ^{2}\left(u\right)\right] +\zeta_{3} \log \left(u\right)\,,
    \end{aligned}
    \end{equation}
\normalsize
where $G$ is a Goncharov polylogarithm. 

To get this compact expression, symbol (denoted as $\mathcal{S}$) is a powerful tool, see  \cite{Goncharov:2010jf, goncharov2009simple}. 
We briefly review the strategy of simplifying remainders with the aid of symbols below.
\begin{enumerate}[leftmargin=*]
    \item To start with, determine terms can not be expressed with classical polylogarithms. If a transcendentality-4 function $f$ satisfies Goncharov criterion \cite{Goncharov:2010jf}
    \begin{equation}
    \label{eq:ConcharovCrit}
        \delta \circ \mathcal{S}(f)\big|_{\Lambda^{2}B_2} \equiv \mathcal{S}_{a b c d}-\mathcal{S}_{b a c d}-\mathcal{S}_{a b d c}+\mathcal{S}_{b a d c}-(a \leftrightarrow c, b \leftrightarrow d)=0\,,
    \end{equation}
then it can be re-written in terms of classical polylogarithms, \emph{i.e.} $\log(x),\text{Li}_{k}(x)$.
Planar remainder density $\mathcal{R}^{\scriptscriptstyle \rm PL}_{123}$ does not pass this criterion and one can use $G(\{1-u,1-u,1,0\},{v})$ to capture terms beyond classical polylogarithms, see \cite{Brandhuber:2014ica} for more details.
    \item In order to determine the classical polylogrithm part, one can use an ansatz as linear combination of all possible polynomial of classical polylogarithms with transcendentality-4 and solve the ansatz at symbol level. 
    \item Finally, one can fix the terms ignored by symbols, which are linear combination of  functions $\zeta_2 \log(x)\log(y)$, $\zeta_2 \text{Li}_2(x)$, $\zeta_3 \log(x)$ and $\zeta_4$ in transcendentality-4 case, for example numerically. 
\end{enumerate}

Let us comment on the definition of remainder at density level. 
In the planar case, one can sum up range-2 and range-3 form factor densities to define remainder density in \eqref{eq:rembasis} as 
\begin{align}
\mathcal{R}^{\pln}_{123} = & \ {\cal I}_{123}^{(2),{\scriptscriptstyle \rm PL}} + {1\over2} \Big( {\cal I}_{12}^{(2),{\scriptscriptstyle \rm PL}} + {\cal I}_{23}^{(2),{\scriptscriptstyle \rm PL}} \Big) - 
{1\over2} \Big[ {1\over2} \big( {\cal I}_{12}^{(1),{\scriptscriptstyle \rm PL}}\big)^2 +  {1\over2} \big({\cal I}_{23}^{(1),{\scriptscriptstyle \rm PL}}\big)^2 + 2{\cal I}_{12}^{(1),{\scriptscriptstyle \rm PL}}{\cal I}_{23}^{(1),{\scriptscriptstyle \rm PL}} \Big] \nonumber
 \\
& -\tilde{f}^{(2)}(\epsilon) \Big( {\cal I}_{12}^{(1),{\scriptscriptstyle \rm PL}}+ {\cal I}_{23}^{(1),{\scriptscriptstyle \rm PL}}\Big) \big|_{\epsilon \rightarrow 2 \epsilon} \,.
\label{eq:planar-density-subtraction}
\end{align}
For the non-planar case, \eqref{eq:bds2lnp} may not hold directly at density level.
Below we would like to introduce the density of remainders for range-3 and range-2 separately. To achieve this, we need to introduce some auxiliary divergent quantities which appear in remainder densities. These auxiliary quantities, in the end, will be shown to cancel in the full form factor, after summing over all densities. 
Such a definition without adding range-3 and range-2 together will make it easier to deal with the non-planar contribution and also help to understand the structure better.

For the convenience of discussion below, we introduce $\mathcal{I}^{(2)}_{\rm BPS,a}(p_1,p_2,p_3)$, or denoted as $\mathcal{I}^{(2)}_{\rm BPS,a}(123)$, to represent  the terms in parenthesis in \eqref{eq:bpsr32new} or  \eqref{eq:bpsr31new}:
\begin{equation}
\label{eq:IBPSa}
    \mathcal{I}_{\text{BPS,a}}^{(2)}(123)=\begin{aligned}
       \includegraphics[width=0.15\linewidth]{fig/tribox2top.eps}
    \end{aligned}
    \hlmath{5}{-\frac{1}{2}}
    \begin{aligned}
       \includegraphics[width=0.15\linewidth]{fig/tritri2top.eps}
    \end{aligned}\,,
\end{equation}
as well as the following ones for range-2:
\begin{equation}
    \mathcal{I}_{\text{BPS,b}}^{(2)}(12)=\begin{aligned}
          \includegraphics[width=0.15\linewidth]{fig/pladderm1top.eps}
        \end{aligned}\hlp{5}\begin{aligned}
          \includegraphics[width=0.135\linewidth]{fig/tritri2pt1top.eps}
        \end{aligned}
        \hskip -8pt ,
        \quad 
        \mathcal{I}_{\text{BPS,c}}^{(2)}(12)=\ \frac{1}{2}\hskip -3pt \begin{aligned}
          \includegraphics[width=0.15\linewidth]{fig/npladdertop.eps}
        \end{aligned}\hlmath{5}{-2}\begin{aligned}
          \includegraphics[width=0.135\linewidth]{fig/tritri2pt1top.eps}
        \end{aligned} .
\end{equation}

\subsubsection*{Full-color remainder densities}

First we provide  a definition for color-stripped remainder density for the range-3 with color factor ${\check{\mathcal{D}}_1}$.  Following \eqref{eq:partsubtract}, we find that
\begin{equation}
	\left(\left[\mathcal{I}^{(2)}({123})\right]_{\check{\mathcal{D}}_1}-\frac{1}{2}\left[ \big(\mathcal{I}^{(1)}\mathcal{I}^{(1)}\big)({123})\right]_{\check{\mathcal{D}}_1}\right)_{\rm div}=U(12)+U(23)\,,
\end{equation}
where $U(ij)$ are the auxiliary divergences mentioned above, and are defined as
\begin{equation}\label{eq:fcan}
	U(ij)= \left(-\frac{\zeta_2}{2\epsilon^2}-\frac{2\zeta_3}{\epsilon}\right)(-s_{ij})^{-2\epsilon } \,.
\end{equation} 
The corresponding remainder density can be defined as 
\begin{equation}\label{eq:Rr3d1}
    \left[\mathcal{R}^{(2)}(123)\right]_{\check{\mathcal{D}}_1}=\left(\mathcal{I}^{(2)}_{\rm BPS,a}(123)-\frac{1}{2}\mathcal{I}^{(1)}(12)\mathcal{I}^{(1)}(23)- U(12) \right) + (p_1 \leftrightarrow p_3)\,,
\end{equation}
where we have used explicit expression in \eqref{eq:bpsr31new}, \eqref{1lsqr3} and \eqref{eq:fcan}.

Similarly, the range-2 density remainder associated with ${\check{\mathcal{D}}_{25}}$ can be defined as 
\begin{equation}\label{eq:Rr2d25}
	 \left[\mathcal{R}^{(2)}(12)\right]_{\check{\mathcal{D}}_{25}}=\mathcal{I}_{\text{BPS,b}}^{(2)}(12)-\frac{1}{2}\left(\mathcal{I}^{(1)}(12)\right)^{2}-(-2)U(12)-f^{(2)}(\epsilon)\big(\mathcal{I}^{(1)}(12)\big)_{\epsilon\shortrightarrow 2\epsilon}\,,
\end{equation}
and it is easy to check that $ \big[\mathcal{R}^{(2)}(12)\big]_{\check{\mathcal{D}}_{25}}=19\zeta_4/2$ is just a pure number. 

Note that the above  remainder densities in \eqref{eq:Rr3d1} and \eqref{eq:Rr2d25} are different from planar definition $\mathcal{R}_{123}^{\pln}$ in \eqref{eq:planar-density-subtraction}, since the latter is a combination of range-3 and range-2 densities. One can relate $\mathcal{R}_{123}^{\pln}$ to the densities in \eqref{eq:Rr3d1} and \eqref{eq:Rr2d25}:
\begin{equation}\label{eq:Rpl}
   \begin{aligned}
	\mathcal{R}^{\pln}_{123}=&\left[\mathcal{R}^{(2)}(123)\right]_{\check{\mathcal{D}}_{1}}+\frac{1}{2} \left[\mathcal{R}^{(2)}(12)\right]_{\check{\mathcal{D}}_{25}}+\frac{1}{2} \left[\mathcal{R}^{(2)}(23)\right]_{\check{\mathcal{D}}_{25}}\\
	=&\bigg[
	\mathcal{I}_{\text{BPS,a}}^{(2)}(123)
    +\frac{1}{2}
    \mathcal{I}_{\text{BPS,b}}^{(2)}(12)
        \\
    &-\frac{1}{2}\left(\mathcal{I}^{(1)}(12)\mathcal{I}^{(1)}(23)+\frac{1}{2}\left(\mathcal{I}^{(1)}(12)\right)^2\right)-\frac{1}{2} f^{(2)}(\epsilon)\big(\mathcal{I}^{(1)}(12)\big)_{\epsilon\shortrightarrow 2\epsilon}\bigg]+(p_1\leftrightarrow p_3)\,.
\end{aligned}
\end{equation}
In this combination one can see  clearly that the $U(12)$ and $U(23)$ cancel between range-2 and range-3.
Also, it is straightforward to get
\begin{equation}
    \left[\mathcal{R}^{(2)}(123)\right]_{\check{\mathcal{D}}_{1}}
    =\mathcal{R}_{\rm basis}(u,v,w)+(u\leftrightarrow v)+20\zeta_4\,.
\end{equation}

Now we turn to the calculation of the remainder associated to non-planar color factors, \emph{i.e.}~$\big[\mathcal{R}^{(2)}\big]_{\check{\mathcal{D}}_{19}}$ and  $\big[\mathcal{R}^{(2)}\big]_{\check{\mathcal{D}}_{27}}$. The range-3 non-planar remainder can be defined as:
\begin{align}
    \left[\mathcal{R}^{(2)}(123)\right]_{\check{\mathcal{D}}_{19}}
    =&\left(\mathcal{I}^{(2)}_{\rm BPS,a}(123)
    +\frac{1}{2}\mathcal{I}^{(1)}(12)\mathcal{I}^{(1)}(23)+U(12)\right) +\text{cyc}(p_1,p_2,p_3) \nonumber\\
    &-\frac{1}{4\epsilon}\log\left({-s_{12}\over- s_{23}}\right)\log\left({-s_{23}\over -s_{13}}\right)\log\left({-s_{13}\over -s_{12}}\right)\,,
    \label{eq:Rr3d19}
\end{align}
where the last term in \eqref{eq:Rr3d19} precisely corresponds to $\itbf{H}^{(2)}_{123}$ in \eqref{nir}, and to cancel the divergence we also add auxiliary function $U(ij)$.
The range-2 remainder is defined as
\begin{equation}\label{eq:Rr2d27}
\begin{aligned}
    \frac{1}{2} \left[\mathcal{R}^{(2)}(12)\right]_{\check{\mathcal{D}}_{27}}
    =&\bigg[
    \mathcal{I}_{\text{BPS,c}}^{(2)}(12)
    -4U(12)
    -f^{(2)}(\epsilon)\big(\mathcal{I}^{(1)}(12)\big)_{\epsilon\shortrightarrow 2\epsilon}
    \bigg]\,,
\end{aligned}
\end{equation}
and $\big[\mathcal{R}^{(2)}\big]_{\check{\mathcal{D}}_{27}}$ is also a pure number equals to $2\times(-17\zeta_4)$. 

Now one may ask the question that: does the non-planar remainder have the same maximal transcendental part as the planar remainder?
Interestingly, the answer is no. 
A simple calculation reveals that $\big[\mathcal{R}^{(2)}(123)\big]_{\check{\mathcal{D}}_{19}}$ can not be given as a linear combination of $\mathcal{R}_{\rm basis}$ function in \eqref{eq:bpsrem} plus pure numbers.  
To compute its analytic form, we follow the similar strategy as described above for the planar remainder. 

We first check the Goncharov criterion \eqref{eq:ConcharovCrit}. Observing that
\begin{equation}
	\delta \circ \mathcal{S}\left( \left[\mathcal{R}^{(2)}\right]_{\check{\mathcal{D}}_{19}}\right) \Big|_{\Lambda^{2}B_2}=-\delta \circ (\mathcal{S}\left(G(\{1-u,1-u,1,0\},v)+\text{cyc}(u,v,w)\right))\Big|_{\Lambda^{2}B_2}\,,
\end{equation}
we know that non-classical part of $\Big[\mathcal{R}^{(2)}\Big]_{\check{\mathcal{D}}_{19}}$ can be chosen as the same $G$ functions as in $\mathcal{R}_{\rm basis}$. 

Inspired by this,  we define\footnote{The $\zeta_4$ term is separated intentionally to get symmetric properties of $\mathcal{R}^{\scriptscriptstyle\text{NP}}_{123}$ in \eqref{eq:symofRNP}.} 
\begin{equation}\label{eq:RNP}
     \mathcal{R}^{\scriptscriptstyle\text{NP}}
     =\left[\mathcal{R}^{(2)}\right]_{\check{\mathcal{D}}_{19}}
     +\mathcal{R}_{\rm basis}(u,v,w)+\mathcal{R}_{\rm basis}(v,w,u)+\mathcal{R}_{\rm basis}(w,u,v)-3\times \frac{ 19}{4}\zeta_4\,,
\end{equation}
as new non-planar (NP) information with the following symmetry property
\begin{equation}\label{eq:symofRNP}
\begin{aligned}
    &\mathcal{R}^{\scriptscriptstyle\text{NP}}(u,v,w)=(-1)^{\sigma}\mathcal{R}^{\scriptscriptstyle\text{NP}}\left(\sigma(u,v,w)\right) \,.
\end{aligned}
\end{equation}
Since $\mathcal{R}^{\scriptscriptstyle\text{NP}}$ passes the Goncharov criterion and  can be expressed as classic polylogarithm consequently. It can be fixed using the ansatz method aforementioned and the final expression takes a very simple form as 
\begin{equation} 
\mathcal{R}^{\scriptscriptstyle\text{NP}}(u,v,w)
	 =\mathcal{R}^{\scriptscriptstyle\text{NP}}_{\rm basis}(u,v,w)+\text{cyc}(u,v,w)\,,
\end{equation}
where we introduce another basis function
\begin{equation}\label{eq:nprembasis}
	\mathcal{R}^{\scriptscriptstyle\text{NP}}_{\rm basis}(u,v,w)=\text{Li}_{3}\left(1-\frac{1}{u}\right)\log\left(\frac{v}{w}\right)+\frac{1}{12}\log(u)^{3}\log\left(\frac{v}{w}\right)+\zeta_2 \log(1-u)\log\left(\frac{v}{w}\right)\,.
\end{equation}
As a result, the range-3 non-planar remainder density can be given as
\begin{equation}
\begin{aligned}
	\left[\mathcal{R}^{(2)}(123)\right]_{\check{\mathcal{D}}_{19}}
	&=\left[-\mathcal{R}_{\rm basis}(u,v,w)+\mathcal{R}^{\scriptscriptstyle\text{NP}}_{\rm basis}(u,v,w)+\frac{19}{4}\zeta_4\right]+\text{cyc}(u,v,w)\,.
\end{aligned}
\end{equation}

With the symmetry of \eqref{eq:symofRNP}, we can also dress color factor to $\mathcal{R}^{\scriptscriptstyle\text{NP}}_{\rm basis}(u,v,w)$. Define 
\begin{equation}
\begin{aligned}
    \itbf{R}^{\scriptscriptstyle\text{NP}}&=\sum_{i<j<k}\itbf{R}^{\scriptscriptstyle \text{NP}}_{ijk}\,, \\
    \itbf{R}^{\scriptscriptstyle \text{NP}}_{123}&=\sum_{\sigma \in S_3} \check{\mathcal{D}}_{\Delta_{2}}(\sigma) \mathcal{R}^{\scriptscriptstyle\text{NP}}_{\rm basis}(\sigma(u,v,w))=\bigg(\sum_{\sigma\in S_3}(-1)^{\sigma}\check{\mathcal{D}}_{\Delta_{2}}(\sigma)\bigg)\mathcal{R}^{\scriptscriptstyle\text{NP}}_{\rm basis}(u,v,w)\,.
\end{aligned}
\end{equation}
We discuss more on this new non-planar contribution in Section~\ref{ssec:hterm}.

\subsection{UV renormalization and dilatation operator}\label{ssec:uv}

For non-BPS form factors, after subtracting the universal IR divergences, the remaining divergences have UV origin. The UV divergences should be cancelled by the renormalization constant, which can be used to determine the dilatation operator. We apply this procedure to the SU(2) form factors. 

We generalize \eqref{prenorm} and introduce the full-color renormalized loop correction $\underline{\itbf{I}}:=\itbf{Z} + \itbf{I}$. 
Up to two loops, one has
\begin{equation}\label{eq:renormupto2l}
\begin{aligned}
    \underline{\itbf{I}}^{(1)}=&\itbf{I}^{(1)}+\itbf{Z}^{(1)} \,, \\
   \underline{\itbf{I}}^{(2)}=&\itbf{I}^{(2)}+\itbf{Z}^{(2)}+\itbf{I}^{(1)}\itbf{Z}^{(1)} \,. 
\end{aligned}
\end{equation}
The one-loop corrections are easy to obtain and are given  in Appendix~\ref{ap:1l}. 
The two-loop corrections have IR structure satisfying \eqref{eq:bds2lnp}. Equivalently, renormalization constant $\itbf{Z}^{(2)}$ can be fixed by requiring the matching of divergent part in both sides in the following equation:
\begin{equation}\label{eq:2lrenorm}
    \itbf{I}^{(2)}=-\left(\itbf{Z}^{(2)}+\itbf{I}^{(1)}\itbf{Z}^{(1)}\right)+\frac{1}{2}\left(\underline{\itbf{I}}^{(1)}\right)^{2}+\itbf{H}^{(2)}_{\rm SU(2)}+\tilde{f}^{(2)}(\epsilon) \left(\underline{\itbf{I}}^{(1)}\right)_{\epsilon\shortrightarrow 2\epsilon}+ \itbf{R}^{(2)}+\mathcal{O}(\epsilon),
\end{equation} 
where $\itbf{H}^{(2)}_{\rm SU(2)}$ is similar to the $\itbf{H}^{(2)}$ terms in \eqref{eq:bds2lnp}. Complete details and comments about these $\itbf{H}^{(2)}_{\rm SU(2)}$ terms are provided in Section~\ref{ssec:hterm}. 

At one-loop level, from \eqref{eq:1lsu2}, UV and IR structures can be separated from each other easily by the bubble and triangle integrals. At two-loop order, generally such an easy separation is not possible. However, for SU(2) form factors, we find it is convenient to group  $\itbf{I}^{(2)}$ in three parts: 
\begin{equation}\label{eq:su2decomp}
    \itbf{I}^{(2)}=\itbf{I}^{(2)}_{\text{BPS}}+\itbf{I}^{(2)}_{\text{fish}}+\itbf{I}^{(2)}_{\text{mix}},
\end{equation}
which manifests the separation of IR and UV divergences.
Let us explain the decomposition of \eqref{eq:su2decomp} in detail: 
\begin{itemize}[leftmargin=*]
    \item[$\square$] $\itbf{I}^{(2)}_{\text{BPS}}$ is the BPS part as discussed in \eqref{eq:isu2tilde}. These integrals, physically, can be regarded as containing IR divergence only. Besides, these are the terms contributing to the maximally transcendental part of remainders.
    
    This BPS part satisfies
    \begin{equation}\label{eq:su2bps}
        \itbf{I}^{(2)}_{\text{BPS}} = \frac{1}{2}\left(\underline{\itbf{I}}^{(1)}_{\text{tri}}\right)^{2}+\tilde{f}^{(2)}(\epsilon)\left(\underline{\itbf{I}}^{(1)}_{\text{tri}}\right)_{\epsilon\shortrightarrow 2\epsilon}+\itbf{H}^{(2)}_{\rm BPS}+\itbf{R}^{(2)}_{\text{a}} \,,
    \end{equation}
    where $\itbf{I}_{\rm tri}$ is the triangle integral part\footnote{Later we also use $\itbf{I}_{\rm bub}$ as bubble integral part and $\underline{\itbf{I}}_{\rm bub}$ as renormalized bubble part. Their corresponding kinematic parts are denoted by simply replacing $\itbf{I}$ with $\mathcal{I}$.} of one-loop correction (see Appendix~\ref{ap:1l}) and $\itbf{H}^{(2)}_{\rm BPS}$ has the same kinematic part as $\itbf{H}^{(2)}$ in \eqref{nir}. Such a subtraction has no difference from our discussion in the previous sections and we will skip this part. 

    \item[$\square$] The second term $\itbf{I}^{(2)}_{\text{fish}}$ is composed of two types of integrals:
    \begin{equation*}
    \mathcal{I}_{\text{SU(2),a}}^{(2)}(p_1,p_2,p_3)=\begin{aligned}
        \includegraphics[width=0.15\linewidth]{fig/bubtrimasstop.eps}
        \end{aligned}\,,\quad \mathcal{I}_{\text{SU(2),b}}^{(2)}(p_1,p_2)=\begin{aligned}
           \includegraphics[width=0.15\linewidth]{fig/bubbub2pttop.eps}
        \end{aligned} \hskip -7pt\,.
    \end{equation*}
    These integrals are special because they do not have divergences in soft region, in other words, they have UV divergence only. Explicit form of these integrals is listed in \eqref{eq:fishsa} and \eqref{eq:fishsb}. The  renormalization constant can be fixed using the following equation
    \begin{equation}\label{eq:su2fish}
           \itbf{I}^{(2)}_{\text{fish}}=-\itbf{Z}^{(2)}-\itbf{I}^{(1)}_{\text{bub}}\itbf{Z}^{(1)}+\frac{1}{2}\left(\underline{\itbf{I}}^{(1)}_{\text{bub}}\right)^{2}+\itbf{R}^{(2)}_{\text{b}} \,.  
    \end{equation}
    \item[$\square$] The remaining part of integrand contains
    \begin{equation*}
    \begin{aligned}
        \mathcal{I}_{\text{SU(2),c}}^{(2)}(p_1,p_2,p_3)&=\hskip -2pt\begin{aligned}
           \includegraphics[height=0.10\linewidth]{fig/tritrimasstop.eps}
        \end{aligned}\hlp{5}\begin{aligned}
           \includegraphics[height=0.10\linewidth]{fig/bubboxtop.eps}
        \end{aligned}\,, \\
        \mathcal{I}_{\text{SU(2),d}}^{(2)}(p_1,p_2)&=\hskip -2pt \begin{aligned}
           \includegraphics[height=0.10\linewidth]{fig/pladdersu2top.eps}
        \end{aligned}
        \hlmath{8}{+}
        \begin{aligned}
           \includegraphics[height=0.10\linewidth]{fig/pladdersu2b.eps}
        \end{aligned}\,.
    \end{aligned}
    \end{equation*}
     This part has both UV and IR divergences but are all canceled by one-loop data:
     \begin{equation}\label{eq:su2mix}
        \itbf{I}^{(2)}_{\text{mix}}= -\itbf{I}^{(1)}_{\text{tri}}\itbf{Z}^{(1)}+
        \underline{\itbf{I}}^{(1)}_{\text{tri}}\underline{\itbf{I}}^{(1)}_{\text{bub}}
        +\tilde{f}^{(2)}(\epsilon)\left(\underline{\itbf{I}}^{(1)}_{\text{bub}}\right)_{\epsilon\shortrightarrow 2\epsilon}+ \itbf{R}^{(2)}_{\text{c}} \,. 
     \end{equation}
$\itbf{R}^{(2)}_{\text{a,b,c}}$ represent finite remainder parts.
\end{itemize}

Equations \eqref{eq:su2bps}-\eqref{eq:su2mix} can be expanded on DDM basis defined in Section~\ref{ssec:nbasis2l} and the subtraction procedure is similar to the previous sections. 
We will not reproduce the details for such calculation.
Rather, it will be helpful to reorganize the results, especially the color factors, and manifest the structure of $\tilde{\itbf{I}}^{(2)}=\itbf{I}^{(2)}_{\rm mix}+\itbf{I}^{(2)}_{\rm fish}$ which is defined in \eqref{eq:isu2tilde} and decomposed in \eqref{eq:su2decomp}. Without loss of generality, we specify fields $YXX$ for sites $123$ and $p_1^{Y}p_2^{X}p_3^{X}$ as external states, and any other fields configuration can be obtained by acting permutations and relabeling. 
The density form of  \eqref{eq:su2fish}-\eqref{eq:su2mix} can be written as
\begin{equation}
\begin{aligned}
  \tilde{\itbf{I}}_{123,\rm fish}^{(2)}=& \ 
    \check{\mathcal{C}}_{\rm fish}^{r=3} \, \mathcal{I}_{\text{SU(2),a}}^{(2)}(p_1,p_2,p_3)+ (a_2,p_2\leftrightarrow a_3,p_3) \,,\\
     \tilde{\itbf{I}}_{12,\rm fish}^{(2)}=& \ 
    \check{\mathcal{C}}_{\rm fish}^{r=2} \,
    \mathcal{I}_{\text{SU(2),b}}^{(2)}(p_1,p_2)
    \label{eq:ifish}\,,
\end{aligned}
\end{equation}
\begin{align}\label{eq:imix}
    \nonumber \tilde{\itbf{I}}_{123,\rm mix}^{(2)}=&\ 
    \check{\mathcal{C}}_{\rm mix,A}^{r=3} \,
    \mathcal{I}_{\text{SU(2),c}}^{(2)}(p_1,p_2,p_3)
        + \check{\mathcal{C}}_{\rm mix,B}^{r=3} \,
        \mathcal{I}_{\text{SU(2),c}}^{(2)}(p_2,p_1,p_3)
       + (a_2,p_2\leftrightarrow a_3,p_3) \,, \\
    \tilde{\itbf{I}}_{12,\rm mix}^{(2)}=& \ 
    \check{\mathcal{C}}_{\rm mix}^{r=2} \,
    \mathcal{I}_{\text{SU(2),d}}^{(2)}(p_1,p_2)
    \,,
\end{align}
where $\check{\mathcal{C}}$ are color factors defined in terms of trivalent graphs as
\begin{equation}\label{eq:su2cfs}
\begin{aligned}
    \check{\mathcal{C}}_{\rm mix,A}^{r=3}&=
    \begin{aligned}
    \includegraphics[width=0.18\linewidth]{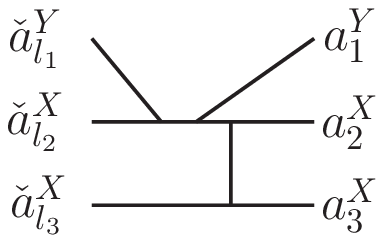}
    \end{aligned}\,, \quad 
    \check{\mathcal{C}}_{\rm mix,B}^{r=3}=
    \begin{aligned}
    \includegraphics[width=0.18\linewidth]{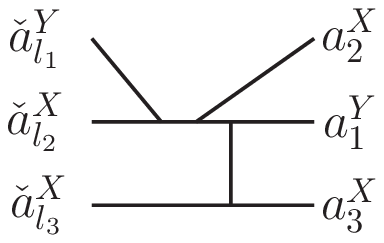}
    \end{aligned}\,,\\
    \check{\mathcal{C}}_{\rm fish}^{r=3}&=
    \begin{aligned}
    \includegraphics[width=0.18\linewidth]{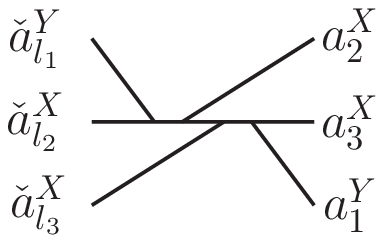}
    \end{aligned}\,, \quad
    \check{\mathcal{C}}_{\rm mix}^{r=2}=\check{\mathcal{C}}_{\rm fish}^{r=2}=
    \begin{aligned}
    \includegraphics[width=0.18\linewidth]{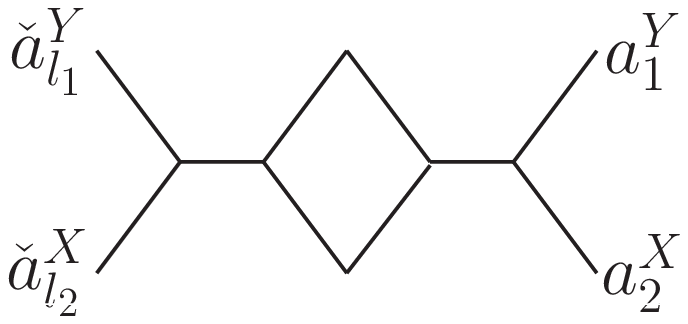}
    \end{aligned}\,.
\end{aligned}
\end{equation}

We will use the above form for subtraction below and the complete derivation of this form can be found in Appendix~\ref{ap:su2}. Technical details about the subsequent subtraction are also included there.

\subsubsection*{Renormalization constant and Dilatation operator}

To obtain the two-loop renormalization constant and dilatation operator, one just needs to focus on \eqref{eq:su2fish}. 
To realize that, we first check the structure of $\itbf{Z}^{(1)}\itbf{I}^{(1)}$ 
\begin{equation}\label{eq:su2compactZIform}
\begin{aligned}
	(\itbf{Z}^{(1)}\itbf{I}^{(1)}_{\rm bub})_{123} &=
 \check{\mathcal{C}}_{\rm fish}^{r=3} \, \mathcal{Z}^{(1)}\mathcal{I}^{(1)}_{\rm bub}(12)  +(a_2,p_2\leftrightarrow a_3,p_3) \,, \\ 
	(\itbf{Z}^{(1)}\itbf{I}^{(1)}_{\rm bub})_{12}&=
	 2\ \check{\mathcal{C}}_{\rm fish}^{r=2} \,  \mathcal{Z}^{(1)}\mathcal{I}^{(1)}_{\rm bub}(12)\,,
\end{aligned}
\end{equation}
where we have used the fact that $\mathcal{Z}^{(1)}$ have no kinematic dependence. The factor 2 can be explained as permuting the two intermediate fields in $\check{\mathcal{C}}_{\rm fish}^{r=2}$|they can be $XY$ or $YX$. 
This fact is easy to understand given the fact that the color factor of both $\itbf{Z}^{(1)}$ and $\itbf{I}^{(1)}_{\rm bub}$ is of $\begin{aligned}
\includegraphics[width=0.12\linewidth]{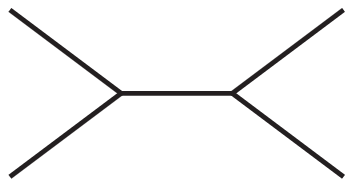}
\end{aligned}$ type such that their connection is $\check{\mathcal{C}}_{\rm fish}$. 
$\Big(\underline{\itbf{I}}_{\rm bub}^{(1)}\Big)^{2}$ have similar structure, although it is completely finite and do not affect UV divergence and renormalization. 

Because of the similar color structure for terms in \eqref{eq:su2fish}, it is possible to expand \eqref{eq:su2fish} on the color factors $\check{\mathcal{C}}_{\rm fish}$ written down in \eqref{eq:ifish} with range $r=2,3$. Therefore, extracting kinematic coefficient of \eqref{eq:su2fish} gives
\begin{align}
 	\text{Range-3: }\quad &
 	\mathcal{I}_{\text{SU(2),a}}^{(2)}(123)
		=-\mathcal{Z}^{(1)}\mathcal{I}_{\rm bub}^{(1)}(12)+\underline{\mathcal{I}}_{\rm bub}^{(1)}(12)\underline{\mathcal{I}}_{\rm bub}^{(1)}(13)-\mathcal{Z}^{(2)}(123)+\mathcal{R}^{(2)}_{\rm b}(123) , \nonumber\\
		\text{Range-2: }\quad &
		\mathcal{I}_{\text{SU(2),b}}^{(2)}(12)
		= -\mathcal{Z}^{(1)}{\mathcal{I}}^{(1)}_{\rm bub}(12)+\frac{1}{2}\left(\underline{\mathcal{I}}_{\rm bub}^{(1)}(12)\right)^{2}-\mathcal{Z}^{(2)}(12)+\mathcal{R}_{\rm b}^{(2)}(12),  \label{eq:su2fishsubtraction}
\end{align}
where the renormalization constant and remainder should satisfy\footnote{Here $\mathcal{Z}^{(2)}$ has no kinematic dependence. The (123) and (12) in argument are just used to distinguish the range.}
\begin{equation}
\begin{aligned}
    &\mathcal{Z}^{(2)}(123)=\itbf{Z}^{(2)}_{123}\big|_{\check{\mathcal{C}}_{\rm fish}^{r=3}}\,, \quad \mathcal{Z}^{(2)}(12)=\itbf{Z}^{(2)}_{12}\big|_{\check{\mathcal{C}}_{\rm fish}^{r=2}}\,, \\
    &\mathcal{R}^{(2)}_{\rm b}(123)=\itbf{R}^{(2)}_{123,\rm b}\big|_{\check{\mathcal{C}}_{\rm fish}^{r=3}}\,, \quad \mathcal{R}^{(2)}_{\rm b}(12)=\itbf{R}^{(2)}_{12,\rm b}\big|_{\check{\mathcal{C}}_{\rm fish}^{r=2}}\,.
\end{aligned}
\end{equation}

Since all the elements in \eqref{eq:su2fishsubtraction} are known, $\mathcal{Z}^{(2)}$ and $\mathcal{R}^{(2)}$ can be  easily solved as
\begin{align}
\nonumber
    \mathcal{Z}^{(2)}(123)& =\frac{1}{2\epsilon^2}-\frac{1}{2\epsilon}\,,\qquad \mathcal{Z}^{(2)}(12)=\frac{1}{2\epsilon^2}-\frac{1}{2\epsilon} ,\\
    \mathcal{R}_{\rm b}^{(2)}(12)& = \frac{7}{2}-\log (-s_{12}) , \label{eq:su2fishres} \\
    \nonumber
    \mathcal{R}_{\rm b}^{(2)}(123)& = \frac{7}{2}- \log (-s_{12})+\frac{1}{2}\left(\log (-s_{123})-2\right)\log \left(\frac{u}{v}\right)    -\Big( \text{Li}_2(1-u)+\frac{1}{2} \log \left(u\right) \log (v)\Big).
\end{align}

To get dilatation operators, we first write down the complete form of renormalized constant
\begin{equation}
\begin{aligned}
	&\itbf{Z}^{(2)}_{123}=\check{\mathcal{C}}_{\rm fish}^{r=3}\left(\frac{1}{2\epsilon^2}-\frac{1}{2\epsilon}\right) +  (a_2,p_2\leftrightarrow a_3,p_3) \,, \\
		&\itbf{Z}^{(2)}_{12}=2\ \check{\mathcal{C}}_{\rm fish}^{r=2} \left(\frac{1}{2\epsilon^2}-\frac{1}{2\epsilon}\right) \,.
\end{aligned}
\end{equation}

And the dilatation operator, derived from
$\mathds{D}^{(2)}=4 \epsilon\Big[\itbf{Z}^{(2)}-\frac{1}{2}\big(\itbf{Z}^{(1)}\big)^{2}\Big]$ with $\big(\itbf{Z}^{(1)}\big)^{2}$ given in  \eqref{eq:su2z1sqgra}, happens to read\footnote{
Recall that this dilatation density is derived under the momenta labeling $p_1^{Y}p_2^{X}p_3^{X}$ acting on 123 sites with Fields $YXX$.}
\begin{equation}
 	(\mathds{D}^{(2)})_{123}=-2 \ \check{\mathcal{C}}^{r=3}_{\rm fish} ,
		\qquad (\mathds{D}^{(2)})_{12}=-4 \ \check{\mathcal{C}}^{r=2}_{\rm fish} .
\end{equation}
Given this, the complete abstract operator form should be
\begin{equation}
\begin{aligned}
	(\mathds{D}^{(2)})_{\rm SU(2)}&=-2\begin{aligned}
		\includegraphics[width=0.17\linewidth]{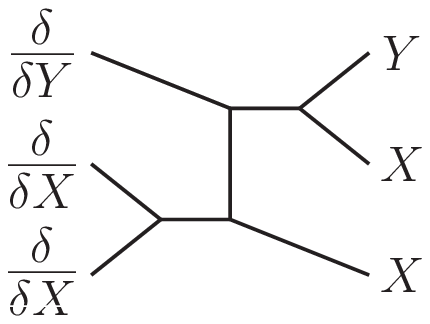}
		\end{aligned} -2  \begin{aligned}
		\includegraphics[width=0.17\linewidth]{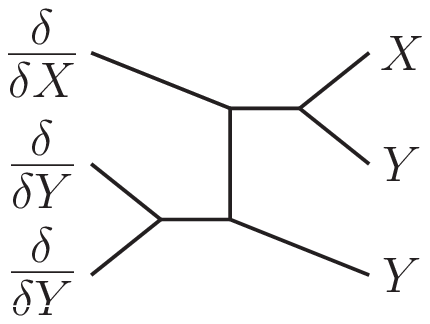}
		\end{aligned} -4\begin{aligned}
		\includegraphics[width=0.17\linewidth]{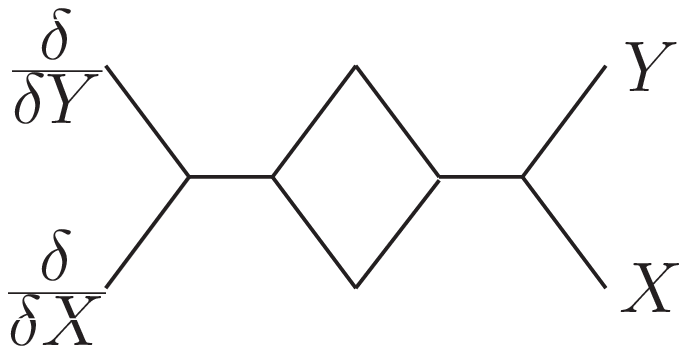}
		\end{aligned}\,,
\end{aligned}
\end{equation}
where the abstract dilatation operator $(\mathds{D}^{(2)})_{\rm SU(2)}$ acts on operators like $\operatorname{tr}(XYXY\cdots)$ directly and bypass the process of momenta labeling of fields. 
This form is equivalent to full-color dilatation operator in SU(2) Sector in \cite{Beisert:2003jj}. 

Finally, let us briefly comment on remainders. $\itbf{R}^{(2)}_{\text{fish}}$ have simple kinematic functions with transcendentality ranging form 2 to 0 as in \eqref{eq:su2fishres}. An interesting fact is that only $\itbf{R}^{(2)}_{\text{fish}}$ have rational pieces (or say transcendentality-0 parts), and rational pieces satisfy
    \begin{equation}
        \mathds{D}^{(2)}=-\frac{4}{7} \itbf{R}^{(2)} \Big|_{\textrm{deg-0}}\,,
    \end{equation}
where $\itbf{R}\big|_{\textrm{deg-$m$}}$ denotes transcendentality-$m$ part of remainders. This can be derived directly from \eqref{eq:su2fishres} that
\begin{equation}
    \mathcal{Z}^{(2)}(123)=\mathcal{Z}^{(2)}(12)=\frac{1}{7}\left(\frac{1}{\epsilon^{2}}-\frac{1}{\epsilon}\right)\mathcal{R}^{(2)}_{\rm b}(123)\Big|_{\textrm{deg-0}}=\frac{1}{7}\left(\frac{1}{\epsilon^{2}}-\frac{1}{\epsilon}\right)\mathcal{R}^{(2)}_{\rm b}(12)\Big|_{\textrm{deg-0}} .
\end{equation}
 
    
\subsubsection*{Divergence cancellation in $\itbf{I}^{(2)}_{\rm mix}$} 

Finally we turn to the $\itbf{I}^{(2)}_{\rm mix}$ part of  our analysis on SU(2) form factor. This part serves to provide a consistency check as well as  the final part of remainder. 
Similar to our previous discussion for $\itbf{I}_{\rm fish}$, we can also expand the subtraction equation \eqref{eq:su2mix} for $\itbf{I}_{\rm mix}$ on the color factors $\check{\mathcal{C}}_{\rm mix}$ in \eqref{eq:imix}.

Hence, it is convenient to write down an equation parallel to \eqref{eq:su2fishsubtraction} as follows
\begin{align}
    \text{Range-3: } \quad &
    \mathcal{I}_{\text{SU(2),c}}^{(2)}(123)
        =\mathcal{I}^{(1)}_{\rm tri}(23)\mathcal{I}^{(1)}_{\rm bub}(12)+V(\epsilon)+\mathcal{R}_{\rm c}(123) \,, \\
        \nonumber
    \text{Range-2: } \quad &
    \mathcal{I}_{\text{SU(2),d}}^{(2)}(12)
        =\mathcal{I}^{(1)}_{\rm tri}(12) \, \mathcal{I}^{(1)}_{\rm bub}(12)+\tilde{f}^{(2)}(\epsilon) \, \underline{\mathcal{I}}_{\rm bub}^{(1)}(12)+2 V(\epsilon)+\mathcal{R}_{\rm c}(12) \,,
\end{align}    
where we have introduced another auxiliary quantity $V\equiv\zeta_2/\epsilon$  similar to \eqref{eq:fcan}. And the final task of the cancellation of auxiliary $V$, beyond density level, if discussed in Appendix~\ref{ap:su2}.

The results of remainders are
\begin{equation}
\begin{aligned}
    \mathcal{R}^{(2)}_{\rm c}(12)=&-4\zeta_2+8\zeta_3 \,, \\
    \mathcal{R}^{(2)}_{\rm c}(123)\Big|_{\textrm{deg-3}}=&
\left[-\mathrm{Li}_{3}\left(-\frac{u}{w}\right)+\log (u) \operatorname{Li}_{2}\left(\frac{v}{1-u}\right)-\frac{1}{2} \log(u) \log (1-u) \log \left(\frac{w^{2}}{1-u}\right)\right. \\
&\left.+\frac{1}{2} \mathrm{Li}_{2}\left(-\frac{u v}{w}\right)+\frac{1}{2} \log (u) \log (v) \log (w)+\frac{1}{12} \log ^{3}(w)+(u \leftrightarrow v)\right]\\
&+\mathrm{Li}_{3}(1-u)-\mathrm{Li}_{3}(v)+\frac{1}{2} \log ^{2}(v) \log \left(\frac{1-v}{u}\right)-\zeta_{2} \log \left(\frac{u^2}{w}\right) - 3\zeta_{3} \,, \\
\mathcal{R}^{(2)}_{\rm c}(123)\Big|_{\textrm{deg-2}}=&-\Big( \mathrm{Li}_{2}(1-u)+\frac{1}{2}\log^{2}(u)\Big)+2\zeta_2\,.
\end{aligned} 
\end{equation}
We observe that these remainders (of degree less than 4) are compatible with the planar remainders in \cite{Loebbert:2015ova, Loebbert:2016xkw} up to some $\zeta_2\times \log$ and $\zeta_3$ terms.  This can be explained by the observation that the full-color density in \eqref{eq:imix}  behaves like the same blocks appearing in planar densities, for example in  \cite{Loebbert:2015ova}, by dressing color factors.  

\subsection{Comment on new features of non-planar subtractions and remainders}\label{ssec:hterm}

Through the discussion in this chapter, we have encountered several new features originating from non-planarity, including the divergent $\itbf{H}^{(2)}$ terms and the finite non-planar remainder $ \itbf{R}^{\scriptscriptstyle\text{NP}}$. Both of these can be explained by the breaking of parity symmetry for the kinematic part of  non-planar color factors, such as $\left[\mathcal{I}^{(2)}\right]_{\check{\mathcal{D}}_{19}}$ for BPS form factors. Comparing \eqref{eq:bpsr31new} and \eqref{eq:bpsr32new}, it is clear that \eqref{eq:bpsr32new} has only cyclic symmetry while  \eqref{eq:bpsr31new} has parity symmetry $(p_1\leftrightarrow p_3)$. The part violating parity symmetry in \eqref{eq:bpsr32new} can be regarded as 
\begin{equation}
    \mathcal{I}^{\textrm{anti-sym}} \equiv \sum_{\sigma \in S_3} (-1)^{\sigma}\hskip -3pt  \begin{aligned}
       \includegraphics[height=0.1\linewidth]{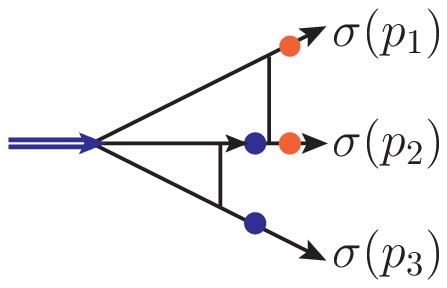}
    \end{aligned}\,.
\end{equation}
One can show that the divergent part of these kinematic integrals is just $\mathcal{H}^{(2)}$, \emph{i.e.}~the kinematic part of $\itbf{H}^{(2)}$, up to certain overall number. Such argument gives us a heuristic reasoning about the origin of new contributions from non-planar calculation, and we will discuss them in detail below. 

We start from $\itbf{H}^{(2)}$ terms first. These $\itbf{H}$ terms should have only $\epsilon^{-1}$ divergence and can be regarded as commutators of one-loop quantities. It is the non-commutative property of full-color one-loop quantities that results in such $\itbf{H}$ terms. 

A clean example is the $\itbf{H}^{(2)}$ in BPS form factors, for example in \eqref{nir}, whose density, $\itbf{H}^{(2)}_{123}$, can be written as a commutator between finite and divergent part of one-loop density
\begin{equation}\label{commutator}
\begin{aligned}
    \itbf{H}^{(2)}_{123}=\frac{1}{2}\big(\itbf{I}^{(1)}_{23}\big|_{\text{fin}}\itbf{I}^{(1)}_{12}\big|_{\text{div}}- \itbf{I}^{(1)}_{12}\big|_{\text{fin}}\itbf{I}^{(1)}_{23}\big|_{\text{div}} \big) +\text{cyc}(p_1,p_2,p_3),
\end{aligned}
\end{equation}
where $\itbf{I}_{ij}$ is one-loop BPS correction in Appendix~\ref{ap:1l}, and subscript `fin' or `div' means taking the finite part or divergent part. The concrete computation  have been given in \eqref{eq:hkine} in Appendix~\ref{ap:bps} and we just highlight the triple-log structure here
\begin{equation}\label{eq:hkinepre}
    \big[\mathcal{H}^{(2)}\big]_{\check{\mathcal{D}}_{19}}(123)=\frac{1}{4\epsilon}\mathrm{log}\left(\frac{-s_{12}}{-s_{23}}\right)\mathrm{log}\left(\frac{-s_{23}}{-s_{13}}\right)\mathrm{log}\left(\frac{-s_{13}}{-s_{12}}\right)\,.
\end{equation}

This commutator expression is compatible with the $\hat{\mathbf{H}}^{(2)}$ operator in \cite{Aybat:2006mz}, where $\hat{\mathbf{H}}^{(2)}$ is also interpreted as a commutator. 
The form of \eqref{commutator} also inspires us that it is possible to deform the one-loop square to a close function as:
\begin{equation}
\label{1lsqrep}
    \frac{1}{2}\left({\underline{\itbf{I}}}^{(1)}\right)^{2} +\itbf{H}^{(2)}= \frac{1}{2}\left(\underline{\itbf{I}}^{(1)}\Big|_{\text{fin}}\right)^{2}+\frac{1}{2}\left({\underline{\itbf{I}}}^{(1)}\Big|_{\text{div}}\right)^{2}+\left({\underline{\itbf{I}}}^{(1)}\Big|_{\text{fin}}\right)\left({\underline{\itbf{I}}}^{(1)}\Big|_{\text{div}}\right) \,,
\end{equation}
such that if we use the RHS to replace $\frac{1}{2}({\underline{\itbf{I}}}^{(1)})^{2}$ in the subtraction formula \eqref{eq:bds2lnp}, then the ${\itbf{H}}^{(2)}$ function is no-longer needed.

As claimed in Section~\ref{ssec:uv}, the $\itbf{H}$ functions for SU(2) form factors are similar but slightly different from BPS form factors. More concretely, 
\begin{equation}
    \itbf{H}^{(2)}_{\rm SU(2)}=\itbf{H}^{(2)}_{\rm BPS}+\itbf{H}^{(2)}_{\rm mix}\,.
\end{equation}
The only point deserving notice for $\itbf{H}^{(2)}_{\rm BPS}$ is that only some configurations for a certain color factor have such triple-log divergence, similar to the discussion in Section~\ref{ssec:2lsu2}. The second part $\itbf{H}^{(2)}_{\rm mix}$ is the commutator $\frac{1}{2}\big[\underline{\itbf{I}}^{(1)}_{\rm tri}\big|_{\rm div},\underline{\itbf{I}}^{(1)}_{\rm bub}\big]$\footnote{$\underline{\itbf{I}}_{\rm bub}$ is finite, and its commutator with finite part of $\underline{\itbf{I}}_{\rm tri}$ is just finite simple $\log$ functions so we omit that.} pointed out in the last subsection. Such commutators have kinematic parts
\begin{equation}
\begin{aligned}
    &\left[\left(\mathcal{H}^{(2)}_{\rm mix}\right)^{\scriptscriptstyle YXX}_{\scriptscriptstyle YXX}(p_1^{Y},p_2^{X},p_3^{X})\right]_{\check{\mathcal{D}}_{19}}=\frac{-2+\log(s_{23})}{\epsilon}\log\left(\frac{-s_{12}}{-s_{13}}\right)\,,\\
    &\left[\left(\mathcal{H}^{(2)}_{\rm mix}\right)^{\scriptscriptstyle XYX}_{\scriptscriptstyle YXX}(p_1^{X},p_2^{Y},p_3^{X})\right]_{\check{\mathcal{D}}_{19}}=\frac{-2+\log(s_{12})}{\epsilon}\log\left(\frac{-s_{13}}{-s_{23}}\right)\,,\\
    &\left[\left(\mathcal{H}^{(2)}_{\rm mix}\right)^{\scriptscriptstyle XYX}_{\scriptscriptstyle YXX}(p_1^{X},p_2^{X},p_3^{Y})\right]_{\check{\mathcal{D}}_{19}}=\frac{-2+\log(s_{13})}{\epsilon}\log\left(\frac{-s_{23}}{-s_{12}}\right)\,.
\end{aligned}
\end{equation}
Besides, because of the commutator structure, the deformation in \eqref{1lsqrep} also seems to be applicable for SU(2) form factors. It would be interesting to check if such a deformation also suits more general form factors in other sectors.

Let us further discuss the new contribution for non-planar remainders  $ \itbf{R}^{\scriptscriptstyle\text{NP}}$ and its component given in \eqref{eq:nprembasis}. 
The different between planar and non-planar can be also understood from the different symmetry at integrand level, as that the planar \eqref{eq:bpsr31new} has parity symmetry $(p_1\leftrightarrow p_3)$ while non-planar case \eqref{eq:bpsr32new} has only cyclic symmetry.
The overall structure is similar to \eqref{eq:hkinepre} discussed above, so most conclusions are also valid here. 
In this case, although we have not found a direct commutator explanation to this remainder, the argument based on parity symmetry is still true. It would be interesting to understand this point better or consider what will happen in other situations. 
We would like to also stress that the discrepancy between planar and non-planar contributions is still consistent with the maximal transcendentality principle first introduced in study of anomalous dimensions in \cite{Kotikov:2002ab, Kotikov:2004er} and later observed for the remainder functions of various form factors \cite{Brandhuber:2012vm, Brandhuber:2014ica, Loebbert:2015ova, Brandhuber:2016fni, Loebbert:2016xkw,  Ahmed:2016vgl, Banerjee:2016kri, Brandhuber:2017bkg,  Banerjee:2017faz, Jin:2018fak, Brandhuber:2018xzk, Brandhuber:2018kqb, Jin:2019ile, Jin:2019opr}. Therefore, we expect that the new non-planar contribution will also be a universal function that appears in generic (non-planar)  gauge theories.

Finally, we discuss the situation when these non-planar corrections contribution in the final result. For $\itbf{H}^{(2)}$ in the BPS case, the color structure of \eqref{nir} is really special, which is
\begin{equation}
    \check{\mathcal{C}}^{\textrm{anti-sym}} \equiv \sum_{\sigma \in S_{3} }(-1)^{\sigma}\check{\mathcal{D}}_{\Delta_{2}}(\sigma)\,.
\end{equation}
Action of $ \check{\mathcal{C}}^{\textrm{anti-sym}}$ on tree-level form factors shows that only length$\geq$6 operators, regardless their trace structures, possibly give non-vanishing contribution under action of $\check{\mathcal{C}}^{ \textrm{anti-sym}}$. 
Such structure certainly relies on the  character of identical fields for  BPS form factors. For example, for $\big[\mathcal{I}^{(2)}\big]_{\check{\mathcal{D}}_{19}}$, only $YXX\rightarrow YXX$ has a BPS part.

Similar observation holds for other quantities for SU(2) form factors, for example the $\itbf{H}$ terms and remainder $\itbf{R}$. Expanding on DDM basis defined in Section~\ref{ssec:nbasis2l}, the coefficient of $\check{\mathcal{D}}_{19}$ receives contribution appearing in BPS form factors only for $YXX\rightarrow YXX$ configuration. Consequently, the color coefficient of triple-log divergence in \eqref{eq:hkinepre}  can appear in form factors of operators with length$\geq 4$, which is not as strict as BPS form factors staring form length-6. 
Likewise, four-point amplitudes in QCD also have $\itbf{H}$ terms contribution, see \cite{Aybat:2006mz,Bern:2004cz}.


\section{Summary and Outlook}\label{sec:disc}

Comparing with the planar case, the study of non-planar sector of gauge theories remains a much harder problem. 
One main complication is that the color factors have to be taken into account. 
The correlation between color and kinematic degrees of freedom typically makes the structures much more intricate. 
Besides, certain powerful symmetries in the planar limit, such as integrability in ${\cal N}=4$ SYM, are no longer applicable. 
In this paper, we consider non-planar form factors of generic high-length operators in ${\cal N}=4$ SYM, 
which are an interesting class of observables that capture the information of both off-shell anomalous dimensions and on-shell amplitudes.
To probe the non-planar sector efficiently, we apply both the on-shell unitarity method and color-kinematics duality. 
Below we recapitulate the main results of this paper.
\begin{itemize}
\item[(1)] 
To simplify the problem, we make a careful analysis on color decomposition, using both trace and trivalent (DDM) basis, to disentangle color and spacetime degrees of freedom. 
Since we consider form factors of high-length operators, to simplify the structure, we also decompose the loop correction in terms of density functions.
These decompositions help to reduce the full computation to a small set of kinematic basis functions. 

    \item[(2)] We develop unitarity method with non-planar cuts to compute kinematic functions of form factors. Explicit two-loop results for BPS and SU(2) form factors are obtained. For the SU(2) case, the supersymmetric Ward identities are used to simplify the calculation.
    
    \item[(3)] Color-kinematics duality, as an alternative way to study non-planar correction, is also considered. In this strategy, the symmetry relations inherited from the color factors play a (bonus) role to simplify the computation of kinematic parts.
    We construct the explicit CK-dual representation for two-loop density corrections of half-BPS form factors. An interesting observation is that certain scaleless integrals are needed in order to satisfy the CK duality. 
    
    \item[(4)] After integration, an analysis on full-color IR structure is performed. In contrast to the planar BDS ansatz, the full-color IR divergence contains a new non-planar contribution at $1/\epsilon$ order (the $\itbf{H}^{(2)}$ term), which provides an independent check of the known dipole formula for two-loop IR divergences.
    
    \item[(5)] As for the finite remainder function, we obtain a new non-planar contribution in the maximally transcendental four part. This new contribution can be expressed in terms of classical polylogarithms and has a similar structure and non-planar origin as the $\itbf{H}^{(2)}$ term in the divergence part.
    
    \item[(6)] Finally, the full-color SU(2) dilatation operator is also derived and is consistent with the previously known result, which also serves as a check for our SU(2) calculation.
\end{itemize}

We hope the above concrete studies and results will open an avenue to explore the non-planar sector in the gauge theory, in particular for the off-shell operators using on-shell amplitudes-inspired techniques. 
Let us mention a few generalizations and other problems that can be considered based on the study of this paper.
First, it would be interesting to generalize the present study to operators of more generic sectors and also to higher loops. For some previous non-planar studies of anomalous dimension and correlation functions, see \emph{e.g.}~\cite{Beisert:2003jj, Beisert:2003tq, Zwiebel:2005er, Xiao:2009pv, Caputa:2010ep, Velizhanin:2009gv, Velizhanin:2010ey, Velizhanin:2014zla, Eden:2012tu, Fleury:2019ydf, Kniehl:2020rip}.
Second, the non-planar data would be useful for the study of possible hidden symmetries in the non-planar sector. 
A promising direction appears to be the the Hexagon program which suggests that the integrability may be still applicable beyond the planar sector for correlation functions \cite{Eden:2017ozn, Bargheer:2017nne, Eden:2018vug, Bargheer:2018jvq, Bargheer:2019kxb}. (See other integrability related studies in \emph{e.g.}~\cite{Carlson:2011hy, Koch:2011jk}.)
Another important structure to explore is the color-kinematics duality, in particular for form factors of general non-BPS operators at two loops. 
Furthermore, the form factors at strong coupling via AdS/CFT have been studied in \cite{Alday:2007he, Maldacena:2010kp, Gao:2013dza}, and more recently, the Pentagon OPE method has been generalized to form factor \cite{Sever:2020jjx}.
It would be certainly interesting to have a non-planar generalization (see a related study for amplitudes \cite{Ben-Israel:2018ckc}).
Finally, although we focus on ${\cal N}=4$ SYM in this paper, the on-shell method is certainly applicable to form factors of high-dimensional operators in more generic gauge theories such as QCD, see \emph{e.g.}~\cite{Jin:2020pwh}.

\acknowledgments

We would like to thank Siyuan Zhang for discussions. 
This work is supported in part by the National Natural Science Foundation of China (Grants No.~11822508, 11947302, 11935013),
and by the Key Research Program of Frontier Sciences of CAS under Grant No.~QYZDB-SSW-SYS014.
We also thank the support of the HPC Cluster of ITP-CAS.

\appendix

\section{Two-loop DDM basis}\label{ap:ddm}

In this appendix, we provide the explicit DDM basis $\check{\mathcal{D}}_{i}$ with $i=1,\ldots,27$ for two-loop form factors that are defined in \eqref{eq:set1}-\eqref{eq:set4} in Section\,\ref{ssec:nbasis2l}: 
\begin{equation}
\begin{aligned}
&
\check{\mathcal{D}}_{1} =  f^{  \check{a}_{l_1} a_1 \text{x}_1} f^{\text{x}_1 a_2  \text{x}_2} f^{ \text{x}_2 \text{x}_3 \check{a}_{l_2}} f^{ \text{x}_3 a_3 \check{a}_{l_3}},
\qquad \check{\mathcal{D}}_{2} =  f^{\check{a}_{l_2} a_1  \text{x}_1} f^{\text{x}_1 a_2  \text{x}_2} f^{ \text{x}_2 \text{x}_3 \check{a}_{l_3}} f^{ \text{x}_3 a_3 \check{a}_{l_1}},\\&
 \check{\mathcal{D}}_{3} =   f^{\check{a}_{l_3} a_1 \text{x}_1} f^{\text{x}_1 a_2 \text{x}_2} f^{ \text{x}_2 \text{x}_3  \check{a}_{l_1}} f^{ \text{x}_3 a_3 \check{a}_{l_2}},
\qquad \check{\mathcal{D}}_{4} =  f^{\check{a}_{l_1} a_1 \text{x}_1}f^{\text{x}_1 a_3  \text{x}_2} f^{\text{x}_2 \text{x}_3 \check{a}_{l_2} } f^{\text{x}_3 a_2 \check{a}_{l_3} },\\&
 \check{\mathcal{D}}_{5} =  f^{\check{a}_{l_2} a_1 \text{x}_1} f^{\text{x}_1 a_3  \text{x}_2} f^{\text{x}_2 \text{x}_3 \check{a}_{l_3}} f^{\text{x}_3 a_2  \check{a}_{l_1}},
\qquad \check{\mathcal{D}}_{6} =  f^{\check{a}_{l_3} a_1 \text{x}_1} f^{\text{x}_1 a_3 \text{x}_2} f^{\text{x}_2 \text{x}_3 \check{a}_{l_1}} f^{\text{x}_3 a_2 \check{a}_{l_2} },\\&
 \check{\mathcal{D}}_{7} =  f^{\check{a}_{l_1} a_2 \text{x}_1} f^{\text{x}_1 a_1 \text{x}_2} f^{\text{x}_2 \check{a}_{l_2} \text{x}_3} f^{\text{x}_3 a_3 \check{a}_{l_3}},
\qquad \check{\mathcal{D}}_{8} = f^{\check{a}_{l_2} a_2 \text{x}_1} f^{\text{x}_1 a_1 \text{x}_2}  f^{\text{x}_2  \text{x}_3  \check{a}_{l_3}} f^{\text{x}_3 a_3  \check{a}_{l_1} },\\&
 \check{\mathcal{D}}_{9} =  f^{\check{a}_{l_3} a_2 \text{x}_1}f^{\text{x}_1a_1 \text{x}_2} f^{ \text{x}_2 \text{x}_3  \check{a}_{l_1}} f^{ \text{x}_3 a_3 \check{a}_{l_2}},
\qquad \check{\mathcal{D}}_{10} =  f^{\check{a}_{l_1} a_2 \text{x}_1} f^{\text{x}_1 a_3 \text{x}_2} f^{\text{x}_2 \text{x}_3 \check{a}_{l_2}} f^{\text{x}_3 a_1 \check{a}_{l_3}},\\&
 \check{\mathcal{D}}_{11} =  f^{\check{a}_{l_2} a_2 \text{x}_1} f^{\text{x}_1 a_3 \text{x}_2} f^{\text{x}_2 \text{x}_3 \check{a}_{l_3}} f^{\text{x}_3 a_1  \check{a}_{l_1} },
\qquad \check{\mathcal{D}}_{12} =  f^{\check{a}_{l_3} a_2 \text{x}_1} f^{\text{x}_1 a_3 \text{x}_2} f^{ \text{x}_3 \text{x}_2 \check{a}_{l_1}} f^{\text{x}_3 a_1 \check{a}_{l_2}},\\&
 \check{\mathcal{D}}_{13} =  f^{\check{a}_{l_1} a_3   \text{x}_1} f^{\text{x}_1 a_1 \text{x}_2} f^{\text{x}_2 \text{x}_3 \check{a}_{l_2}} f^{\text{x}_3 a_2 \check{a}_{l_3}},
\qquad \check{\mathcal{D}}_{14} =  f^{\check{a}_{l_2} a_3  \text{x}_1} f^{\text{x}_1 a_1 \text{x}_2} f^{\text{x}_2 \text{x}_3 \check{a}_{l_3} } f^{\text{x}_3 a_2  \check{a}_{l_1}},\\&
 \check{\mathcal{D}}_{15} =  f^{\check{a}_{l_3} a_3  \text{x}_1} f^{\text{x}_1 a_1 \text{x}_2} f^{\text{x}_2 \text{x}_3 \check{a}_{l_1}} f^{\text{x}_3 a_2 \check{a}_{l_2}},
\qquad \check{\mathcal{D}}_{16} =  f^{ \check{a}_{l_1} a_3  \text{x}_1}f^{\text{x}_1 a_2 \text{x}_2}f^{\text{x}_2 \text{x}_3 \check{a}_{l_2}}f^{\text{x}_3 a_1 \check{a}_{l_3} },\\&
 \check{\mathcal{D}}_{17} =  f^{\check{a}_{l_2} a_3 \text{x}_1} f^{\text{x}_1 a_2 \text{x}_2} f^{\text{x}_2 \text{x}_3 \check{a}_{l_3}} f^{\text{x}_3 a_1  \check{a}_{l_1}},
\qquad \check{\mathcal{D}}_{18} =  f^{\check{a}_{l_3} a_3  \text{x}_1} f^{\text{x}_1 a_2 \text{x}_2} f^{\text{x}_2 \text{x}_3 \check{a}_{l_1}} f^{\text{x}_3 a_1 \check{a}_{l_2} },\\&
 \check{\mathcal{D}}_{19} =  f^{a_1  \check{a}_{l_1} \text{x}_1}f^{a_2 \check{a}_{l_2} \text{x}_2}f^{a_3 \check{a}_{l_3} \text{x}_3}f^{\text{x}_1 \text{x}_2\text{x}_3},
\qquad \check{\mathcal{D}}_{20} =  f^{a_1  \check{a}_{l_1} \text{x}_1}f^{a_3 \check{a}_{l_2} \text{x}_2}f^{a_2 \check{a}_{l_3} \text{x}_3}f^{\text{x}_1 \text{x}_2 \text{x}_3},\\&
 \check{\mathcal{D}}_{21} =  f^{a_2  \check{a}_{l_1}\text{x}_1}f^{a_1 \check{a}_{l_2} \text{x}_2}f^{a_3 \check{a}_{l_3} \text{x}_3}f^{\text{x}_1 \text{x}_2 \text{x}_3},
\qquad \check{\mathcal{D}}_{22} =  f^{a_2  \check{a}_{l_1} \text{x}_1}f^{a_3 \check{a}_{l_2}\text{x}_2}f^{a_1 \check{a}_{l_3} \text{x}_3}f^{\text{x}_1 \text{x}_2 \text{x}_3},\\&
 \check{\mathcal{D}}_{23} =  f^{a_3  \check{a}_{l_1} \text{x}_1}f^{a_1 \check{a}_{l_2} \text{x}_2}f^{a_2 \check{a}_{l_3}\text{x}_3}f^{\text{x}_1 \text{x}_2 \text{x}_3},
\qquad \check{\mathcal{D}}_{24} =  f^{a_3  \check{a}_{l_1} \text{x}_1}f^{a_2 \check{a}_{l_2} \text{x}_2}f^{a_1 \check{a}_{l_3} \text{x}_3}f^{\text{x}_1 \text{x}_2\text{x}_3}\\&
 \check{\mathcal{D}}_{25} =    f^{\text{x}_{1}\check{a}_{l_2}\text{x}_{2}}f^{\text{x}_{2} \check{a}_{l_1}\text{x}_{3}}f^{\text{x}_{3} a_{1}\text{x}_{4}}f^{\text{x}_{4} a_{2} \text{x}_{1}},
\qquad \check{\mathcal{D}}_{26} =    f^{\text{x}_{1}\check{a}_{l_2}\text{x}_{2}}f^{\text{x}_{2} \check{a}_{l_1}\text{x}_{3}}f^{\text{x}_{3} a_{2}\text{x}_{4}}f^{\text{x}_{4} a_{1} \text{x}_{1}},\\&
 \check{\mathcal{D}}_{27} =    f^{\text{x}_{1}\check{a}_{l_2}\text{x}_{2}}f^{\text{x}_{2}a_{2}\text{x}_{3}}f^{\text{x}_{3} a_{1}\text{x}_{4}}f^{\text{x}_{4}  \check{a}_{l_1} \text{x}_{1}} .
\end{aligned}
\end{equation}

Below we briefly comment on how to relate the BPS form factor from  CK-dual form in Section~\ref{sec:CKBPS} to the result in Section~\ref{ssec:2lbps}.
Consider the CK-dual represeduntation \eqref{eq:compck}, the permutations of internal lines $\sigma^i$ action on $\check{\mathcal{C}}_{1}$ belong to $S_3$, while  in the DDM basis \eqref{eq:set1} one has only $\sigma^i \in \mathbb{Z}_3$. 
The new permutations in CK form, 
namely $\sigma_{i}\in S_3 \backslash \mathbb{Z}_3$,
can be expressed in terms of DDM basis \eqref{eq:set1}. For example, using color Jacobi relation, one has
\begin{equation}\label{eq:d1md1a}
    \check{\mathcal{D}}_{\Delta_1(\mathbf{1},\mathbf{1})}- \check{\mathcal{D}}_{\Delta_1(\mathbf{1},\mathbf{1})} |_{\text{refl.}}=\check{\mathcal{D}}_{\Delta_2(\mathbf{1})}\,,
\end{equation}
where ``refl." represents the reflection that maps $\{ \{l_1,l_2,l_3\},\{p_1,p_2,p_3\}\}$ to $\{ \{l_3,l_2,l_1\},\{p_3,p_2,p_1\}\}$, with which one can get the permutations  $\sigma_{i}\in S_3 \backslash \mathbb{Z}_3$.
\eqref{eq:d1md1a} can be also given  in a diagrammatic form
\begin{equation}\label{eq:d1md1}
    \begin{aligned}
    \includegraphics[width=0.55\linewidth]{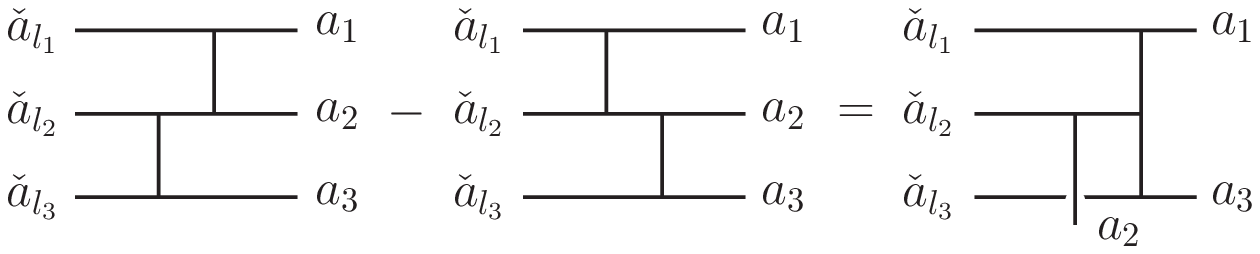}
    \end{aligned}\,.
\end{equation}
With such equations, one can expand the CK-dual form \eqref{eq:compck} on the DDM basis defined in Section~\ref{ssec:nbasis2l} and then compare it with the result in Section~\ref{ssec:2lbps}. 

\subsubsection*{Two-loop DDM basis of next-to-minimal form factors}

\begin{figure}
    \centering
    \subfigure[Cubic Graph: $\check{\mathcal{D}}_{\Delta_1} $]{
        \includegraphics[width=0.23\linewidth]{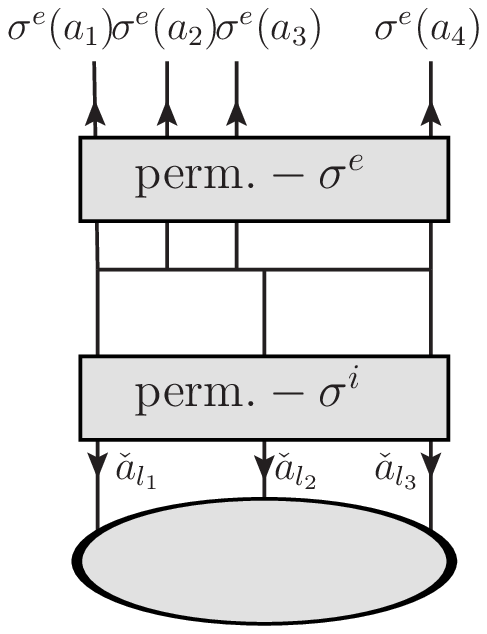}
         \label{fig:ddm2lnm1}
        }
     \centering
   \subfigure[Cubic Graph: $\check{\mathcal{D}}_{\Delta_2} $]{
        \includegraphics[width=0.23\linewidth]{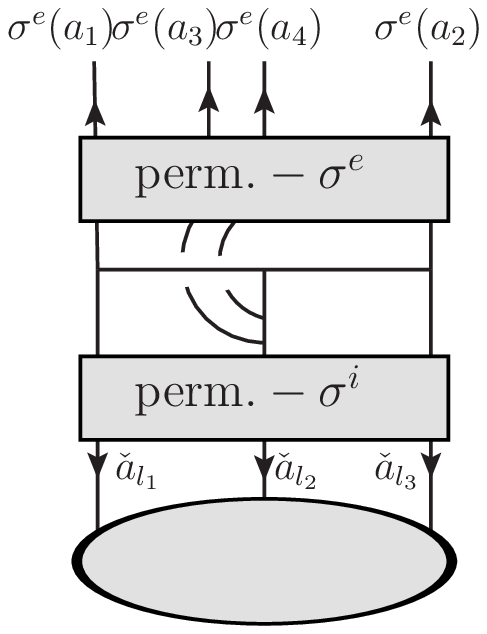}
        \label{fig:ddm2lnm2}
        }
    \subfigure[Cubic Graph: $\check{\mathcal{D}}_{\Delta_3} $]{
    \includegraphics[width=0.22\linewidth]{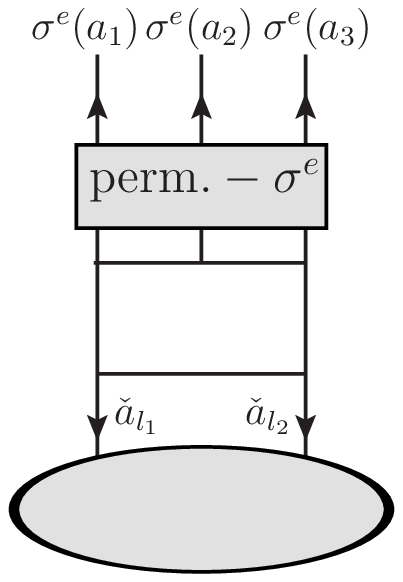}
    \label{fig:ddm2lnm3}
    }
    \subfigure[Cubic Graph: $\check{\mathcal{D}}_{\Delta_4}$]{
    \includegraphics[width=0.22\linewidth]{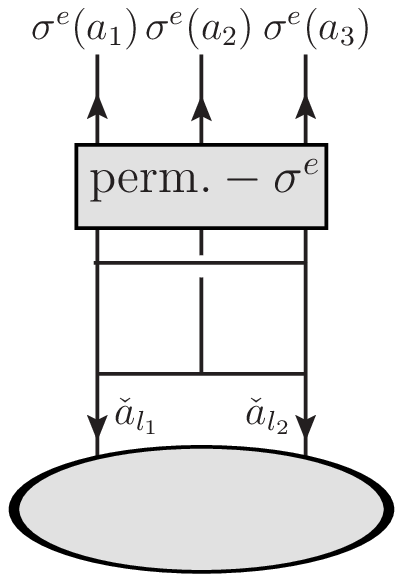}
    \label{fig:ddm2lnm4}
    }
    \caption{Cubic Graphs for next-to-minimal two-loop correction densities.}
    \label{fig:tribas2ln}
\end{figure}

As a generalization to the discussion in Section\,\ref{ssec:nbasis2l}, we consider further the two-loop next-to-minimal trivalent basis. For the range-3 density, we have the following two sets of trivalent color structures:
\begin{equation}\label{cf2lddmnr3}
\begin{aligned}
    &\text{Set}_{1}=\left\{\check{\mathcal{D}}_{\Delta_1}(\sigma^{e},\sigma^{i})\right\}_{\sigma^{i}\in S_{3},\sigma^{e}\in S_{4}/\mathbb{Z}_{2}}=\left\{\sigma^{i}\sigma^{e}\cdot f^{\check{a}_{l_1}a_{1}\text{x}_{1}}f^{\text{x}_{1}a_{2}\text{x}_{2}}f^{\text{x}_{2}a_{3}\text{x}_{3}}f^{\text{x}_{3}\check{a}_{l_2}\text{x}_{4}}f^{\text{x}_{4}a_{4}\check{a}_{l_3}}\right\},\\
    &\text{Set}_{2}=\left\{\check{\mathcal{D}}_{\Delta_2} (\sigma^{e},\sigma^{i})\right\}_{\sigma^{i}\in S_3,\sigma^{e}\in \mathbb{Z}_{2}\times\mathbb{Z}_{4}}=\left\{\sigma^i\sigma^{e}\cdot f^{\check{a}_{l_1}a_{1}\text{x}_{1}}f^{\check{a}_{l_2}a_{2}\text{x}_{2}}f^{\check{a}_{l_3}a_{3}\text{x}_{4}}f^{\text{x}_{4}a_4\text{x}_{3}}f^{\text{x}_{1}\text{x}_{2}\text{x}_{3}}\right\},
\end{aligned}
\end{equation}
where $S_{4}/\mathbb{Z}_{2}$ means permutation of $\{a_1,a_2,a_3,a_4\}$ mod reflection, \emph{i.e.}~$\{1,2,3,4\}\leftrightarrow \{4,3,2,1\}$, $\mathbb{Z}_{2}\times \mathbb{Z}_{4}$ represents cyclic permutaions of $\{a_1,a_2,a_3,a_4\}$ time exchanging the last two elements in the set, namely, exchanging the middle two lines in Figure~\ref{fig:ddm2lnm2}. 
Both of these choices are based on the symmetry of topology of diagrams. In this color basis, similar to the planar case, all color-stripped form factors in Set$_1$ can be one-to-one mapped to certain planar matrix elements.
As a check, the number of above range-3 basis is $3!\times 4!/2 + 3! \times 8 =5!$,  which is the same as the dimension of DDM basis for seven-point amplitude $\itbf{A}_{7}^{(0)}$.

Similarly for the range-2 color basis, one has 
\begin{equation}\label{cf2lddmnr2}
\begin{aligned}
    \text{Set}_{3}&=\left\{\check{\mathcal{D}}_{\Delta_3} (\sigma_{e})\right\}_{\sigma_{e}\in S_{3}}=\left\{\sigma_{e}\cdot  f^{\text{x}_1\check{a}_{l_1}\text{x}_{2}}f^{\text{x}_{2}\check{a}_{l_2}\text{x}_{3}}f^{\text{x}_{3} {a}_{1}\text{x}_{4}}f^{\text{x}_{4}{a}_{2}\text{x}_{5}}f^{\text{x}_{5}{a}_{3}\text{x}_{1}}\right\},\\
    \text{Set}_{4}&=\left\{\check{\mathcal{D}}_{\Delta_4} (\sigma_{e})\right\}_{\sigma_{e}\in S_{3}}=\left\{\sigma_{e}\cdot   f^{\text{x}_1\check{a}_{l_1}\text{x}_{2}}f^{\text{x}_{2}{a}_{3}\text{x}_{3}}f^{\text{x}_{3}\check{a}_{l_2}\text{x}_{4}}f^{\text{x}_{4} {a}_{1}\text{x}_{5}}f^{\text{x}_{5}{a}_{2}\text{x}_{1}}\right\}.
\end{aligned}
\end{equation}
The diagrammatic representation of \eqref{cf2lddmnr3}-\eqref{cf2lddmnr2} are given in Figure~\ref{fig:tribas2ln}. We can also check the counting of range-2 basis as $2\times 3!=4!/2$, which is equal to the number of elements in trivalent basis of one-loop five-point amplitudes.

\section{Non-planar features of unitarity}\label{ap:npuni}

In this appendix, we consider some further details on non-planar unitarity following the discussion in Section\,\ref{sec:unitarity}. 
We will first address the question about how to classify the cut channels contributing to certain color factor $\mathcal{T}_{k}$, and then we discuss some subtle points about adding cut contribution from different cut channels. 

\subsubsection*{Channels from different ordering}

To classify cut channels corresponding to certain color factors, one can in principle consider all possible orderings of each tree blocks and sew them together. 
Below we will employ a procedure separately considering internal cut legs and external uncut legs . 

\begin{figure}
     \centering
    \subfigure[Crossing number=0]{
         \includegraphics[width=0.25\linewidth]{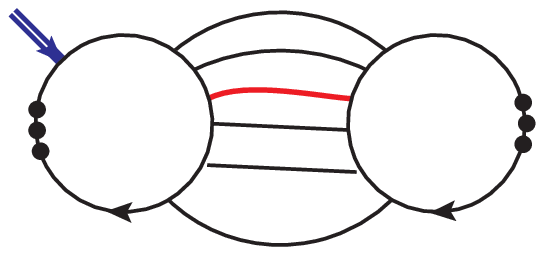}
          \label{fig:crnum0}
         }
     \centering
     \subfigure[Crossing number=1]{
         \includegraphics[width=0.25\linewidth]{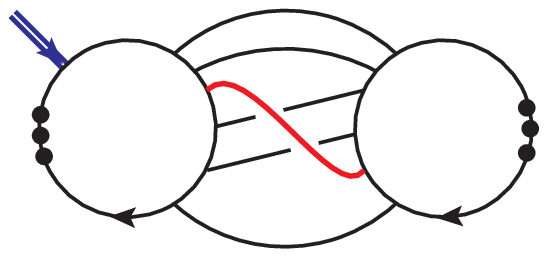}
          \label{fig:crnum1}
         }
      \centering
   \subfigure[Crossing number=2]{
         \includegraphics[width=0.25\linewidth]{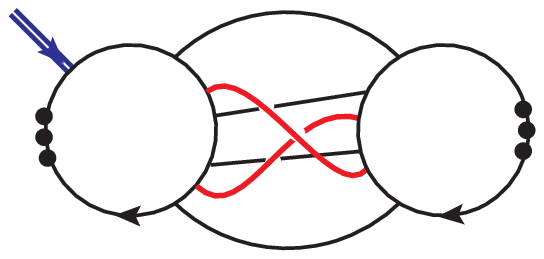}
         \label{fig:crnum2}
         }
     \caption{Examples of connections with different crossing numbers.  The circle with arrow represents a trace, and the black dots represent possible insertion of external legs.}
     \label{fig:crnum}
\end{figure}

In the first step, we consider the sewing of internal cut legs (and ignore the external legs). 
The crossing of internal lines can lead to $1/N_c^2$ suppression. For the convenience of discussion,  we introduce the \emph{crossing number}, which is the minimal number of lines that have to be adjusted to reach a planar topology. For example, Figure~\ref{fig:crnum0} is a planar example which has crossing number 0; Figure~\ref{fig:crnum1} shows a connection with crossing number 1, which means one can move one side of the red line to get a planar topology (although seemingly the red line cross two other cut lines);  Figure~\ref{fig:crnum2} shows a diagram with crossing number 2, since it is necessary to adjust two red lines.

The crossing number defined above has a direct meaning in terms of color factors.
 Roughly speaking, the crossing number determines the number of traces in color factors.
Here one can regard $N_{\mathrm{c}}=\operatorname{tr}(\mathds{1})$ as a trace.
Then the crossing number determines the number of trace plus the power of $N_{\mathrm{c}}$ (after simplifying the color factor). 
For connections with the same crossing number, the power of $N_{ c}$ plus the number of traces at leading order are the same.
The number of traces is reduced by 2$j$ when increasing the crossing number by $j$. 

After fixing the sewing of internal legs, we next consider the distribution of external legs. For the case with crossing number 0, there is always a one-to-one correspondence between distributions of external legs and color factors $\mathcal{T}_{k}$. For higher crossing numbers, in contrast, there are possibilities that different ways of distributing external legs can contribute to the same color factor.

\begin{figure}[t]
    \centering
    \subfigure{
        \includegraphics[width=0.225\linewidth]{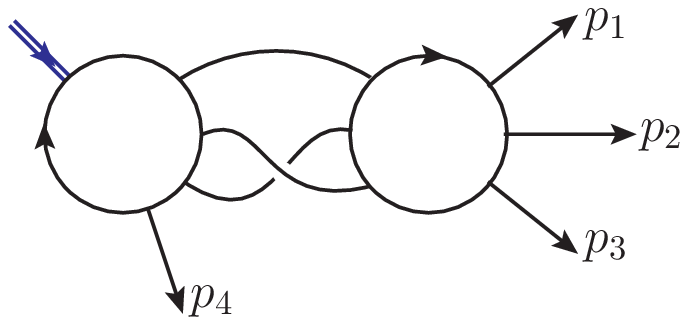}
         \label{fig:t4eg1b}
        }
    \centering
    \subfigure{
        \includegraphics[width=0.225\linewidth]{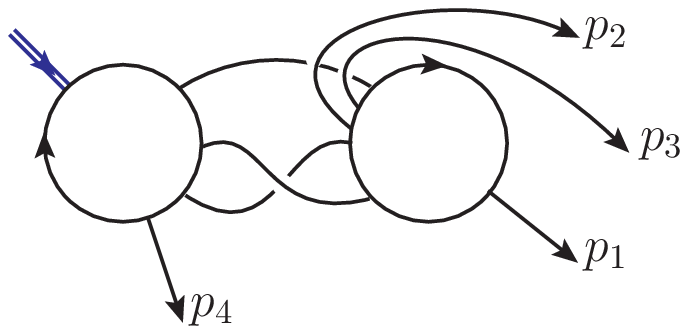}
         \label{fig:t4eg2b}
        }
     \centering
   \subfigure{
        \includegraphics[width=0.225\linewidth]{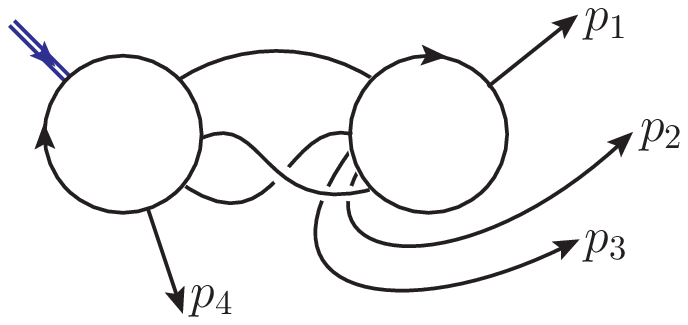}
        \label{fig:t4eg3b}
        }
        \centering
    \subfigure{
        \includegraphics[width=0.225\linewidth]{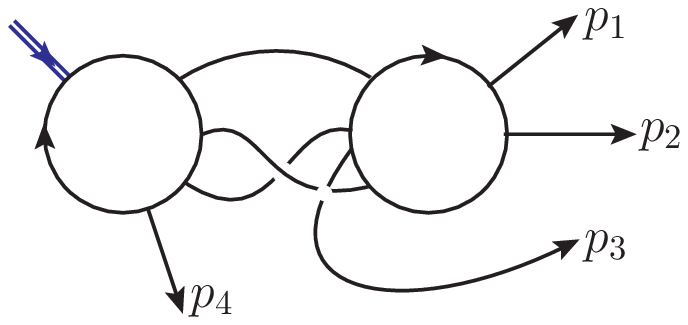}
         \label{fig:t4eg4b}
        }
    \centering
    \subfigure{
        \includegraphics[width=0.225\linewidth]{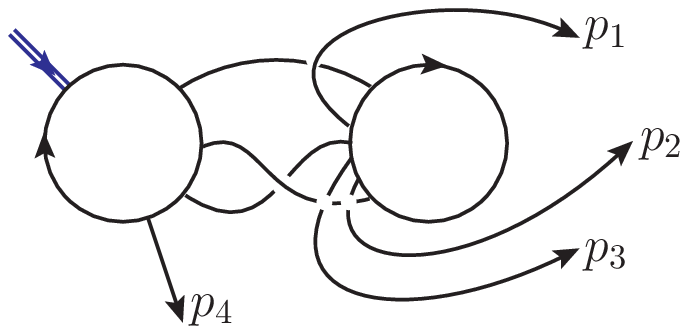}
         \label{fig:t4eg5b}
        }
     \centering
   \subfigure{
        \includegraphics[width=0.225\linewidth]{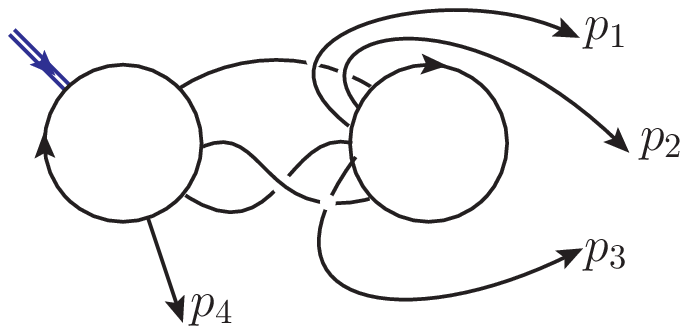}
        \label{fig:t4eg6b}
        }
    \subfigure{
        \includegraphics[width=0.225\linewidth]{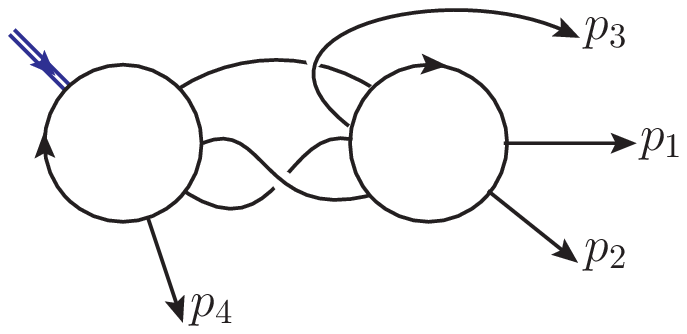}
        \label{fig:t4eg7b}
        }
        \centering
    \subfigure{
        \includegraphics[width=0.225\linewidth]{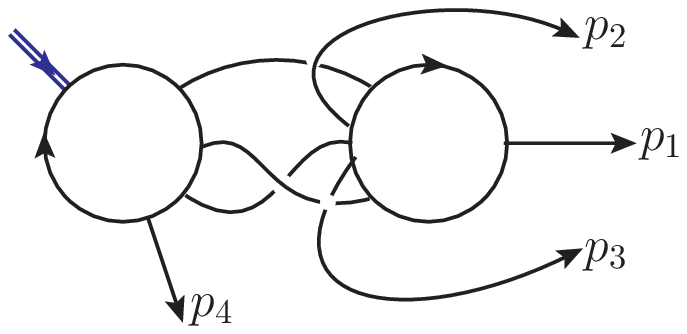}
         \label{fig:t4eg8b}
        }
    \centering
    \subfigure{
        \includegraphics[width=0.225\linewidth]{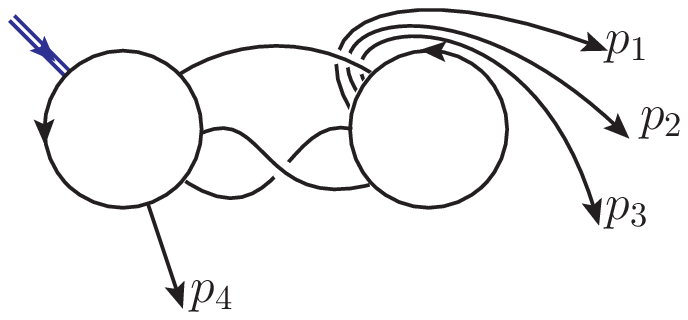}
         \label{fig:t4eg9b}
        }
     \centering
   \subfigure{
        \includegraphics[width=0.225\linewidth]{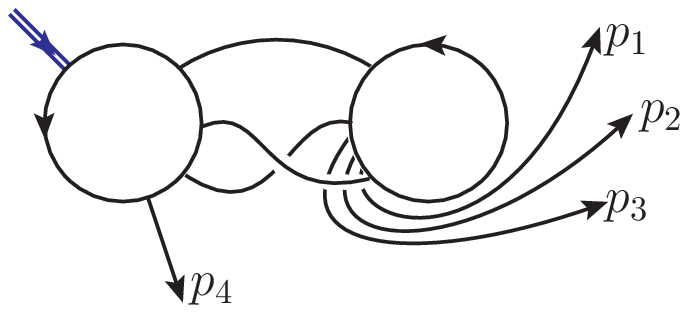}
        \label{fig:t4eg10b}
        }
    \caption{Triple-cut for two-loop minimal form factor $\mathcal{O}=\operatorname{tr}(XXXX)$ with crossing number 1 as in  Figure~\ref{fig:crnum}. These 10 figures correspond to different distributions of external legs $\{p_1,p_2,p_3\}$ (\emph{i.e.}~oP$_{3}\{p_1,p_2,p_3\}$). }
    \label{fig:t4eg}
\end{figure}

Consider the following example: a $s_{123}$-triple-cut of two-loop minimal form factor of a length-four operator, say $\mathcal{O}=\operatorname{tr}(XXXX)$, and our targeting color factor is $\mathcal{T}_c=\operatorname{tr}(a_1a_2a_3a_4)$.  By carefully inspecting all possibilities, there are 10 configurations contribute to the same color factor 
$\mathcal{T}_c$, as listed in Figure~\ref{fig:t4eg}. 
To label the partitions of external legs within each block, we introduce $\text{oP}_{k} S$ as \emph{ordered} partitions of set $S$ into $k$ subsets. In the above example, we need $\{D,F,E\}\in\text{oP}_{3}\{p_1,p_2,p_3\}$ explicitly given as
\begin{align}
\big\{ 
& \big\{ \{p_1,p_2,p_3\}, \emptyset, \emptyset \big\}, \ \big\{ \emptyset, \{p_1,p_2,p_3\}, \emptyset \big\}, \ \big\{ \emptyset, \emptyset, \{p_1,p_2,p_3\} \big\}, \nonumber \\
& \big\{ \{p_1,p_2\}, \{p_3\}, \emptyset\big\}, \ \big\{ \{p_1,p_2\}, \emptyset, \{p_3\}\big\}, \ \big\{ \emptyset, \{p_1,p_2\}, \{p_3\}\big\},  \nonumber\\
& \big\{ \{p_1\}, \{p_2,p_3\}, \emptyset\big\}, \ \big\{ \{p_1\}, \emptyset, \{p_2,p_3\}\big\}, \ \big\{ \emptyset, \{p_1\}, \{p_2,p_3\}\big\},  \ \big\{ \{p_1\}, \{p_2\}, \{p_3\}\big\}
\big\} \,.
\end{align}
Since the above partitions contribute the same color factor, one can add them together in the cut integrand, in which one can consider applying amplitude identities (such as KK relations) to simplify the computation.

\subsubsection*{Comments on non-planar cut integrands}

In the remaining part of this appendix, we consider a few features that are related to the non-planar cut integrands.

When constructing non-planar cut integrand via unitarity method, a complication is to  identify integral topologies from the cut integrand. 
When the orderings of cut legs do not match with each other in different blocks, one typically encounters certain twisted topologies. 
It is then important to keep track the twisting ordering for integrands. 

\begin{figure}
    \centering
    \includegraphics[width=0.55\linewidth]{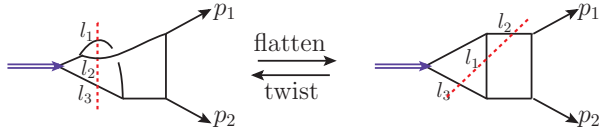}
    \caption{An example of twisted diagrams.}
    \label{fig:pladdert1}
\end{figure}

For example, consider the form factor of $\mathcal{O}=\operatorname{tr}(XX\cdots X)$ in the following cut channel:
\begin{equation}\label{npintegrand2}
\begin{aligned}
     &\int\mathrm{dPS}_{3,l}\hat{\mathcal{F}}_{L+1}(-l_1,-l_2,A,-l_3,B)\hat{\mathcal{A}}_{5}(l_3,l_1,l_2,p_1,p_2) \\
     =&\int\mathrm{dPS}_{3,l}\left(\cdots+ \frac{s_{12}^{2}}{s_{1 l_2}s_{l_1l_2}s_{l_1l_3}}+\cdots\right) \,.
\end{aligned}
\end{equation}
The term written down explicitly on the RHS in \eqref{npintegrand2} corresponds to the cut integral representation as shown in Figure~\ref{fig:pladdert1}.

In  the left diagram in Figure~\ref{fig:pladdert1}, we preserve the twisting of cut legs so that the ordering of legs match that in planar tree blocks, for example, the ordering in $\mathcal{A}_5$ is $\{l_3,l_1,l_2,p_1,p_2\}$.
This integral, after being flattened, is equivalent to a planar ladder integral in the RHS of  Figure~\ref{fig:pladdert1}.
We would like to stress that in the practical computation, it is important to keep track of twist ordering (as in the LHS of  Figure~\ref{fig:pladdert1}) when we consider summarizing contribution from different cut channels. 
To illustrate this point, as a non-trivial example, we discuss in more detail the cuts of Figure~\ref{fig:d27a5} and Figure~\ref{fig:d27a52} below. 

The cut channel of Figure~\ref{fig:d27a5} has been discussed in Section\,\ref{ssec:2lbps}, and the cut contribution has been given explicitly in \eqref{eq:ID27}.
Parallel to the discussion of Figure~\ref{fig:d27a5}, for the cut shown in Figure~\ref{fig:d27a52}, one has 
\begin{equation}
\begin{aligned}
	\left[\mathcal{F}^{(2)}\right]_{\mathcal{T}_{t_3}}\Big|_{s_{12}-\textrm{DC}} =&\int \hat{\mathrm{dPS}}_{3,l}\hat{\mathcal{F}}^{(0)}_{L+1,\scriptscriptstyle \rm dens}(p_5,-l_2,-l_3,p_6)\hat{\mathcal{A}}_{5}^{(0)}(p_1,l_1,p_2,l_3,l_2)\\
	&+ \big(\text{kinematic dependence on  $p_3$ and $p_4$}\big) . 
\end{aligned}
\end{equation}
After subtracting contribution from range-3 density, the detected cut contribution to the density of ${\check{\mathcal{D}}_{27}}$ is similar to  \eqref{eq:ID27} as
\begin{equation}
\label{eq:ID27n}
\begin{aligned}
    \left[\mathcal{I}^{(2)}(12)\right]_{\check{\mathcal{D}}_{27}}=& \begin{aligned}
        \includegraphics[width=0.17 \linewidth]{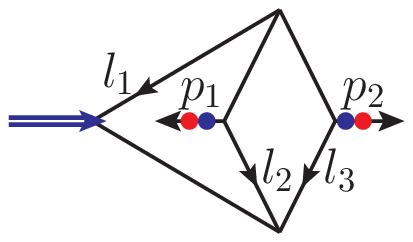}
    \end{aligned}
    \hlm{5}\Big(
   \hskip +2pt\begin{aligned}
       \includegraphics[width=0.15\linewidth]{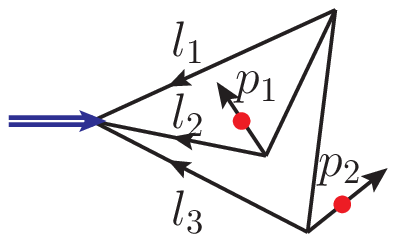}
    \end{aligned} \hlmath{5}{+ \frac{1}{2} } \begin{aligned}
       \includegraphics[width=0.15\linewidth]{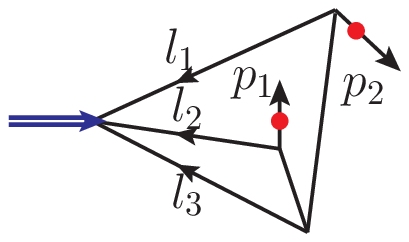}
    \end{aligned}\hlmath{5}{+\frac{1}{2}} \begin{aligned}
       \includegraphics[width=0.15\linewidth]{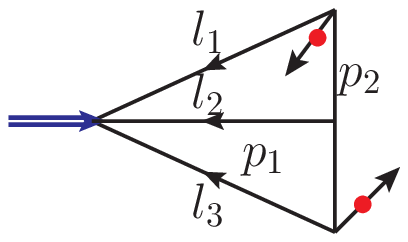}
    \end{aligned}\hskip -5pt \Big) \\
    &+ \big(\text{terms detected in other channels}\big).
\end{aligned}
\end{equation}
Let us compare \eqref{eq:ID27} and \eqref{eq:ID27n}:
\begin{itemize}
    \item The first term in \eqref{eq:ID27} and \eqref{eq:ID27n} respectively correspond to the same integral with identical twisting. This twisted integral can be cut in the following two ways:
     \begin{equation}\label{eq:npladdertwo}
    \text{Cut}_1:
    \begin{aligned}
    \includegraphics[width=0.2\linewidth]{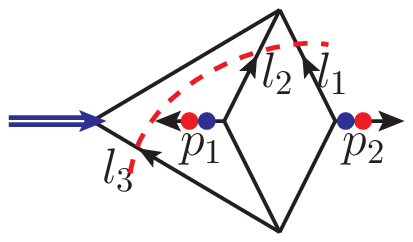}
    \end{aligned},\quad
    \text{Cut}_2:
    \begin{aligned}
    \includegraphics[width=0.2\linewidth]{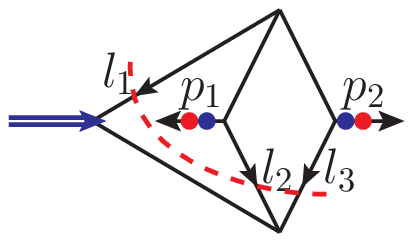} 
    \end{aligned} \,,
    \end{equation}
 contributing to cut-channels of \eqref{eq:ID27} and \eqref{eq:ID27n}, respectively. In this case, the two cut channels detect the same integral.
    
As an illuminating analogy, one can recall the planar calculation, for example in in \eqref{eq:bpspl2}: the ``planar-ladder" integral has two cuts in the $s_{12}$ triple-cut  channel:
   \begin{equation}\label{pladdertwo}
    \text{Cut}_1:
    \begin{aligned}
    \includegraphics[width=0.2\linewidth]{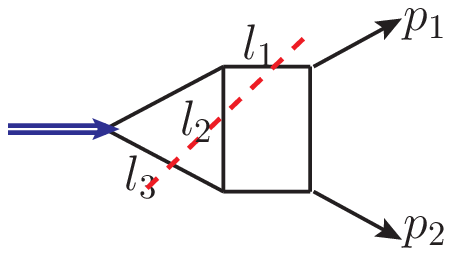}
    \end{aligned},\quad
    \text{Cut}_2:
    \begin{aligned}
    \includegraphics[width=0.2\linewidth]{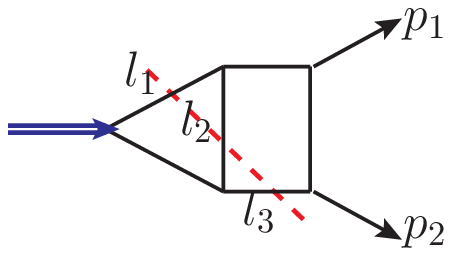} 
    \end{aligned} \,,
    \end{equation}
     but both two cuts emerge in the same cut channel (with the same  ordering).
  
    \item In contrast to the first term, the other terms in the parenthesis in \eqref{eq:ID27} and \eqref{eq:ID27n} need to be treated differently.  After being flattened, they are all equivalent to the same integral 
      $\begin{aligned}
          \includegraphics[width=0.12\linewidth]{fig/tritri2pt1top.eps}
        \end{aligned}$.
    However,  as shown in \eqref{eq:ID27} and \eqref{eq:ID27n}, they have different twisting structures. 
    Therefore, they should be taken as different integral contributions. 
    In other words, one should sum up their coefficients to construct the full form factor integrand, and this explains the total coefficient (-4) of this integral in \eqref{eq:sid27}.
    
\end{itemize}
Similar argument also holds for other cuts that we have used before. The point is that we must treat integrals with different twisting as different contributions and add them up. Another instance for illustrating this point is the factor 3 in \eqref{eq:finalcheck}. 


\section{Six-point Amplitudes}\label{ap:amp}
In this appendix, we provide the explicit expressions of six-scalar tree amplitudes which are needed in the unitarity construction in Section\,\ref{sec:2l}. Some of them have been given in \cite{Loebbert:2015ova}. Here we also give a few new elements that are need for full non-planar construction. 
\begin{align}
    \mathcal{A}_{6}^{(0)}(p_1^{{X}},& p_2^{X},p_3^{X},q_3^{\bar{X}},q_2^{\bar{X}},q_{1}^{\bar{X}}) 
    =-\frac{s_{p_1 p_2 p_3}}{s_{p_1 q_1} s_{p_3 q_3}}+\frac{s_{p_2 p_3} s_{q_1 q_2}}{s_{p_1 q_1} s_{p_3 q_3} s_{p_2 p_3 q_3}}+\frac{s_{p_1 p_2} s_{q_2 q_3}}{s_{p_1 q_1}
   s_{p_3 q_3} s_{p_3 q_2 q_3}}, \\
    \mathcal{A}_{6}^{(0)}(q_1^{\bar{X}},& p_1^{X},p_2^{X},q_2^{\bar{X}},p_3^{X},q_{3}^{\bar{X}})  \nonumber\\
    =&-\frac{s_{p_1 p_2}}{s_{p_2 q_2} s_{p_3 q_3}}-\frac{s_{p_1 p_3}}{s_{p_1 q_1}s_{p_2 q_2}}-\frac{s_{p_2 p_3}}{s_{p_1 q_1} s_{p_3 q_3}}+\frac{s_{p_1 p_2 p_3}}{s_{p_1 q_1} s_{p_2 q_2}}+\frac{s_{p_1 p_2 p_3}}{s_{p_1 q_1} s_{p_3 q_2}}+\frac{s_{p_1 p_2 p_3}}{s_{p_1 q_1} s_{p_3 q_3}}\nonumber\\
     &+\frac{s_{q_1q_3} s_{p_1 p_2}}{s_{p_2 q_2} s_{p_1 p_2 q_2} s_{p_3 q_3}}-\frac{s_{q_2 q_3} s_{p_1 p_2}}{s_{p_1 q_1} s_{p_1 p_2 q_1} s_{p_3 q_2}}-\frac{s_{q_2 q_3}s_{p_1 p_2}}{s_{p_1 q_1} s_{p_1 p_2 q_1} s_{p_3 q_3}}-\frac{s_{p_2 p_3} s_{q_1 q_3}}{s_{p_1 q_1} s_{p_3 q_2} s_{p_2 p_3 q_2}}\nonumber\\
     &-\frac{s_{p_2 p_3}s_{q_1 q_3}}{s_{p_1 q_1} s_{p_2 q_2} s_{p_2 p_3 q_2}}+\frac{s_{p_1 p_3} s_{p_1 p_2}}{s_{p_1 q_1}s_{p_2 q_2} s_{p_3 q_3}}+\frac{s_{p_2 p_3} s_{p_1 p_2}}{s_{p_1 q_1} s_{p_2 q_2} s_{p_3 q_3}}+\frac{s_{p_3 q_2} s_{p_1 p_2}}{s_{p_1 q_1} s_{p_2 q_2} s_{p_3 q_3}}\nonumber\\
     &+\frac{s_{p_2 p_3} s_{p_1 q_3}}{s_{p_1 q_1} s_{p_2 q_2}s_{p_3 q_3}}+\frac{s_{p_1 p_3} s_{p_2 q_1}}{s_{p_1 q_1} s_{p_2 q_2}s_{p_3 q_3}}+\frac{s_{p_1 p_3} s_{p_2 p_3}}{s_{p_1 q_1} s_{p_2 q_2} s_{p_3 q_3}} , \\
     \mathcal{A}_{6}^{(0)}(p_1^{{Y}},& p_2^{X},p_3^{X},q_3^{\bar{X}},q_2^{\bar{X}},q_{1}^{\bar{Y}})\nonumber\\
     =&-\frac{s_{p_2p_3} s_{q_1q_2}}{s_{p_1q_1} s_{p_3q_3} s_{p_2p_3q_3}}-\frac{s_{p_1p_2} s_{q_2q_3}}{s_{p_1q_1} s_{p_3q_3} s_{p_3q_2q_3}}+\frac{s_{p_1p_2p_3}}{s_{p_1q_1} s_{p_3q_3}}
     \nonumber\\
     &
     -\frac{s_{p_2p_3}}{s_{p_3q_3} s_{p_2p_3q_3}}-\frac{s_{q_2 q_3}}{s_{q_3p_3} s_{p_3q_2q_3}}+\frac{1}{s_{p_3q_3}} , \\
     \mathcal{A}_{6}^{(0)}(p_1^{{Y}},& p_2^{X},p_3^{X},q_1^{\bar{Y}},q_2^{\bar{X}},q_{3}^{\bar{X}}) 
     =-\frac{1}{s_{p_1p_2q_3}} , \\
     \mathcal{A}_{6}^{(0)}(p_1^{Y},& p_2^{X},p_3^{X},q_3^{\bar{X}},q_{1}^{\bar{Y}},q_2^{\bar{X}}) 
     =\frac{s_{p_2p_3}}{s_{p_3q_3} s_{p_2p_3q_3}}+\frac{s_{q_1q_3}}{s_{p_3q_3} s_{p_3q_3q_1}}-\frac{1}{s_{p_3q_3}}+\frac{1}{s_{p_3q_3q_1}} , \\
     \mathcal{A}_{6}^{(0)}(q_1^{\bar{Y}},& p_{1}^{Y}, p_2^{X},q_2^{\bar{X}},q_{3}^{\bar{X}},p_3^{X})\nonumber\\
     &=-\frac{s_{p_1 p_2}}{s_{p_2 q_2} s_{p_1 p_2 q_2}}+\frac{s_{p_2 p_3}}{s_{p_2 q_2} s_{p_2 p_3 q_2}}-\frac{s_{p_3 q_2}}{s_{p_1 p_2 q_1} s_{p_3 q_3}}+\frac{s_{p_3 q_1}}{s_{p_1 p_2 q_2} s_{p_3 q_3}}\nonumber\\
     & \quad +\frac{s_{p_2 p_3}}{s_{p_3 q_2}s_{p_2 p_3 q_2}}-\frac{s_{p_3 q_3}}{s_{p_3 q_2}s_{p_3 q_2 q_3}}-\frac{2}{s_{p_3 q_2 q_3}}-\mathcal{A}_{6}^{(0)}(q_1^{\bar{X}}, p_1^{X},p_2^{X},q_2^{\bar{X}},p_3^{X},q_{3}^{\bar{X}}) , \\
     \mathcal{A}_{6}^{(0)}(q_1^{\bar{Y}},& p_{2}^{X},p_{1}^{Y},q_2^{\bar{X}},q_{3}^{\bar{X}},p_3^{X}) 
     =\frac{s_{p_3 q_1}}{s_{p_1 p_2 q_2} s_{p_3 q_3}}-\frac{s_{p_3 q_2}}{s_{p_1 p_2 q_1} s_{p_3 q_3}}+\frac{1}{s_{p_3 q_1 q_3}}-\frac{1}{s_{p_3 q_2 q_3}} .
\end{align}
Other amplitudes can be obtained by the SU(2) Super-Ward Identities and KK relations.


\section{One-loop BPS and SU(2) form factors}\label{ap:1l}

In this appendix we discuss one-loop full-color BPS and SU(2) form factors, which are used in the two-loop subtraction.
Besides, one loop analysis is also a simple test ground of the full-color method that we have developed. 

From the analysis in Section~\ref{ssec:cfddm}, 
one loop minimal loop corrections have only two basis given in \eqref{eq:1lbasis}:
\begin{equation*}
    \Big\{\check{\mathcal{D}}_{1}=\tilde{f}^{\check{a}_{{l}_1}a_{1}\text{x}}\tilde{f}^{\text{x} a_{2}\check{a}_{{l}_2}}, \qquad \check{\mathcal{D}}_{2}= \tilde{f}^{\check{a}_{{l}_1} a_{2} \text{x}}\tilde{f}^{\text{x} a_{1}\check{a}_{{l}_2}} \Big\}\,.
\end{equation*}
Also, one just needs planar matrix elements to reconstruct $\big[\mathcal{I}^{(1)}(12)\big]_{\check{D}_{i}}$. 
The required planar elements can be given explicitly as:
\begin{equation}
\begin{aligned}
    &\left(\mathcal{I}^{(1),\scriptscriptstyle\text{PL}}(12)\right)_{XX}^{XX}=-\text{I}_{\rm tri}[s_{12}],\\
    &\left(\mathcal{I}^{(1),\scriptscriptstyle\text{PL}}(12)\right)_{XY}^{XY}=-\text{I}_{\rm tri}[s_{12}]-\text{I}_{\rm bub,12}[1], \quad  \left(\mathcal{I}^{(1),\scriptscriptstyle\text{PL}}(12)\right)_{XY}^{YX}=\text{I}_{\rm bub,12}[1] .
\end{aligned}
\end{equation}
It is straightforward to use a double cut to fix these integands \cite{Loebbert:2015ova}.
By dressing the color factors, the full-color loop correction density can be given as
\begin{equation}\label{eq:1lsu2}
\begin{aligned}
	\itbf{I}^{(1)}_{12}&=-\begin{aligned}
		\includegraphics[width=0.18\linewidth]{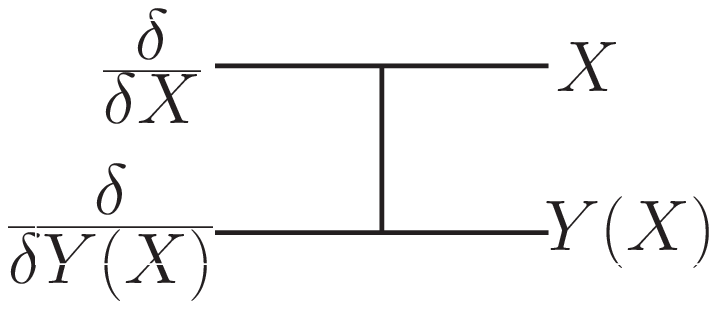}
		\end{aligned}\text{I}_{\rm tri}[s_{12}] -\begin{aligned}
		\includegraphics[width=0.15\linewidth]{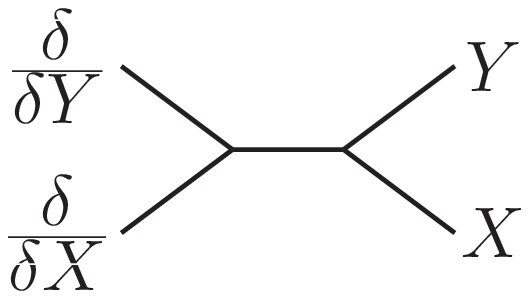}
		\end{aligned} \text{I}_{\rm bub,12}[1] , \\
	\itbf{Z}^{(1)}_{12}&= \frac{1}{\epsilon}\begin{aligned}
		\includegraphics[width=0.15\linewidth]{fig/colorxy1.eps}
		\end{aligned}\,,
\end{aligned}
\end{equation}
where $\frac{\delta}{\delta Y(X)}$ means variation acting on Y fields or X fields. \eqref{eq:1lsu2} contains both one-loop BPS and SU(2) loop corrections. We also call the first part in $\itbf{I}$ as $\itbf{I}_{\rm tri}$ and the second part $\itbf{I}_{\rm bub}$. 

\begin{figure}
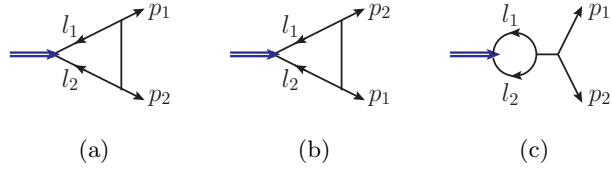

    \centering
    \subfigure[]{\includegraphics[width=0.18\linewidth]{fig/trinew.eps}}
     \subfigure[]{\includegraphics[width=0.18\linewidth]{fig/trinewb.eps}}
    \subfigure[]{\includegraphics[width=0.18\linewidth]{fig/bubnew.eps}}
    \caption{Trivalent topologies for one-loop non-BPS form factors. 
    Note that we treat two triangle topology (a) and (b) differently, because the fields related to $l_1,l_2,p_1,p_2$ may be distinct for general form factor and an overall tree-level factors may not be  factorized out.
    }
    \label{fig:ck1lb}
\end{figure}

As mentioned in Section~\ref{sec:CKBPS}, a CK-dual form for SU(2) form factor can also be constructed. 
One difference between the BPS and SU(2) case is that we have to distinguish external particles in the latter situation. 
As a result, the diagrams in Figure~\ref{fig:ck1l} have to be extended to Figure~\ref{fig:ck1lb}.
With this fact in mind, we simply write down
\begin{equation}
    \itbf{I}^{(1)}_{12,\rm SU(2)}=\tilde{f}^{\check{a}_{l_1}a_{1}\text{x}_1}\tilde{f}^{\check{a}_{l_2}a_{2}\text{x}_1}s_{2l_1}\begin{aligned}
    \includegraphics[width=0.14\linewidth]{fig/trinew.eps}
    \end{aligned}+\tilde{f}^{\check{a}_{l_1}a_{2}\text{x}_1}\tilde{f}^{\check{a}_{l_2}a_{1}\text{x}_1}s_{2l_1}\begin{aligned}
    \includegraphics[width=0.14\linewidth]{fig/trinewb.eps}
    \end{aligned}\,.
\end{equation}
And the CK duality check is straight forward: we have $N_{\rm tri,a}=N_{\rm tri,b}=s_{2l_1}$ and $N_{\rm bub}=0$. This fact ($N_{\rm tri,a}=N_{\rm tri,b}$ and $N_{\rm bub}=0$) can be generalized to more general form factors, usually including non-local numerator, \emph{i.e.}~numerator with $s_{12}$ pole.  More concrete discussion will be covered in another article.

From the above discussion, one loop correction and renormalization constant have a very nice form which will be helpful for our understanding and calculation of the production of one-loop quantities in \eqref{eq:2lrenorm}. 
For example, to get densities of $\big(\itbf{Z}^{(1)}\big)^{2}$, one simply connects two $\itbf{Z}^{(1)}$ together
\begin{equation}
\label{eq:su2z1sqgra}
	\text{Range-3}\sim\frac{1}{\epsilon^2}\begin{aligned}
		\includegraphics[width=0.27\linewidth]{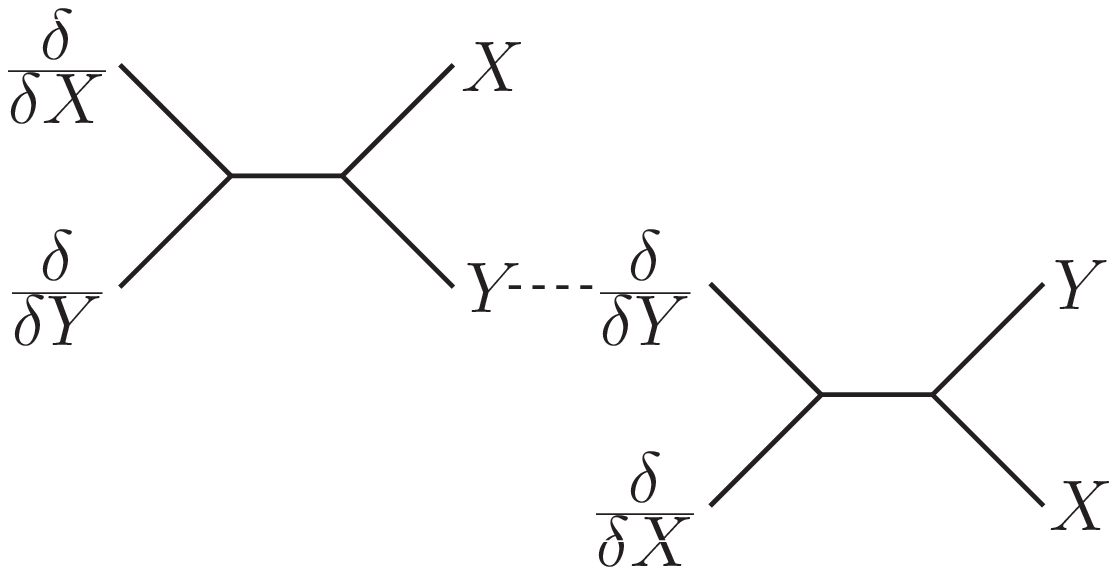}
		\end{aligned} , \qquad 
	\text{Range-2}\sim\frac{2}{\epsilon^2}\begin{aligned}
		\includegraphics[width=0.27\linewidth]{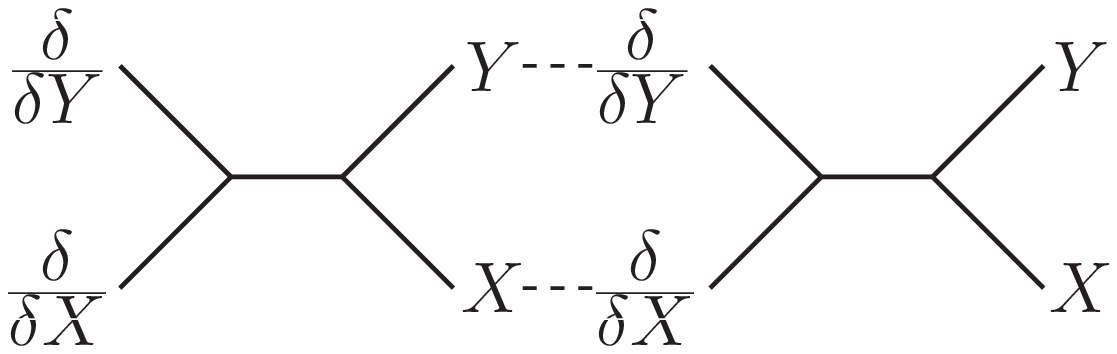}
		\end{aligned}\,,
\end{equation}
where the factor 2 in range-2 density comes from permuting two internal contracted fields. 

As for integrals appearing above, the one loop triangle integral is 
\begin{equation}\label{eq:triint}
    \text{I}_{\rm tri,12}[1]=\left(-s_{12}\right)^{-\epsilon-1}\left(-\frac{1}{\epsilon^2}+\frac{\zeta_2}{2}+O(\epsilon)\right).
\end{equation}
And the one loop bubble integral is
\begin{equation}\label{eq:bubint}
    \text{I}_{\rm bub,12}[1]=\left(-s_{12}\right)^{-\epsilon}\left[\frac{1}{\epsilon}+2+\mathcal{O}\left(\epsilon\right)\right]\,,
\end{equation}
where our convention of integrals is
\begin{equation}
    \text{I}^{(L)}\left[N\left(l_{i}, p_{j}\right)\right]=e^{L \epsilon \gamma_{\mathrm{E}}} \int \frac{d^{D} l_{1}}{i \pi^{\frac{D}{2}}} \cdots \frac{d^{D} l_{L}}{i \pi^{\frac{D}{2}}} \frac{N\left(l_{i}, p_{j}\right)}{\prod_{j} D_{j}} .
\end{equation}

For completeness, two fish integrals used in Section~\ref{ssec:uv} are
\begin{subequations}
\begin{align}
    \mathcal{I}_{\text{SU(2),b}}^{(2)}(p_1,p_2)=&(-s_{12})^{-2\epsilon}\left[\frac{1}{2 \epsilon ^2}+\frac{5}{2\epsilon }+\frac{19}{2}-\frac{\pi^2}{12}+\mathcal{O}(\epsilon)\right]\label{eq:fishsa} , \\
   \nonumber
     \mathcal{I}_{\text{SU(2),a}}^{(2)}(p_1,p_2,p_3)=&(-s_{12})^{-2\epsilon}\bigg[\frac{1}{2 \epsilon ^2}+\frac{5}{2\epsilon }+\frac{19}{2} -\frac{\pi ^2}{12}
     -\left(\text{Li}_2\left(1-u\right)+\frac{1}{2} \log^2 (u)\right)+\mathcal{O}(\epsilon)\bigg] .
   \label{eq:fishsb}
\end{align}
\end{subequations}

\section{Cancellation of two-loop non-planar integrals}

In this appendix we give a proof to \eqref{eq:npr3detop}, following a similar discussion in \cite{Aybat:2006mz}.

To derive \eqref{eq:npr3detop}, we need a useful relation, which implies an implicit parity symmetry:
\begin{equation}
    \mathcal{I}_{\rm 3, np}^{\prime}  =\mathcal{I}_{\rm 3, np}\,,
\end{equation}
where
\begin{equation}
    \mathcal{I}_{\rm 3, np}^{\prime}  \equiv  \mathcal{I}_{\rm 3, np}\big|_{\{k_i\}\rightarrow \{k_{i-1}\}}=\left(s_{13} ~ 2k_3\cdot p_2 + s_{12} ~ 2k_1\cdot p_3 + s_{23} ~ 2k_2\cdot p_1\right) 
    \begin{aligned}
        \includegraphics[width=0.12\linewidth]{fig/pirade.eps}
    \end{aligned} .
\end{equation}
To prove this, we calculate $\Delta\mathcal{I}_{\rm 3,np} = \mathcal{I}_{\rm 3,np}-\mathcal{I}_{\rm 3,np}^{\prime}$ with a new parametrization. 
\begin{equation}\label{eq:i3np-i3npprime}
  \Delta  \mathcal{I}_{\rm 3,np}=e^{2 \epsilon \gamma_{E}}\int {\mathrm{d}^{D}k_1 \over i \pi^{D\over 2} } {\mathrm{d}^{D}k_2 \over i \pi^{D\over 2} } {s_{13} ~ 2(2k_1+k_2)\cdot p_2 + s_{12} ~ 2(k_2-k_1)\cdot p_3 - s_{23} ~ 2(2k_2+k_1)\cdot p_1 \over k_1^2 k_2^2 k_3^2 (k_1-p_1)^{2} (k_2-p_2)^{2} (k_3-p_3)^{2}}\,.
\end{equation}
Follow the discussion in \cite{Aybat:2006mz}, we write down 
\begin{equation}
    k_i^{\mu}={p_1^{\mu} \over p_1\cdot p_2}(p_2\cdot k_i)+{p_2^{\mu} \over p_1\cdot p_2}(p_1\cdot k_i)+k_{i,T}\,,
\end{equation}
and define 
\begin{equation}
    \xi_{i}={p_1\cdot p_3 \over p_1\cdot p_2}(p_2\cdot k_i),\quad \eta_{i}={p_2\cdot p_3 \over p_1\cdot p_2}(p_1\cdot k_i)\,.
\end{equation}
With these variables, we can re-write measure as
\begin{equation}
\begin{aligned}
    \mathrm{d}^{D}k_{i}&=\mathrm{d}k_{i}^{+}\mathrm{d}k_{i}^{-}\mathrm{d}^{D-2}k_{i,T}\\
    &=\frac{p_1 \cdot p_2}{\left(p_1 \cdot p_3\right)\left(p_2 \cdot p_3\right)} d \xi_{i} d \eta_{i} \mathrm{d}^{D-2}k_{i,T}\,,
\end{aligned}
\end{equation}
and 
\begin{equation}
    k_{i}^{2}=2 \frac{\left(p_1 \cdot k_{i}\right)\left(p_2 \cdot k_{i}\right)}{p_1 \cdot p_2}-k_{i, T}^{2}=2 \frac{p_1 \cdot p_2}{\left(p_1 \cdot p_3\right)\left(p_2 \cdot p_3\right)} \xi_{i} \eta_{i}-k_{i, T}^{2}\,.
\end{equation}
If we exchange $\xi_{1}\leftrightarrow \eta_{2}$,  $\xi_{2}\leftrightarrow \eta_{1}$ and $k_{1,T}\leftrightarrow k_{2,T}$, then $(k_1+p_1)^{2}\leftrightarrow (k_2+p_2)^{2}$ and $k_1^{2}\leftrightarrow k_2^{2}$. The denominator of \eqref{eq:i3np-i3npprime} is symmetric under $\xi_{1}\leftrightarrow \eta_{2}$,  $\xi_{2}\leftrightarrow \eta_{1}$ and $k_{1,T}\leftrightarrow k_{2,T}$. And the numerator, which can be written as
\begin{equation}
\begin{aligned}
   &s_{13} ~ 2(2k_1+k_2)\cdot p_2 + s_{12} ~ 2(k_2-k_1)\cdot p_3 - s_{23} ~ 2(2k_2+k_1)\cdot p_1\\
   =&\xi_{1}-\xi_{2}+\eta_{1}-\eta_{2}-p_{3, T} \cdot\left(k_{1}-k_{2}\right)_{T}+\xi_{1}+2 \xi_{2}-2 \eta_{1}-\eta_{2} ,
\end{aligned}
\end{equation}
is anti-symmetric under the transformation. So actually \eqref{eq:i3np-i3npprime} gives zero. Namely, $\mathcal{I}_{\rm 3,np}=\mathcal{I}_{3,np}^{\prime}={\mathcal{I}_{\rm 3,np}+\mathcal{I}_{3,np}^{\prime} \over 2}$ which equals to  the following form using momentum conservation 
\begin{equation}
    -\left(s_{13} ~ k_2\cdot p_2 + s_{12} ~ k_3\cdot p_3 + s_{23} ~ k_1\cdot p_1\right)
    \begin{aligned}
        \includegraphics[width=0.14\linewidth]{fig/pirade.eps}
    \end{aligned} .
\end{equation}

\section{Details for IR and UV cancellation}
\label{app:IRUVcancel}

In this appendix we provide some details of IR and UV subtractions for two-loop form factors.

\subsection{Details for divergence cancellation in \eqref{eq:bds2lnp}}\label{ap:bps}

Performing subtractions in \eqref{eq:bds2lnp} requires a full-color generalization of production of one-loop quantities, such as the one-loop-square. To fully illustrate the full-color one-loop-square, we take a little detour and consider a general case. Concretely, we consider two arbitrary one-loop quantities $\itbf{P}^{(1)}$ and $\itbf{Q}^{(1)}$, with color decomposition $\itbf{P}^{(1)}=\sum_{\sigma\in S_2} \sigma \cdot \left( \tilde{f}^{\check{a}_{{l}_1}a_{1}\text{x}}\tilde{f}^{\text{x} a_{2}\check{a}_{{l}_2}} \mathcal{P}^{(1)}\right)$, and their product $\itbf{P}^{(1)}\itbf{Q}^{(1)}$. 
To dress the one loop product with color, we impose the condition that the planar component of full-color one-loop square should recover planar results, because it has been proven to be effective in the planar case. It turns out that the product reads\footnote{The permutation $\sigma$ act on all the external labels, including momenta, color indices and the field configurations.}
\begin{align}\label{eq:abfc}
&\itbf{P}^{(1)}\itbf{Q}^{(1)}=\sum_{\sigma\in S_3}\sigma \cdot 
\bigg\{\hskip -3pt \begin{aligned}
\includegraphics[width=0.75\linewidth]{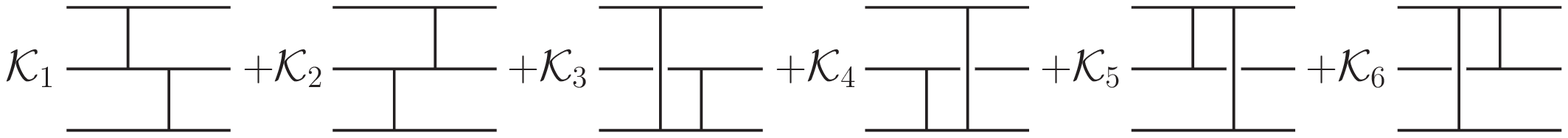}
\end{aligned}\bigg\}\,,
\end{align}
where we omit the labels for color factors represented with graphs as 
$\begin{aligned}
	\includegraphics[width=0.15\linewidth]{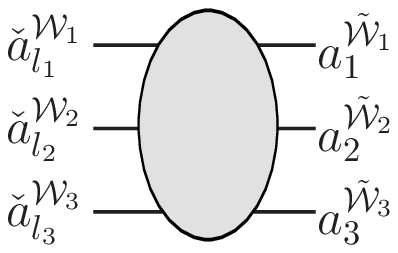}
\end{aligned}$|inside 

\vskip -12pt \noindent  the gray oval can be any color factors in \eqref{eq:abfc}.

The kinematic factors in \eqref{eq:abfc} are
\begin{equation}\label{eq:pq}
\begin{aligned}
	& \mathcal{K}_1 =\big(\mathcal{P}^{(1)}\big)_{\mathcal{W}_{I}\mathcal{W}_{3}}^{\widetilde{\mathcal{W}}_{2} \widetilde{\mathcal{W}}_{3}}({ p_2 , p_3}) \big(\mathcal{Q}^{(1)}\big)_{\mathcal{W}_{1}\mathcal{W}_{2}}^{ \widetilde{\mathcal{W}}_{1}\mathcal{W}_{I}}({ p_1 , p_2})\,,\ \mathcal{K}_2 =\big(\mathcal{Q}^{(1)}\big)_{\mathcal{W}_{2}\mathcal{W}_{3}}^{\mathcal{W}_{I} \widetilde{\mathcal{W}}_{3}}({ p_2 , p_3})  \big(\mathcal{P}^{(1)}\big)_{\mathcal{W}_{1}\mathcal{W}_{2}}^{ \widetilde{\mathcal{W}}_{1} \widetilde{\mathcal{W}}_{2}}({ p_1 , p_2})\,,\\
	& \mathcal{K}_3 =\big(\mathcal{P}^{(1)}\big)_{\mathcal{W}_{2}\mathcal{W}_{I}}^{ \widetilde{\mathcal{W}}_{2} \widetilde{\mathcal{W}}_{3}}({ p_2 , p_3})  \big(\mathcal{Q}^{(1)}\big)_{\mathcal{W}_{1}\mathcal{W}_{3}}^{ \widetilde{\mathcal{W}}_{1}\mathcal{W}_{I}}({ p_1 , p_3})\,,\ \mathcal{K}_4 =\big(\mathcal{Q}^{(1)}\big)_{\mathcal{W}_{2}\mathcal{W}_{3}}^{ \widetilde{\mathcal{W}}_{2}\mathcal{W}_{I}}({ p_2 , p_3})  \big(\mathcal{P}^{(1)}\big)_{\mathcal{W}_{1}\mathcal{W}_{I}}^{ \widetilde{\mathcal{W}}_{1} \widetilde{\mathcal{W}}_{3}}({ p_1 , p_3})\,,\\
	& \mathcal{K}_5 =\big(\mathcal{P}^{(1)}\big)_{\mathcal{W}_{I}\mathcal{W}_{3}}^{ \widetilde{\mathcal{W}}_{1} \widetilde{\mathcal{W}}_{3}}({ p_1 , p_3})  \big(\mathcal{Q}^{(1)}\big)_{\mathcal{W}_{1}\mathcal{W}_{2}}^{\mathcal{W}_{I} \widetilde{\mathcal{W}}_{2}}({ p_1 , p_2})\,, \ \mathcal{K}_6 =\big(\mathcal{Q}^{(1)}\big)_{\mathcal{W}_{1}\mathcal{W}_{3}}^{\mathcal{W}_{I} \widetilde{\mathcal{W}}_{3}}({ p_1 , p_3})  \big(\mathcal{P}^{(1)}\big)_{\mathcal{W}_{I}\mathcal{W}_{2}}^{ \widetilde{\mathcal{W}}_{1} \widetilde{\mathcal{W}}_{2}}({ p_1 , p_2})\,,
\end{aligned}
\end{equation}
where $\mathcal{W}_{I}$ is intermediate fields: for our discussion for SU(2) form factors, it can be $X$ or $Y$ according to specific situation. 

As an example of \eqref{eq:pq}, one can take $\itbf{H}^{(2)}$ function as a commutator, whose origination is discussed in \eqref{commutator}:
\begin{align}
    \itbf{H}^{(2)}_{123}=&\frac{1}{2}\big(\itbf{I}^{(1)}_{23}\big|_{\text{fin}}\itbf{I}^{(1)}_{12}\big|_{\text{div}} - \itbf{I}^{(1)}_{12}\big|_{\text{fin}}\itbf{I}^{(1)}_{23}\big|_{\text{div}}\big) +\text{cyc}(p_1,p_2,p_3)\nonumber\\
    \nonumber=&\mathcal{K}_1 \bigg\{
    \begin{aligned}
    \includegraphics[width=0.18\linewidth]{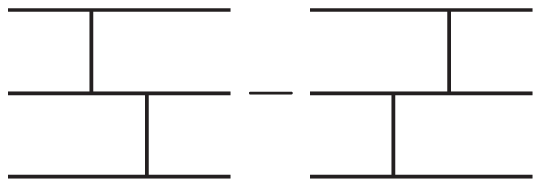}
    \end{aligned}
    \bigg\}+\mathcal{K}_2 \bigg\{
    \begin{aligned}
    \includegraphics[width=0.18\linewidth]{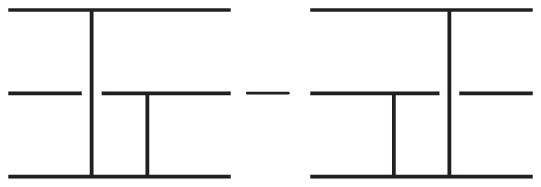}
    \end{aligned}
    \bigg\}+\mathcal{K}_3 \bigg\{
    \begin{aligned}
    \includegraphics[width=0.18\linewidth]{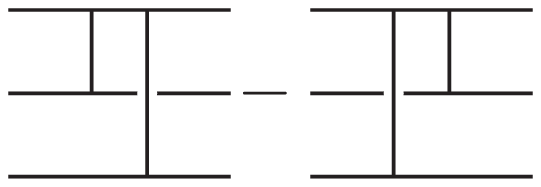}
    \end{aligned}
    \bigg\}+\cdots .
\end{align}
In this case, we take the difference between 
$\itbf{P}=\itbf{I}\big|_{\rm fin}, \itbf{Q}=\itbf{I}\big|_{\rm div}$ and $\itbf{Q}=\itbf{I}\big|_{\rm fin}, \itbf{P}=\itbf{I}\big|_{\rm div}$, \emph{i.e.} the commutator. 
This example clearly shows that the order in full-color case is extremely important and must be paid extra attention to.
Below we only explicitly list terms contribute to $\check{\mathcal{D}}_{19}$ and the label of external lines of color diagrams are always $123$ in these terms. The $\mathcal{K}_i$ are 
\begin{equation}
\begin{aligned}
    \mathcal{K}_1&=\frac{1}{2}\mathcal{I}^{(1)}(12)\Big|_{\rm div}\mathcal{I}^{(1)}(23)\Big|_{\rm fin}-\frac{1}{2}\mathcal{I}^{(1)}(23)\Big|_{\rm div}\mathcal{I}^{(1)}(12)\Big|_{\rm fin} \,, \\
    \mathcal{K}_2&=\frac{1}{2}\mathcal{I}^{(1)}(23)\Big|_{\rm div}\mathcal{I}^{(1)}(13)\Big|_{\rm fin}-\frac{1}{2}\mathcal{I}^{(1)}(13)\Big|_{\rm div}\mathcal{I}^{(1)}(23)\Big|_{\rm fin} \,, \\
    \mathcal{K}_3&=\frac{1}{2}\mathcal{I}^{(1)}(13)\Big|_{\rm div}\mathcal{I}^{(1)}(12)\Big|_{\rm fin}-\frac{1}{2}\mathcal{I}^{(1)}(12)\Big|_{\rm div}\mathcal{I}^{(1)}(13)\Big|_{\rm fin} \,.
\end{aligned}
\end{equation}
Interestingly, all of the color-coefficient of $\mathcal{K}_i$ is $-\check{\mathcal{D}}_{19}$, leading to
\begin{equation}\label{eq:hkine}
    \big[\mathcal{H}^{(2)}\big]_{\check{\mathcal{D}}_{19}}(123)=-\sum_{i=1}^{3}\mathcal{K}_i=\frac{1}{4\epsilon}\mathrm{log}\left(\frac{-s_{12}}{-s_{23}}\right)\mathrm{log}\left(\frac{-s_{23}}{-s_{13}}\right)\mathrm{log}\left(\frac{-s_{13}}{-s_{12}}\right)\,.
\end{equation}
Now we can say that the production of one-loop quantities is clear, and the derivation of one-loop-square in  \eqref{1lsqr2} and  \eqref{1lsqr3} is trivial. 

Afterwards, we can perform subtractions and define remainders as in Section~\ref{ssec:2lrem}. Recall that to define remainder at density level, we introduce auxiliary divergences $U$ which do not appear in IR subtraction formula \eqref{eq:bds2lnp}. So we check the cancellation of these ${U}$s. Collecting all the $U$ together, we have
\begin{align}
    &\itbf{U}=\nonumber\\
    &\sum_{i<j<k}\bigg\{\sum_{\substack{\sigma^{i}\in \mathbb{Z}_3\\ \sigma^{e}\in S_3}} \check{\mathcal{D}}_{\Delta_{1}}(\sigma^{i},\sigma^{e}) \left(\sigma^{e}\cdot \left(U(ij)+U(jk)\right)\right) - \sum_{ \sigma^{e}\in S_3} \check{\mathcal{D}}_{\Delta_{2}}(\sigma^{e}) \left(U(ij)+U(jk)+U({ik})\right)\bigg\} \nonumber\\
    &+2 \sum_{i<j} \sum_{\sigma^{e}\in S_2}\Big(\check{\mathcal{D}}_{\Delta_{3}}(\sigma^{e})U(ij) -2 \check{\mathcal{D}}_{\Delta_{4}}(\mathbf{1}) U(ij)\Big)\,.
\end{align}
$\itbf{U}$ has the similar structure as  \eqref{eq:redistribution} and it is easy to check the following identity beyond density level 
\begin{equation}
    \itbf{U} \cdot \itbf{F}^{(0)}=0,\qquad \forall \itbf{F}^{(0)}\,.
\end{equation}

Finally, we also comment on the $N_{ c}$ factor in the definition of $\tilde{f}^{(2)}$. Such factor is natural to expect because we use coupling $g_{YM}^{2}$ rather than $g_{YM}^{2}N_c$ in the planar calculation. In our definition of $\left[\mathcal{R}^{(2)}\right]_{\check{\mathcal{D}}_{i}}$ with $i=25,26,27$ in \eqref{eq:Rr2d25} and \eqref{eq:Rr2d27}, we already include $\tilde{f}^{(2)}$ factor, namely, we have expand the term $\tilde{f}^{(2)}\itbf
{I}^{(1)}(2\epsilon)$ in \eqref{eq:bds2lnp} with range-2 color basis. More precisely, we have used the following color identity
\begin{equation}
    \check{\mathcal{D}}_{25}+\check{\mathcal{D}}_{26}+2\check{\mathcal{D}}_{27}=2 N_c f^{ \check{a}_{l_1} a_1 \text{x}_1} f^{\text{x}_1 {a}_2 \check{a}_{l_2} } \,,
\end{equation}
where $ f^{ \check{a}_{l_1} a_1 \text{x}_1} f^{\text{x}_1 {a}_2 \check{a}_{l_2} }$ is the color factor of full-color one-loop density $\itbf{I}^{(1)}$. This explains the $f^{(2)}$ term, for example, in \eqref{eq:Rr2d25}.

\subsection{Analysis for SU(2) form factors}\label{ap:su2}

Below we provide further  details for the SU(2) form factors. We first reorganize the two-loop results in a simpler form, and then discuss the cancellation of  divergences.

\subsubsection*{Simplified full-color density}

The SU(2) density function ${\itbf{I}}$ given in Section\,\ref{ssec:2lsu2} can be simplified if we choose another set of basis, which is motivated by rendering various symmetry manifest.

First we observe that $\tilde{\itbf{I}}_{\rm SU(2)}$ (as defined in \eqref{eq:isu2tilde}) should be symmetric if exchanging two internal or external $X$ fields. So we have%
\footnote{Note that although the form has better symmetric property, the color basis for range-3 density are not all independent, providing us freedom for adjusting the kinematic parts which we use below.}
\begin{align}\label{eq:newsu2dens}
    \nonumber 
    \tilde{\itbf{I}}_{\rm SU(2)}^{(2),r=3}(p_1^{Y},p_2^{X},p_3^{X})=&\bigg(
    \sum_{\sigma_{a}\in \mathbb{Z}_3}\sum_{\sigma_{b}\in \mathbb{Z}_{3}}\bigg\{
	 \begin{aligned}
		\includegraphics[width=0.17\linewidth]{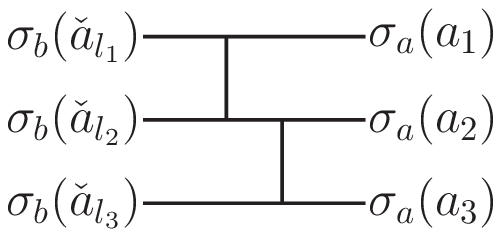}
	\end{aligned}  \left(\tilde{\mathcal{I}}_{\rm SU(2)}(\sigma_{a}(p_1,p_2,p_3)\right)^{\sigma_{a}(\scriptscriptstyle{YXX})}_{\sigma_{b}(\scriptscriptstyle{YXX})}) \bigg\}\\
	&\hskip 2.2cm +(\check{a}_{l_2}^{X}\leftrightarrow \check{a}_{l_3}^{X})\bigg) + (a_2,p_2\leftrightarrow a_3,p_3)\,,
\end{align}
leaving us 9 functions $\left(\tilde{\mathcal{I}}_{\rm SU(2)}\right)^{\sigma_{a}(\scriptscriptstyle{YXX})}_{\sigma_{b}(\scriptscriptstyle{YXX})}$. 

Second, we still have SU(2) relations satisfied by these functions
\begin{equation}
\begin{aligned}
    &\sum_{\tau_{b}} \left(\tilde{\mathcal{I}}_{\rm SU(2)}\right)^{\tau_{a}(\scriptscriptstyle{YXX})}_{\tau_{b}(\scriptscriptstyle{YXX})}=0,\quad \forall \tau_{a}\in S_{3}\,;
\qquad     \sum_{\tau_{a}} \left(\tilde{\mathcal{I}}_{\rm SU(2)}\right)^{\tau_{a}(\scriptscriptstyle{YXX})}_{\tau_{b}(\scriptscriptstyle{YXX})}=0,\quad \forall \tau_{b}\in Z_{3}\,.
\end{aligned}
\end{equation}
These SU(2) relations are identities at function level and the difference between permutation $\tau$ and $\sigma$ can be found in Section~\ref{ssec:2lsu2}.  
These six equations are not independent and from them one finds four basis functions for $\tilde{\mathcal{I}}_{\rm SU(2)}$
\begin{equation}\label{eq:foursu2basis}
	\mathcal{B}_1= \left(\tilde{\mathcal{I}}_{\rm SU(2)}\right)_{\scriptscriptstyle YXX}^{\scriptscriptstyle YXX},\ 
    \mathcal{B}_2= \left(\tilde{\mathcal{I}}_{\rm SU(2)}\right)_{\scriptscriptstyle YXX}^{\scriptscriptstyle XXY},\
	\mathcal{B}_3= \left(\tilde{\mathcal{I}}_{\rm SU(2)}\right)^{\scriptscriptstyle YXX}_{\scriptscriptstyle XXY},\ 
	 \mathcal{B}_4= \left(\tilde{\mathcal{I}}_{\rm SU(2)}\right)_{\scriptscriptstyle XXY}^{\scriptscriptstyle XXY}\,. 
\end{equation}

One can further  set $\mathcal{B}_3=\mathcal{B}_4=0$ in \eqref{eq:foursu2basis}. At this stage, this is nothing but a choice and we will explain this choice later. 
Other non-vanishing kinematic matrix elements are  
\begin{equation}
\begin{aligned}
	&\left(\tilde{\mathcal{I}}_{\rm SU(2)}\right)_{\scriptscriptstyle YXX}^{\scriptscriptstyle XYX}=-\mathcal{B}_{1}, \qquad 
	\left(\tilde{\mathcal{I}}_{\rm SU(2)}\right)_{\scriptscriptstyle XYX}^{\scriptscriptstyle YXX}=-\mathcal{B}_{1}-\mathcal{B}_{2},\\
	&\left(\tilde{\mathcal{I}}_{\rm SU(2)}\right)_{\scriptscriptstyle XXY}^{\scriptscriptstyle XYX}=-\mathcal{B}_{2}, \qquad
	\left(\tilde{\mathcal{I}}_{\rm SU(2)}\right)_{\scriptscriptstyle XYX}^{\scriptscriptstyle XYX}=\mathcal{B}_{1}+\mathcal{B}_{2}\,.
\end{aligned}
\end{equation}

Comparing this ansatz form \eqref{eq:newsu2dens} with the unitarity results obtained in Section\,\ref{ssec:2lsu2}, we see that such basis functions is nothing but the integrals in Section~\ref{ssec:uv}
\begin{equation}\label{eq:su2newres}
    \mathcal{B}_1(p_1,p_2,p_3)=
    \mathcal{I}_{\text{SU(2),c}}^{(2)}(p_1,p_2,p_3)
        ,\quad 
    \mathcal{B}_2(p_1,p_2,p_3)=
    \mathcal{I}_{\text{SU(2),a}}^{(2)}(p_1,p_2,p_3)
        \,.
\end{equation}
This also justifies the rationale of our previous choice about $\mathcal{B}_3=\mathcal{B}_4=0$.

Using \eqref{eq:foursu2basis} and \eqref{eq:su2newres}, we can get \eqref{eq:imix} and \eqref{eq:ifish}, by collecting color factors for the same kinematic functions. For example, consider $\left(\tilde{\mathcal{I}}_{\rm SU(2)}\right)_{\scriptscriptstyle YXX}^{\scriptscriptstyle XYX}$ and $\left(\tilde{\mathcal{I}}_{\rm SU(2)}\right)_{\scriptscriptstyle YXX}^{\scriptscriptstyle YXX}$. They all have $\mathcal{B}_1(p_1,p_2,p_3)$ kinematic factor, and the color coefficients of such kinematic part are 
$\begin{aligned}
    \includegraphics[height=0.07\linewidth]{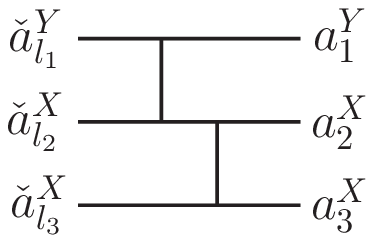}
\end{aligned}$ and 
$\begin{aligned}
    \includegraphics[height=0.07\linewidth]{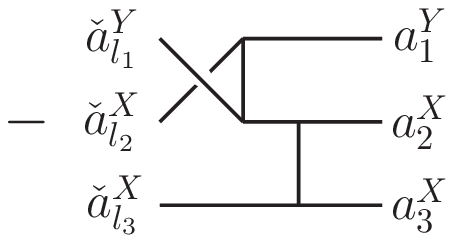}
   \end{aligned}$ 
respectively. These two color factors satisfy
\begin{equation}
     \begin{aligned}
    \includegraphics[width=0.55\linewidth]{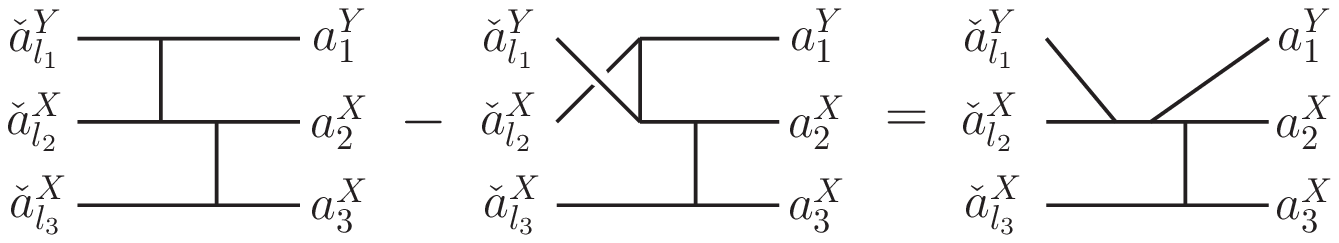}
   \end{aligned} ,
\end{equation}
which recovers $\check{\mathcal{C}}_{\rm mix,A}^{r=3}$ in \eqref{eq:imix}.

Finally, one also considers a check with Feynman diagram computation. For $\begin{aligned}
\includegraphics[width=0.12\linewidth]{fig/bubboxtop.eps}
\end{aligned}$ and $\begin{aligned}
\includegraphics[width=0.12\linewidth]{fig/bubtrimasstop.eps}
\end{aligned}$, the color factors in \eqref{eq:ifish} and \eqref{eq:imix} seem to be natural. The only remaining problem is that from Feynman diagrams, the color factor of $\begin{aligned}
\includegraphics[width=0.12\linewidth]{fig/tritrimasstop.eps}
\end{aligned}$should look like $\begin{aligned}
    \includegraphics[width=0.15\linewidth]{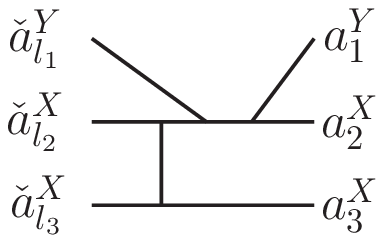}
\end{aligned}$. 
This problem can be resolved as one can generate $\check{\mathcal{C}}_{\rm mix}^{r=3}$ using the identity in the following pattern
\begin{equation}
    \begin{aligned}
\includegraphics[width=0.5\linewidth]{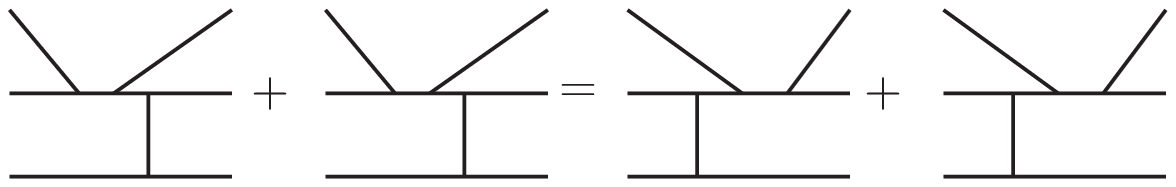}
\end{aligned}\,.
\end{equation}

\subsubsection*{Details for divergence cancellation}

The subtraction of SU(2) form factors in Section \ref{ssec:uv} is separated into three parts. The UV part, \emph{i.e.}~$\itbf{I}_{\rm fish}$ contributing to the most important information of dilatation operator, has been discussed in detail there. The subtractions and remainders of $\itbf{I}_{\rm BPS}$ have no difference from the BPS subtraction in Section~\ref{ssec:ir} and \ref{ssec:2lrem}, and we will comment on $\itbf{H}$ terms later in Section~\ref{ssec:hterm}. Here we provide some further discussion for $\itbf{I}_{\rm mix}$. 

One should be cautious about the difference between \eqref{eq:su2mix} and \eqref{eq:2lrenorm}|they are not strictly equivalent.
The problem comes from the order of $\underline{\itbf{I}}^{(1)}_{\rm bub}$ and  $\underline{\itbf{I}}^{(1)}_{\rm tri}$. Directly from \eqref{eq:2lrenorm}, the proper order of $\underline{\itbf{I}}^{(1)}_{\rm tri}$ and $\underline{\itbf{I}}^{(1)}_{\rm bub}$ is
$\frac{1}{2}\big\{\underline{\itbf{I}}^{(1)}_{\rm tri},\underline{\itbf{I}}^{(1)}_{\rm bub}\big\}$, which have a commutator  $\frac{1}{2}\big[\underline{\itbf{I}}^{(1)}_{\rm tri},\underline{\itbf{I}}^{(1)}_{\rm bub}\big]$ difference from $\underline{\itbf{I}}^{(1)}_{\rm tri}\underline{\itbf{I}}^{(1)}_{\rm bub}$ in \eqref{eq:su2mix}. This commutator, with $\epsilon^{-1}$ divergence, can be regarded being cancelled by another $\itbf{H}^{(2)}_{\rm mix}$, as will be discussed in Section~\ref{ssec:hterm}. Nevertheless, it is convenient to just write down the subtraction \eqref{eq:su2mix} and obtain the remainders in Section~\ref{ssec:uv}.

After discussing the subtraction equation \eqref{eq:su2mix}, the subtraction procedure is mostly depicted in Section~\ref{ssec:uv} with one small remaining problem, which is the cancellation of auxiliary divergence $V=\zeta_2 /\epsilon$. Such auxiliary quantity just show up in subtractions of $\itbf{I}_{\rm mix}$, whose color factors (those $\check{C}_{\rm mix}$) are presented in \eqref{eq:su2cfs}. For example, if the external states are $p_1^{Y},p_2^{X},p_3^{X}$, then the contributions, with color factors, proportional to $V=\zeta_2/\epsilon$ are
\begin{equation}
   \hskip -3pt  {\itbf{V}}_{123}=
    V\left(\check{\mathcal{C}}_{\rm mix,A}^{r=3}+ \check{\mathcal{C}}_{\rm mix,B}^{r=3}+ (a_2\leftrightarrow a_3)\right)
   , \quad  
    {\itbf{V}}_{12}= 2 V
    \check{\mathcal{C}}_{\rm mix}^{r=2},\quad  
    {\itbf{V}}_{13}= 2 V
    \check{\mathcal{C}}_{\rm mix}^{r=2}\Big|_{a_2\shortrightarrow a_3}.
\end{equation}
Summing up all these auxiliary divergence, we find out that they cancel beyond density level
\begin{equation}
    \itbf{V} \cdot \itbf{F}^{(0)}=\Big(\sum_{i<j<k}  \itbf{V}_{ijk} + \sum_{i<j}  \itbf{V}_{ij} \Big) \cdot  \itbf{F}^{(0)}=0 \,.
\end{equation}
Note that both $\itbf{V}_{ijk}$ and $\itbf{F}^{(0)}$ depend on specific operators, so we have to analysis case by case. Check performed up to length-6 operators with arbitrary distribution of $X,Y$ supports this cancellation. It would be interesting to have a general proof of such conclusion. 



\providecommand{\href}[2]{#2}\begingroup\raggedright\endgroup

\end{document}